\documentclass[hyperpdf,nobind, sftitles, twoside]{hepthesis}

\usepackage{emptypage}

\usepackage[T1]{fontenc}
\usepackage{charter}
\usepackage[expert]{mathdesign}

\usepackage{morefloats,subfig,afterpage}
\usepackage{mathrsfs} 
\usepackage{verbatim}
\usepackage{amsmath}
\usepackage{xcolor}
\usepackage[
            paperwidth=17cm,    
            paperheight=24cm,   
            inner=2.25cm,
            outer=1.75cm,
            bottom=2cm,
            textheight=19.5cm]{geometry}
\usepackage[english]{babel}
\usepackage{cite}
\usepackage{bm}
\usepackage{capt-of}   
\usepackage{array}
\usepackage{booktabs}
\newcolumntype{C}[1]{>{\centering\let\newline\\\arraybackslash\hspace{0pt}}m{#1}}
\usepackage{enumerate}
\usepackage[shortlabels]{enumitem}

\usepackage{stackrel}

\definecolor{cerulean}{rgb}{0.0, 0.48, 0.65}

\definecolor{ao(english)}{rgb}{0.0, 0.5, 0.0}

\definecolor{goldenrod}{rgb}{0.85, 0.65, 0.13}

\definecolor{blue-violet}{rgb}{0.54, 0.17, 0.89}

\definecolor{fandango}{rgb}{0.71, 0.2, 0.54}

\usepackage{anyfontsize}
\newcommand{\myparagraph}[1]{\noindent \textbf{{\fontsize{11}{11}\selectfont #1}} --}

\usepackage[ddmmyyyy]{datetime}
\usepackage{braket}
\usepackage{slashed}
\usepackage{graphicx}
\usepackage{fontawesome}
\usepackage{placeins}
\usepackage[export]{adjustbox}
\usepackage{lipsum}

\usepackage[subfigure]{tocloft}
\addtocontents{toc}{\vspace{-4ex}} 
\addtocontents{preface}{\vspace{-4ex}} 
\setkomafont{subsubsection}{\normalfont}
\makeatletter
\renewcommand\subsubsection{\@startsection{subsubsection}{3}{\z@}%
                                     {-3.25ex\@plus -1ex \@minus -.2ex}%
                                     {-1.5ex \@plus -.2ex}
                                     {\normalfont\normalsize\bfseries}}
\makeatother

\usepackage{listings}
\lstdefinestyle{mystyle}{
    commentstyle=\color{black},
    keywordstyle=\color{blue},
    numberstyle=\tiny\color{gray},
    stringstyle=\color{purple},
    basicstyle=\ttfamily\footnotesize,
    breakatwhitespace=false,         
    breaklines=true,                                     
    keepspaces=true,                 
    numbers=none,                    
    numbersep=5pt,                  
    showspaces=false,                
    showstringspaces=false,
    showtabs=false,                  
    tabsize=2,
    frame=single,
    rulecolor=\color{black},
    literate={\$}{{\textcolor{blue}{\$}}}1, 
}

\lstdefinelanguage{yaml}{
    keywords={true,false,null,yes,no},
    keywordstyle=\color{blue}\bfseries,
    comment=[l]{\#},
    commentstyle=\color{blue},
    morestring=[b]',
    morestring=[b]",
    stringstyle=\color{brown},
    literate = *{:}{{\textcolor{orange}{:}}}{1}%
                {,}{{\textcolor{orange}{,}}}{1}%
                {\{}{{\textcolor{orange}{\{}}}{1}%
                {\}}{{\textcolor{orange}{\}}}}{1}%
                {[}{{\textcolor{orange}{[}}}{1}%
                {]}{{\textcolor{orange}{]}}}{1}%
}

\definecolor{delim}{RGB}{20,105,176}
\definecolor{numb}{RGB}{106, 109, 32}

\lstdefinelanguage{json}{
    numbers=none,
    frame=single,
    rulecolor=\color{black},
    showspaces=false,
    showtabs=false,
    breaklines=true,
    postbreak=\raisebox{0ex}[0ex][0ex]{\ensuremath{\color{gray}\hookrightarrow\space}},
    breakatwhitespace=true,
    upquote=true,
    morestring=[b]",
    literate=
     *{0}{{{\color{numb}0}}}{1}
      {1}{{{\color{numb}1}}}{1}
      {2}{{{\color{numb}2}}}{1}
      {3}{{{\color{numb}3}}}{1}
      {4}{{{\color{numb}4}}}{1}
      {5}{{{\color{numb}5}}}{1}
      {6}{{{\color{numb}6}}}{1}
      {7}{{{\color{numb}7}}}{1}
      {8}{{{\color{numb}8}}}{1}
      {9}{{{\color{numb}9}}}{1}
      {\{}{{{\color{delim}{\{}}}}{1}
      {\}}{{{\color{delim}{\}}}}}{1}
      {[}{{{\color{delim}{[}}}}{1}
      {]}{{{\color{delim}{]}}}}{1},
}

\usepackage{dsfont}
\usepackage{amssymb}
\newcommand{\gsim}{\raisebox{-0.13cm}{~\shortstack{$>$ \\[-0.07cm]
      $\sim$}}~}
\newcommand{\lsim}{\raisebox{-0.13cm}{~\shortstack{$<$ \\[-0.07cm]
$\sim$}}~}

\usepackage{bibentry}

\usepackage{booktabs}
\usepackage{multirow}
\usepackage{tabularx}
\usepackage{longtable}
\usepackage{comment}

\usepackage{wrapfig}
\usepackage{makecell}

\usepackage{neuralnetwork}
\newcommand{\lag}{\mathcal{L}}

\newcommand{\mcO}{\mathcal{O}}

\newcommand{\chm}{\checkmark}

\newcommand{\mycite}[1]{\underline{\hspace{-0.5pt}\cite{#1}}}

\newcommand{\smefit}{{\sc\small SMEFiT}}
\newcommand{\matchTOfit}{{\sc \small match2fit}}

\DeclareOldFontCommand{\rm}{\normalfont\rmfamily}{\mathrm}
\DeclareOldFontCommand{\sf}{\normalfont\sffamily}{\mathsf}
\DeclareOldFontCommand{\tt}{\normalfont\ttfamily}{\mathtt}
\DeclareOldFontCommand{\bf}{\normalfont\bfseries}{\mathbf}
\DeclareOldFontCommand{\it}{\normalfont\itshape}{\mathit}
\DeclareOldFontCommand{\sl}{\normalfont\slshape}{\@nomath\sl}
\DeclareOldFontCommand{\sc}{\normalfont\scshape}{\@nomath\sc}
\newcommand{\be}{\begin{equation}}
\newcommand{\ee}{\end{equation}}
\newcommand{\bea}{\begin{eqnarray}}
\newcommand{\eea}{\end{eqnarray}}
\newcommand{\bi}{\begin{itemize}}
\newcommand{\ei}{\end{itemize}}
\newcommand{\ben}{\begin{enumerate}}
\newcommand{\een}{\end{enumerate}}

\newcommand{\lc}{\left[}
\newcommand{\rc}{\right]}
\newcommand{\lp}{\left(}
\newcommand{\rp}{\right)}

\usepackage{floatrow}


\usepackage{caption}
\usepackage{sidecap}
\usepackage{lscape}
\usepackage{makecell}
\usepackage{fancyvrb}

\usepackage{tikz}
\usetikzlibrary{fit}
\usetikzlibrary{positioning}
\usetikzlibrary{shapes.geometric, arrows}
\usetikzlibrary{arrows.meta}
\usepackage{varwidth}
\usepackage{xcolor}
\definecolor{mtplotlib1}{HTML}{1f77b4}
\definecolor{mtplotlib2}{HTML}{ff7f0e}
\definecolor{mtplotlib3}{HTML}{2ca02c}
\definecolor{mtplotlib4}{HTML}{d62728}

\tikzset{%
  >={Latex[width=2mm,length=2mm]},
            base/.style = {rectangle, rounded corners, draw=black,
                           minimum width=4cm, minimum height=1cm,
                           text centered}, 
            mystyle/.style={rectangle, rounded corners, draw=black,
            minimum width=12cm, minimum height=1cm,
            text centered}, 
    col0/.style = {base, fill=white!30},
    col1/.style = {base, fill=mtplotlib1!30},
    col11/.style = {mystyle, fill=mtplotlib1!30},
    col2/.style = {base, fill=mtplotlib2!30},
    col3/.style = {base, fill=mtplotlib3!30},
    col4/.style = {base, minimum width=2.5cm, fill=mtplotlib4!15,}
}

\makeatletter
\@ifpackageloaded{hyperref}{%
\hypersetup{%
  pdftitle = {Fingerprinting New Physics with Effective Field Theories},
  pdfsubject = {Jaco ter Hoeve's PhD thesis},
  pdfkeywords = {effective field theories, Machine Learning, neural networks},
  pdfauthor = {\textcopyright\ Jaco ter Hoeve}
}}{}
\makeatother

\begin{document}

\begin{frontmatter}
\thispagestyle{empty}
\begin{center}
\vspace*{3cm}
\Large{\textbf{Fingerprinting New Physics \\with Effective Field Theories}}\\
\vspace{10cm}
\Large{Jaco ter Hoeve}
\end{center}

\newpage
\thispagestyle{empty}

\vspace*{\fill}

\noindent Printed by Ridderprint, The Netherlands.

\noindent ISBN: 978-94-6506-553-3

\noindent http://doi.org/10.5463/thesis.896

\noindent Copyright \copyright 2024 by J.J. ter Hoeve, Edinburgh, United Kingdom.

\noindent This work is financed by the Netherlands Organisation for Scientific Research (NWO) and was carried out at Nikhef and Vrije Universiteit Amsterdam.

\begin{table}[b]
  \centering
  \begin{minipage}{0.25\textwidth}
    \centering
    \includegraphics[width=\textwidth]{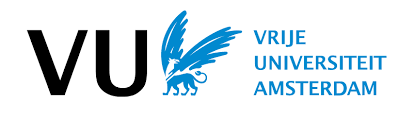}
  \end{minipage}%
  \hspace{0.05\textwidth} 
  \begin{minipage}{0.25\textwidth}
    \centering
    \includegraphics[width=\textwidth]{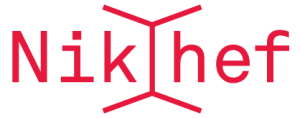}
  \end{minipage}%
  \hspace{0.05\textwidth}
  \begin{minipage}{0.25\textwidth}
    \centering
    \includegraphics[width=.5\textwidth]{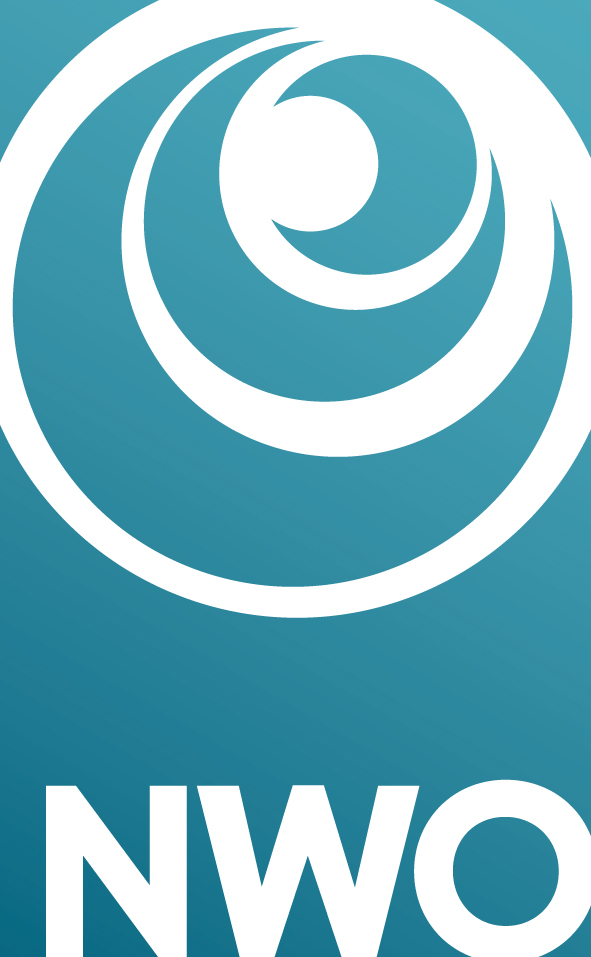}
  \end{minipage}
\end{table}

\newpage
\thispagestyle{empty}
\begin{center}
VRIJE UNIVERSITEIT
\\ \vspace{2cm}
\textbf{{\large F}INGERPRINTING NEW PHYSICS WITH EFFECTIVE FIELD THEORIES} 
\\ \vspace{5cm}
ACADEMISCH PROEFSCHRIFT
\\ \vspace{1cm}
ter verkrijging van de graad Doctor of Philosophy aan \\
de Vrije Universiteit Amsterdam, \\
op gezag van de rector magnificus \\
prof.dr. J.J.G. Geurts, \\
in het openbaar te verdedigen \\
ten overstaan van de promotiecommissie \\
van de Faculteit der Bètawetenschappen \\
op woensdag 8 januari 2025 om 13.45 uur \\
in een bijeenkomst van de universiteit, \\
De Boelelaan 1105
\end{center}
\vspace*{\fill}
\begin{center}
door \\
Jongejan Jacobus ter Hoeve \\
geboren te Apeldoorn
\end{center}

\newpage
\thispagestyle{empty}
\begin{table}
  \begin{flushleft}
  \begin{tabular}{ll} 
promotor: & prof.dr. J. Rojo  \\
& \\
& \\
copromotor: & prof.dr. W. Verkerke \\ 
& \\
& \\
promotiecommissie: &
prof.dr. H.G. Raven\\
&prof.dr. E.L.M.P. Laenen\\
&prof.dr. F. Maltoni\\
&dr. F. de Almeida Dias\\
&dr. L. Brenner\\

\end{tabular}
\end{flushleft}
\end{table}

\clearpage

\vspace*{1.25cm}
\tableofcontents
\newpage
\markboth{}{}
\thispagestyle{plain}

\begin{preface}
\phantomsection
\addcontentsline{toc}{chapter}{Preface}
What enables us to make progress in Particle Physics? We have mapped out Nature up to the tiniest length scales, all the way down to the proton's substructure. Yet, we are potentially blind to new high-energetic phenomena beyond our current experimental reach. How can our theories be so incredibly successful at one scale, while remaining completely agnostic about whatever new physics might exist beyond it?  

This thesis revolves around this question and aims to systematically organise our ignorance through the framework of Effective Field Theories (EFTs) in order to go beyond the current best model in Particle Physics, the Standard Model (SM). The SM falls short to adequately explain various observed phenomena, and leaves many questions unanswered. For example, it is unclear whether the Higgs boson discovered at the Large Hadron Collider (LHC) in 2012 is fundamental or composite \cite{ATLAS:2012yve, CMS:2012qbp, DIMOPOULOS1982206, KAPLAN1984183, KAPLAN1984187}, 
what mechanism accounts for the smallness of neutrino masses \cite{Super-Kamiokande:1998kpq}, and whether a dark matter particle exists as supported by astrophysical and cosmological measurements \cite{Planck:2018nkj}. These shortcomings, among quite a few others, strongly suggest the existence of physics beyond the SM (BSM).

Numerous efforts have been made to extend the SM in the TeV range, although none have resulted in concrete experimental evidence so far \cite{ATLAS:2024fdw}. In this context, the Standard Model Effective Field Theory (SMEFT) provides an alternative strategy to search for BSM physics by offering a robust theoretical framework that relies on a minimal set of model assumptions. It encodes the low-energy \textit{fingerprints} due to high-energetic phenomena that are beyond our current experimental reach, making us simultaneously sensitive to a wide range of possible SM extensions. Its model independence is the key advantage of the SMEFT, and explains why it has received a lot of attention in the past few years, see Refs.~\cite{Grzadkowski:2010es, Brivio:2017vri, Isidori:2023pyp} and references therein. 

A considerable effort of the particle physics' community concerns testing whether current collider data favours the SMEFT over of the SM \cite{DeBlas:2019ehy,Ethier:2021bye,Ethier:2021ydt, Ellis:2020unq,Bissmann:2020mfi,Bruggisser:2021duo,deBlas:2021wap}. Ultimately, the hope is to find a clear correlation pattern among the parameters characterising the SMEFT that may point to an explicit SM extension. This endeavour comes with a number of significant challenges though. First, ensuring model independence comes at the cost of a large parameter space that is hard to explore - without additional assumptions, the SMEFT is sensitive to $\mathcal{O}\lp 10^3 \rp$ parameters to first non-trivial order, while the SM depends on $\mathcal{O}\lp 10 \rp$ free parameters. In addition, constraining this parameter space requires experimental data that covers a wide range of production and decay processes, with corresponding theoretical calculations that are non-trivial to obtain numerically. Finally, methodological advances are often necessary to realise the full potential of the SMEFT, for instance through Machine Learning (ML). 

In this thesis, we face these challenges and present a combined state-of-the art SMEFT analysis of the top, Higgs, diboson and electroweak sectors. We also analyse how the EFT picture that emerges connects to explicit BSM models, and develop methodological improvements based on ML to optimise sensitivity to possible BSM signals. We also present the improvement we may expect from the High-Luminosity LHC (HL-LHC) upgrade, as well as from two proposed future circular colliders, the electron-positron Future Circular Collider (FCC-ee) and the Circular Electron Positron Collider (CEPC).

The outline and my contributions are as follows. Chapter \ref{chap:smeft} is introductory and lays the foundation for the rest of this thesis. It briefly presents the SM, the framework of EFTs, and in particular the SMEFT. An introduction to the  methodologies and tools adopted and developed in this thesis is given in Chapter \ref{chap:methdologies}. These first two chapters have been written specifically for this thesis. 

Chapter \ref{chap:smefit3}, based on Ref.~\mycite{Celada:2024mcf}, presents a combined SMEFT interpretation of Higgs, top quark, and diboson data from the LHC complemented by electroweak precision observables (EWPOs) from the Large Electron-Positron Collider (LEP) and the SLAC Large Detector (SLD) within the {\sc\small SMEFiT} framework \cite{Hartland:2019bjb, Ethier:2021ydt, Ethier:2021bye, Giani:2023gfq,terHoeve:2023pvs,Celada:2024mcf}. We account for next-to-leading (NLO) order corrections in quantum chromodynamics (QCD) to the LHC cross-sections in the SMEFT and include corrections up to quadratic order in the EFT expansion. Including recent Run II data based on the full integrated luminosity, we set bounds on 45 (50) Wilson coefficients in the linear (quadratic) fit. I have produced all plots and fits, implemented an independent treatment of the EWPOs, implemented new LHC datasets, computed the corresponding SMEFT theory predictions while extending existing ones, and contributed to the writing. 

Moving on, Chapter \ref{chap:smefit_uv} is based on my work in Ref.~\mycite{terHoeve:2023pvs} and provides an automatised framework to set constraints on the couplings and masses of SM ultraviolet-completions (UV-completions) by matching them onto the SMEFT, either at tree-level or one-loop. Leveraging new functionalities in the {\sc\small SMEFiT} code, we present constraints on one-particle SM extensions matched at tree-level, as well as at one-loop, and furthermore demonstrate the interplay present in multi-particle SM extensions. The framework presented in this chapter can be used to constrain any user defined SM extension up to some mild conditions. I have run all fits, contributed to the writing and produced all plots. 

Chapter \ref{chap:ml4eft} presents the {\sc\small ML4EFT} framework, an open-source software framework I developed and presented in Ref.~\mycite{GomezAmbrosio:2022mpm}. Based on ML techniques, we construct unbinned multivariate observables that provide optimal sensitivity in the parameter space of the SMEFT. Compared to traditional approaches that adopt binned distributions, a significant gain in sensitivity to the EFT coefficients may be achieved. For this work, I developed a new open-source software framework, produced the accompanying website and tutorial, contributed to the writing, and produced all plots and fits. 

Chapter \ref{chap:smefit_fcc} analyses the impact of future colliders on the SMEFT parameter space and is based on my work in Ref.~\mycite{Celada:2024mcf}.  With the first runs of the HL-LHC upgrade scheduled in 2029, it is relevant to assess its expected impact in the context of the SMEFT. Beyond hadron colliders, proposals of new lepton colliders are actively being discussed, such as circular colliders like the FCC-ee and the CEPC. In this context, it is especially timely to quantify the information the expected measurements can bring into the parameter space of the SMEFT. For the purpose of this chapter, I developed a numerical projection module that extrapolates existing LHC Run II data to the HL-LHC, run all fits, contributed to the writing, and created all plots. 

Finally, we summarise and conclude our main findings in Chapter \ref{chap:conclusions} and outline our future visions. 
\end{preface}

\newpage
\phantomsection
\addcontentsline{toc}{chapter}{List of Publications}
\pagestyle{plain}
\section*{List of Publications}
This thesis is based on the following publications:

\begin{itemize}
    \item E.~Celada, T.~Giani, \textbf{J.~ter Hoeve}, L.~Mantani, J.~Rojo, A.~N. Rossia, M.~O.~A.
    Thomas, and E.~Vryonidou, \textit{{Mapping the SMEFT at high-energy
    colliders: from LEP and the (HL-)LHC to the FCC-ee}},  {\em JHEP} \textbf{09}
    (2024) 091,
    [\href{http://arxiv.org/abs/2404.12809}{{\texttt{arXiv:2404.12809}}}].
    
    \item \textbf{J.~ter Hoeve}, G.~Magni, J.~Rojo, A.~N. Rossia, and E.~Vryonidou, \textit{{The automation of SMEFT-assisted constraints on UV-complete models}},  {\em JHEP} \textbf{01} (2024) 179, [\href{http://arxiv.org/abs/2309.04523}{{\texttt{arXiv:2309.04523}}}].
    
    \item R.~Gomez~Ambrosio, \textbf{J.~ter Hoeve}, M.~Madigan, J.~Rojo, and V.~Sanz, \textit{{Unbinned multivariate observables for global SMEFT analyses from machine learning}},  {\em JHEP} \textbf{03} (2023) 033, [\href{http://arxiv.org/abs/2211.02058}{{\texttt{arXiv:2211.02058}}}].
    
\end{itemize}
During my PhD, I have further contributed to the following publications:

\begin{itemize}

    \item \textbf{J.~ter Hoeve}, E.~Laenen, C.~Marinissen, L.~Vernazza, and G.~Wang, \textit{{Region analysis of QED massive fermion form factor}},  {\em JHEP} \textbf{02} (2024) 024, [\href{http://arxiv.org/abs/2311.16215}{{\texttt{arXiv:2311.16215}}}].
    
    \item H.~La, A.~Brokkelkamp, S.~van~der Lippe, \textbf{J.~ter Hoeve}, J.~Rojo, and S.~Conesa-Boj, \textit{{Edge-induced excitations in Bi2Te3 from spatially-resolved electron energy-gain spectroscopy}},  {\em Ultramicroscopy} \textbf{254} (2023) 113841, [\href{http://arxiv.org/abs/2305.03752}{{\texttt{arXiv:2305.03752}}}].
    
    \item A.~Brokkelkamp, \textbf{J.~ter Hoeve}, I.~Postmes, S.~E. van Heijst, L.~Maduro, A.~V. Davydov, S.~Krylyuk, J.~Rojo, and S.~Conesa-Boj, \textit{{Spatially Resolved Band Gap and Dielectric Function in Two-Dimensional Materials from Electron Energy Loss Spectroscopy}},  {\em J. Phys. Chem. A} \textbf{126} (2022), no.~7 1255--1262, [\href{http://arxiv.org/abs/2202.12572}{{\texttt{arXiv:2202.12572}}}].
    
\end{itemize}

\pagestyle{headings}
\thispagestyle{empty}
\end{frontmatter}

\RedeclareSectionCommand[
  beforeskip=0pt,
  afterskip=2\baselineskip]{chapter}
\RedeclareSectionCommand[
  beforeskip=\baselineskip,
  afterskip=.5\baselineskip]{section}
\RedeclareSectionCommand[
  beforeskip=.75\baselineskip,
  afterskip=.5\baselineskip]{subsection}

\begin{mainmatter}

  \cleardoublepage 
  \chapter{The Standard Model as an Effective Field Theory}
\label{chap:smeft}
\vspace{-1cm}
\begin{center}
  \begin{minipage}{1.\textwidth}
      \begin{center}
          \textit{This chapter is based on reviews from Refs.~\cite{Isidori:2023pyp, Cohen:2019wxr, Brivio:2017vri, Manohar:2018aog} and my work in Ref.~\mycite{terHoeve:2023pvs}}.
      \end{center}
  \end{minipage}
  \end{center}
In this first chapter, we start by reviewing the SM that describes the elementary building blocks of Nature together with their interactions. The SM is arguably the most successful theory ever constructed by humankind and has been verified by experiment to incredible precision. Yet, open questions remain both within the SM and beyond related to theoretical and experimental observations that the SM falls short to explain.  

After defining the SM particle content and their interactions in Sect.~\ref{sec:sm_intro}, we highlight some of these shortcomings, which will motivate much of the rest of this thesis. As will be argued in Sect.~\ref{sec:smeft_intro}, these can be partially described by rephrasing the SM in the language of the SMEFT, an introduction to which will be given in Sects.~\ref{sec:efts_intro}-\ref{sec:smeft_intro}. The SMEFT provides a handle to search for new physics in a model agnostic way, with explicit models entering only at a later stage via matching, a procedure that will be introduced in Sect.~\ref{sec:matching_intro}. We end this introductory chapter in Sect.~\ref{sec:smeft_intro_pheno} by giving an overview of existing work and tools in the field of SMEFT phenomenology.

\section{The Standard Model of Particle Physics}
\label{sec:sm_intro}
The SM is built on the core principle of local gauge invariance under the following symmetry group,
\begin{equation}
  \mathcal{G}_{\rm SM} = SU(3)_c \times SU(2)_L \times U(1)_Y .
  \label{eq:sm_gauge_groups}
\end{equation}
Its field content can be categorised into the fermionic fields ($q_i$, $u_i$, $d_i$, $\ell_i$, $e_i$) with generation indices $i, j=1,2,3$, the gauge fields ($G$, $W$, $B$) associated to the three gauge groups in Eq.~\eqref{eq:sm_gauge_groups} and the Higgs doublet $\varphi$ that is responsible for electroweak symmetry breaking (EWSB) from $SU(2)_L \times U(1)_Y$ down to the gauge group $U(1)_{\rm em}$ of electromagnetism. In this notation, $q_i$ and $\ell_i$ denote respectively the left-handed quark and lepton $SU(2)_L$ doublets; $u_i$ and $d_i$ the up-type and down-type right-handed quark singlets, while $e_i$ denotes the right-handed lepton singlet. The third generation quark doublet and singlet we sometimes also denote as $Q$ and $t$ respectively in this thesis. The fields transform under $\mathcal{G}_{\rm SM}$ according to the group representations specified in Table~\ref{tab:field_content_sm}. The matter fields transform in the fundamental representation of the gauge group, while the gauge fields ($G$, $W$, $B$) transform in the adjoint representation. 

\begin{table}[t]
  \centering
  \caption{The field content of the SM and their transformation properties under the SM gauge group in Eq.~\eqref{eq:sm_gauge_groups} split into matter fields ($q_i$, $u_i$, $d_i$, $\ell_i$, $e_i$), the Higgs doublet $\varphi$ and the local gauge fields ($G$, $W$, $B$) before electroweak symmetry breaking. The first and second row list the corresponding representation under $SU(3)_c$ and $SU(2)_L$ respectively, while the bottom row displays the hypercharge assignments $y$.   }
  \label{tab:field_content_sm}
  \begin{tabular}{|c| c c c c c| c | c c c |} 
   \hline
    & $q_i$ & $u_i$ & $d_i$ & $\ell_i$ & $e_i$ & $\varphi$ & $G$ & $W$ & $B$\\ [0.5ex] 
   \hline
   $SU(3)_c$& $3$ & $3$ & $3$ & $1$ & $1$ & $1$ & $8$ & $1$ & $1$ \\ [0.5ex] 
   \hline
   $SU(2)_L$& $2$  & $1$ & $1$ & $2$ & $1$ & $2$ & $1$ & $3$ & $1$ \\ [0.5ex] 
   \hline
   $U(1)_Y$& $1/6$ &$2/3$  &$-1/3$ & $-1/2$ & $-1$& $1/2$ & $0$  &$0$ & $0$\\ [0.5ex] 
   \hline
   \end{tabular}
\end{table}

The dynamics of the SM fields is governed by the SM Lagrangian $\mathcal{L}_{\rm SM}$ that constitutes all operators built out of derivatives and fields at mass-dimension four, respecting Lorentz and local gauge invariance, 
\begin{align}
  \nonumber
  \mathcal{L}_{\rm SM} = &-\frac{1}{4}G_{\mu\nu}^A G^{A\mu\nu}  -\frac{1}{4}W_{\mu\nu}^I W^{I\mu\nu} - \frac{1}{4}B_{\mu\nu} B^{\mu\nu} \\
  \nonumber
  & + i \lp \bar{\ell}_i\slashed{D}\ell_i + \bar{e}_i\slashed{D}e_i + \bar{q}_i\slashed{D}q_i + \bar{u}_i\slashed{D}u_i + \bar{d}_i\slashed{D}d_i \rp \\
  \nonumber
  & + \lp D_\mu \varphi \rp^\dagger \lp D^\mu \varphi \rp + m^2 \varphi^\dagger \varphi - \frac{\lambda}{2}\lp \varphi^\dagger \varphi\rp^2 \\
  & - \lp Y_{e, ij}\bar{\ell}_i e_j \varphi + Y_{u, ij}\bar{q}_i u_j \tilde{\varphi} + Y_{d, ij}\bar{q}_i d_j \varphi + {\rm h.c.}\rp .
  \label{eq:sm_lag}
\end{align}
In principle, Eq.~\eqref{eq:sm_lag} may contain additional terms that contribute to topological effects. As these enter through total derivatives, we shall not consider these further in this work. The first line in Eq.~\eqref{eq:sm_lag} describes the kinematics of the gauge fields through the following field strength tensors,
\begin{align}
G_{\mu\nu}^A &= \partial_\mu G_\nu^A - \partial_\nu G_\mu^A + g_s f^{ABC}G_\mu^B G_\nu^C \, , \\
W_{\mu\nu}^I &= \partial_\mu W_\nu^I - \partial_\nu W_\mu^I + g \epsilon^{IJK}W_\mu^B W_\nu^C \, , \\
B_{\mu\nu} &= \partial_\mu B_\nu - \partial_\nu B_\mu \, 
\end{align}
where $A, B, C=1,\dots, 8$ and $I, J, K=1,2,3$ denote the $SU(3)$ and $SU(2)$ indices in the adjoint representation, with the corresponding fully antisymmetric structure constants $f^{ABC}$ and $\epsilon^{IJK}$. The corresponding gauge couplings are denoted $g_s$ and $g$. The second line in Eq.~\eqref{eq:sm_lag} encodes the fermionic kinematic terms and their interactions to the gauge bosons through the covariant derivative, which acts as 
\begin{equation}
  D_\mu = \partial_\mu - ig_s T^A G_\mu^A - i g t^I W_\mu^I - i g' y B_\mu.
  \label{eq:cov_der}
\end{equation}
In Eq.~\eqref{eq:cov_der}, $T^A=\lambda^A/2$ and $t^I = \tau^I/2$ are the generators of $SU(3)$ and $SU(2)$ in the fundamental representation with $\lambda^A$ and $\tau^I$ denoting the Gell-Mann and Pauli matrices respectively, while $y$ denotes the hypercharge of the $U(1)_Y$ gauge group with coupling $g'$ as indicated in Table~\ref{tab:field_content_sm}. Finally, the third and fourth line describe the dynamics of the Higgs boson through the Higgs potential and its interactions to the other SM matter and gauge fields. After EWSB, these terms give rise to the masses of the physical electroweak bosons, and the fermionic masses through the Yukawa interactions as written on the fourth line with Yukawa matrices $Y_e$, $Y_u$ and $Y_d$.

Even though the SM has been extremely successful, it fails to explain some pressing questions that drive much of today's research, for example:
\begin{enumerate}
  \item \textbf{Gravity:} The SM make no reference to gravity, and it is theoretically incompatible with General Relativity.
  \item \textbf{Neutrino masses:} Experimental evidence for non-zero neutrino masses observed through neutrino oscillations \cite{Super-Kamiokande:1998kpq} is inconsistent with the SM as formulated in Eq.~\eqref{eq:sm_lag} due to the absence of a neutrino mass term. Extensions of the SM exist that aim to incorporate neutrino masses, such as the Seesaw model \cite{King:2003jb}.
  \item \textbf{Dark matter and dark energy:} Galactic observations suggest that about $27\%$ of matter is composed of dark matter \cite{Bertone:2004pz}. The SM offers no such dark matter particle candidates. Dark energy makes up an additional $68\%$ of the universe's energy and accounts for the expansion of the universe. The SM also does not provide sufficient vacuum energy to be compatible with experimental observations \cite{Planck:2018vyg}. 
  \item \textbf{Matter anti-matter asymmetry:} It is unclear what explains the large abundance of ordinary matter over anti-matter. The SM does not provide a large enough source of Charge-Parity (CP) violation to account for this imbalance, as required by the Sakharov conditions \cite{Sakharov_1991}. 
\end{enumerate}
Accommodating these shortcomings demands BSM physics. However, the absence of new resonances around the electroweak scale suggests that new physics might in fact be heavy, which motivates the introduction of EFTs, to which we turn next. 
 
\section{Effective Field Theories}
\label{sec:efts_intro}
Here we motivate and present the principles underlying EFTs. We assume first a generic EFT, before specialising to the SMEFT in Sect.~\ref{sec:smeft_intro} as a probe to search for BSM physics.

\subsection{Power counting and decoupling}
\label{subsec:powercounting}
EFTs are perfectly valid quantum field theories (QFTs) that are both local and renormalisable except they may only be applied up to some finite scale.  
Conceptually, we are used to this in everyday life where day-to-day dynamics can be described without explicit reference to the existence of quarks and gluons. Another example is celestial motion, which can be accurately modelled in terms of Newtonian mechanics up to small relativistic corrections that are power suppressed by the speed of light. Put more technically, physics in the ultraviolet (UV) is said to decouple from physics in the infrared (IR), as formalised in the Appelquist-Carazzone decoupling theorem \cite{PhysRevD.11.2856}.

The above can be made more concrete by considering the example of two real scalar fields $\phi_H$ and $\phi_L$ with associated masses $M$ and $m \ll M$ respectively. Suppose we are interested in describing dynamics in a regime at energies $E\sim m$, then the heavy fields $\phi_H$ may be integrated out from the full theory $\mathcal{L}_{\rm full}$, 
\begin{equation}
  \int\mathcal{D}\phi_L\exp\left[i\int d^4x \mathcal{L}_{\rm eft}(\phi_L)\right] \equiv \int\mathcal{D}\phi_L\mathcal{D}\phi_H\exp\left[i \int d^4x \mathcal{L}_{\rm full}(\phi_L, \phi_H)\right] \, ,
\end{equation}
to arrive at an effective description $\mathcal{L}_{\rm eft}$ that only depends on the light fields $\phi_L$. The effective Lagrangian can be expressed as an infinite tower of local higher dimensional operators $\mathcal{O}_i^{(d)}$ of mass-dimension $d$ through the operator product expansion,
\begin{equation}
  \mathcal{L}_{\rm eft} = \mathcal{L}_{d \le 4} + \sum_{d=5}^\infty \sum_{i=1}^{n_{\rm eft}}\frac{c_i^{(d)}}{\Lambda^{d-4}}\mathcal{O}_i^{(d)} \, .
  \label{eq:lag_eft}
\end{equation}
In Eq.~\eqref{eq:lag_eft}, $\mathcal{O}_i^{(d)}$ is constructed exclusively out of the light fields $\phi_L$ and gets associated a dimensionless Wilson coefficient $c_i^{(d)}$ that becomes increasingly power suppressed by the cut-off scale $\Lambda \sim M$ towards higher mass-dimensions. The Wilson coefficients encode low-energy imprints due to interactions at high energies involving the heavy fields.  It is important to stress that Eq.~\eqref{eq:lag_eft} is only valid below the cut-off scale, where predictions can be made arbitrarily precise by including sufficient inverse power corrections of the heavy scale. In practice, predictions need only be accurate up to some predefined order in the EFT power expansion, meaning that higher order terms may be safely omitted. The power counting parameter keeps track of this, which is a small dimensionless quantity, and depending on the nature of the EFT, this may either be a ratio of a light mass to a heavy mass as discussed above, or a low velocity to the speed of light for example.

In case the full theory is known, an EFT valid at lower scales may be constructed following the \textit{top-down} approach. This amounts to integrating out heavy fields such that only light fields remain, which simplifies calculations significantly by dealing with only one scale at a time. In this approach, the Wilson coefficients are expressed explicitly in terms of parameters of the full theory via \textit{matching} to make sure that the full theory and the EFT agree in the IR. Another advantage shows up at loop-level, where the presence of two widely separated scales such as $m$ and $M$ may lead to large logarithms of the form $(\alpha \log(m^2 /M^2))^n$ that spoil the perturbative convergence even for small couplings $\alpha$. The EFT is able to resum these large logarithms to all orders in perturbation theory via matching and running, as will be further discussed in the next section. 

In contrast, a \textit{bottom-up} approach applies whenever the full theory is unknown. In this case, the value of the Wilson coefficients is inferred indirectly from measurements at low energies through (global) fits. In order to stay model agnostic about the UV, all higher dimensional operators compatible with symmetry requirements must be included. This leads to an unfeasibly large number of operators in practice, such that additional symmetry requirements must be imposed, e.g. flavour assumptions as will be discussed in Sect.~\ref{subsec:flavour_assumption}. Another simplification involves the inclusion of only specific subsets of measurements. At the end of the day, the philosophy behind the bottom-up approach is to determine the Wilson coefficients once, and connect them to multiple theories in the UV through matching. This makes it a highly efficient approach to search for BSM physics. 

\subsection{Renormalisation and matching}
\label{subsec:rg_matching_eft}
The earliest example in hindsight of what we call an EFT today is given by Fermi's theory of weak decays. It was originally constructed before the discovery of the $W/Z$ bosons \cite{Fermi:1933jpa}, which demonstrates that EFTs do not need explicit reference to their UV completion to constitute valid theories. Consider the matrix element of neutron to proton decay via tree-level $W$-boson exchange in the regime of small momentum transfer $p$ in comparison to $M_W$, 
  \begin{align}
    \nonumber
    i \mathcal{M} &= \lp \frac{-ig}{\sqrt{2}} \rp^2 V_{du}\lp\bar{d}\gamma^\mu P_L u\rp\frac{-i g_{\mu\nu}}{p^2-M_W^2}\lp\bar{e}\gamma^\nu P_L\nu_e\rp \\
    &\approx \frac{i}{M_W^2}\lp\frac{-ig}{\sqrt{2}} \rp^2 V_{du}\lp\bar{d}\gamma^\mu P_L u\rp\lp\bar{e}\gamma^\nu P_L\nu_e\rp + \mathcal{O}\lp M_W^{-4}\rp \, ,
    \label{eq:fermi_w_decay}
  \end{align}
  where on the last line we have expanded in the small ratio of scales $p^2/M_W^2 \ll 1$ and $V_{du}$ denotes the relevant Cabibbo-Kobayashi-Maskawa (CKM) element. The low-energy expansion of Eq.~\eqref{eq:fermi_w_decay} demonstrates that the same matrix element can in fact be obtained up to $\mathcal{O}\lp M_W^{-2}\rp$ through the following effective Lagrangian,
  \begin{equation}
    \mathcal{L}_{\rm eff} = - \frac{4 G_F}{\sqrt{2}}V_{du}\lp\bar{d}\gamma^\mu P_L u\rp\lp\bar{e}\gamma^nu P_L\nu_e\rp \, ,
    \label{eq:eft_fermi}
  \end{equation} 
  after identifying $G_F = \sqrt{2} g^2/8 M_W^2$. The effect of the $W$-boson on the low-energy degrees of freedom has now been encoded into Fermi's constant $G_F$, which provides an early example of matching. 
   
The addition of higher dimensional operators such as the one in Eq.~\eqref{eq:eft_fermi} breaks renormalisability in the traditional sense, meaning that an infinite number of counterterms would be needed to absorb all UV divergences. This can be understood starting from imagining a double dim-5 insertion in a one-loop integral. Absorbing its divergence requires a dim-6 counterterm, but two insertions of this dim-6 counterterm lead to yet another divergence that would require a counterterm of dim-8, etcetera. An infinite pattern arises where new counterterms keep introducing others.

However, an EFT requires only a finite number of counterterms as infinities that arise at higher orders may be omitted by power counting; EFTs are renormalisable order by order in the power expansion. This perspective provides the modern view of renormalisability, and consequently, Wilson coefficients are subject to renormalisation group (RG) evolution. Starting from the observation that the bare Wilson coefficients are independent of the $\overline{\rm MS}$ renormalisation scale $\tilde{\mu}$, the RG equation (RGE) for the renormalised Wilson coefficients $c_{i}^{(d)}$ is given by \cite{Manohar:2018aog}
\begin{equation}
  \frac{d}{d\log\tilde{\mu}^2}c_{i}^{(d)} = \gamma_{ij_1j_2\dots j_k}c_{j_1}^{(d_1)}\dots c_{j_k}^{(d_k)} \, ,
  \label{eq:rge_eft}
\end{equation}
with $d-4 = \sum_i (d_i - 4)$ and $\gamma_{ij_1j_2\dots j_k}$ denoting the anomalous dimension matrix. It is worth to point out several important aspects related to Eq.~\eqref{eq:rge_eft}.
\begin{itemize}
  \item First, the RGE in the EFT are coupled non-linear differential equations, as opposed to QFT in four spacetime dimensions. This can be understood by considering Eq.~\eqref{eq:rge_eft} up to $d=6$, 
\begin{align}
  \frac{d}{d\log\tilde{\mu}^2}c_{i}^{(5)} &= \gamma_{ij}^{(5)}c_{j}^{(5)} \\
  \frac{d}{d\log\tilde{\mu}^2}c_{i}^{(6)} &= \gamma_{ij}^{(6)}c_{j}^{(6)} + \gamma_{ijk}c_{j}^{(5)}c_{k}^{(5)} \, ,
\end{align} 
where the non-linear term $c_{j}^{(5)}c_{k}^{(5)}$ in the second line derives from the fact that amplitudes with two insertions at dim-5 require a dim-6 counterterm.
  \item Second, the RGE governs the evolution of the renormalised Wilson coefficients from some high scale $\mu_H$ to a lower scale $\mu_L$ or vice versa. In case of non-zero off-diagonal elements $\gamma_{ij}^{(d)}$ with $i\neq j$, the Wilson coefficients undergo mixing, meaning that $c_i \neq 0$ and $c_j = 0$ at $\mu_H$ may induce non-zero values $c_j \neq 0$ at $\mu_L$. 
\item Third, the RGE enables the summing of large logarithms. In case of two widely separated scales $m$ and $M$ with $m \ll M$, loop calculations involving both scales introduce large logarithms of the form $\log(m^2/ M^2)$ that spoil the perturbative convergence, causing perturbation theory to break down. However, one of the main benefits of the EFT is that it enables to resum these large logarithms through the RGE, thereby restoring perturbation theory. This is done by introducing a third intermediate scale  $\tilde{\mu}_M$, the so-called matching scale, with $\tilde{\mu}_L < \tilde{\mu}_M < \tilde{\mu}_H$ that splits up the large logarithm into two contributions $\log(\tilde{\mu}^2/M^2)$ and $\log(m^2/\tilde{\mu}^2)$. The first can be resummed with the RGE of the full theory up to the matching scale $\tilde{\mu}_M$, while the second can be resummed with the RGE in the EFT with the matching equation serving as boundary condition. 
\end{itemize}

\subsection{A minimal basis}
In order to stay model agnostic about the UV, all operators consistent with symmetry requirements and power counting must be included in Eq.~\eqref{eq:lag_eft}. Naively, one might expect to construct all operators $\mathcal{O}_i^{(d)}$ by writing down all permutations of field operators and derivatives that add up to mass-dimension $d$. This would, however, introduce redundancies as we illustrate in this section. We consider the following EFT Lagrangian up to dim-6 built out of a single real scalar field $\phi$, 
\begin{equation}
  \mathcal{L}_{\rm eft} \supset \frac{1}{2}\partial_\mu \phi \partial^\mu\phi - \frac{m^2}{2}\phi^2-\frac{\lambda}{4!}\phi^4 + \frac{c_1}{\Lambda^2}\frac{\phi^3\Box\phi}{3!} + \frac{c_2}{\Lambda^2}\frac{\phi^2\lp\partial_\mu\phi\rp\lp\partial^\mu\phi\rp}{4} \, , 
  \label{eq:eft_red_toy}
\end{equation}
where we have only written higher dimensional operators that will turn out to be redundant. Redundancies can be made apparent by one of the following techniques.
\begin{enumerate}
  
 \item{\textbf{Integration by parts}:
This first technique exploits the fact that operators corresponding to total derivatives vanish when integrated over spacetime. In Eq.~\eqref{eq:eft_red_toy}, this lets us remove the operator associated to coefficient $c_2$,
\begin{equation}
  \partial_{\mu}(\phi^3\partial^\mu\phi) = 3\phi^2 \lp \partial_\mu \phi\rp \lp \partial^\mu \phi\rp + \phi^3 \Box \phi \, ,
  \label{eq:op_obp}
\end{equation}
and trade it for the operator associated to $c_1$. One could also have done the reverse, the choice of operators in the final set is ultimately a matter of convenience.}

\item{\textbf{Field redefinitions:}
 In contrast to ordinary QFTs where fields may be redefined under linear mappings, EFTs must only be renormalisable by power counting and therefore allow for non-linear field transformations
\begin{equation}
  \phi \rightarrow \phi'= \phi + \frac{f(\phi)}{\Lambda^n} \, ,
  \label{eq:field_redef}
\end{equation}
where $f(\phi)$ performs a non-linear map on the field $\phi$. The effective Lagrangian in Eq.~\eqref{eq:eft_red_toy} transforms under Eq.~\eqref{eq:field_redef} according to 
\begin{equation}
  \mathcal{L}_{\rm eft} \rightarrow \mathcal{L}_{\rm eft} - \frac{1}{\Lambda^n}f(\phi)\lp \Box + m^2 + \frac{\lambda}{3!}\phi^2 \rp \phi + \mathcal{O}\lp\Lambda^{-(n+2)}\rp \, 
\end{equation}
which produces a shift at order $\Lambda^{-n}$ proportional to the classical equations of motion (EOM). This demonstrates that the effect of operators proportional to the classical EOM can be completely moved to higher orders in the EFT power expansion by a suitable choice of $f(\phi)$. Indeed, choosing $f(\phi) = c_1\phi^3/3!$ removes the operator associated to $c_1$ in the original effective Lagrangian Eq.~\eqref{eq:eft_red_toy}.
}
\item{\textbf{Fierz identities:} The last technique applies to operators composed of a chain of Dirac spinors $\psi$ and allows one to reshuffle their order at the cost of introducing or removing Dirac matrices $\gamma^\mu$ or products thereof. For example, one can relate 
\begin{equation}
  [\bar{\psi}_{L_1}\psi_{R_2}][\bar{\psi}_{R_3}\psi_{L_4}] = \frac{1}{2}[\bar{\psi}_{L_1}\gamma^\mu\psi_{L_4}][\bar{\psi}_{R_3}\gamma_\mu\psi_{R_2}] \, ,
\end{equation}
where $L/R$ denotes the chirality of the Dirac spinor.
Fierz identities derive from completeness relations, e.g. the set $\{\mathds{1}, \gamma_5, \gamma^\mu, \gamma_5\gamma^\mu, \sigma^{\mu\nu}\}$ with $\mu > \nu$ provides a basis of 16 elements that span the space of all $4\times4$ matrices in spinor space \cite{Nieves:2003in, Nishi:2004st}.  
}
\end{enumerate}

\section{Going beyond the SM: the SMEFT framework}
\label{sec:smeft_intro}
We now specialise to the EFT that we shall adopt throughout the rest of this thesis. Assuming new physics lives far above the electroweak scale $v\sim 246$ GeV, the SMEFT provides a systematic expansion of the theory space around the SM through the addition of higher dimensional operators $\mathcal{O}_i^{(d)}$ suppressed by inverse powers of the scale of new physics $\Lambda$,
\begin{equation}
  \mathcal{L}_{\rm SMEFT} = \mathcal{L}_{\rm SM} + \sum_{d=5}^\infty \sum_{i=1}^{n_{\rm eft}}\frac{c_i^{(d)}}{\Lambda^{d-4}}\mathcal{O}_i^{(d)} \, ,
  \label{eq:smeft_lag}
\end{equation}
with the following properties:
\begin{itemize}
  \item The operators $\mathcal{O}_i^{(d)}$ depend only on SM fields plus its derivatives, and respect Lorentz invariance as well as the $SU(3)_c \times U(2)_L \times U(1)_Y$ SM gauge group symmetries. At order $\mathcal{O}\lp\Lambda^{4-d}\rp$, each operator $O_i^{(d)}$ must be of mass-dimension $d$ consistent with power counting.
  \item EWSB is linearly realised, identically to the SM. The Higgs Effective Field Theory (HEFT) as a generalisation of the SMEFT allows for non-linear EWSB mechanisms in addition \cite{Brivio:2017vri}. 
  \item Operators $\mathcal{O}_i^{(d)}$ with $d > 4$ introduce either existing SM-like vertices with modified couplings or new Lorentz structures. Based on dimensional analysis, an amplitude $\mathcal{A}^{(6)}$ at dim-6 probed at energy $\sim E$ may scale according to $\mathcal{A}^{(6)} \sim (v^2, vE, E^2)$. The quadratic scaling with energy is especially relevant at the LHC, as it provides a sensitive probe to search for new physics in the high-energy tails of differential cross-section distributions.
\end{itemize}
In the rest of this section, we elaborate more on aspects of Eq.~\eqref{eq:smeft_lag} relevant for LHC phenomenology. Sect.~\ref{subsec:operator_basis} discusses the operator basis, while its impact on cross-section predictions is discussed in Sect.~\ref{subsec:smeft_observables}. We end by discussing additional symmetry considerations in the flavour sector in Sect.~\ref{subsec:flavour_assumption}. 

\subsection{SMEFT operator bases}
\label{subsec:operator_basis}
At mass-dimension five, only one operator exists, known as the Weinberg operator \cite{Weinberg:1979sa} that generates a Majorana mass term for right-handed neutrinos after EWSB, and violates lepton number by two units. The associated cut-off scale $\Lambda$ has been strongly constrained to $10^{14}\:{\rm TeV} < \Lambda < 10^{15}\:\rm {TeV}$ for $\mathcal{O}(1)$ Wilson coefficients due to the small neutrino masses \cite{Isidori:2023pyp}, hence we do not consider its effect in the rest of this thesis. 

The first complete basis up to dim-6 was constructed in Ref.~\cite{Buchmuller:1985jz} in 1986, which, however, turned out to include redundant operators. The final list of non-redundant operators was established in 2010 and is referred to as the Warsaw basis, see Ref.~\cite{Grzadkowski:2010es} for its definition. Other bases exist that are tailored towards certain phenomenological applications, such as the strongly-interacting light Higgs (SILH) basis relevant for UV-complete composite Higgs models \cite{Giudice:2007fh}, the Hagiwara-Ishihara-Szalapski-Zeppenfeld (HISZ) basis relevant for electroweak precision constraints \cite{PhysRevD.48.2182}, and the Green's basis \cite{Gherardi:2020det}. The latter extends the Warsaw basis by operators that are redundant through the EOMs, which is convenient for matching UV-complete models onto the SMEFT, as will be discussed in Sect.~\ref{sec:matching_intro}. It is worth pointing out that the SILH and HISZ bases do not qualify as actual bases as they do not represent a complete list at dim-6 operators \cite{Brivio:2017bnu}.

We now classify the operators in the Warsaw basis, which we adopt in this thesis, following the logic of Ref.~\cite{Grzadkowski:2010es}. In general, a dim-6 operator must be of the form,
\begin{equation}
  O_i^{(6)} =  \varphi ^{N_{\varphi}} D^{N_{D}} X^{N_{X}} \lp \bar{\psi} \psi \rp^{N_{\psi}}
\end{equation}
with $\varphi$ denoting the Higgs doublet, $D$ the covariant derivative, $X$ the field strength tensor and $\bar{\psi}\psi$ a bilinear of fermionic fields, all raised to appropriate powers to add up to mass-dimension 6:
\begin{equation}
  \left[ O_i^{(6)} \right] = N_\varphi + N_D + 2 N_X + 3N_{\psi} = 6 \, .
\end{equation}
Let us first focus on purely bosonic operators characterised by $N_\psi=0$. In this case, the Warsaw basis consists of the following classes of operators, 
\begin{equation}
  \{X^3, X^2\varphi^2, X^2D^2, X\varphi^2D^2, \varphi^6, \varphi^4D^2, \varphi^2D^4\} \, ,
  \label{eq:bosonic_ops_warsaw_1}
\end{equation}
where the class $X\varphi^4$ does not contribute by the antisymmetry of $X$, and $XD^4$ is related to $X^2D^2$ via contractions. As argued in Sect.~\ref{subsec:operator_basis}, the operators $\varphi^2D^4$, $\varphi^2 X D^2$, $X^2D^2$ can be reduced by the EOMs either to operators containing fermions, or to $X^3$, $X^2\varphi^2$, $\varphi^6$ or $\varphi^4D^2$, such that the set \eqref{eq:bosonic_ops_warsaw_1} further reduces to
\begin{equation}
  \{X^3, X^2\varphi^2, \varphi^6, \varphi^4D^2 \} \, .
  \label{eq:bosonic_ops_warsaw_2}
\end{equation}
Examples of operators in \eqref{eq:bosonic_ops_warsaw_2} are
\begin{align}
  \nonumber
  X^3 \qquad : \qquad & \mathcal{O}_{W} = \epsilon_{IJK}W_{\mu\nu}^I W^{J, \nu\rho}W^{K, \mu}_\rho \\
  \nonumber
  X^2\varphi^2 \qquad : \qquad & \mathcal{O}_{\varphi G} = \lp \varphi^\dagger \varphi\rp G^{A, \mu\nu} G^A_{\mu\nu}\\
  \nonumber
  \varphi^6 \qquad : \qquad & \mathcal{O}_{\varphi} = \lp \varphi^\dagger\varphi \rp^3\\ 
  \nonumber
  \varphi^4 D^2 \qquad : \qquad & \mathcal{O}_{\varphi D} = \lp \varphi^\dagger D^\mu \varphi \rp^\dagger \lp \varphi^\dagger D_\mu \varphi \rp \, , 
\end{align}
where $\mathcal{O}_{\varphi G}$ introduces effective gluonic-Higgs interactions, $X^3$ contributes to the anomalous triple gauge couplings and multiboson interactions, $\mathcal{O}_\varphi$ modifies the Higgs potential and $\mathcal{O}_{\varphi D}$ modifies the interactions between the Higgs and the electroweak bosons.

Next, switching on $N_\psi=2$ gives rise to the following additional classes of operators,
\begin{equation}
  \{\psi^2D^3, \psi^2\varphi D^2, \psi^2 XD, \psi^2\varphi^3, \psi^2 X \varphi, \psi^2\varphi^2 D\} \, ,
  \label{eq:2_fermionic_ops_warsaw_1}
\end{equation}
which can be further reduced by the EOMs to
\begin{equation}
  \{\psi^2X\varphi, \psi^2\varphi^2 D, \psi^2\varphi^3\}.
  \label{eq:2_fermionic_ops_warsaw_2}
\end{equation}
The operators in Eq.~\eqref{eq:2_fermionic_ops_warsaw_2} are known as dipole, current and Yukawa operators respectively. The latter category modifies the Yukawa interactions in the SM. Finally, the case $N_\psi=4$ leads to four-fermionic operators that can be distinguished based on their chirality $L$ or $R$: $(\bar{L}L)(\bar{L}L), (\bar{R}R)(\bar{R}R), (\bar{L}L)(\bar{R}R)$. The number of operators in this class is especially high due to combinatorics - 2205 baryon/lepton number preserving four-fermion operators exist at dim-6 for three generations. The anomalous dimension matrix at dim-6 in the Warsaw basis was constructed in Ref.~\cite{Alonso:2013hga}. 

Non-redundant bases beyond dim-6 have also been established. For dim-7 we refer to Ref.~\cite{Liao:2016hru}, while dim-8 and dim-9 are tabulated in Refs.~\cite{PhysRevD.104.015026, Murphy:2020rsh} and Refs.~\cite{Li:2020xlh, Liao:2020jmn} respectively. A generalised procedure to count the number of non-redundant operators has been obtained via the Hilbert Series in Refs.~\cite{Henning:2015alf, Marinissen:2020jmb}.



\subsection{Observables and input schemes}
\label{subsec:smeft_observables}
Armed with an operator basis, next we translate how Eq.~\eqref{eq:smeft_lag} manifests itself at the level of (differential) cross-sections measurements. Up to $\mathcal{O}\lp\Lambda^{-4}\rp$, the matrix element in the SMEFT is given by
\begin{align}
  \mathcal{M}(\boldsymbol{c}) &= \mathcal{M}^{(\rm sm)} + \sum_{i=1}^{n_{\rm{eft}}}\frac{c_i^{(d_6)}}{\Lambda^2}\mathcal{M}^{(\rm d_6)}_i  
  + \sum_{i=1}^{n_{\rm{eft}}}\frac{c_i^{(d_8)} }{\Lambda^4} \mathcal{M}^{(\rm d_8)}_i + \sum_{i,j=1}^{n_{\rm eft}} \frac{c_i^{(d_6)}c_j^{(d_6)}}{\Lambda^4}\mathcal{M}_{ij}^{(d_8)'} \, ,
  \label{eq:mem_smeft}
\end{align}
where dim-5 and dim-7 operators have been omitted because they violate either lepton or baryon number conservation \cite{Helset:2019eyc, Kobach:2016ami}. These are already tightly constrained in comparison to effects entering at even mass-dimensions as a result of proton stability and the smallness of neutrino masses. Studies that include odd-dimensional operators can be found in Refs.~\cite{Felkl:2021uxi, Deppisch:2020oyx, Li:2019fhz, Cirigliano:2018yza, MINKOWSKI1977421, Deppisch:2017ecm, Fridell:2023rtr}. Furthermore, the last term in Eq.~\eqref{eq:mem_smeft} contains contributions from double insertions of dim-6 operators. These mix with dim-8 operators under field redefinitions, whose effect we denote by a prime. Since the cross-section $\sigma_m(\boldsymbol{c})$ in some generic kinematic bin $m$ is related to its matrix element squared, we obtain 
\begin{align}
  \nonumber
  \sigma_m(\boldsymbol{c}) &\sim \left| \mathcal{M}^{\rm(sm)}\right|^2 + \frac{1}{\Lambda^2}\sum_{i=1}^{n_{\rm eft}}2 {\rm Re}\left[\mathcal{M}^{(\rm sm)\dagger}\mathcal{M}^{(\rm d_6)}_i\right]c_i^{d_6} +
  \frac{1}{\Lambda^4}\sum_{i,j=1}^{n_{\rm eft}} \mathcal{M}_i^{\rm (d_6)\dagger}\mathcal{M}_j^{\rm (d_6)}c_{i}^{d_6}c_j^{d_6} \\
  & + \frac{1}{\Lambda^4}\sum_{i=1}^{n_{\rm eft}} 2 {\rm Re}\left[ \mathcal{M}^{(\rm sm)\dagger}\mathcal{M}^{(\rm d_8)}_i\right]c_{i}^{d_8} 
  + \frac{1}{\Lambda^4}\sum_{i,j=1}^{n_{\rm eft}} 2 {\rm Re}\left[ \mathcal{M}^{(\rm sm)\dagger}\mathcal{M}^{'(\rm d_8)}_i\right]c_{i}^{d_6}c_j^{d_6} \, .
  \label{eq:mem_smeft_squared}
\end{align}
In this thesis, we restrict ourselves to real valued Wilson coefficients, thus imposing CP symmetry, and we furthermore neglect the $\mathcal{O}\lp\Lambda^{-4}\rp$ contributions from dim-8 operators on the second line of Eq.~\eqref{eq:mem_smeft_squared}. This is justified mostly on practical grounds; global EFT fits require a simultaneous treatment of all operators that enter a given process, which becomes unfeasible in practice due to the large number of operators at dim-8 \footnote{For three generations, and including baryon- and lepton-number violating operators, as well as CP violating ones, dim-8 contains $44807$ operators, while dim-6 contains $3045$ operators \cite{Henning:2015alf}.}. However, effects of subsets of dim-8 operators in specific processes have been studied in Refs.~\cite{Ellis:2023zim, Hays:2018zze, Corbett:2021iob, Asteriadis:2022ras, Corbett:2021eux, Alioli:2020kez, Boughezal:2021tih, Kim:2022amu, Degrande:2023iob, Corbett:2023qtg}. 

With these considerations, and performing the phase-space integral, we are left with the following modified cross-section prediction that we adopt in this thesis, 
\begin{equation}  
\sigma_m(\boldsymbol{c}) =
  \sigma^{(\rm sm)}_m + \sum_{i=1}^{n_{\rm{eft}}}
  \sigma_{m,i}^{(\rm eft)}c_i
  +\sum_{i=1, j \ge i}^{n_{\rm{eft}}}
  \sigma_{m, ij}^{(\rm eft)}c_i c_j \, ,
  \label{eq:smeft_xsec}
\end{equation} 
where we have dropped the superscripts $d_i$ on the Wilson coefficients as all of them are of dim-6. In Eq.~\eqref{eq:smeft_xsec}, we have redefined the Wilson coefficients $c_i$ to absorb the cut-off scale $\Lambda$, i.e. the coefficients $c_i$ are to be understood as $\tilde{c}_i/\Lambda^2$ with $\tilde{c}_i$ the original Wilson coefficient entering the SMEFT Lagrangian Eq.~\eqref{eq:smeft_lag}. Looking at Eq.~\eqref{eq:smeft_xsec}, the SM cross-section $\sigma_m^{(\rm sm)}$ gets complemented by a linear correction  $\sigma_{m,i}^{(\rm sm)}$ at $\mathcal{O}\lp\Lambda^{-2}\rp$ arising from interference between the SM and EFT amplitudes with a single dim-6 $c_i$ insertion, and a pure quadratic EFT correction $\sigma_{m, ij}^{(\rm eft)}$ at $\mathcal{O}\lp\Lambda^{-4}\rp$, which we recall here does not include dim-8 corrections. 


\myparagraph{Input schemes} In addition to operators entering amplitudes directly, operators may also come in indirectly through the input parameters that define the theory. As an example we consider the muon decay rate used to determine Fermi's constant $\hat{G}_F$,
\begin{equation}
  \Gamma(\mu \rightarrow e \bar{\nu}_e\nu_\mu) = \hat{G}_F^2\frac{m_\mu^5}{192 \pi^3} \, ,
\end{equation}
where the hat signifies quantities that are determined directly from data. 
%
However, in the SMEFT, the muon decay rate receives corrections from dim-6 operators,
\begin{equation}
  \Gamma(\mu \rightarrow e \bar{\nu}_e\nu_\mu) = \underbrace{\lp \bar{G}_F + \frac{\sqrt{2}}{4}\lp c_{\ell\ell} - c_{\varphi\ell_1}^{(3)}- c_{\varphi\ell_2}^{(3)}\rp \rp^2}_{\hat{G}_F}\frac{m_\mu^5}{192 \pi^3} \, ,
  \label{eq:muon_decay_rate_smeft}
\end{equation}
such that possibly non-zero effects of $c_{\ell\ell}$, $c_{\varphi\ell_1}^{(3)}$ and $c_{\varphi\ell_1}^{(3)}$ get absorbed in the numerical value of $\hat{G}_F$, with $\bar{G}_F$ following the usual SM relation. Consequently, any derived quantity that depends on $\bar{G}_F$, such as the fine structure constant $\alpha_{\rm ew}$, receives indirect corrections from these operators. Explicitly, the SM relation
\begin{equation}
  \bar{G}_F &= \frac{\pi \alpha_{\rm ew}}{\sqrt{2}m_W^2(1-m_W^2/m_Z^2)}  \, ,
\end{equation}
implies together with Eq.~\eqref{eq:muon_decay_rate_smeft}
\begin{equation}
  \alpha_{\rm ew} = \lp \hat{G}_F - \frac{\sqrt{2}}{4}\lp c_{\ell\ell} - c_{\varphi\ell_1}^{(3)} - c_{\varphi\ell_2}^{(3)} \rp\rp \frac{\sqrt{2}m_W^2(1-m_W^2/m_Z^2)}{\pi} \, ,
  \label{eq:alpha_ew_inputs}
\end{equation}
which shows that any electroweak (EW) process will always depend on $c_{\ell\ell}$, $c_{\varphi\ell_1}^{(1)}$ and $c_{\varphi\ell_1}^{(2)}$ indirectly through the definition of the input parameter $\hat{G}_F$.

Two common input parameter schemes in the EW sector exist. The $\alpha_{\rm ew}$ scheme fixes $\{\hat{\alpha}_{\rm ew}, \hat{m}_Z, \hat{G}_F\}$, meaning that $m_W$ becomes a derived quantity, while the $m_W$-scheme fixes $\{m_W, m_Z, G_F\}$ with $\alpha_{\rm ew}$ a derived quantity, as indicated explicitly in Eq.~\eqref{eq:alpha_ew_inputs}. As $\alpha_{\rm ew}$ enters amplitudes via numerators, as opposed to $m_W$ that enters via propagators, we adopt the $m_W$-scheme in this thesis. 

\subsection{Flavour assumptions}
\label{subsec:flavour_assumption}
Imposing baryon and lepton number conservation, the Lagrangian in Eq.~\eqref{eq:lag_eft} contains $2499$ dim-6 operators for three generations \cite{Alonso:2013hga}. To reduce this number, and make practical analyses feasible, one may adopt one of the following flavour assumptions:
\begin{enumerate}
  \item $U(3)^5$: this is the maximal symmetry that can be imposed and assumes equal treatment of all fermionic generations. 
  \item MFV: Minimal Flavour Violation assumes that the Yukawa couplings and the CKM phase provide the only source of CP violation and assumes that this continues to hold in the SMEFT. This is obtained by imposing a $U(3)^5$ symmetry on the fermionic fields, while spurions invariant under $U(3)^5$ are inserted in the currents that lead to flavour violating effects \cite{Brivio:2020onw}.
  \item \verb|top|: assumes a $U(2)^3$ symmetry in the quark sector, i.e. only the first two generations are treated equally, while top and bottom are treated separately. In the lepton sector, the maximal symmetry $U(3)_\ell \times U(3)_e$ is imposed.
  \item  \verb|topU3l|: similar to \verb|top|, except that the lepton sector now assumes the less restrictive $\left[ U(1)_\ell \times U(1)_e \right]^3$ symmetry. 
  \item {\sc\small SMEFiT} flavour assumptions: $U(2)_q \times U(2)_u \times U(3)_d \times \left[ U(1)_\ell \times U(1)_e \right]^3$. This is similar to \verb|topU3l| except that the bottom-quark is no longer singled out. This set of flavour of assumptions is consistent with the implementation of \verb|SMEFTatNLO| \cite{Degrande:2020evl}, which is the automatised Monte-Carlo tool that will allow us to obtain theory predictions in the SMEFT at next-to-leading order (NLO) in the QCD perturbative expansion in Chapter \ref{chap:methdologies}.
\end{enumerate}
Throughout the rest of this thesis, we will adopt the {\sc\small SMEFiT} flavour assumptions.


\section{Matching onto the SMEFT}
\label{sec:matching_intro}
In this section, we describe in more detail the concept of matching UV-complete models onto the SMEFT. This is adapted from my work in Ref.~\mycite{terHoeve:2023pvs} and serves as introduction to Chapter \ref{chap:smefit_uv}.

Matching an EFT, such as 
the SMEFT, with the associated UV-complete model can be performed by means of two well-established techniques, as well as with another recently developed method based on on-shell amplitudes~\cite{DeAngelis:2023bmd}.
The first of these matching techniques is known as the functional method and is based on the manipulation of the path integral, the action, and the Lagrangian~\cite{Fuentes-Martin:2016uol,Ellis:2017jns,Kramer:2019fwz,Angelescu:2020yzf,Ellis:2020ivx,Summ:2020dda,Banerjee:2023iiv,Chakrabortty:2023yke}.
It requires to specify the UV-complete Lagrangian, the heavy fields, and the matching scale, 
while the EFT Lagrangian is part of the result although not necessarily in the desired basis.
The second technique is the diagrammatic method, based on equating off-shell Green's functions computed in both the EFT and the UV model, and therefore it requires the explicit form of both Lagrangians from the onset~\cite{deBlas:2017xtg,Carmona:2021xtq}.
Both methods provide the same final results and allow for both tree-level and one-loop matching computations. 
The automation of this matching procedure up to 
the one-loop level is mostly solved in the case of the diagrammatic technique~\cite{Carmona:2021xtq}, and is well advanced in the functional method case~\cite{DasBakshi:2018vni,Fuentes-Martin:2022jrf}.

Let us illustrate the core ideas underlying
this procedure by reviewing the matching to 
the SMEFT at tree and one-loop level of a 
specific benchmark UV-complete model.
This is taken to be the single-particle
extension of the SM resulting from adding a new heavy scalar boson, $\phi$. This scalar transforms under the SM gauge group in the same manner as the Higgs boson, i.e. $\phi\sim\left(1,2\right)_{1/2}$, where we denote the irreducible representations under the SM gauge group as $(\text{SU}(3)_{\text{c}},\text{SU}(2)_{\text{L}})_{\text{U}(1)_{\text{Y}}}$. 
Following the notation of~\cite{deBlas:2017xtg}, the Lagrangian of this model reads
\begin{equation}
\begin{split}
 \lag_{\rm UV}= &\lag_{\rm SM} + |D_{\mu}\phi|^2 - m_{\phi}^2 \phi^{\dagger}\phi - \left((y_{\phi}^{e})_{ij}\, \phi^{\dagger} \bar{e}_{R}^{i} \ell_{L}^{j} + (y_{\phi}^{d})_{ij}\, \phi^{\dagger} \bar{d}_{R}^{i} q_{L}^{j} \right.\\
	&\left. + (y_{\phi}^{u})_{ij} \, \phi^{\dagger} i\sigma_2  \bar q_{L}^{T,i} u_{R}^{j} + \lambda_{\phi}\, \phi^{\dagger}\varphi |\varphi|^2+\text{h.c.}\right) - {\rm scalar~potential}\, ,
\end{split}	
\label{eq:Lag_UV_full_Varphi_model}
\end{equation}
with $\lag_{\rm SM}$ being the SM
Lagrangian and  $\varphi$ the SM Higgs
doublet.
We do not write down explicitly the complete form of the scalar potential in Eq.~(\ref{eq:Lag_UV_full_Varphi_model}), of which $\lambda_{\phi}\, \phi^{\dagger}\varphi |\varphi|^2$ is one of the components, since it has no further effect on the matching outcome as long as it leads to an expectation value satisfying  $\langle\phi\rangle=0$,
such that $m_\phi^2>0$ corresponds to the pole mass. 

This heavy doublet $\phi$ interacts with the SM fields via the Yukawa couplings $(y_{\phi}^{u,d,e})_{ij}$, the scalar coupling $\lambda_\phi$, and the electroweak gauge couplings.
In the following, we consider as  ``UV couplings''
exclusively those couplings between UV and SM particles that are not gauge couplings.
The model described by Eq.~(\ref{eq:Lag_UV_full_Varphi_model}) corresponds to the two-Higgs doublet model (2HDM) in the decoupling limit~\cite{Gunion:2002zf,Bernon:2015qea}.
For simplicity, we assume that all the couplings between the SM and the heavy particle are real
and satisfy $(y_{\phi}^{\psi})_{ij} = \delta_{i,3}\delta_{j,3}\, (y_{\phi}^{\psi})_{33}$ for $\psi=u,\,d,\,e$, and the only SM Yukawa couplings that we consider as non-vanishing are the ones of the third-generation fermions. 
 
\myparagraph{Tree-level matching} The matching of UV-complete models to dim-6 SMEFT operators at tree-level has been fully tackled in~\cite{deBlas:2017xtg}, which considers all possible UV-completions  with particles of spin up to $s=1$  generating non-trivial Wilson coefficients. 
These results can be reproduced with the automated codes {\sc\small MatchingTools}~\cite{Criado:2017khh} and {\sc\small MatchMakerEFT}~\cite{Carmona:2021xtq}
based on the diagramatic approach.
At tree-level, the diagrammatic method requires computing the tree-level Feynman diagrams contributing to multi-point Green's functions with only light external particles. 
Then, the covariant propagators $\Delta_i$ must be expanded to a given order in 
inverse powers of the heavy masses. 
The computation of the Feynman diagrams in the EFT can be performed in a user-defined operator basis.

The outcome of matching the model defined by the Lagrangian in Eq.~\eqref{eq:Lag_UV_full_Varphi_model} to 
the SMEFT at tree-level is provided in~\cite{deBlas:2017xtg}.
%
%
A representative subset of the resulting tree-level matching expressions is given by

\begin{alignat}{4}
        \nonumber \frac{\lp c_{qd}^{(1)}\rp_{3333}}{\Lambda^2} &=-\frac{\lp y_{\phi}^{d}\rp_{33}^{2}}{6\,m_\phi^2}, \quad &&   \frac{\lp c_{qd}^{(8)}\rp_{3333}}{\Lambda^2} &&= -\frac{\lp y_{\phi}^{d}\rp_{33}^{2}}{ m_{\phi}^2}, \quad && \frac{\lp c_{d\varphi}\rp_{33}}{\Lambda^2} = \frac{\lambda_{\phi}\,\left(y_{\phi}^{d}\right)_{33}}{m_\phi^2}, \\
        \nonumber \frac{\lp c_{qu}^{(1)}\rp_{3333}}{\Lambda^2} &=-\frac{\lp y_{\phi}^{u}\rp_{33}^2}{6\, m_{\phi}^2}, \quad &&   \frac{\lp c_{qu}^{(8)}\rp_{3333}}{\Lambda^2} &&= -\frac{\lp y_{\phi}^{u}\rp_{33}^2}{ m_{\phi}^2}, \quad && \frac{\lp c_{u \varphi}\rp_{33}}{\Lambda^2} =-\frac{\lambda_{\phi}\,\left(y_{\phi}^{u}\right)_{33}}{m_\phi^2},  \\
         \frac{c_{\varphi}}{\Lambda^2} &=  \frac{\lambda_{\phi}^2}{ m_{\phi}^2}, \qquad && \frac{\lp c_{\varphi q}^{(3)}\rp_{33}}{\Lambda^2} &&= 0 \, . && 
      \label{eq:matching_result_example_tree}
\end{alignat}

Eq.~(\ref{eq:matching_result_example_tree}) showcases the type of constraints on the EFT coefficients that tree-level matching can generate.
First of all, the relations between UV couplings
and Wilson coefficients will be in general non-linear.
Second, some coefficients such as $\left(c_{\varphi q}^{(3)}\right)_{33}$ 
are set to zero by the matching relations. Third, other coefficients acquire a well-defined
sign, such as $c_\varphi$ and $\left(c_{qd}^{(1)}\right)_{3333}$ which become positive-definite and negative-definite after matching, respectively.

When considering multi-particle UV scenarios, rather
than single-particle extensions such 
as the model defined by Eq.~\eqref{eq:Lag_UV_full_Varphi_model},
non-vanishing EFT coefficients generally consist of the sum of several rational terms. 
For example, let us add
to the model of Eq.~\eqref{eq:Lag_UV_full_Varphi_model}
a second heavy scalar with gauge charges $\Phi\sim\left(8,2\right)_{1/2}$ and with mass $m_{\Phi}$ which couples to the SM  fields by means of
\begin{equation}
    \lag_{\rm UV}\supset - \left(y_{\Phi}^{qu}\right)_{ij}\,\Phi^{A\dagger}\,i\sigma_2\,\bar{q}_{L,i}^{T} T^{A} u_{R,j}+\text{h.c.} \, .
\label{eq:Lag_UV_add_1part}
\end{equation}
Integrating out this additional heavy
scalar field modifies
two of the tree-level matching relations listed in Eq.~\eqref{eq:matching_result_example_tree}
as follows
\begin{equation}
    \frac{\lp c_{qu}^{(1)} \rp_{3333}}{\Lambda^2} =-\frac{\lp y_{\phi}^{u}\rp_{33}^2}{6\, m_{\phi}^2}-\frac{2\left(y_{\Phi}^{qu}\right)_{33}^2}{9\,m_{\Phi}^2}\,, \quad \frac{\lp c_{qu}^{(8)} \rp_{3333}}{\Lambda^2} = -\frac{\lp y_{\phi}^{u}\rp_{33}^2}{ m_{\phi}^2}+\frac{\left(y_{\Phi}^{qu}\right)_{33}^2}{6\,m_{\Phi}^2}\,.
\end{equation}
%
In this context, we observe that many of the conditions on the EFT coefficients imposed by assuming a certain UV completion are non-linear and hence lead to non-Gaussian effects as we will encounter in Chapter \ref{chap:smefit_uv}. 

The tree-level matching results discussed up to now do not comply with the {\sc\small SMEFiT} flavour assumptions as introduced in Sect.~\ref{subsec:flavour_assumption}, namely U$(2)_{q}\times$U$(2)_{u}\times$U$(3)_{d}\times(\text{U}(1)_{\ell}\times\text{U}(1)_{e})^3$. 
This would cause ambiguities, since for example the coefficient $\lp c_{qd}^{(1)}\rp_{33ii}$ has the same value for $i=1,\,2,\,3$ in the {\sc SMEFTiT} flavour assumptions, while the matching result instead gives a non-vanishing coefficient only for $i=3$. 
In this specific case, the flavour symmetry can be respected after tree-level matching by further imposing
\begin{equation}
\left( y_{\phi}^{e} \right)_{33} = \left(y_{\phi}^{d}\right)_{33}=0 \, ,
\end{equation}
and leaving $\lambda_\phi$ and $\left(y_{\phi}^{u}\right)_{33}$ as the only non-vanishing UV couplings. Notice that this implies that the heavy new particle interacts only with the Higgs boson and the top quark, a common situation in well-motivated UV models. 
%

%

\myparagraph{One-loop matching} Extending the diagrammatic matching technique to the one-loop case is conceptually straightforward, and requires the computation of one-loop diagrams in the UV model with off-shell external light particles and at least one heavy-particle propagator inside the loop.
From the EFT side, diagrammatic matching
at one-loop involves the calculation of the diagrams with the so-called Green's basis, which includes also those operators that are redundant by EOMs. 
The dim-6  and dim-8 Green's bases in the
SMEFT have been computed in \cite{Gherardi:2020det,Chala:2021cgt,Ren:2022tvi}. 
Further technicalities such as evanescent operators acquire special relevance at this order~\cite{Fuentes-Martin:2022vvu}.
The automation of one-loop matching with the diagrammatic technique is provided by  {\sc\small MatchMakerEFT}~\cite{Carmona:2021xtq} for any renormalisable UV model with heavy scalar bosons and spin-$1/2$ fermions.
The equivalent automation
applicable to models containing heavy spin-1 bosons is work in progress.

In this thesis we use {\sc\small MatchMakerEFT} (v1.1.3) to evaluate the one-loop matching of 
a selected UV model. 
When applied to the Lagrangian of Eq.~\eqref{eq:Lag_UV_full_Varphi_model}, this procedure generates  
additional non-vanishing operators in comparison
to those arising at tree-level
%

An example of the results of one-loop matching corrections to
the EFT coefficients is provided for 
$c_{Qt}^{(8)}$, for which the tree-level
matching relation in Eq.~\eqref{eq:matching_result_example_tree}
is now extended as follows
\begin{eqnarray} 
\nonumber
\label{eq:oneloop_logs_example}
    \frac{c_{Qt}^{(8)}}{\Lambda^2} &=& - \lc \frac{25 g_1^2}{1152\pi^2}+\frac{3g_2^2}{128\pi^2} - \frac{3 \lp y_{t}^{\rm SM}\rp^2}{16 \pi^2} + \frac{g_3^2}{16\pi^2}\left(1-\log \lp \frac{m_\phi^2}{\mu^2}\rp\right) \rc \frac{\lp y_{\phi}^{u}\rp_{33}^{2}}{m_\phi^2} \\&&-\frac{\lp y_{\phi}^{u}\rp_{33}^{2}}{m_\phi^2 } + \frac{3}{64\pi^2} \lc 1-2\log \lp \frac{m_\phi^2}{\mu^2}\rp \rc \frac{\lp y_{\phi}^{u}\rp_{33}^4}{m_{\phi}^2} \, ,
\end{eqnarray}
with $\mu$ being the matching scale,
$g_i$ the SM gauge coupling constants,
and $y_{t}^{\rm SM}$ the top Yukawa coupling
in the SM. 
To estimate the numerical impact of
loop corrections to matching 
in the specific case of Eq.~(\ref{eq:oneloop_logs_example}), one can substitute the corresponding values of the SM couplings.
One finds that the term proportional to $\left(y_{\phi}^{u}\right)_{33}^2$
receives a correction at the few-percent
level from one-loop matching effects. 

The logarithms in the matching scale $\mu$
appearing in Eq.~(\ref{eq:oneloop_logs_example}) are  generated by the running of the couplings and Wilson
coefficients between the heavy particle mass $m_\phi$ and $\mu$. 
Since in this thesis we neglect RG running effects~\cite{Aoude:2022aro}, the matching expressions may be simplified by choosing the scale $\mu$ to be equal to the mass of the integrated-out UV heavy field such that the logarithms vanish.
As compared to the tree-level matching relations, common features of the one-loop contributions are the appearance of terms proportional to the UV couplings at the fourth order and the presence
of the SM gauge couplings. 
%
%

It is also possible to find EFT coefficients that are matched to the sum of a piece depending only on SM couplings and the UV mass and a piece proportional to the UV couplings, e.g.,
\begin{equation}
    \frac{c_{\varphi Q}^{(3)}}{\Lambda^2} = -\frac{g_2^4}{3840\pi^2 m_\phi^2} - \frac{\left(y_{t}^{\rm SM}\right)^2 \left(y_{\phi}^{u}\right)_{33}^2}{192\pi^2 m_\phi^2} + \frac{ g_2^2 \left(y_{\phi}^{u}\right)_{33}^2}{1152\pi^2 m_\phi^2}.
\end{equation} 
This kind of relations could in principle favour non-vanishing UV couplings even when the EFT
coefficients are very tightly constrained, provided that the gauge-coupling term is of similar size to the other terms in
the matching relation.


\section{Phenomenology of the SMEFT}
\label{sec:smeft_intro_pheno}
Switching again to a bottom-up approach, the cross-section predictions in the SMEFT, Eq.~\eqref{eq:smeft_xsec}, can be fitted to LHC, LEP and SLD data in order to constrain and correlate the corresponding Wilson coefficients. Many efforts in this direction have been made in the past couple of years, featuring increasingly wider datasets, more precise theory predictions, as well as more advanced analyses techniques. The aim of the remainder of this chapter is to give an overview of the existing efforts, highlighting especially how the main results presented in this thesis compare. We also list some of the available tools relevant for SMEFT phenomenology. 

On the experimental side, the ATLAS collaboration published a first internal global SMEFT analysis to Higgs and EW data in Ref.~\cite{ATLAS:2022xyx}, which is simultaneously sensitive to 28 operators and presents bounds at linear and quadratic order in the EFT expansion. It adopts the STXS 1.2 framework in the Higgs sector \cite{Monti:2803606} and includes 5 Higgs-decay channels ($ZZ$, $WW$, $\gamma\gamma$, $bb$, $\tau\tau$) for all production modes, which was updated recently to include in addition the $\mu\mu$ and $Z\gamma$ decay channels \cite{ATLAS:2024lyh}. The EW data span LHC diboson differential measurements, four-lepton production (which includes lepton-pairs produced via virtual photons), and $Z$-boson production in Vector Boson Fusion (VBF), as well as the the EW precision observables measured at LEP and SLD \cite{ALEPH:2005ab}. Fits are performed in a rotated basis with respect to the standard Warsaw basis in order to maximise sensitivity and remove ill-constrained directions from the fit.  

The CMS collaboration published a standalone EFT interpretation of the top-sector in Ref.~\cite{CMS:2023xyc} where 26 operators were fitted to all major associated top quark production modes ($t\bar{t}H$, $t\bar{t}W$, $t\bar{t}Z$, $tZq$, $tHq$, $t\bar{t}t\bar{t}$). Singling out top processes is motivated by the large mass of the top-quark and its Yukawa coupling around unity. Regarding the Higgs sector, CMS presented a fit to 7 operators in the STXS framework (up to stage 1.1) based on $6$ decay channels ($\gamma\gamma$, $ZZ$, $WW$, $\tau\tau$, $bb$, $\mu\mu$) in Ref.~\cite{CMS:2020gsy}. On the production side, measurements of $ggH$, VBF, $VH$ and $ttH$ were included, as well as the single-top induced processes $tH$, $tHW$ and $tHq$.    

Moving on to efforts performed within the the theory community, we include in Tables \ref{tab:eft_existing_efforts_1} and \ref{tab:eft_existing_efforts_2} an overview of the existing global SMEFT fitting efforts. For each collaboration, we indicate, from top to bottom, the measurement sectors entering the fit, whether linear or quadratic EFT corrections are included, the order in perturbative QCD, the number of Wilson coefficients, the operator basis and whether individual or marginal bounds are reported. In individual fits, coefficients are fitted one at a time with all other coefficients set to zero. Continuing, we list the EW input scheme, the flavour assumptions, whether RG effects are included, the statistical model, the fitting methodology, and finally, the corresponding publication references.

From Tables \ref{tab:eft_existing_efforts_1} and \ref{tab:eft_existing_efforts_2} we observe that the most extensive global SMEFT fit to date is performed by the {\sc\small SMEFiT} collaboration, as presented in Ref.~\mycite{Celada:2024mcf}. It includes the most updated dataset comprising 445 measurements from the combined Higgs, top, diboson, and electroweak sectors and is simultaneously sensitive to 50 Wilson coefficients up to quadratic order in the EFT expansion. It adopts SMEFT theory predictions accurate up to NLO in the QCD perturbative expansion for the LHC processes. Other global efforts include work done by the \verb|Fitmaker| collaboration that performed a combined interpretation of the same sectors, being simultaneously sensitive to 34 Wilson coefficients up to linear order in the EFT expansion \cite{Ellis:2020unq}. Its dataset was updated later to account for the anomalous $m_W$ measurement by CDF \cite{CDF:2022hxs} in the $\alpha_{\rm ew}$ input scheme \cite{Bagnaschi:2022whn}. The {\sc\small SFitter} collaboration performed a combined analysis of the Higgs, top and diboson sectors in Ref.~\cite{Brivio:2019ius, Biekotter:2018ohn, Butter:2016cvz, Corbett:2015ksa, Lopez-Val:2013yba}, while a dedicated analysis based on public experimental likelihoods in the top sector was performed in Ref.~\cite{Elmer:2023wtr}. RG effects to account for measurements defined at different scales have been included so far in Refs.~\cite{Castro:2016jjv, Grunwald:2023nli, Bartocci:2023nvp} and more recently in Ref.~\cite{Allwicher:2023shc}. The latter adopts a $U(2)^5$ flavour symmetry among the first two generations resulting in 124 Wilson coefficients constrained in individual fits including linear EFT corrections. Another interesting direction concerns the simultaneous determination of parton distribution functions (PDFs) and Wilson coefficients. Often, hadronic theory predictions in the SMEFT assume SM PDFs, while non-vanishing EFT effects might in fact be present in the PDFs themselves. This motivates their combined treatment, as performed by {\sc\small SIMUnet} developed and used in Refs.~\cite{Kassabov:2023hbm, Costantini:2024xae}. 

In addition to the fitting tools outlined above, an impressive collection of dedicated tools to perform SMEFT phenomenology has been developed: 
\begin{itemize}
 \item \textbf{Matching:} As mentioned in Sect.~\ref{sec:matching_intro}, the automatised matching of explicit BSM models onto the SMEFT is provided by the packages {\sc\small MatchMakerEFT} \cite{Carmona:2021xtq} and {\sc\small Matchete} \cite{Fuentes-Martin:2022jrf}. Automatised bounds can be obtained using {\sc\small SMEFiT} as developed in Ref.~\mycite{terHoeve:2023pvs}, which will be covered extensively in Chapter \ref{chap:smefit_uv}. 
 \item \textbf{Observables:} Regarding the computation and collection of observables, \verb|SMEFTsim| \cite{Brivio:2020onw} interfaced to \verb|Madgraph5_aMC@NLO| \cite{Aoude:2022aro} enables automated computations of observables in the SMEFT at LO and accommodates generic flavour assumptions. NLO accurate predictions can be obtained with \verb|SMEFTatNLO| \cite{Degrande:2020evl}, although it has embedded more restrictive flavour assumptions. The Mathematica package {\sc\small HighPT} provides automatised SMEFT corrections in the high $p_T$ tails of Drell-Yan (DY), while {\sc\small flavio} offers hundreds of implemented flavour observables. The latter feed into {\sc\small smelli} \cite{Aebischer:2018iyb} to construct the corresponding global likelihood in the SMEFT parameter space. Another tool in this direction is {\sc\small HEPfit} \cite{DeBlas:2019ehy, DeBlas:2019qco, deBlas:2019rxi} that incorporates around a thousand SM observables complemented by explicit SM extensions as well as several EFT scenarios. Optimal observables based on Machine Learning techniques that exploit the full multivariate nature of the final state are also actively being developed, such as {\sc\small Madminer} \cite{Brehmer:2019xox} and the package {\sc\small ML4EFT} that we developed in Ref.~\mycite{GomezAmbrosio:2022mpm}.
 \item \textbf{RG running:}  Connecting flavour and high $p_T$ observables requires automatised RG evolution, which has been implemented in public codes such as {\sc\small Wilson} \cite{Aebischer:2018bkb}, \verb|Madgraph5_aMC@NLO| \cite{Aoude:2022aro} and {\sc\small DsixTools} \cite{Celis:2017hod, Fuentes-Martin:2020zaz}. 
\end{itemize}

\begin{table}[t]
  \renewcommand{\arraystretch}{3}
  \scriptsize
  \centering
  \caption{ Overview of the existing SMEFT fitting efforts, continued in Table \ref{tab:eft_existing_efforts_2}. From top to bottom, we list the measurement sectors used as input to the fit, the order in the EFT expansion, the order in the perturbative QCD expansion, the number of EFT parameters ($n_{\rm eft}$), the fitting basis, whether the bounds are individual (Ind.) or marginal (Marg.), the EW input scheme, the flavour assumptions, whether RG effects are included, the statistical model and fitting methodology, and, finally, the corresponding publication references.}
  \label{tab:eft_existing_efforts_1}
  \begin{tabular}{ c | c | c | c | c  }  
    & {\sc\small SMEFiT} & \verb|Fitmaker| & {\sc SFitter} & HEPfit \\ [0.5ex]
   \hline
   \hline
    \textbf{Data} & \makecell{EW + Higgs + \\ diboson + top} & \makecell{EW + Higgs + \\ diboson + top}& \makecell{EW + Higgs + \\ diboson, top}&  \makecell{EW + Higgs, \\ flavour} \\ [0.5ex] 
   \hline
    \textbf{Order} EFT& Quadratic& Linear & Quadratic & Linear \\ [0.5ex] 
    \hline
    \textbf{Order} QCD& NLO & NLO & NLO & LO  \\ [0.5ex] 
    \hline
    $n_{\rm eft}$ & 50 & 34 & 22 & 28 \\
    \hline
    \textbf{Basis}& Warsaw & Warsaw & HISZ, Warsaw & Higgs, Warsaw  \\ [0.5ex] 
   \hline
   \textbf{Ind./Marg.} & Both & Both& Both& Both \\
   \hline
    \textbf{Input scheme}& $m_W$& $\alpha_{\rm ew}$& $\alpha_{\rm ew}$& $\alpha_{\rm ew}$ \\ [0.5ex] 
    \hline
   \textbf{Flavour}&\makecell{$U(2)_q \times U(2)_u \times$ \\ $U(3)_d \times $ \\ $\left[ U(1)_\ell \times U(1)_e \right]^3$}& \makecell{$U(2)_q \times U(2)_u \times $\\$\times U(3)_d \times U(3)_\ell$\\ $ \times U(3)_e$} & \makecell{$U(2)_q\times U(2)_u$ \\ $\times U(2)_d$}& \makecell{$U(2)_q \times U(2)_u \times U(2)_d$ \\ $\times \left[ U(1)_\ell \times U(1)_e \right]^3$}\\ [0.5ex] 
   \hline
   \textbf{RG effects}& No & No & No & No \\ [0.5ex] 
   \hline
   \textbf{Stat. model} & Gaussian & Gaussian & \makecell{Poisson, Flat,\\ Gaussian}& Gaussian  \\ [0.5ex]
   \hline
   \textbf{Methodology}& Bayesian & Bayesian & Frequentist&Bayesian \\ [0.5ex]
   \hline
   \textbf{References} & \mycite{Hartland:2019bjb, Ethier:2021ydt, Ethier:2021bye, Giani:2023gfq,terHoeve:2023pvs,Celada:2024mcf} & \cite{Ellis:2020unq, Bagnaschi:2022whn}& \cite{Brivio:2019ius, Biekotter:2018ohn, Butter:2016cvz, Corbett:2015ksa, Lopez-Val:2013yba, Elmer:2023wtr}&\cite{DeBlas:2019ehy, DeBlas:2019qco, deBlas:2019rxi} \\
   \end{tabular}
\end{table}


  \begin{table}[t]
    \renewcommand{\arraystretch}{3}
    \scriptsize
    \centering
    \caption{Same as Table \ref{tab:eft_existing_efforts_1}, continued.}
    \label{tab:eft_existing_efforts_2}
    \begin{tabular}{ c | c | c | c | c }  
     & TopFitter & EFTfitter & Mainz group & Zurich group \\ [0.5ex]
     \hline
     \hline
      \textbf{Data} & top & \makecell{top + DY + \\ flavour} & \makecell{EW + Higgs + \\ top + flavour + \\ $jj\gamma$ + parity violation + \\ lepton scattering }& \makecell{EW + flavour + \\ DY + Jet}\\ [0.5ex] 
     \hline
      \textbf{Order} & Linear & Quadratic & Linear& Linear \\ [0.5ex] 
      \hline
      \textbf{Order} QCD& NLO & LO & NLO & NLO \\ [0.5ex] 
      \hline
      $n_{\rm eft}$ &  14& 14& 41& 124\\
      \hline
      \textbf{Basis}& Warsaw & Warsaw& Warsaw & Warsaw  \\ [0.5ex] 
     \hline
     \textbf{Ind./Marg.} & Both& Both& Both& Ind.\\
     \hline
      \textbf{Input scheme}& $\times$ & $\times$ & $\alpha_{\rm ew}$& $\alpha_{\rm ew}$  \\ [0.5ex] 
      \hline
     \textbf{Flavour}& \makecell{$U(2)_q\times U(2)_u$\\ $\times U(2)_d$}& MFV & $U(3)^5$& $U(2)^5$\\ [0.5ex] 
     \hline
     \textbf{RG effects}& No & Yes & Yes & Yes \\ [0.5ex] 
     \hline
     \textbf{Stat. model} & Gaussian & Gaussian& Gaussian & Gaussian \\ [0.5ex]
     \hline
     \textbf{Methodology}& Frequentist &Bayesian & Frequentist & Frequentist  \\ [0.5ex]
     \hline
     \textbf{References} & \cite{Brown:2018gzb, Buckley:2015lku, Buckley:2015nca}& \cite{Castro:2016jjv, Grunwald:2023nli}& \cite{Bartocci:2023nvp} & \cite{Allwicher:2023shc} \\
     \end{tabular}
  \end{table}


   \chapter{Methodologies}
\label{chap:methdologies}
\vspace{-1cm}
  
In this chapter, we introduce the two main methodological frameworks that were used to obtain the results
presented in Refs.~\mycite{GomezAmbrosio:2022mpm, terHoeve:2023pvs,Celada:2024mcf}. After briefly describing the full experimental likelihood in Sect.~\ref{sec:exp_likelihood}, we introduce {\sc\small SMEFiT} in Sect.~\ref{sec:smefit_framework}, which is a flexible open-source software package to fit a large class of experimental datasets within the context of the SMEFT as previously introduced in Chapter \ref{chap:smeft}. In Sect.~\ref{sec:ml4eft_framework}, we introduce the statistical preliminaries leading to the {\sc\small ML4EFT} framework that we developed in Ref.~\mycite{GomezAmbrosio:2022mpm} to construct and integrate unbinnend multivariate observables into global SMEFT fits. This will be the subject of Chapter \ref{chap:ml4eft}.

\section{The full experimental likelihood}
\label{sec:exp_likelihood}
The key concept at the heart of any statistical analysis is the likelihood function. It encodes the probability of observed data given a particular hypothesis specified by a set of parameters entering a statistical model. These are usually split up in a parameter of interest (POI) - denoted $\mu$ - that one would like to determine, and nuisance parameters $\boldsymbol{\theta}$ that are not of interest directly, but are needed to fully characterise the model. Nuisance parameters generally enter to model systematic uncertainties, such as those related to efficiencies, energy scales, the luminosity and missing higher order corrections. 

A typical LHC measurement consists of a list of event counts in a set of bins, which are usually grouped into $N_c$ separate channels $k$ (or categories) to improve signal to background ratios. The number of events $n_{i, k}$ in each bin $i$ inside category $k$ follows a Poisson distribution with an expected yield $\nu_{i, k}$ that can be decomposed into a nominal signal rate $S_{i, k}$ and background $B_{i, k}$ contribution,
\begin{equation}
  \nu_{i, k}(\mu, \boldsymbol{\theta}) = \mu \epsilon_{i, k}(\boldsymbol{\theta})S_{i, k}(\boldsymbol{\theta}) + B_{i, k}(\boldsymbol{\theta}) \, ,
\end{equation}
with efficiency $\epsilon_{i, k}$ and a parameter of interest $\mu$ describing the overall rate. The expected event yield carries an explicit dependence on $\boldsymbol{\theta}$ as the effect of the nuisances is to distort the efficiencies, as well as the signal and background rates.  

Each event yield $n_i$ receives contributions from individual events $j$ characterised by a set of kinematics $\boldsymbol{x}_{ij}$, such as invariant masses or angles, which follow a distribution $p_i$. The nuisance parameters $\boldsymbol{\theta}$ may be determined through auxiliary measurements $\boldsymbol{y}$ obtained in a separate control region. As the control region provides independent measurements, the overall probability $p$ of the collection $\{\boldsymbol{n}, \boldsymbol{x}, \boldsymbol{y}\}$ may then be modelled as a product of probabilities, 
\begin{equation}
  p(\boldsymbol{n}, \boldsymbol{x}, \boldsymbol{y }|\mu, \boldsymbol{\theta}) =
  \prod_{k=1}^{N_c}\prod_{i=1}^{n_{\rm bins}}\left[{\rm Pois}\left(n_{i, k}\left.\right\rvert \nu_{i, k}(\mu, \boldsymbol{\theta})\right) \prod_{j=1}^{n_{i,k}}p_i(\boldsymbol{x}_{ij}|\mu, \boldsymbol{\theta})\right]\prod_{j=1}^{n_{\rm syst}}\mathcal{C}(\boldsymbol{y}|\theta_j) \, ,
  \label{eq:exp_likelihood}
\end{equation}
which returns the full experimental likelihood when evaluated on the observed data \cite{Cranmer:2021urp}. 

Although Eq.~\eqref{eq:exp_likelihood} provides only a schematic representation, it contains many of the building blocks that enter an actual experimental analysis. In practice, experimental likelihoods contain $\mathcal{O}\lp 1000 \rp$ nuisance parameters, which makes them complex objects to massage and, more importantly, communicate outside the experimental collaborations. So far, only a limited number of publicly available likelihoods exist, which, for example, have been interpreted in a SMEFT context in Ref.~\cite{Elmer:2023wtr}. It is more common instead to have outside access either to the experimental covariance matrix, or separate statistical and systematic uncertainties, in which case the theory community cannot do much but assume the data is Gaussianly distributed, as will be discussed in the next section. This provides a good approximation as was recently demonstrated explicitly in case of STXS measurements in Ref.~\cite{ATLAS:2024lyh}.



\section{The SMEFiT framework}
\label{sec:smefit_framework}
In order to interpret the experimental likelihood, Eq.~\eqref{eq:exp_likelihood}, in the language of the SMEFT, a robust framework is needed that is able to analyse whether current data is compatible with the SM or prefers another point in the SMEFT parameter space instead. The {\sc\small SMEFiT} framework was developed with this particular goal in mind and was used in Refs.~\cite{Hartland:2019bjb, Ethier:2021bye,Ethier:2021ydt, Giani:2023gfq} and my work in Refs.~\mycite{terHoeve:2023pvs,Celada:2024mcf}. 

In this section, we present {\sc\small SMEFiT}, focussing on its treatment of experimental data, theory predictions and methodological tools. We also describe the procedure that {\sc\small SMEFiT} adopts to extrapolate existing LHC Run II measurements to the HL-LHC conditions. We end by providing instructions on how to install the {\sc\small SMEFiT} code.

\subsection{Treatment of experimental data in SMEFiT}
\label{subsec:smefit_exp_data}
Rather than providing the full experimental likelihood, Eq.~\eqref{eq:exp_likelihood}, experimentalists commonly provide measurements in the following format:
\begin{itemize}
  \item The central value $\sigma_i$ of the measurement in kinematic bin $i$, or $\sigma$ in case of an inclusive cross-section; 
  \item Uncorrelated additive uncertainties $s_{i, \theta}^{\rm (uncorr)}$ pertaining to source $\theta$, such as statistical uncertainties. Multiple uncorrelated uncertainties are added in quadrature to obtain $s_{i}^{\rm (uncorr)}$ ;
  \item Correlated additive uncertainties $s_{i, \theta}^{\rm (corr)}$ from source $\theta$;
  \item Correlated multiplicative uncertainties $\delta_{i, \theta}^{\rm (corr)}$ originating from source $\theta$ such as the uncertainty on the total integrated luminosity. The corresponding absolute uncertainties can be obtained upon multiplying with the central value $\sigma_i$.
\end{itemize}
These various sources of uncertainty form the ingredients that enter the experimental covariance matrix, which should be constructed consistently depending on the type of uncertainty.

\myparagraph{Experimental covariance matrix}
In case the measurement comes with separate sources of systematics, we construct the experimental covariance between bins $i$ and $j$ as the sum of uncorrelated uncertainties and correlated ones,
\begin{equation}
  {\rm cov}_{ij}^{\rm (exp)} = s_{i}^{\rm (uncorr)} s_{j}^{\rm (uncorr)} \delta_{ij} + \sum_{\theta} s_{i, \theta}^{\rm (corr)} s_{j, \theta}^{\rm (corr)} \, ,
  \label{eq:fit_covmat}
\end{equation}
where the sum runs over the correlated systematic sources. In case the breakdown into separate sources of systematics is not reported, the total systematic error $s_i^{\rm (tot)}$ is reported instead. This is often accompanied by a correlation matrix pertaining to the total error such that the covariance is constructed as
\begin{equation}
  {\rm cov}_{ij}^{\rm (exp)} = {\rm cor}_{ij}^{\rm (exp)} s_i^{\rm (tot)} s_j^{\rm (tot)}.
  \label{eq:cov_from_cor}
\end{equation}
Whenever a separate correlation (or covariance) matrix is provided, we decompose it into so-called artificial systematics $s_i^{\rm (art)}$ such that when plugged into Eq.~\eqref{eq:fit_covmat} we retrieve the full covariance from Eq.~\eqref{eq:cov_from_cor} \cite{NNPDF:2014otw}. This is done by performing the eigenvector decomposition, 
\begin{align}
  {\rm cov}_{ij}^{\rm (exp)} &= \sum_{k,l} u_{ik} \lp \lambda_{k}\delta_{kl}\rp  u_{jl} = \sum_l u_{il} \sqrt{\lambda_l} \sqrt{\lambda_l} u_{jl} \equiv \mathbf{s}^{\rm (art)} \cdot  \mathbf{s}^{\rm (art)} \, ,
  \label{eq:covmat_decomp}
\end{align}
where $u_{ik}$ is a square matrix whose $k^{\rm th}$ column contains the $k^{\rm th}$ eigenvector with eigenvalue $\lambda_k$. In Eq.(\ref{eq:covmat_decomp}), we have furthermore defined the artificial systematic $s_i^{\rm (art)} = u_{il} \sqrt{\lambda_l}$ of measurement $i$, with no sum on $l$.  

\myparagraph{Asymmetric uncertainties}
Another common aspect that enters the construction of the experimental covariance are asymmetric uncertainties $[s_{i, \theta}^+, s_{i, \theta}^-]$, with $s_{i, \theta}^+$ and $s_{i, \theta}^-$ positively defined. Since we work in the Gaussian approximation, any asymmetric uncertainty must be symmetrised prior to serving as input to the fit. Here we adopt the approach as proposed in Ref.~\cite{DAgostini:2004kis}, which consists of i) shifting the central value $\sigma_i$ by the semi-difference between $s_{i, \theta}^+$ and $s_{i, \theta}^-$ for each source,
\begin{equation}
  \sigma_i \rightarrow \sigma_i + \frac{1}{2}\sum_\theta  \lp s_{i, \theta}^+ - s_{i, \theta}^-\rp \, ,
\end{equation}
and ii) defining a symmetrised uncertainty
\begin{equation}
  s_{i,{\rm sym}}^{\rm (\theta)} \equiv \sqrt{\lp \frac{s_{i, \theta}^+ + s_{i, \theta}^-}{2} \rp^2 + 2 \lp \frac{s_{i, \theta}^+ - s_{i, \theta}^-}{2}\rp^2}.
\end{equation}
\myparagraph{d'Agostini bias}
In the presence of multiplicative uncertainties, we apply the so-called $t_0$-prescription in order to avoid the d'Agostini bias that would show up otherwise whenever the covariance matrix as published by the experimentalists is used \cite{Ball:2009qv}. This consists of converting the multiplicative uncertainties into additive ones by multiplying them against the SM prediction, i.e. the $t_0$ prediction.


\subsection{Theory predictions in SMEFiT}
\label{subsec:smefit_theory}
We now move on to describing how theory predictions are obtained in {\sc\small SMEFiT}. For this, we first recall from Eq.~\eqref{eq:smeft_xsec} the form of the  the modified cross-section predictions in the SMEFT for a kinematic bin $m$, 
\begin{equation}  
  \sigma_m(\boldsymbol{c}) =
    \sigma^{(\rm sm)}_m + \sum_{i=1}^{n_{\rm{eft}}}
    \sigma_{m,i}^{(\rm eft)}c_i
    +\sum_{i=1, j \ge i}^{n_{\rm{eft}}}
    \sigma_{m, ij}^{(\rm eft)}c_i c_j .
    \label{eq:smeft_xsec_2}
\end{equation} 
In {\sc\small SMEFiT}, we adopt SM predictions  $\sigma^{(\rm sm)}_m$ accurate up to next-to-next-to-leading order (NNLO) in the QCD perturbative expansion plus NLO electroweak corrections. These are obtained either with the Monte Carlo (MC) generator {\sc\small mg5\_aMC@NLO} \cite{Alwall:2014hca} complemented by SM NNLO K-factors, or with \verb|MATRIX| \cite{Grazzini:2017mhc} directly at NNLO. We refer to Ref.~\cite{Ethier:2021bye} for a complete overview. 

Regarding the linear and quadratic SMEFT corrections, $\sigma_{m,i}^{(\rm eft)}$ and $\sigma_{m, ij}^{(\rm eft)}$, we use \verb|MG5_aMC@NLO| interfaced to \verb|SMEFT@NLO| \cite{Degrande:2020evl} to obtain NLO QCD accurate predictions. Whenever a specific process cannot be computed directly at NLO in the SMEFT, as for instance in case of $t\bar{t}Z$ production, we use SM NLO K-factors to rescale the leading order (LO) predictions, see Table 3.9 in Ref.~\cite{Ethier:2021bye}. We furthermore adopt the $\{\hat{m}_W, \hat{m}_Z, \hat{G}_F\}$ input-scheme and use the NNPDF4.0 NNLO no-top \cite{NNPDF:2021njg} PDF set in order to avoid potential overlap between PDF and EFT effects in the top-sector \cite{Costantini:2024xae}. The renormalisation and factorisation scales are set to representative scales depending on the process considered, e.g. to the $W$-mass in case of single top production in association with a $W$ boson. The SMEFT corrections to the EWPOs from LEP and SLD \cite{ALEPH:2005ab} include LO electroweak corrections and are obtained analytically, except for diboson production, which is computed numerically with {\sc\small mg5\_aMC@NLO}. 

\myparagraph{Numerical precision}
Extra care is taken to ensure that the SMEFT corrections  $\sigma_{m,i}^{(\rm eft)}$ and $\sigma_{m, ij}^{(\rm eft)}$ are numerically precise and do not suffer from MC fluctuations, which is relevant especially towards tails of distributions. 

In case the SMEFT correction makes up more than $5\%$ of the corresponding SM prediction, we demand the relative MC uncertainty to be below $1\%$. This is realised by running at optimised values of the EFT parameters so that the relevant SMEFT contribution is artificially enhanced with respect to the total SMEFT prediction. Indeed, dropping the label $m$ for simplicity, the linear correction associated to EFT parameter $c_i$ is optimised by maximising the ratio
\begin{equation}
  \frac{c_i \sigma^{\rm (eft)}_i}{\sigma^{\rm (sm)} + c_i\sigma_i^{\rm(eft)}+ c_i^2 \sigma^{\rm (eft)}_{ii}} \rightarrow c_i^{\rm (lin, opt)} = \sqrt{\frac{\sigma^{\rm (sm)}}{\sigma^{\rm (eft)}_{ii}}} \qquad\text{(no sum on $i$)}\, ,
  \label{eq:optimised_lin} 
\end{equation}
while the quadratic interference correction $\sigma_{ij}^{\rm (eft)}$ between EFT parameters $c_i$ and $c_j$ is optimised by maximising 
\begin{equation}
  \frac{c_i c_j \sigma^{\rm (eft)}_{ij}}{c_i^2 \sigma^{\rm (eft)}_{ii} + c_j^2 \sigma^{\rm (eft)}_{jj} + c_i c_j \sigma^{\rm (eft)}_{ij}} \rightarrow c_i^{\rm (quad, opt)} = c_j \sqrt{\frac{\sigma^{\rm (eft)}_{jj}}{\sigma^{\rm (eft)}_{ii}}}  \qquad\text{(no sum on $i, j$)} .
  \label{eq:optimised_quad_int}
\end{equation}
In Eqs.~\eqref{eq:optimised_lin} and \eqref{eq:optimised_quad_int}, $\sigma^{\rm (eft)}_{ii}$ may be taken from a previous (non-optimised) run. Note also how Eq.~(\ref{eq:optimised_quad_int}) still depends on $c_j$, which may be fixed to $c_j=1000$ to let the quadratic corrections dominate over the linear ones. Finally, it is straightforward to see that the diagonal quadratic corrections $\sigma^{\rm (eft)}_{ii}$ can be made arbitrarily large by setting $c_i$ sufficiently large.

\myparagraph{Theory covariance matrix} Regarding theoretical uncertainties on the SM predictions, we include missing higher order uncertainties (MHOU) through scale variations, as well as PDF uncertainties \cite{NNPDF:2021njg} and statistical uncertainties. All sources are added in quadrature to construct the theory covariance matrix ${\rm cov}_{ij}^{\rm (th)}$ between bins $i$ and $j$, which combined with the experimental covariance from Eq.~(\ref{eq:fit_covmat}) gives the total covariance matrix
\begin{equation}
  {\rm cov}_{ij} = {\rm cov}_{ij}^{\rm (exp)} + {\rm cov}_{ij}^{\rm (th)} \, .
  \label{eq:full_covmat}
\end{equation}
Eq.~\eqref{eq:full_covmat} encodes the total covariance between measurement $i$ and $j$, accounting for both theory and experimental sources of uncertainties.

\subsection{Optimisation}
\label{subsec:optimisation}
Given the experimental measurements $\sigma_i^{\rm (exp)}$ with theory predictions $\sigma_i^{\rm (th)}(\boldsymbol{c})$, the figure of merit that is minimised to find the best-fit values and uncertainties on the EFT parameters $\mathbf{c}$ is defined as
\begin{equation}
  \chi^2(\mathbf{c}) = \frac{1}{n_{\rm dat}} \sum_{i,j=1}^{n_{\rm dat}} \lp\sigma_i^{\rm (exp)} - \sigma_i^{\rm (th)}(\mathbf{c})\rp {\rm cov}_{ij}^{-1}\lp\sigma_j^{\rm (exp)} - \sigma_j^{\rm (th)}(\mathbf{c})\rp \, , 
  \label{eq:smefit_chi2}
\end{equation}
where the covariance matrix is the one defined in Eq.~\eqref{eq:full_covmat}, accounting for both experimental and theory uncertainties. 

Two complementary methods exists to minimise Eq.~(\ref{eq:smefit_chi2}) depending on the order in the EFT expansion. 
In case of only linear corrections, the best-fit point can be evaluated analytically, thus bypassing the need for numerical optimisation making it the preferred fitting mode. 
If quadratics corrections are present, no such analytic solution exists, and one needs to resort to numerical methods. 

\myparagraph{Linear optimisation} In the absence of quadratic EFT corrections, Eq.~\eqref{eq:smeft_xsec} simplifies to
\begin{align}
  \sigma_m^{\rm (th)}(\boldsymbol{c}) = \sigma_m^{\rm (sm)} + \sum_{i=1}^{n_{\rm dat}}c_i\sigma_{m,i}^{\rm (eft)} \, ,\qquad i = 1, \dots, n_{\rm eft} \, ,
  \label{eq:linear_eft}
\end{align}
 which upon substitution into Eq.~\eqref{eq:smefit_chi2} gives,
\begin{equation}
  \chi^2(\mathbf{c}) = \frac{1}{n_{\rm dat}}\lp \sigma_i^{\rm (exp)} - \sigma_i^{\rm (sm)} - c_k\sigma_{i,k}^{\rm (eft)}\rp {\rm cov}^{-1}_{ij} \lp \sigma_j^{\rm (exp)} - \sigma_j^{\rm (sm)} - c_l\sigma_{j,l}^{\rm (eft)}\rp \, ,
  \label{eq:chi2_lin}
\end{equation}
where repeated indices are summed over. Minimising Eq.~\eqref{eq:chi2_lin} with respect to $c_i$ results into the best fit point $\hat{c}_i$
\begin{equation}
  \hat{c}_i = \lp \sigma^{\rm (eft)}_{j, i} {\rm cov}^{-1}_{jk}\sigma^{\rm (eft)}_{k, l} \rp^{-1} \lp \sigma^{\rm (eft)}_{m, l}{\rm cov}^{-1}_{m,n}\lp \sigma^{\rm(exp)}_n - \sigma^{\rm (sm)}_{n}\rp \rp .
  \label{eq:best_fit_linear}
\end{equation}
Identifying the first term in parentheses in Eq.~(\ref{eq:best_fit_linear}) as the Fisher information matrix,
\begin{equation}
  I_{ij} = \sigma^{\rm (eft)}_{j, i} {\rm cov}^{-1}_{jk}\sigma^{\rm (eft)}_{k, l} \, , 
  \label{eq:fisher_lin_analytic}
\end{equation} 
we rewrite Eq.~(\ref{eq:chi2_lin}) using Eq.~\eqref{eq:best_fit_linear} as a quadratic polynomial centred around the best fit point with minimum $\chi^2_{\rm min}$,
\begin{equation}
  \chi^2(\mathbf{c}) = \chi^2_{\rm min} + \frac{1}{n_{\rm dat}}\lp c - \hat{c}\rp_i I_{ij} \lp c - \hat{c}\rp_j.
  \label{eq:chi2_lin_shifted}
\end{equation}
Eq.~(\ref{eq:chi2_lin_shifted}) demonstrates that the covariance of the best fit point is given by the inverse of the Fisher information matrix. This means that the EFT parameters are distributed according to a multivariate Gaussian distribution $\mathcal{N}(\hat{c}, I^{-1})$ centred around $\hat{c}$ with covariance $I^{-1}$.

\myparagraph{Quadratic optimisation} In the presence of quadratic EFT corrections, the figure of merit $\chi^2$ in Eq.~(\ref{eq:smefit_chi2}) becomes a quartic polynomial in the EFT parameters such that no analytic solution exists. In this case, one has to resort to numerical methods, such as Nested Sampling, which we describe next. We follow largely the discussion in Refs.~\cite{10.1214/06-BA127, Feroz:2013hea, 10.1111/j.1365-2966.2009.14548.x}.

Given a set of hypotheses parametrised by $\mathbf{c}$ and some observed dataset denoted $D$, Bayes's theorem lets one update one's prior believe $\pi(\mathbf{c})$ by the likelihood $\mathcal{L}(\mathbf{c}) \equiv p(D|\mathbf{c})$ of observing $D$ given $\mathbf{c}$, 
\begin{equation}
  p(\mathbf{c}) = \frac{\mathcal{L}(\mathbf{c})\pi(\mathbf{c})}{\mathcal{Z}} \, ,
  \label{eq:bayes_rule}
\end{equation} 
and normalises it by the Bayesian evidence $\mathcal{Z}$ defined as, 
\begin{equation}
  \mathcal{Z} = \int \mathcal{L}(\mathbf{c})\pi(\mathbf{c}) d\mathbf{c} \, ,
  \label{eq:bayesian_evidence}
\end{equation}
to end up with the posterior distribution $p(\mathbf{c})$. However, the Bayesian evidence is hard to evaluate in practice due to the high dimensionality of $\mathbf{c}$. Nested Sampling \cite{10.1111/j.1365-2966.2009.14548.x} evaluates $\mathcal{Z}$ by mapping the high dimensional integral in Eq.~(\ref{eq:bayesian_evidence}) to a single dimensional one that is easier to solve. As a by-product of this method one obtains samples from the posterior distribution $p(\mathbf{c})$, which can then be used to evaluate averages, correlations and other derived quantities of the EFT parameters.

\begin{figure}
  \centering
  \includegraphics[width=.7\linewidth]{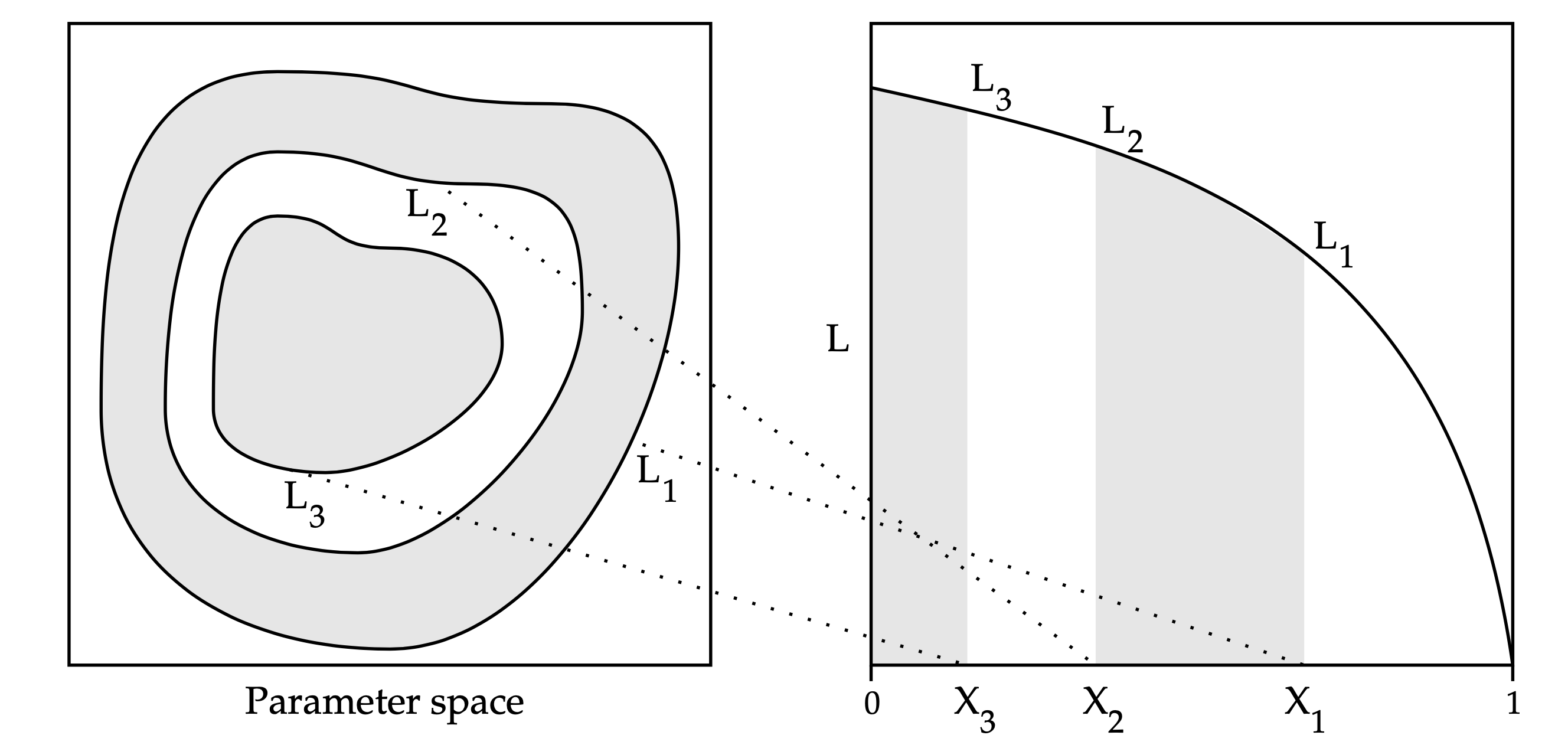}
  \caption{Likelihood contours at $L_i$ in parameter space (left) are sorted by their enclosed prior mass $X_i$ (right). Figure taken from \cite{10.1214/06-BA127}.}
  \label{fig:NS_shells}
\end{figure}

The starting point of the Nested Sampling algorithm is to reparametrise Eq.~(\ref{eq:bayesian_evidence}) in terms of (nested) likelihood shells sorted by their enclosed prior mass. For this, we define the survival function $X(\lambda)$ as the amount of prior mass with likelihood greater than some threshold value $\lambda$,
\begin{equation}
  X(\lambda) = \int_{\{\mathbf{c}:\mathcal{L}(\mathbf{c})>\lambda\}} \pi(\mathbf{c})d\mathbf{c}.
  \label{eq:survival_fct}
\end{equation}
Note how the enclosed prior mass $X$ decreases as $\lambda$ increases. Specifically, we have $X(0) = 1$ and $X(L_{\rm max}) = 0$ at maximum likelihood $L_{\rm max}$. Next, using that expectation values of positive random variables can be evaluated by integrating their corresponding survival function, we rewrite $\mathcal{Z}$ as
\begin{equation}
  \mathcal{Z} = \int_0^\infty X(\lambda)d\lambda.
  \label{eq:averaged_survival_fct}
\end{equation}
Furthermore, as Eq.~(\ref{eq:survival_fct}) is monotonically decreasing, we may rewrite Eq.~(\ref{eq:averaged_survival_fct}) in terms of its inverse $L(X)$, 
\begin{equation}
  \mathcal{Z} = \int_0^1 L(X)dX,
  \label{eq:evidence_1d}
\end{equation}
which demonstrates the earlier statement that the evidence $\mathcal{Z}$ is evaluated in terms of sorted likelihood-shells, as also illustrated by Fig.~\ref{fig:NS_shells}.

Now, if $L(X)$ were known exactly, we could simply evaluate Eq.~(\ref{eq:evidence_1d}) with standard numerical integration methods as
\begin{equation}
  \mathcal{Z} \approx \sum_{i=1}^N w_i L_i \, ,
\end{equation}
where the weight $w_i$ belonging to point $i$ may be set according to the trapezium rule $w_i = \lp X_{i-1} - X_{i+1}\rp/2$. In practice, we adopt MC methods to estimate $L(X_i)$. This follows an iterative procedure that starts with drawing  $N_{\rm live}$ points from the full prior $\pi(\mathbf{c})$ at iteration $i=0$ with $X_0=1$. Then we sort the live-points by their likelihood and remove the point associated with the lowest likelihood and record it as $L_i$. The discarded live-point is replaced by a new sample drawn from the prior subject to the constraint that its likelihood is larger than $L_i$. Finally, we shrink the prior volume down from $X_i$ to $X_{i-1}$ with $X_i = \exp(-i/N_{\rm live})$. 

By repeating this iterative procedure, the algorithm gradually traverses the entire prior volume until the evidence has been determined up to some specified precision. As a by-product we obtain posterior samples by associating to each discarded live point an importance weight $p_i = L_i w_i / \mathcal{Z}$. The samples can subsequently be unweighted by accepting $\boldsymbol{c}_i$ with probability $p_i/K$ where $K \geq {\rm max}(p_i)$. 

In this thesis, we adopt a flat prior on the EFT parameters. Its range should be chosen carefully so as not to exclude any physical solutions, or unnecessarily slow down the the convergence of the fit. We follow a semi-automatic approach based on the outcome of fits to one EFT parameter at a time (individual fits) that are comparably quick. As a rule of thumb, we take the prior in the full EFT parameter space four times as wide as the individual bounds at $95\%$ CI. 

\subsection{Analysis tools}
\label{subsec:smefit_analyse}
The output of either the linear or quadratic fit as described in Sect.~\ref{subsec:optimisation} is a set of posterior samples drawn from the posterior distribution $p(\mathbf{c})$ in Eq.~(\ref{eq:bayes_rule}). In order to further interpret these, we present here a couple of statistical tools that help in this regard and will also be used in the rest of this thesis.

A metric that quantifies the amount of information that datasets bring into the EFT parameter space is given by the Fisher information matrix, see Ref.~\cite{Brehmer:2016nyr} for a review. It is related to the second derivative of the log-likelihood as a measure of its curvature, and defined as
\begin{equation}
  I_{ij} \equiv -E\left[\frac{\partial^2\log \mathcal{L}(\mathbf{c})}{\partial c_i \partial c_j}|\mathbf{c}\right] \, ,
  \label{eq:fisher_def}
\end{equation}
 where the expectation value is evaluated with respect to the distribution that describes the data. Large (small) eigenvalues of $I_{ij}$ signal a relatively strongly (weakly) curved likelihood surface at $\mathbf{c}$ along the associated eigenvector, and thus quantify the degree of sensitivity to the EFT parameters. Null eigenvectors correspond to flat directions that cannot be constrained.

In the SMEFT, the log-likelihood at linear order in the EFT expansion follows directly from substituting $\mathcal{L}(\mathbf{c}) = \exp[-\chi^2/2]$ into Eq.~(\ref{eq:fisher_def}), with $\chi^2$ defined in Eq.~(\ref{eq:chi2_lin}), 
\begin{equation}
  I_{ij} = \sum_{k,l}^{n_{\rm dat}} \sigma_{k,i}^{\rm (eft)}{\rm cov}^{-1}_{kl}\sigma_{l,j}^{\rm (eft)} \, . 
  \label{eq:fisher_def_2}
\end{equation}
Note how Eq.~\eqref{eq:fisher_def_2} reproduces Eq.~\eqref{eq:fisher_lin_analytic} and is independent of the values of the EFT parameters. We further note that the covariance in Eq.~\eqref{eq:fisher_smeft_lin} may only contain experimental sources of uncertainty as theoretical uncertainties do not affect the distribution of the data. The diagonal entries of Eq.~\eqref{eq:fisher_def_2} simplify to
\begin{equation}
  I_{ii} = \sum_{k=1}^{n_{\rm dat}} \frac{\lp \sigma_{k, i}^{\rm (eft)}\rp^2}{\delta_{k, \rm exp}^2} \, ,
  \label{eq:fisher_smeft_lin}
\end{equation}
where $\delta_{k, \rm exp}^2$ represents the total experimental uncertainty. We find that the Fisher information of EFT parameter $c_i$ grows with the size of the corresponding linear SMEFT correction as well as with increased experimental precision. 

Rather than performing a fit with respect to the standard basis $\{c_i\}$, which might suffer from flat or poorly constrained directions, one may fit in the eigenbasis of the Fisher information $I_{ij}$, Eq.~\eqref{eq:fisher_def_2}. This (locally) decorrelates the EFT parameters such that flat directions no longer propagate to the other degrees of freedom. Based on this eigendecomposition, one may additionally want to fit in a basis of reduced dimensionality by keeping only directions with eigenvalues above a certain threshold, which is also known as principle component analysis (PCA).


\subsection{Projections for the HL-LHC}
\label{sec:hl_lhc_projections}


Here we describe the procedure that {\sc\small SMEFiT} adopts in order to extrapolate existing LHC Run II measurements to the HL-LHC conditions, which follows largely the discussion presented in our work in Ref.~\mycite{Celada:2024mcf}.
Assuming the SM as underlying theory, these projections will be used to quantify  the impact of Higgs, top quark, and diboson production measurements at the HL-LHC on the SMEFT parameter space in Chapter \ref{chap:smefit_fcc}.
We note that dedicated projections for HL-LHC pseudo-data are available, e.g.~\cite{deBlas:2019rxi,deBlas:2022ofj,DeBlas:2019qco,Durieux:2022cvf} and references therein.
Here we adopt instead a different strategy to ensure consistency with the {\sc\small SMEFiT} analysis settings, based on generating pseudo-data for future LHC runs by means of extrapolating from available Run II datasets.

To this end, we use a new {\sc\small SMEFiT} module generating projections of pseudo-data for future experiments.
This projection is based on extrapolating existing datasets, acquired at the same center-of-mass energy but lower luminosities, for the same underlying process.
The same strategy was adopted  in~\cite{Ethier:2021ydt,Greljo:2021kvv} for the SMEFT impact projections of vector-boson scattering and high-mass Drell-Yan data at the HL-LHC,  as well as in the PDF projections at the  HL-LHC \cite{AbdulKhalek:2018rok} 
and at the Forward Physics Facility of~\cite{Cruz-Martinez:2023sdv,Feng:2022inv}.

This module starts by considering a given available measurement from the LHC Run  II, composed of $n_{\rm bin}$ data points, and with the corresponding SM predictions given by $\mathcal{O}_i^{{\rm (th)}}$.
 The central values for the pseudo-data, denoted by $\mathcal{O}_i^{{\rm (exp)}} $, are obtained
 by fluctuating these theory predictions
 by the fractional statistical ($\delta_i^{\rm (stat)}$) and systematic ($\delta_{k,i}^{\rm (sys)}$)
 uncertainties,
 \begin{equation}
  \label{eq:pseudo_data_v2}
  \mathcal{O}_i^{{\rm (exp)}}
  = \mathcal{O}_i^{{\rm (th)}}
    \left( 1+ r_i \delta_i^{\rm (stat)}
    + \sum_{k=1}^{n_{\rm sys}}
    r_{k,i} \delta_{k,i}^{\rm (sys)}
    \right) \,
    , \qquad i=1,\ldots,n_{\rm bin} \, ,
 \end{equation}
where $r_i$ and $r_{k,i}$ are univariate random Gaussian numbers, whose distribution is such as to reproduce the experimental covariance matrix of the data, and the index $k$ runs over the individual sources of correlated systematic errors. 
We note that theory uncertainties are not included in the pseudo-data generation, and enter only the calculation of the $\chi^2$.

Since one is extrapolating from an existing measurement, whose associated statistical and systematic errors are denoted by $\tilde{\delta}_i^{\rm (stat)}$ and 
$\tilde{\delta}_{k,i}^{\rm (sys)}$, one needs to account for the increased statistics and the expected reduction of the systematic uncertainties for the HL-LHC data-taking period.
The former follows from the increase in integrated luminosity,
\begin{equation}
\delta_i^{\rm (stat)} = \tilde{\delta}_i^{\rm (stat)} \sqrt{\frac{\mathcal{L}_{\rm Run2}}{\mathcal{L}_{\rm HLLHC}}} \,, \qquad i=1,\ldots, n_{\rm bin} \, ,
\end{equation}
while the reduction of systematic errors is estimated by means of an overall rescaling factor
\begin{equation}
\delta_{k,i}^{\rm (sys)} = \tilde{\delta}_{k,i}^{\rm (sys)}\times f_{\rm red}^{(k)} \,, \qquad i=1,\ldots, n_{\rm bin} \, ,\quad k=1,\ldots, n_{\rm sys} \, . 
\end{equation}
with $f_{\rm red}^{(k)}$ indicating a correction estimating improvements in the experimental performance, in many cases possible thanks to the larger available event sample.
Here for simplicity we adopt the optimistic scenario considered in the HL-LHC projection studies~\cite{Cepeda:2019klc}, namely $f_{\rm red}^{(k)}=1/2$ for all the datasets.
For datasets without the breakdown of statistical and systematic errors, 
Eq.~(\ref{eq:pseudo_data_v2}) is replaced by 
\begin{equation}
  \label{eq:pseudo_data_v3}
  \mathcal{O}_i^{{\rm (exp)}}
  = \mathcal{O}_i^{{\rm (th)}}
    \left( 1+ r_i \delta_i^{\rm (tot)}
    \right) \,
    , \qquad i=1,\ldots,n_{\rm bin} \, ,
 \end{equation}
 with the total error being reduced
 by a factor $\delta_i^{\rm (tot)}=f_{\rm red}^{{\rm tot}} \times \tilde{\delta}_i^{\rm (tot)}$ with $f_{\rm red}^{{\rm tot}}\sim 1/3$, namely the average of the expected reduction of statistical and systematic uncertainties as compared to the baseline Run II measurements.
 For such datasets, the correlations are neglected in the projections due to the lack of their  breakdown.

 The main benefit of our approach is the possibility to extrapolate the full set of processes entering the current SMEFT global fit, to make sure that all relevant directions in the parameter space are covered in these projections, as well as bypassing the need to evaluate separately the SM and SMEFT theory predictions. 
 One drawback is that it does not account for the possible increase in kinematic range covered by HL-LHC measurements, for example with additional bins in the high-energy region, which can only be considered on a case-by-case basis. 


When generating pseudo-data based on some known underlying law, the total $\chi^2$ values will depend on the random seed used for the pseudo-data generation.
Statistical consistency of the procedure demands that the empirical distribution of these $\chi^2$ values over a large number of different sets of synthetic pseudo-data follows a $\chi^2$ distribution with the appropriate number of degrees of freedom. We have verified this property and found excellent agreement. 

\subsection{Code and installation}
The {\sc\small SMEFiT} code is an open source {\sc\small Python} package made available via the Python Package Index (pip) and can be installed by running:
\begin{lstlisting}[language=bash]
  $ pip install smefit
\end{lstlisting}
Alternatively, it can be installed from source by cloning the {\sc\small SMEFiT} GitHub repository and running the install script: 
\begin{lstlisting}[language=bash]
  $ git clone https://github.com/LHCfitNikhef/smefit_release.git
  $ ./install.sh -n <env_name='smefit_installation'>
  $ conda activate <env_name='smefit_installation'>
\end{lstlisting}
To start a fit with Nested Sampling, one should run:
\begin{lstlisting}[language=bash]
  $ smefit NS <path/to/runcard.yaml>
\end{lstlisting}
The analytic solution can be obtained instead with:
\begin{lstlisting}[language=bash]
  $ smefit A <path/to/smefit_runcard.yaml>
\end{lstlisting}
The latter only applies to linear fits as discussed in Sect.~\ref{subsec:optimisation}. An example of a \verb|smefit_runcard.yaml| is given below: 
\begin{lstlisting}[language=yaml]
  # name to give to fit
  result_ID: hello_smefit
  
  # path where results are stored
  result_path: ./results
  
  # path to data
  data_path: ./smefit_database/commondata
  
  # path to theory tables
  theory_path: ./smefit_database/theory
  
  # perturbatve QCD order (LO or NLO)
  order: NLO

  # SMEFT expansion order 
  use_quad: False
  
  # include theory uncertainties and the t0-prescription
  use_theory_covmat: True
  use_t0: True
  
  # number of samples (for analytic solution)
  n_samples: 20000
  
  # Datasets to include
  datasets:
  
    - CMS_ttZ_13TeV
    - ATLAS_ttZ_13TeV_pTZ
  
  # Coefficients to fit
  coefficients:
  
    O81qq: { 'min': -2, 'max': 2 }  # prior range
    O83qq: { 'min': -2, 'max': 2 }
\end{lstlisting}
The datasets and EFT coefficients to be fitted need to be specified after the \verb|datasets| and \verb|coefficients| key respectively. We provide all datasets available within {\sc\small SMEFiT} and its corresponding theory predictions on a dedicated {\sc\small SMEFiT} database that can be downloaded by:
\begin{lstlisting}[language=bash]
  $ git clone https://github.com/LHCfitNikhef/smefit_database.git
\end{lstlisting}
Realistic {\sc\small SMEFiT} runcards that were used to obtain the results in this thesis are linked on the {\sc\small SMEFiT} website, 
\begin{center}
  \url{https://lhcfitnikhef.github.io/smefit_release} \, ,
\end{center}
where we also include a tutorial, which runs on Google Colab, to get familiar with the code.

\section{The ML4EFT framework}
\label{sec:ml4eft_framework}
The {\sc\small SMEFiT} framework as introduced in Sect.~\ref{sec:smefit_framework} assumes a multivariate Gaussian likelihood and includes either inclusive or differential cross-section measurements. The latter are reported as central values plus corresponding (asymmetric) experimental uncertainties in each bin, where the bin typically refers to either one or two kinematic variables in case of a single or doubly differential distribution respectively. In Chapter \ref{chap:ml4eft}, we will argue that this choice is unnecessarily restrictive for EFT analyses as it neglects the continuous multivariate nature of the particles' phase space by projecting onto a subset of (non-optimal) kinematics. Here we lay out the statistical foundation of the {\sc\small ML4EFT} framework that we developed in Ref.~\mycite{Gomez-Ambrosio:2018pnl} to remedy this. 

{\sc\small ML4EFT} is an open source {\sc\small Python} code designed to facilitate the integration of unbinned multivariate observables into fits of Wilson coefficients in the SMEFT. It is based on  Machine Learning regression and classification techniques to parameterise high-dimensional likelihood ratios as required to carry out parameter inference in the context of global SMEFT analyses. In this section, we start by briefly reviewing binned likelihoods in Sect.~\ref{subsec:binned_likelihoods}. We then introduce in Sect.~\ref{subsec:unbinned_likelihood} the unbinned likelihood counterpart, which is smoothly connected to the binned likelihood in the infinitely narrow bin limit, and point out where Machine Learning enters. Sect.~\ref{subsec:ml4eft_code} provides information on how to use and install the {\sc\small ML4EFT} code.

\subsection{Binned likelihoods}
\label{subsec:binned_likelihoods}
Let us consider a dataset $\mathcal{D}$.
The corresponding theory prediction $\mathcal{T}$ will in general depend
on $n_{\rm p}$ model parameters, denoted by $\boldsymbol{c} = \{ c_1, c_2, \ldots, c_{n_{\rm p}}\}$,
and hence we write these predictions as $\mathcal{T}(\boldsymbol{c})$.
The likelihood function is defined as the
probability to observe the dataset $\mathcal{D}$ assuming that the corresponding
underlying law is described by the theory predictions $\mathcal{T}(\boldsymbol{c})$
associated to the specific set of parameters $\boldsymbol{c}$,
\begin{equation} 
\label{eq:likelihood}
\mathcal{L}( \boldsymbol{c}) = P\lp \mathcal{D} | \mathcal{T}(\boldsymbol{c}) \rp \, .
\end{equation} 
This likelihood function makes it possible to discriminate between different theory
hypotheses and to determine, within a given theory hypothesis $\mathcal{T}(\boldsymbol{c})$,
the preferred values and confidence level (CL) intervals for
a given set of model parameters.
In particular,
the best-fit values of the parameters $\boldsymbol{\hat{c}}$ are then
determined from the maximisation of the likelihood function $\mathcal{L}( \boldsymbol{c})$,
with contours of fixed likelihood determining their CL intervals.

The most common manner of presenting the information contained in the dataset $\mathcal{D}$
is by binning the data in terms of specific values of selected variables characteristic of each
event, such as the final state kinematics.
In this case, the individual events are combined into $N_{b}$ bins.
Let us denote by $n_i$ the number of observed events in the $i$-th bin
and by $\nu_i(\boldsymbol{c})$ the corresponding theory prediction for the
model parameters $\boldsymbol{c}$.
For a sufficiently large number of events $n_i$ per bin (typically taken to be $n_i\gsim 30$)
one can rely on the Gaussian approximation.  Hence, the likelihood
to observe $\boldsymbol{n} = (n_1, \dots, n_{N_{b}})$ events in each bin, given the
theory predictions $\boldsymbol{\nu}(\boldsymbol{c})$,
is given by
\begin{equation}
  \label{eq:binned_gaussian_likelihood}
	\mathcal{L}(\boldsymbol{n}; \boldsymbol{\nu}(\boldsymbol{c})) = \prod_{i=1}^{N_b} \exp\left[-\frac{1}{2}\frac{(n_i-\nu_i(\boldsymbol{c}))^2}{\nu_i(\boldsymbol{c})}\right] \, ,
\end{equation}
where we consider only statistical uncertainties and neglect possible sources of
correlated systematic errors in the measurement (uncorrelated systematic errors can be
accounted for in the same manner as the statistical counterparts).
This approximation is justified since in Chapter \ref{chap:ml4eft} we will focus on
statistically-limited  observables, e.g. the high-energy tails
of differential distributions.
The binned Gaussian likelihood Eq.~(\ref{eq:binned_gaussian_likelihood}) can also be expressed
as
\begin{equation}
   \label{eq:binned_gaussian_likelihood_chi2}
   -2\log  \mathcal{L}(\boldsymbol{n}; \boldsymbol{\nu}(\boldsymbol{c})) =
    \sum_{i=1}^{N_b} \frac{(n_i-\nu_i(\boldsymbol{c}))^2}{\nu_i(\boldsymbol{c})}\, ,
\end{equation}
that is, as the usual $\chi^2$ corresponding to Gaussianly distributed binned measurements.
The most likely values of the parameters $\boldsymbol{\hat{c}}$ given the theory hypothesis
$\mathcal{T}(\boldsymbol{c})$ and the measured dataset $\mathcal{D}$ are obtained
from the minimisation of Eq.~(\ref{eq:binned_gaussian_likelihood_chi2}).

The Gaussian binned likelihood, Eq.~(\ref{eq:binned_gaussian_likelihood_chi2}), is not appropriate
whenever the number of events in some bins becomes too small.
%
Denoting by $n_{\rm tot}$ the total number of observed events and $\nu_{\rm tot}(\boldsymbol{c})$
the corresponding theory prediction, the corresponding likelihood is the product
of Poisson and multinomial distributions:
\begin{equation}
  \label{eq:likelihood_binned_poisson}
  \mathcal{L}(\boldsymbol{n}; \boldsymbol{\nu}(\boldsymbol{c}))  = \frac{\lp \nu_{\mathrm{tot}}(\boldsymbol{c})\rp^{n_\mathrm{tot}}e^{-\nu_{\mathrm{tot}}(\boldsymbol{c})}}{n_\mathrm{tot}!}\frac{n_\mathrm{tot}!}{n_1!\dots n_{N_b}!}\prod_{i=1}^{N_b}
\left(\frac{\nu_i (\boldsymbol{c})}{\nu_{\mathrm{tot}}(\boldsymbol{c}))}\right)^{n_i} \, ,
\end{equation}
where the total number of observed events (and the corresponding theory prediction)
is equivalent to the sum over all bins,
\begin{equation} 
\nu_{\mathrm{tot}}(\boldsymbol{c}) = \sum_{i=1}^{N_b} \nu_i(\boldsymbol{c})\, , \qquad n_{\mathrm{tot}} = \sum_{i=1}^{N_b} n_i \, .
\end{equation} 
When imposing these constraints, Eq.~(\ref{eq:likelihood_binned_poisson}) simplifies to
\begin{equation}
\mathcal{L}(\boldsymbol{n}; \boldsymbol{\nu}(\boldsymbol{c})) = \prod_{i=1}^{N_b}\frac{\nu_i(\boldsymbol{c})^{n_i}}{n_i!}e^{-\nu_i(\boldsymbol{c})} \, ,
\label{eq:joint_pdf_simp}
\end{equation}
which is equivalent to the likelihood of a binned measurement  in which the number of events $n_i$ in each bin follows an independent Poisson distribution with mean $\nu_i(\boldsymbol{c})$.
As in the Gaussian case, Eq.~(\ref{eq:binned_gaussian_likelihood_chi2}), one often
considers the negative log-likelihood,
and for the Poissonian likelihood of Eq.~(\ref{eq:joint_pdf_simp})
this translates into
\begin{equation}
  -2 \log \mathcal{L}(\boldsymbol{n}; \boldsymbol{\nu}(\boldsymbol{c}))  = -2 \sum_{i=1}^{N_b} \lp 
  n_i\log \nu_i(\boldsymbol{c})-\nu_i(\boldsymbol{c}) \rp,
	\label{eq:likelihood_binned_2}
\end{equation}
where we have dropped the $c$-independent terms.
In the limit of large number of events per bin, $n_i\gg 1$, it can be shown
that the Poisson log-likelihood, Eq.~(\ref{eq:likelihood_binned_2}), reduces
to its Gaussian counterpart, Eq.~(\ref{eq:binned_gaussian_likelihood_chi2}).
Again, the most likely values of the model parameters, $\boldsymbol{\hat{c}}$,
    are those obtained from the minimisation of Eq.~(\ref{eq:likelihood_binned_2}).

\subsection{Unbinned likelihoods}
\label{subsec:unbinned_likelihood}

The previous discussion applies to binned observables, and leads
to the standard Gaussian and Poisson likelihoods, Eqns.~(\ref{eq:binned_gaussian_likelihood})
and~(\ref{eq:joint_pdf_simp}) respectively, in the case of statistically-dominated
measurements.
Any binned measurement entails some information loss by construction,
since the information provided by individual events falling into the same bin
is being averaged out.
To eliminate the effects of this information loss, one can
construct unbinned likelihoods that reduce to their binned
counterparts Eqns.~(\ref{eq:binned_gaussian_likelihood})
and~(\ref{eq:joint_pdf_simp}) in the appropriate limits.

Instead of collecting the $N_{\rm ev}$ measured events into $N_b$ bins,
when constructing unbinned observables one treats  each event individually.
We denote now the dataset under consideration as
\begin{equation} 
\label{eq:dataset_unbinned_definition}
\mathcal{D}=\{ \boldsymbol{x}_i\}\, \qquad
\boldsymbol{x}_i= \lp x_{i,1}, x_{i,2},\ldots, x_{i,n_{k}} \rp \, ,\qquad  i=1, \ldots, N_{\rm ev} \, ,
\end{equation} 
with $\boldsymbol{x}_i$ denoting the
array indicating
the values of the $n_{k}$ final-state variables that are being measured.
Typically the array $\boldsymbol{x}_i$ will contain the values of the transverse
momenta, rapidities, and azimuthal angles of the measured final state particles, but
could also be composed of higher-level variables such as in jet substructure measurements.
Furthermore, the same approach can be applied to detector-level quantities,
in which case the array $\boldsymbol{x}_i$ contains information
such as energy deposits in the calorimeter cells.

As in the binned case, we assume that this process is described
by a theoretical framework $\mathcal{T}(\boldsymbol{c})$
depending on the $n_p$ model parameters  $\boldsymbol{c} = \{ c_1, c_2, \ldots, c_{n_{\rm p}}\}$.
The kinematic variables of the events constituting the
dataset Eq.~(\ref{eq:dataset_unbinned_definition})
are independent and identically distributed random variables following a given distribution,
which we denote by $f_\sigma\lp \boldsymbol{x}, \boldsymbol{c} \rp $, where the notation
reflects that this probability density will be given, in the cases we are interested in,
by the differential cross-section evaluated using the null hypothesis (theory
$\mathcal{T}(\boldsymbol{c})$ in this case).
For such an unbinned measurement, the likelihood factorises into the contributions
from individual events such that
\begin{equation}
\mathcal{L}( \boldsymbol{c}) = \prod_{i=1}^{N_{\rm ev}} f_\sigma\lp \boldsymbol{x}_i, \boldsymbol{c} \rp \, .
\label{eq:likelihood_unbinned_fixednev}
\end{equation}
It is worth noting that in Eq.~(\ref{eq:likelihood_unbinned_fixednev}) the
data enters as the experimentally observed values of the kinematic variables
$\boldsymbol{x}_i$ for each event, while the theory predictions
enter at the level of the model adopted $f_\sigma\lp \boldsymbol{x}, \boldsymbol{c} \rp $
for the underlying probability density.

By analogy with the binned Poissonian case, the likelihood can be generalised
to the more realistic case where the measured number of events $N_{\rm ev}$ is not
fixed but rather distributed according to a Poisson with mean $\nu_{\rm tot}(\boldsymbol{c})$,
namely the total number of events predicted by the theory $\mathcal{T}(\boldsymbol{c})$,
see also Eq.~(\ref{eq:likelihood_binned_poisson}).
The likelihood Eq.~(\ref{eq:likelihood_unbinned_fixednev}) then receives an
extra contribution to account for the random size of the dataset $\mathcal{D}$ which reads
\begin{equation}
\mathcal{L}( \boldsymbol{c}) = \frac{\nu_{\rm tot}(\boldsymbol{c})^{N_{\rm ev}}}{N_{\rm ev}!}e^{-\nu_{\rm tot}(\boldsymbol{c})}\prod_{i=1}^{N_{\rm ev}} f_\sigma\lp \boldsymbol{x}_i, \boldsymbol{c} \rp \, .
\label{eq:likelihood_extended}
\end{equation}
Eq.~(\ref{eq:likelihood_extended}) defines the extended unbinned likelihood,
with corresponding log-likelihood given by
\begin{align}
  \nonumber
  \log \mathcal{L}( \boldsymbol{c}) &= -\nu_{\rm tot}(\boldsymbol{c}) + N_{\rm ev}\log \nu_{\rm tot}(\boldsymbol{c}) +
  \sum_{i=1}^{N_{\rm ev}} \log f_\sigma\lp \boldsymbol{x}_i, \boldsymbol{c} \rp \\
  &=  -\nu_{\rm tot}(\boldsymbol{c}) +
  \sum_{i=1}^{N_{\rm ev}} \log  \lp  \nu_{\rm tot}(\boldsymbol{c})  f_\sigma\lp \boldsymbol{x}_i, \boldsymbol{c} \rp \rp \, ,
\label{eq:log_likelihood_extended}
\end{align}
where again we have  dropped all terms that do not depend on the theory
parameters $\boldsymbol{c} $ since these are not relevant to determine
the maximum likelihood estimators and confidence level intervals.
The unbinned log-likelihood Eq.~(\ref{eq:log_likelihood_extended}) can also be obtained from the Poissonian
binned likelihood Eq.~(\ref{eq:likelihood_binned_2}) in the infinitely narrow bin limit,
that is, when taking $n_i\rightarrow 1\,( \forall i)$ and $N_{b}\to N_{\rm ev}$, where $\nu_i(\boldsymbol{c})  \rightarrow
\nu_{\rm tot}(\boldsymbol{c})f_\sigma\lp \boldsymbol{x}_i, \boldsymbol{c} \rp $.
Indeed, in this limit one has that
\begin{align}
&\log \mathcal{L}(\boldsymbol{n}; \boldsymbol{\nu}(\boldsymbol{c}))\Big|_{\rm binned}  = \sum_{i=1}^{N_b} \lp 
  n_i\log \nu_i(\boldsymbol{c})-\nu_i(\boldsymbol{c})   \rp
  \label{eq:likelihood_binned_2_limit}\\
 \nonumber&\to  \sum_{i=1}^{N_{\rm ev}} \lc 
  \log \lp \nu_{\rm tot}(\boldsymbol{c})f_\sigma\lp \boldsymbol{x}_i, \boldsymbol{c} \rp\rp -\nu_{\rm tot}(\boldsymbol{c})f_\sigma\lp \boldsymbol{x}_i, \boldsymbol{c} \rp   \rc \\
  \nonumber&\hspace{2cm}= -\nu_{\rm tot}(\boldsymbol{c}) +
  \sum_{i=1}^{N_{\rm ev}} 
  \log \lp \nu_{\rm tot}(\boldsymbol{c})f_\sigma\lp \boldsymbol{x}_i, \boldsymbol{c} \rp\rp   \nonumber \, ,
\end{align}
  as expected, 
  where we have used the normalisation condition for the probability density
  \begin{equation} 
   \sum_{i=1}^{N_{\rm ev}}f_\sigma\lp \boldsymbol{x}_i, \boldsymbol{c} \rp  = 1 \, .
   \end{equation} 
   Hence one can smoothly interpolate between the (Poissonian) binned and unbinned
   likelihoods by reducing the bin size until there is at most one event per bin.
   Again, we ignore correlated systematic errors in this derivation.

As mentioned above, the probability density associated to the events
that constitute the dataset Eq.~(\ref{eq:dataset_unbinned_definition}) and enter
the corresponding likelihood Eq.~(\ref{eq:log_likelihood_extended}) is, in the case
of high-energy collisions, given by the  normalised
differential cross-section
\begin{equation}
f_\sigma\lp \boldsymbol{x}, \boldsymbol{c} \rp= \frac{1}{\sigma_{\rm fid}( \boldsymbol{c})}\frac{d\sigma(\boldsymbol{x}, \boldsymbol{c})}{d\boldsymbol{x}} \, ,
\label{eq:dsigma_dx}
\end{equation}
with $\sigma_{\rm fid}(\boldsymbol{c})$ indicating the total fiducial cross-section corresponding
to the phase space region in which the kinematic variables $\boldsymbol{x}$ that describe
the event are being measured.
By construction, Eq.~(\ref{eq:dsigma_dx}) is normalised as should be the case
for a probability density.
We can now use Eq.~(\ref{eq:log_likelihood_extended}) together with Eq.~(\ref{eq:dsigma_dx})
in order to evaluate the unbinned profile likelihood ratio,
\begin{equation}
  q_{\boldsymbol{c}} \equiv   -2 \log \frac{\mathcal{L}(\boldsymbol{c})}{\mathcal{L}(\boldsymbol{\hat{c}})}=
  2\left[\nu_{\rm tot}(\boldsymbol{c}) - \nu_{\rm tot}(\boldsymbol{\hat{c}}) -\sum_{i=1}^{N_{\rm ev}} \log \lp 
	  \frac{d\sigma(\boldsymbol{x}_i, \boldsymbol{c})}{d\boldsymbol{x}} \Bigg{/}\frac{d\sigma(\boldsymbol{x}_i, \boldsymbol{\hat{c}})}{d\boldsymbol{x}} \rp \right].
	\label{eq:plr_4}
\end{equation}
For a given significance level $\alpha$, the profile likelihood ratio can be used to determine the endpoints of the $100(1-\alpha)\%$ CL interval by imposing the condition
\begin{equation}
  \label{eq:confidence_level}
	p_{\boldsymbol{c}} \equiv 1 - F_{\chi^2_{n_{\rm p}}}(q_{\boldsymbol{c}}) = \alpha
\end{equation}
on the p-value $p_{c}$, where $F_{\chi^2_k}(y)$ is the cumulative distribution function of the $\chi^2_k$ distribution with $k$ degrees of freedom. 
Continuing from Eq.~\eqref{eq:plr_4}, let us define for convenience of notation
\begin{equation}
  r_\sigma( \boldsymbol{x}_i, \boldsymbol{c}, \boldsymbol{\hat{c}}) \equiv
  \frac{d\sigma(\boldsymbol{x}_i, \boldsymbol{c})}{d\boldsymbol{x}} \Bigg{/}\frac{d\sigma(\boldsymbol{x}_i, \boldsymbol{\hat{c}})}{d\boldsymbol{x}}  \qquad \text{and}\qquad  r_\sigma( \boldsymbol{x}_i, \boldsymbol{c})\equiv  r_\sigma( \boldsymbol{x}_i, \boldsymbol{c}, \boldsymbol{0}) \, .
  \label{eq:ratio_rsigma_definition}
\end{equation}
The latter is especially useful in cases such as the SMEFT, where
the alternative
hypotheses corresponds to the vanishing of all the theory
parameters (the EFT Wilson coefficients), and $\mathcal{T}$ reduces
to the SM.
In terms of this notation, we can then express the profile likelihood ratio for the unbinned
observables Eq.~(\ref{eq:plr_4}) as
\begin{equation}
q_{\boldsymbol{c}}= 2\left[\nu_{\rm tot}(\boldsymbol{c})-\sum_{i=1}^{N_{\rm ev}} \log  r_\sigma( \boldsymbol{x}_i, \boldsymbol{c})\right] -  2\left[\nu_{\rm tot}(\boldsymbol{\hat{c}})-\sum_{i=1}^{N_{\rm ev}} \log  r_\sigma( \boldsymbol{x}_i, \boldsymbol{\hat{c}})\right] \, .
	\label{eq:plr_5}
\end{equation}
One can then use either Eq.~(\ref{eq:plr_4}) or Eq.~(\ref{eq:plr_5}) to derive confidence level intervals associated
to the theory parameters $\boldsymbol{c}$ in the same manner as in the binned case, namely
by imposing Eq.~(\ref{eq:confidence_level}) for a given choice of the CL range.

Provided double counting is avoided, binned and unbinned observables can simultaneously be used
in the context of parameter limit setting.
In this general case one assembles a joint likelihood which
accounts for the contribution of all available types of observables,
namely
\begin{equation} 
\mathcal{L}( \boldsymbol{c}) = \prod_{k=1}^{N_{\mathcal{D}}}\mathcal{L}_k( \boldsymbol{c}) =
\prod_{k=1}^{N_{\mathcal{D}}^{\rm (ub)}}\mathcal{L}_k^{\rm (ub)}( \boldsymbol{c})
\prod_{j=1}^{N_{\mathcal{D}}^{\rm (bp)}}\mathcal{L}_j^{\rm (bp)}( \boldsymbol{c})
\prod_{\ell=1}^{N_{\mathcal{D}}^{\rm (bg)}}\mathcal{L}_\ell^{\rm (bg)}( \boldsymbol{c}) \, ,
\end{equation} 
where we have $N_{\mathcal{D}} = N_{\mathcal{D}}^{\rm (ub)}+N_{\mathcal{D}}^{\rm (bp)} + N_{\mathcal{D}}^{\rm (bg)}$
datasets classified into unbinned (ub), binned Poissonian (bp), and binned Gaussian (bg)
datasets, where the corresponding likelihoods are given by Eq.~(\ref{eq:likelihood_extended})
for unbinned, Eq.~(\ref{eq:joint_pdf_simp}) for binned Poissonian, and
Eq.~(\ref{eq:binned_gaussian_likelihood}) for binned Gaussian observables.
The associated log-likelihood function is then
\begin{equation} 
\label{eq:log_likelihood_global}
\log \mathcal{L}( \boldsymbol{c}) = \sum_{k=1}^{N_{\mathcal{D}}} \log \mathcal{L}_k( \boldsymbol{c}) =
\sum_{k=1}^{N_{\mathcal{D}}^{\rm (ub)}}\log \mathcal{L}_k^{\rm (ub)}( \boldsymbol{c}) +
\sum_{j=1}^{N_{\mathcal{D}}^{\rm (bp)}}\log \mathcal{L}_j^{\rm (bp)}( \boldsymbol{c}) + 
\sum_{\ell=1}^{N_{\mathcal{D}}^{\rm (bg)}} \log \mathcal{L}_\ell^{\rm (bg)}( \boldsymbol{c}) \, ,
\end{equation} 
which can then be used to construct the profile likelihood ratio Eq.~(\ref{eq:plr_4})
in order to test the null hypothesis and determine confidence level intervals
in the theory parameters $\boldsymbol{c}$. 

The main challenge for the integration of unbinned observables in global fits
using the framework summarised by Eq.~(\ref{eq:log_likelihood_global}) is that the evaluation of $\mathcal{L}_k^{\rm (ub)}( \boldsymbol{c})$
is in general costly, since the underlying probability density is not known in closed
form and hence needs to be computed numerically using Monte Carlo methods. This is plagued by an exponential scaling with the dimension of $\boldsymbol{x}$.
In Chapter \ref{chap:ml4eft} we discuss how to bypass this problem by adopting Machine Learning
techniques to parametrise this probability density (the differential cross-section)
and hence assemble unbinned observables that are fast and efficient to evaluate,
as required for their integration into a global SMEFT analysis.

\subsection{Code and installation}
\label{subsec:ml4eft_code}

{\sc\small ML4EFT} is made available via the Python Package Index (pip) and can be installed 
by running:
\begin{lstlisting}[language=bash]
  $ pip install ml4eft
\end{lstlisting}
Alternatively, the code can be downloaded from its public GitHub repository and installed from source:
\begin{lstlisting}[language=bash]
  $ git clone https://github.com/LHCfitNikhef/ML4EFT.git
  $ cd code
  $ pip install -e . 
\end{lstlisting}
The unbinned likelihood $\mathcal{L}(\boldsymbol{c})$ from Eq.~\eqref{eq:likelihood_extended} can be obtained by training a multivariate classifier using {\sc\small ML4EFT} as follows:
\begin{lstlisting}[language=python]
  import ml4eft.core.classifier as classifier

  classifier.Fitter(json_path = "./path/to/ml4eft_runcard.json", 
                    mc_run = 0, # Monte Carlo replica number
                    c_name = "name_of_wilson_coefficient",
                    output_dir = "./path/to/model/output")
\end{lstlisting}
An example of a \verb|ml4eft_runcard.json| is provided below:

\begin{lstlisting}[language=json]
  {
    "process_id": "tt", # process (top-quark pair production)
    "epochs": 2000, # max number (nr.) of epochs
    "lr": 0.0001,  # learning rate
    "n_batches": 50,  # nr. of mini-batches
    "output_size": 1,  # nr. of output nodes
    "hidden_sizes": # architecture of the hidden nodes
    [ 
      100,
      100,
      100
    ],
    "n_dat": 100000,  # total nr. of data points
    "event_data": "./path/to/training_data/",
    "features": # final-state kinematic features
    [  
      "pt_l1",
      "pt_l2",
      "DeltaPhi_ll",
      "DeltaEta_ll",
      "m_bb",
      "..."
    ],
    "loss_type": "CE",  # adopt cross entropy loss
    "patience": 200,  # nr. of epochs the loss may keep increasing
    "val_ratio": 0.2,  # validation set ratio
    "c_train": # value of the EFT coefficients of the training data
    {
      "ctu8": 10  
    }
  }
\end{lstlisting}
The framework is documented in greater detail on a dedicated website
\begin{center}
  \url{https://lhcfitnikhef.github.io/ML4EFT},
\end{center}
where, in addition, one can find a self-standing tutorial (which can also
be run in Google Colab) where the user is guided step by step in how
unbinned multivariate observables can be constructed given a choice of
EFT coefficients and of final-state kinematic features.

In the same website we also provide links to the main results
 presented in Chapter \ref{chap:ml4eft},
including the likelihood ratio parametrisations that have been obtained
for the two processes ($t\bar{t}$ and $hZ$) considered in Ref.~\mycite{GomezAmbrosio:2022mpm}.
We also include there animations demonstrating the 
training of the neural networks, such as e.g.
\begin{center}
\url{https://lhcfitnikhef.github.io/ML4EFT/sphinx/build/html/results/ttbar_analysis_parton2.html#overview}
\end{center}
Additional unbinned multivariate observables to be constructed in the future using
our framework will be added to the same page.
Furthermore, we also plan to tabulate the neural network
outputs using fast grid techniques so that these observables can be  stand-alone integrated
in global fits without the need to link the actual {\sc\small ML4EFT} code.

  \chapter{Global EFT interpretation of Higgs, top, diboson and EWPO data}
\label{chap:smefit3}
\vspace{-1cm}
\begin{center}
\begin{minipage}{1.\textwidth}
    \begin{center}
        \textit{This chapter is based on my results that are presented in Ref.~\mycite{Celada:2024mcf}} 
    \end{center}
\end{minipage}
\end{center}
 
\myparagraph{Introduction} In this chapter, we present {\sc\small SMEFiT3.0}, a global SMEFT analysis of Higgs, top quark, and diboson data from the LHC complemented by EWPOs from LEP and SLD.
This analysis, carried out within the {\sc\small SMEFiT2.0} framework~\cite{Hartland:2019bjb,Ethier:2021ydt,Ethier:2021bye,vanBeek:2019evb,Giani:2023gfq}~\mycite{terHoeve:2023pvs}, considers  recent inclusive and differential measurements from the LHC Run II, in several cases based on its full integrated luminosity, alongside with a new implementation of the EWPOs.
In contrast to {\sc\small SMEFiT2.0} \cite{Ethier:2021bye}, the EWPO implementation is based on independent calculations of the relevant EFT contributions, and ensures consistent theory settings between the $e^+e^-$ and hadron collider processes, and is benchmarked with previous studies in the literature. 
Constraining 45~(50) independent directions in the parameter space within linear (quadratic) SMEFT fits, this analysis provides a state-of-the-art set of EFT bounds enabled by available LEP and LHC data.

\myparagraph{Motivation} A key component of the LEP and SLD legacy are the EWPOs~\cite{ALEPH:2005ab} from electron-positron collisions at the $Z$-pole and beyond.
The high precision achieved by the measurements of EWPOs imposes stringent stress tests of the SM~\cite{Altarelli:1991fk,Baak:2012kk} and indirectly constrains a wide range of BSM models.
Indeed, as compared to measurements at the LHC, these EWPOs provide complementary, and in many cases dominant, sensitivity to a wealth of BSM scenarios.
In this respect, the EWPOs supplement uniquely the direct information on the properties of the Higgs and electroweak sectors obtained at the LHC~\cite{Dawson:2018dcd,Butter:2016cvz,Azatov:2019xxn}.

In the language of the SMEFT, EWPOs provide information on several directions in the EFT parameter space~\cite{Brivio:2017bnu}, some of them of great relevance when matching onto compelling UV-completions~\mycite{terHoeve:2023pvs}.
However, the implementation of EWPOs within a SMEFT fit must deal with several theoretical subtleties, and ensure that the underlying settings such as the choice of electroweak input scheme, flavour assumptions, and operator basis are consistent with those adopted in the associated interpretation of LHC measurements.

\myparagraph{Outline and overview of the next chapters} The outline of this chapter is as follows.
First, in Sect.~\ref{sec:ewpos}, we describe the new implementation of the EWPOs in {\sc\small SMEFiT}.
Sect.~\ref{sec:updated_global_fit} presents the {\sc\small SMEFiT3.0} global analysis, including the LHC Run II measurements alongside the new EWPO implementation.
We summarise our findings of the current chapter in Sect.~\ref{sec:summary_smefit3}.

Moving on, in Chapter \ref{chap:smefit_uv}, we will shift perspective and present constraints on UV-complete models directly that have been matched onto the SMEFT. This ultimately bridges the energy gap between the generic EFT landscape and the wide range of explicit SM extensions \mycite{terHoeve:2023pvs}. Chapter \ref{chap:ml4eft} will introduce the {\sc\small ML4EFT} framework that we developed to integrate unbinned multivariate observables into global SMEFT fits based on Machine Learning techniques \mycite{GomezAmbrosio:2022mpm}. Finally, building on the results presented in Chapter \ref{chap:smefit3}, we will then present in Chapter \ref{chap:smefit_fcc} the impact of the HL-LHC and future colliders on the SMEFT parameter space, as well as on UV-complete models obtained via matching \mycite{Celada:2024mcf}.

\section{Electroweak precision observables in SMEFiT}
\label{sec:ewpos}

Here we describe the new implementation and validation
of EWPOs in the {\sc\small SMEFiT} analysis framework.
In the next section, we compare our results with those obtained with the previous approximation.

\subsection{EWPOs in the SMEFT}
\label{subsec:EWPOs_recap}

For completeness and to set up the notation, we provide a concise overview of how SMEFT operators affect the EWPOs measured at electron-positron colliders operating at the $Z$-pole and beyond such as LEP and SLD.
We work in the $\{\hat{m}_W, \hat{m}_Z, \hat{G}_F\}$ input electroweak scheme.
In the following, physical quantities which are either measured or
derived from measurements
are indicated with a hat, while canonically normalised Lagrangian parameters are denoted with a bar.
Throughout this thesis, Wilson coefficients follow the definitions and conventions of the Warsaw basis~\cite{Grzadkowski:2010es} and we use a U$(2)_q\times$U$(3)_d\times$U$(2)_u\times(\text{U}(1)_\ell\times\text{U}(1)_e)^3$ flavour assumption.
The operators are defined in App.~B of Ref.~\mycite{Celada:2024mcf}. 

In the presence of dim-6 SMEFT operators and adapting \cite{Brivio:2017bnu} to our conventions, the SM values of Fermi's constant and the electroweak boson masses are shifted as follows:
\begin{align}
\delta G_F &= \frac{1}{2 \hat{G}_F} \Bigg( c_{\varphi \ell_1}^{(3)}+c_{\varphi \ell_2}^{(3)} -  c_{\ell\ell}  \Bigg) \, , \nonumber\\
\frac{\delta m_Z^2}{\hat{m}_Z^2} &= \frac{1}{2\sqrt{2}\hat{G}_F}c_{\varphi D} + \frac{\sqrt{2}}{\hat{G}_F}\frac{\hat{m}_W}{\hat{m}_Z}\sqrt{1-\frac{\hat{m}_W^2}{\hat{m}_Z^2}}c_{\varphi WB} \, , \label{eq:input_EW_shifts}\\
\frac{\delta m_W^2}{\hat{m}_W^2} &= 0 \, . \nonumber
\end{align}
In the following, we adopt a notation in which the new physics cut-off scale $\Lambda$ has been reabsorbed into the Wilson coefficients in Eq.~(\ref{eq:input_EW_shifts}), which therefore here and in the rest of the section should be understood to be dimensionful and with mass-energy units of $\lc c\rc = -2$.
We note that in this notation $\delta G_F$ is dimensionless, and hence indicates a relative shift.

These SMEFT-induced shifts in the electroweak input parameters defining the $\{\hat{m}_W, \hat{m}_Z, \hat{G}_F\}$ scheme modify the  interactions of the electroweak gauge bosons.
Specifically, the vector (V) and axial (A) couplings $g_{V,A}$ of the $Z$-boson are shifted in comparison
to the SM reference $\bar{g}_{V, A}$ (recall that the bar indicates renormalised Lagrangian parameters)
according to the following relation:
\begin{align}
\label{eq:VA-shifts}
g_{V, A}^x = \bar{g}_{V, A}^x + \delta g_{V, A}^x \, , \qquad x=\{\ell_i, u_i, d_i, \nu_i\} \, ,
\end{align}
where the superscript $x$ denotes the
fermion to which the $Z$-boson couples: either a charged (neutral) lepton $\ell_i$ ($\nu_i$), an up-type quark $u_i$ or a down-type quark $d_i$, respectively.
The flavour index $i=1,2,3$ runs over fermionic generations.
The SM couplings in Eq.~(\ref{eq:VA-shifts})
are given in the adopted notation by
\begin{equation}
\bar{g}_V^x = T_3^x/2 - Q^xs_{\hat{\theta}}^2  \, ,\qquad \bar{g}_{A}^{x} = T_3^x/2 \, ,
\end{equation}
\begin{equation}
Q^x = \{-1, 2/3, -1/3, 0\} \, ,\quad T_3^x = \{-1/2, 1/2, -1/2, 1/2\} \, ,\quad s_{\hat{\theta}}^2=1-\frac{\hat{m}_W^2}{\hat{m}_Z^2} \, ,
\end{equation}
where $ s_{\hat{\theta}}\equiv \sin \hat{\theta}$.
This shift in the SM couplings of the $Z$-boson arising from the dim-6 operators
in Eq.~(\ref{eq:VA-shifts})
can be further decomposed as
\begin{equation}
\label{eq:Zcoupling_EFT_shifts}
\delta g_V^x = \delta\bar{g}_Z\,\bar{g}_V^x + Q^x\delta s_\theta^2 + \Delta_V^x, \qquad \delta g_A^x = \delta\bar{g}_Z\,\bar{g}_A + \Delta_A^x \, ,
\end{equation}
for the vector and axial couplings respectively.
In Eq.~(\ref{eq:Zcoupling_EFT_shifts}) we have defined the (dimensionless) shifts in terms of the Wilson coefficients in the Warsaw basis
\begin{align}
\nonumber \delta \bar{g}_Z &= -\frac{1}{\sqrt{2}}\delta G_F - \frac{1}{2}\frac{\delta m_Z^2}{\hat{m}_Z^2} + \frac{s_{\hat{\theta}}c_{\hat{\theta}}}{\sqrt{2}\hat{G}_F}c_{\varphi WB}\\
&= -\frac{1}{4\sqrt{2} \hat{G}_F}\left(c_{\varphi D} + 2 c_{\varphi \ell_1}^{(3)} + 2 c_{\varphi \ell_2}^{(3)} - 2c_{\ell\ell}\right) \, ,\\
\delta s_{\theta}^2 &= \frac{1}{2\sqrt{2}\hat{G}_F}\frac{\hat{m}_W^2}{\hat{m}_Z^2}c_{\varphi D} + \frac{1}{\sqrt{2}\hat{G}_F}\frac{\hat{m}_W}{\hat{m}_Z}\sqrt{1-\frac{\hat{m}_W^2}{\hat{m}_Z^2}}c_{\varphi WB} \, ,
\end{align}
where the cosine of the weak mixing angle is given by $ c_{\hat{\theta}}\equiv \cos \hat{\theta} = \hat{m}_W/\hat{m}_Z$.

In this notation, the contributions to the shifts $\delta g_V^x$ and $\delta g_A^x$ which are not proportional to either $\bar{g}_{V,A}^x$ or $Q^x$  are  denoted as $\Delta^{x}_{V,A}$ and are given by
\begin{align}
\nonumber\Delta^{\ell_i}_V &= -\frac{1}{4\sqrt 2 \hat{G}_F}\left(c_{\varphi\ell_i}+c_{\varphi\ell_i}^{(3)}+c_{\varphi e/\mu/\tau}\right) \, ,&
\Delta^{\ell_i}_A &= -\frac{1}{4\sqrt 2 \hat{G}_F}\left(c_{\varphi \ell_i}+c_{\varphi\ell_i}^{(3)}-c_{\varphi e/\mu/\tau}\right),\\
\nonumber\Delta^{\nu_i}_V &=-\frac{1}{4\sqrt 2 \hat{G}_F}\left(c_{\varphi\ell_i}-c_{\varphi\ell_i}^{(3)} \right) \, ,&
\Delta^{\nu_i}_A &= -\frac{1}{4\sqrt 2 \hat{G}_F}\left(c_{\varphi \ell_i}-c_{\varphi\ell_i}^{(3)}\right) \, ,\\
\nonumber\Delta^{u_j}_V &= -\frac{1}{4\sqrt2 \hat{G}_F}\left(c_{\varphi q}^{(1)}-c_{\varphi q}^{(3)}+c_{\varphi u}\right) \, ,&
\Delta^{u_j}_A &= -\frac{1}{4\sqrt2 \hat{G}_F}\left(c_{\varphi q}^{(1)}-c_{\varphi q}^{(3)}-c_{\varphi u}\right),\\
\Delta^{d_j}_V &= -\frac{1}{4\sqrt2 \hat{G}_F}\left(c_{\varphi q}^{(1)}+c_{\varphi q}^{(3)}+c_{\varphi d}\right) \, ,&
\Delta^{d_j}_A &= -\frac{1}{4\sqrt2 \hat{G}_F}\left(c_{\varphi q}^{(1)}+c_{\varphi q}^{(3)}-c_{\varphi d}\right) 
\end{align}
where $i=1,2,3$ for the leptonic generations and $j=1,2$ runs over the two light quark generations. 
Note that in the above equations there is some ambiguity in the definition of $\ell_i$: $\Delta^{\ell_i}$ refers to the shift for the charged leptons, while $\ell_i$ in the coefficient names refers to the left-handed lepton doublet. 
For the heavy third-generation quarks ($j=3$) we have instead:
\begin{align}
\nonumber\Delta^{t}_V &= -\frac{1}{4\sqrt2 \hat{G}_F}\left(c_{\varphi Q}^{(1)}-c_{\varphi Q}^{(3)}+c_{\varphi t}\right) \, ,&
\Delta^{t}_A &= -\frac{1}{4\sqrt2 \hat{G}_F}\left(c_{\varphi Q}^{(1)}-c_{\varphi Q}^{(3)}-c_{\varphi t}\right),\\
\Delta^{b}_V &= -\frac{1}{4\sqrt2 \hat{G}_F}\left(c_{\varphi Q}^{(1)}+c_{\varphi Q}^{(3)}+c_{\varphi d}\right) \, ,&
\Delta^{b}_A &= -\frac{1}{4\sqrt2 \hat{G}_F}\left(c_{\varphi Q}^{(1)}+c_{\varphi Q}^{(3)}-c_{\varphi d}\right).
\end{align}
Concerning the SMEFT-induced shifts to the $W$-boson couplings, these are as follows:
\begin{align}
\nonumber g_{V, A}^{W_\pm, \ell_i} &= \bar{g}_{V, A}^{W_\pm, \ell_i} + \delta\left(g_{V, A}^{W_\pm, \ell_i}\right) \, , \\
g_{V, A}^{W_\pm, q} &= \bar{g}_{V, A}^{W_\pm, q} + \delta\left(g_{V, A}^{W_\pm, q}\right),
\label{eq:VA-shifts-W}
\end{align}
where the SM values are given by $\bar{g}_{V, A}^{W_\pm, \ell_i} = \bar{g}_{V, A}^{W_\pm, q} = 1/2$ and $\ell_i$ refers again to the lepton doublet. 
The SMEFT-induced shifts are given by
\begin{align}
\delta\left(g_{V, A}^{W_\pm, \ell_i}\right) &= \frac{1}{2\sqrt{2}\hat{G}_F}c_{\varphi \ell_i}^{(3)} - \frac{\delta G_F}{2\sqrt{2}}\, ,\\
\delta\left(g_{V, A}^{W_\pm, q}\right) &= \frac{1}{2\sqrt{2}\hat{G}_F}c_{\varphi q}^{(3)} - \frac{\delta G_F}{2\sqrt{2}} \, .
\label{eq:wshift}
\end{align}
where Eq.~\eqref{eq:wshift} applies only to the first two quark generations.

The corrections derived in this section for the couplings of leptons and quarks to the  $Z$ ($g_{V,A}^x$) and to the $W$ $\lp g_{V, A}^{W_\pm, x}\rp$ can be constrained by measurements of $Z$-pole observables at LEP and SLD together with additional electroweak measurements, as discussed below. 

\subsection{Approximate implementation}
\label{subsec:approx-ewpo}

The previous implementation of the EWPOs in the {\sc\small SMEFiT} analysis as presented in \cite{Ethier:2021bye} relied on the assumption that measurements at LEP and SLD were precise enough (compared to LHC measurements), and in agreement with the SM, to constrain the SMEFT-induced shifts modifying the $W$- and $Z$-boson couplings to fermions to be exactly zero.

This assumption results in a series of linear combinations of EFT coefficients appearing in Eqns.~\eqref{eq:VA-shifts} and~\eqref{eq:VA-shifts-W} being set to zero, inducing a number of relations between the relevant coefficients. 
Accounting for the three leptonic generations, this corresponds to 14 constraints parametrised in terms of 16 independent Wilson coefficients such that $14$ of them can be expressed in terms of the remaining two.
 For instance, it was chosen in \cite{Ethier:2021bye} to include $c_{\varphi WB}$ and $c_{\varphi D}$ as the two independent fit parameters and then to parametrise the other 14 coefficients entering in the EWPOs in terms of them as follows:
\begin{flalign}
	\left(
 \def\arraystretch{1.3}
\begin{array}{c}
c_{\varphi \ell_i}^{(3)} \\
 c_{\varphi \ell_i} \\
 c_{\varphi e/\mu/\tau} \\
 c_{\varphi q}^{(-)} \\
 c_{\varphi q}^{(3)} \\
 c_{\varphi u} \\
 c_{\varphi d} \\
 c_{\ell\ell} \\
\end{array}
\right)
= 
\left(
\def\arraystretch{1.3}
\begin{array}{cc}
 -\frac{1}{t_{\hat{\theta}}} & -\frac{1}{4 t_{\hat{\theta}}^2} \\
 0 & -\frac{1}{4} \\
 0 & -\frac{1}{2} \\
 \frac{1}{t_{\hat{\theta}}} & \frac{1}{4 s_{\hat{\theta}}^2}-\frac{1}{6} \\
 -\frac{1}{t_{\hat{\theta}}} & -\frac{1}{4 t_{\hat{\theta}}^2} \\
 0 & \frac{1}{3} \\
 0 & -\frac{1}{6} \\
 0 & 0 \\
\end{array}
\right)
\left(
\begin{array}{c}
	c_{\varphi WB}\\ c_{\varphi D}
\end{array}
\right) \, ,
\label{eq:2independents}
\end{flalign}
where $i=1,2,3$, and $t_{\hat{\theta}}=s_{\hat{\theta}}/c_{\hat{\theta}}$ indicates the tangent of the weak mixing angle. 

We refer to the linear system of equations defined by Eq.~(\ref{eq:2independents}) as the ``approximate'' implementation of the EWPOs used in previous {\sc\small SMEFiT} analyses,
meaning that only $c_{\varphi WB}$ and $c_{\varphi D}$ enter as independent degrees of freedom in the fit, while all other Wilson coefficients in the LHS of Eq.~(\ref{eq:2independents}) are then determined from those two rather than being constrained separately from the data.
Likewise, whenever theory predictions depend on some of these 14 dependent coefficients, for example in LHC processes, they can be reparametrised in terms of  only $c_{\varphi WB}$ and $c_{\varphi D}$.

\subsection{Exact implementation}

The approximate implementation of EWPO constraints as described by Eq.~(\ref{eq:2independents}) encodes a two-fold assumption.
First, it assumes that EWPO measurements coincide with the SM expectations, which in general is not the case.
Second, it also implies that the precision of LEP and SLD measurements is infinite compared to the LHC measurements, which is not necessarily true as demonstrated by LHC diboson production~\cite{Grojean:2018dqj,Banerjee:2018bio}.

To  bypass these two assumptions, which also prevent a robust use of matching results between SMEFT and UV-complete models~\mycite{terHoeve:2023pvs}, here we implement an exact treatment of the EWPOs and include the LEP and SLD measurements in the global fit alongside with the LHC observables.
That is, all 16 Wilson coefficients appearing in Eq.~\eqref{eq:2independents} become independent degrees of freedom, and are constrained by experimental data
from LEP/SLD and LHC sensitive to the shifts in the weak boson couplings given by Eqns.~\eqref{eq:VA-shifts}-(\ref{eq:VA-shifts-W}).
As a consequence, we had to also recompute the dependence of all the observables included in the global fit on these $16$ Wilson coefficients. 

Here we present an overview of the EWPOs included in the fit and discuss the computation of the corresponding theory predictions.
We consider the LEP and SLD legacy measurements specified in Table~\ref{tab:ew-datasets}.
They consist of 19 $Z$-pole observables from LEP-1, 21 bins in $\cos(\theta)$ for various centre of mass energies of Bhabha scattering ($e^+e^- \to e^+e^-$) from LEP-2, the weak coupling $\alpha_{\mathrm{EW}}$ as measured at $m_Z$, the three $W$ branching ratios to all generations of leptons, and 40 bins in $\cos(\theta)$ for four centre of mass energies of four-fermion production mediated by $W$-pairs at LEP-2.
To facilitate comparison with previous results, we adopt the same bin choices as in Table 9 of~\cite{Berthier:2015gja} in the case of Bhabha scattering, which provides an independent constraint on $\alpha_{\mathrm{EW}}$ as this is not fixed by the inputs. 
The information provided by $\alpha_{\mathrm{EW}}$ and by Bhabha scattering is equivalent from the point of view of constraining the SMEFT parameter space, and here we include for completeness both datasets to increase the precision of the resulting fit. 

\begin{table}[t]
\small
\renewcommand{\arraystretch}{1.8}
\centering
\resizebox{\textwidth}{!}{
\begin{tabular}{c|c|c|c|c}
\toprule
Input & Observables & Central values & Covariance & SM predictions \\
\midrule
\multirow{4}{*}{$Z$-pole EWPOs}&$\Gamma_Z$, $\sigma_{\text{had}}^0$, $R_e^0$,
$R_\mu^0$, $R_\tau^0$, 
$A_{FB}^{0,e}$, $A_{FB}^{0,\mu}$, $A_{FB}^{0,\tau}$&  \cite{ALEPH:2005ab} (Table $2.13$) & \cite{ALEPH:2005ab} (Table $2.13$)&\multirow{4}{*}{\cite{Corbett:2021eux} (Table $1$), \cite{Awramik:2003rn,Freitas:2014hra}} \\
&$R_b^0$, $R_c^0$, $A_{FB}^{0,b}$, $A_{FB}^{0,c}$, $A_b$, $A_c$                            & \cite{ALEPH:2005ab} (Table $5.10$) & \cite{ALEPH:2005ab} (Table $5.11$)        \\
&$A_\tau$ ($\mathcal{P}_\tau$), $A_e$ ($\mathcal{P}_\tau$)                                                               & \cite{ALEPH:2005ab} (Table $4.3$)  & n/a          \\
&$A_e$ (SLD), $A_\mu$ (SLD), $A_\tau$ (SLD) & \cite{ALEPH:2005ab} (Table $3.6$)  & \cite{ALEPH:2005ab}  (Table $3.6$)   \\
 \midrule
 \multirow{2}{*}{Bhabha scattering}& $d\sigma/d\cos{\theta}$ ($n_{\rm dat}=21$)
 &\multirow{2}{*}{ \cite{LEP-2} (Tables 3.11-12)}&\multirow{2}{*}{\cite{LEP-2} (App. B.3)}& \multirow{2}{*}{ \cite{LEP-2} (Tables 3.11-12)}\\
 & $\sqrt{s}=189, 192, 196, 200, 202, 205, 207\:\mathrm{GeV}$&&\\
 \midrule
  \multirow{1}{*}{$\alpha_{\mathrm{EW}}$}&$\alpha^{-1}_{\mathrm{EW}}(m_Z)$&\cite{PDG}&\cite{PDG}& \cite{Awramik:2003rn,Corbett:2021eux,PDG} (See text)\\
  \midrule
   \multirow{3}{*}{$W$ branching ratios}&Br($W\rightarrow e \nu_e$)&\multirow{3}{*}{\cite{LEP-2} (Table 5.5) }&\multirow{3}{*}{\cite{LEP-2} (Table E.6) }&\multirow{3}{*}{\cite{Efrati:2015eaa} (Table $2$)}\\
   &Br($W\rightarrow \mu \nu_{\mu}$)&&\\
   &Br($W\rightarrow \tau \nu_{\tau}$)&&\\
    \midrule
 \multirow{2}{*}{$W^+W^-$ production}& $d\sigma/d\cos{\theta}$ ($n_{\rm dat}=40$)
 &\multirow{2}{*}{\cite{LEP-2}}&\multirow{2}{*}{n/a} &\multirow{2}{*}{\cite{LEP-2} (Figure $5.4$)}\\
  & $\sqrt{s}=182, 189, 198, 206\:\mathrm{GeV}$&&\\
 \bottomrule
\end{tabular}}
\caption{Overview of the EWPOs
from LEP and SLD considered in this
work.
For completeness, we also indicate the differential $WW$ measurements
from LEP entering the fit and which
were already included in~\cite{Ethier:2021bye}.
For each measurement, we indicate
the observables considered and
the corresponding 
references for the experimental
central values and covariance matrix. 
In the absence of a covariance matrix,
measurements are assumed to be uncorrelated.
}
\label{tab:ew-datasets}
\end{table}

\myparagraph{Theoretical calculations}
We discuss now the corresponding theory implementation of the observables reported in Table~\ref{tab:ew-datasets}, i.e. $Z$-pole data, $W$ branching ratios, Bhabha scattering, $\alpha_{\mathrm{EW}}$, and $WW$ production.
As mentioned above, we adopt the $\{\hat{m}_W, \hat{m}_Z, \hat{G}_F\}$ input scheme, with the following numerical values of the input electroweak parameters:
\begin{equation}
G_F = 1.1663787\cdot 10^{-5}\;\mathrm{GeV}^{-2} \,, \: m_Z=91.1876\;\mathrm{GeV}\, , \: m_W=80.387\;\mathrm{GeV} \, .
\label{eq:inputparam}
\end{equation}
Concerning flavour assumptions, we adopt the U(2)$_q\times$ U(2)$_u\times $U(3)$_d\times [$U(1)$_\ell\times $U(1)$_e]^3$ flavour symmetry.
Starting with the $Z$-pole observables, we adopt the following definitions:
\begin{align}
\label{eq:gammaZ}
    \Gamma_Z &= \sum_{i=1}^3 \Gamma_{\ell_i} + \Gamma_{\mathrm{
had}} + \Gamma_{\mathrm{inv}}, &\Gamma_{\mathrm{had}} &= \sum_{i=1}^2\Gamma_{u_i} + \sum_{i=1}^3\Gamma_{d_i},  &\Gamma_{\mathrm{inv}} &= \sum_{i=1}^3\Gamma_{\nu_i},\\
\sigma_{\mathrm{had}}^0 &= \frac{12 \pi}{\hat{m}_Z^2}\frac{\Gamma_e\Gamma_{\mathrm{had}}}{\Gamma_Z^2},
&R_{\ell_i}^0 &= \frac{\Gamma_{\mathrm{had}}}{\Gamma_{\ell_i}}\,, &R^0_{b,c} &= \frac{\Gamma_{b,c}}{\Gamma_{\mathrm{had}}},
\\
A_f &= \frac{2g_V^{f}g_A^{f}}{\lp g_V^{f}\rp^2 + \lp g_A^{f}\rp^2},
&A_{FB}^{0, \ell_i} &= \frac{3}{4}A_e A_{\ell_i},
&A_{FB}^{0, b/c} &= \frac{3}{4}A_e A_{b/c}
\label{eq:Rbc}
\end{align}
where $\ell_i=\{e, \mu, \tau\}$ and where the partial decay widths of the $Z$ boson to (massless) quarks and leptons are expressed in terms of their electroweak couplings as
\begin{equation} 
\Gamma_i = \frac{\sqrt{2}\hat{G}_F\hat{m}_Z^3 C}{3\pi}\lp \left|g_V^i\right|^2 + \left|g_A^i\right|^2 \rp
\end{equation} 
where $C=3 \, (1)$ for quarks (leptons) is a colour normalisation factor. 
Substituting the SMEFT-induced shifts to the $Z$-boson couplings Eq.~(\ref{eq:VA-shifts}) into the  $Z$-pole observables Eqns.~(\ref{eq:gammaZ})-(\ref{eq:Rbc}) and expanding up to quadratic order in the EFT expansion, i.e. $\mathcal{O}\lp\Lambda^{-4}\rp$, one obtains the corresponding EFT theory predictions. 

One can proceed in the same manner concerning the $W$-boson branching ratios. 
The starting point is
\begin{align}
\label{eq:Wmass_BR}
\Gamma_W &= \sum_{i=1}^3 \Gamma_{W, \ell_i} + \Gamma_{W, u} + \Gamma_{W, c} \, , \\
\Gamma_{W, i} &= \frac{\sqrt{2}\hat{G}_F \hat{m}_W^3 C} {3\pi}|g_{V,A}^{W,i}|^2\, , 
\end{align}

We then expand $\Gamma_{W,i}/\Gamma_W$ up to quadratic order to end up with the EFT theory predictions for the $W$ branching ratios.
Note that no exotic decays of the $W$-boson are allowed.

The tree-level theoretical expressions for Bhabha scattering in the SMEFT were obtained analytically.
%
We generated all tree-level diagrams with up to one insertion of SMEFT operators for $e^+e^-\rightarrow e^+e^-$ using FeynArts~\cite{Hahn:2000kx} and then obtained the expressions for the cross-section $\sigma({e^+e^-\rightarrow e^+e^-})$ up to order $\mathcal{O}\lp\Lambda^{-4}\rp$ using Feyncalc~\cite{Mertig:1990an,Shtabovenko:2016sxi,Shtabovenko:2020gxv}. We cross-checked our SM expressions with Table $9$ of Ref.~\cite{Berthier:2015gja} and our SMEFT predictions with those obtained using the SMEFT@NLO~\cite{Degrande:2020evl} model in {\sc\small mg5\_aMC@NLO} \cite{Alwall:2014hca}, finding agreement in both cases.

Concerning the EW coupling constant $\alpha_{\mathrm{EW}}$, this is a derived quantity in the \\$\{\hat{m}_W, \hat{m}_Z, \hat{G}_F\}$ input scheme, which can be expressed in terms of the input parameters as follows
\begin{equation}
\bar{\alpha}_{\mathrm{EW}} = \frac{\bar{e}^2}{4\pi} = \frac{\left(\hat{e}-\delta e\right)^2}{4\pi}
=\frac{\hat{e}^2}{4\pi}\left(1-2\frac{\delta e}{\hat{e}}+\left(\frac{\delta e}{\hat{e}}\right)^2+\dots\right),
\label{eq:alpha_ew}
\end{equation}
where the ellipsis indicates higher-order corrections. 
In Eq.~(\ref{eq:alpha_ew}), the SMEFT-induced shift in the electric charge is given by \cite{Brivio:2017bnu}
\begin{equation}
\delta e = \hat{e}\left(-\frac{\delta G_F}{\sqrt{2}}+\frac{\delta m_Z^2}{2\hat{m}_Z^2}\frac{\hat{m}_W^2}{\hat{m}_W^2 - \hat{m}_Z^2}-\frac{\hat{m}_W s_{\hat{\theta}}}{\sqrt{2}\hat{G}_F\hat{m}_Z}c_{\varphi WB}\right) \, ,
\end{equation}
with the measured value of the electric charge given in this electroweak scheme by 
$
\hat{e} = 2^{5/4}\hat{m}_W\sqrt{\hat{G}_F}s_{\hat{\theta}}
$.
We expand Eq.~(\ref{eq:alpha_ew}) up to quadratic order to obtain the sought-for EFT theory predictions for $a_{\mathrm{EW}}$. 
The SM prediction is obtained by solving the on-shell expression for $m_W$ from \cite{Awramik:2003rn} for $\Delta \alpha$ and using
\begin{equation}
    \alpha_{EW}(m_Z) = \frac{\alpha_{EW}}{1-(\Delta \alpha + 0.007127)} \, ,
\end{equation}
 where $\alpha_{EW}$ is the fine-structure constant at zero energy, $0.007127$ represents the conversion factor between the on-shell and $\overline{\rm MS}$ renormalisation schemes \cite{PDG}, and we substitute in the input parameters from \eqref{eq:inputparam} along with those from \cite{Corbett:2021eux}.

Regarding the theory calculations for $WW$-production at LEP-2, we compute linear and quadratic SMEFT contributions to four-fermion production mediated by charged currents using the SMEFT@NLO model in {\sc\small mg5\_aMC@NLO}. 
Only semileptonic final states where a $W$-boson decays to either a $e\nu$ or $\mu\nu$ pair were considered. We computed the angular distribution in $\cos{\theta}$, where $\theta$ is the angle formed by the momentum of the $W^-$ and the incoming $e^-$, and we applied a kinematic cut on the charged lepton angle $\theta_{\ell}$, $|\cos(\theta_{\ell})|< 0.94$, corresponding to the detector acceptance of $20^{\circ}$ around the beam. The four distributions, corresponding to four luminosity-weighted values of centre of mass energy $\sqrt s =182.66,\, 189.09,\, 198.38,\,$ and $ 205.92\:\mathrm{GeV}$, were divided into 10 bins. 
SMEFT corrections to the $W$-boson decays were added {\it a posteriori} following an analogous approach to Ref.~\cite{Brivio:2019myy}. 

In addition to the LEP and SLD datasets listed in Table ~\ref{tab:ew-datasets}, new theory predictions were also computed for LHC processes sensitive to operators entering in the EWPOs and hence in the fit as new independent degrees of freedom.
For this, we used {\sc\small mg5\_aMC@NLO} interfaced to SMEFT@NLO to evaluate linear and quadratic EFT corrections at NLO QCD whenever available. 
In these calculations, in order to avoid any possible overlap between datasets entering simultaneously PDF and EFT fits~\cite{Carrazza:2019sec,Kassabov:2023hbm,Greljo:2021kvv}, we used NNPDF4.0 NNLO no-top~\cite{NNPDF:2021njg} as input PDF set. We refer to Tables 3.1-3.7 in~\cite{Ethier:2021bye} for an overview of the datasets that we include on top of those already presented in Table~\ref{tab:ew-datasets}. 
Furthermore, in comparison to the LHC datasets in~\cite{Ethier:2021bye}, we now include additional datasets from Run II, described in Sect.~\ref{subsec:new_datasets}.

\section{The SMEFiT3.0 global analysis}
\label{sec:updated_global_fit}

Here we present {\sc\small SMEFiT3.0},
an updated version of the global SMEFT analysis from~\cite{Ethier:2021bye,Giani:2023gfq}.
The major differences as compared with these previous analyses are two-fold.
First, the improved treatment of EWPOs as described in Sect.~\ref{sec:ewpos}.
Second, the inclusion of recent measurements of Higgs, diboson, and top quark production data from the LHC Run II, several of them based on its full integrated luminosity of $\mathcal{L}=139$ fb$^{-1}$.
In this section, we start by describing the main features of the new LHC Run II datasets added to the global fit (Sect.~\ref{subsec:new_datasets}). Afterwards, we quantify the impact of the new LHC data and of the  updated implementation of the EWPOs at the level of EFT coefficients (Sect.~\ref{subsec:impact_ewpos}).
\subsection{Experimental dataset}
\label{subsec:new_datasets}

Firstly, we describe the new LHC datasets from Run II which enter the updated global SMEFT analysis and which complement those already included in~\cite{Ethier:2021bye,Giani:2023gfq}.
For consistency with previous studies, and to ensure that QCD calculations at the highest available accuracy can be deployed, for top quark and Higgs boson production we restrict ourselves to parton-level measurements.
For diboson production, we consider instead particle-level distributions, for which NNLO QCD predictions are available for the SM~\cite{Grazzini:2019jkl}.

In the case of top quark production observables, we include the same datasets as in the recent EFT and PDF analysis of the top quark sector from the  {\sc\small PBSP} collaboration~\cite{Kassabov:2023hbm}.
These top quark measurements are extended
with additional datasets that have become available since the release of that study.
Theoretical higher-order QCD calculations and EFT cross-sections for these top quark production datasets are also taken from~\cite{Kassabov:2023hbm}, extended when required to the wider operator basis considered here. 

\begin{table}[t]
  \centering
  \footnotesize
   \renewcommand{\arraystretch}{1.30}
  \begin{tabular}{|C{2.5cm}|C{5cm}|C{1.7cm}|C{1.7cm}|}
  \toprule
 \multirow{2}{*}{Category}   &  \multirow{2}{*}{Processes}    &  \multicolumn{2}{|c|}{$n_{\rm dat}$}     \\
    &     &  {\sc\small SMEFiT2.0} &  {\sc\small SMEFiT3.0}     \\
    \toprule
    \multirow{7}{*}{Top quark}   &  $t\bar{t}+X$    &  94  &115\\
    \multirow{7}{*}{production}&  $t\bar{t}Z$, $t\bar{t}W$    & 14& 21\\
    &  $t\bar{t}\gamma$    & -& 2 \\
    &   single top (inclusive)   & 27 &28\\
    &  $tZ, tW$   &  9&13\\
    &  $t\bar{t}t\bar{t}$, $t\bar{t}b\bar{b}$    & 6 &12\\
    &  {\bf Total}    & {\bf 150 } &{\bf 191 }  \\
    \midrule
    \multirow{3}{*}{Higgs production} & Run I signal strengths  &22 & 22 \\
    \multirow{3}{*}{and decay} & Run II  signal strengths  & 40 & 36 (*)\\
    & Run II, differential distributions \& STXS  & 35 & 71\\
    &  {\bf Total}    & {\bf 97} & {\bf 129}\\
    \midrule
    \multirow{2}{*}{Diboson} & LEP-2 &40 & 40 \\
    \multirow{2}{*}{production} & LHC & 30& 41 \\
    &  {\bf Total}    & {\bf 70} & {\bf 81}  \\
      \midrule
    \multirow{1}{*}{EWPOs} & LEP-2 & - & 44  \\
    \bottomrule
   Baseline dataset     & {\bf Total}      & {\bf 317} & {\bf 445}\\
\bottomrule
  \end{tabular}
  \caption{\small The number of data points $n_{\rm dat}$ in the baseline dataset
    for each of the categories of processes considered in this work.
    We compare the values in the current analysis ({\sc\small SMEFiT3.0}) with those with its predecessor {\sc\small SMEFiT2.0}~\cite{Ethier:2021bye,Giani:2023gfq}.
    Recall that, in  {\sc\small SMEFiT2.0}, the EWPOs were accounted for in an approximate manner. 
    (*) 4 data points from the CMS Run II Higgs dataset were removed because they cannot be described by a multi-Gaussian distribution.
 \label{eq:table_dataset_overview}
}
\end{table}


Table~\ref{eq:table_dataset_overview} indicates the number of data points $n_{\rm dat}$ in the baseline dataset for each of the categories of processes considered here.
We compare these values in the current analysis ({\sc\small SMEFiT3.0}) with those with its predecessor {\sc\small SMEFiT2.0}~\cite{Ethier:2021bye,Giani:2023gfq}.
From  this overview, one observes that the current analysis has $n_{\rm dat}=445$, up from $n_{\rm dat}=317$ in the previous fit.
The processes that dominate this increase in input cross-sections are top quark production ($n_{\rm dat}$ increasing by 41 points), Higgs production (by 32) and the EWPOs, which in SMEFiT2.0 were accounted for in an approximate manner.

We briefly describe the main features of the new Higgs boson, diboson and top quark datasets included here and the settings of the associated theory calculations.
These are summarised in Table~\ref{eq:input_datasets_higgs}, where we indicate the naming convention, the centre-of-mass energy and integrated luminosity, details on the production and decay channels involved, the fitted observables, the number of data points and the corresponding publication reference. 

\begin{landscape}
\begin{table}[t]
  \centering
  \scriptsize
   \renewcommand{\arraystretch}{1.3}
  \begin{tabular}{C{4cm}|c|c|c|c|c|c}
   \toprule
 Dataset   &  $\sqrt{s}$ (TeV) & $\mathcal{L}$ $\lp\rm{fb}^{-1}\rp$  & Info  &  Observables  & $n_{\rm dat}$ & ref.    \\
\midrule
  \multirow{3}{*}{ {\tt ATLAS\_STXS\_RunII\_13TeV\_2022} }  &\multirow{3}{*}{13} &\multirow{3}{*}{139}  &\multirow{3}{*}{
  $gg$F, VBF, $Vh$, $t\bar{t}h$, $th$} & $d\sigma/dp_T^h$    &   \multirow{3}{*}{36}    &  \multirow{3}{*}{ \cite{ATLAS:2022vkf} } \\
     &  &  &  & $d\sigma/dm_{jj}$&     &    \\
     &  &  &  & $d\sigma/dp_T^V$&     &    \\
     \midrule
     \midrule
   {\tt CMS\_WZ\_pTZ\_13TeV\_2022}  & 13 & 137  &
  $WZ$, fully leptonic & $1/\sigma d\sigma/dp_T^Z$    &    10   &   \cite{CMS:2021icx}  \\
     \midrule
     {\tt     CMS\_tt\_13TeV\_ljets\_inc}  & 13 & 137&
   $\ell+\rm{jets}$ & $\sigma(t\bar{t})$    &    1   &   \cite{CMS:2021vhb}  \\
   {\tt     CMS\_tt\_13TeV\_Mtt}  & 13 & 137&
   $\ell+\rm{jets}$ & $1/\sigma d\sigma/dm_{t\bar{t}}$    &    14   &   \cite{CMS:2021vhb}  \\
    \midrule
    \midrule
   {\tt     CMS\_tt\_13TeV\_asy}  & 13 & 138 &
  $\ell+\rm{jets}$ & $A_C$    &    3   &   \cite{CMS:2022ged}  \\
   {\tt     ATLAS\_tt\_13TeV\_asy\_2022}  & 13 & 139 &
  $\ell+\rm{jets}$ & $A_C$    &    5   &   \cite{ATLAS:2022waa}  \\
       \midrule
     \midrule
   {\tt     ATLAS\_Whel\_13TeV}  & 13 & 139 &
  $W$-helicity fraction & $F_0, F_L$    &    2   &   \cite{ATLAS:2022rms}  \\
       \midrule
    \midrule
   {\tt    ATLAS\_ttZ\_13TeV\_pTZ}  & 13 & 139 & $t\bar{t} Z$
   & $d\sigma/dp_T^Z$    &    7   &   \cite{ATLAS:2021fzm}  \\
      \midrule
    \midrule
   {\tt    ATLAS\_tta\_8TeV}  & 8 & 20.2 &
  Inclusive & $\sigma\lp t\bar{t}\gamma\rp$    &    1   &   \cite{ATLAS:2017yax}  \\
   {\tt    CMS\_tta\_8TeV}  & 8 &  19.7&
  Inclusive & $\sigma\lp t\bar{t}\gamma\rp$    &    1   &   \cite{CMS:2017tzb}  \\
        \midrule
    \midrule
   {\tt    ATLAS\_tttt\_13TeV\_slep\_inc}  & 13 & 139 &
  single-lepton & $\sigma_{\rm tot}\lp t\bar{t}t\bar{t}\rp$    &    1   &   \cite{ATLAS:2021kqb}  \\
   {\tt   CMS\_tttt\_13TeV\_slep\_inc}  & 13 & 35.8 &
 single-lepton & $\sigma_{\rm tot}\lp t\bar{t}t\bar{t}\rp$    &    1   &   \cite{CMS:2019jsc}  \\
 {\tt ATLAS\_tttt\_13TeV\_2023}&13&$139$&multi-lepton&$\sigma_{\rm tot}(t\bar{t}t\bar{t})$&1&\cite{ATLAS:2023ajo}\\
{\tt CMS\_tttt\_13TeV\_2023}&13&$139$&same-sign or multi-lepton&$\sigma_{\rm tot}(t\bar{t}t\bar{t})$&1&\cite{CMS:2023ftu}\\
   {\tt   CMS\_ttbb\_13TeV\_dilepton\_inc}  & 13 & 35.9 &
  dilepton & $\sigma_{\rm tot}\lp t\bar{t}b\bar{b}\rp$    &    1   &   \cite{CMS:2020grm}  \\
   {\tt     CMS\_ttbb\_13TeV\_ljets\_inc}  & 13 & 35.9 &
 $\ell+\rm{jets}$ & $\sigma_{\rm tot}\lp t\bar{t}b\bar{b}\rp$    &    1   &   \cite{CMS:2020grm}  \\
     \midrule
    \midrule
   {\tt     ATLAS\_t\_sch\_13TeV\_inc}  & 13 & 139 &
  $s$-channel & $\sigma_{\rm tot}\lp t + \bar{t}\rp$    &    1   &   \cite{ATLAS:2022wfk}  \\
      \midrule
    \midrule
   {\tt     CMS\_tZ\_13TeV\_pTt}  & 13 & 138 &
  dilepton& $d\sigma_{\rm fid}(tZj)/dp_T^t$    &    3   &   \cite{CMS:2021ugv}  \\
   {\tt     CMS\_tW\_13TeV\_slep\_inc}  & 13 & 36 &
 single-lepton & $\sigma_{\rm tot}(tW)$    &    1   &   \cite{CMS:2021vqm}  \\
    \bottomrule
    \end{tabular}
  \caption{\footnotesize Description of the new Higgs boson production and decay, diboson and top-quark datasets added to the global SMEFT fit presented in this work.
For each dataset, we indicate the naming convention, its center-of-mass energy and integrated luminosity, details on the production and decay channels involved, the fitted observables, the number of data points $n_{\rm dat}$, and the publication reference.
  See Sect.~\ref{sec:hl_lhc_projections} for a discussion of which Run II datasets enter the projections for the HL-LHC.
     \label{eq:input_datasets_higgs}
  }
\end{table}
\end{landscape}

\myparagraph{Higgs production and decay}
We include the recent Simplified Template Cross Section (STXS) measurements from ATLAS~\cite{ATLAS:2022vkf}, based on the full Run II luminosity.
All relevant production modes accessible at the LHC (Run II) are considered: $gg$F, VBF, $Vh$, $t\bar{t}h$, and $th$, each of them in all available decay modes.
This Higgs production and decay dataset, which comes with the detailed breakdown of correlated systematic errors (both experimental and theoretical), adds $n_{\rm dat}=36$ data points to the global fit dataset.
The SM cross-sections are taken from the same ATLAS publication~\cite{ATLAS:2022vkf} while we evaluate the linear and quadratic EFT cross-sections using {\sc\small mg5\_aMC@NLO}~\cite{Alwall:2014hca} interfaced to {\sc\small SMEFT@NLO}~\cite{Degrande:2020evl}, with consistent settings with the rest of the observables considered in the fit.
This Higgs dataset is one of the inputs for the most extensive EFT interpretation of their data carried out by ATLAS to date~\cite{ATL-PHYS-PUB-2022-037,ATLAS:2024lyh}. 

\myparagraph{Diboson production}
We include the CMS measurement of the $p_T^Z$ differential distribution in $WZ$ production at $\sqrt{s}=13$ TeV presented in~\cite{CMS:2021icx} and based
on the full Run II luminosity of $\mathcal{L}=137$ fb$^{-1}$.
The measurement is carried out in the 
fully leptonic final state ($\ell^+\ell^- \ell' \bar{\nu}_{\ell'}$) and consists of $n_{\rm dat}=11$ data points. 
The SM theory calculations include NNLO QCD and NLO electroweak corrections and are taken from~\cite{CMS:2021icx}, while the same settings as for Higgs production are employed for the EFT cross-sections. 

\myparagraph{Top quark production}
As mentioned above, here
 we consider the same top quark production datasets entering the analysis of~\cite{Kassabov:2023hbm}, in most cases corresponding to the full Run II integrated luminosity and extended when required to measurements that have become available after the release of that analysis.
As listed in the dataset overview of Table~\ref{eq:input_datasets_higgs}, 
we include the normalised differential $m_{t\bar{t}}$ distribution from CMS in the lepton+jets final state~\cite{CMS:2021vhb}; the charge asymmetries $A_C$ from ATLAS and CMS in the $\ell$+jets final state~\cite{CMS:2022ged,ATLAS:2022waa}; 
the $W$ helicity fractions from ATLAS~\cite{ATLAS:2022rms}; 
the $p_T^Z$ distribution in $t\bar{t}Z$
associated production from ATLAS~\cite{ATLAS:2021fzm}; the inclusive
$t\bar{t}\gamma$ cross-sections from ATLAS and CMS~\cite{ATLAS:2017yax,CMS:2017tzb}; 
the four-top cross-sections from ATLAS and CMS in the single-lepton and multi-lepton final states~\cite{ATLAS:2021kqb,CMS:2019jsc,ATLAS:2023ajo,CMS:2023ftu}; the $t\bar{t}b\bar{b}$ cross-sections from CMS in the dilepton and $\ell$+jets channels~\cite{CMS:2020grm,CMS:2020grm};
the $s$-channel single-top cross-section from ATLAS~\cite{ATLAS:2022wfk}; and finally the single-top associated production cross-sections for $tZ$ and $tW$ from CMS~\cite{CMS:2021ugv,CMS:2021vqm}.
In all cases, state-of-the-art SM and EFT cross-sections from~\cite{Kassabov:2023hbm} are used,  extended whenever required to the additional directions in the EFT parameter space considered here. 

\subsection{Results}
\label{subsec:impact_ewpos}
\label{sec:impact_newdata}

We now present the results of {\sc\small SMEFiT3.0}, which provides the baseline for the subsequent studies with HL-LHC, as well as for the FCC-ee projections introduced later in Chapter \ref{chap:smefit_fcc}. 
We study its main properties, including the data versus theory agreement and its consistency with the SM expectations.
We assess the stability of the results with respect to the order in the EFT expansion adopted (linear versus quadratic), compare individual (one parameter) versus global (marginalised) bounds on the EFT coefficients, map the correlation patterns, quantify the impact of the new data added in comparison with {\sc\small SMEFiT2.0}, and investigate the fit stability with respect to the details of the EWPO implementation. 
All the presented results are based on the Bayesian inference module of {\sc\small SMEFiT} implemented via the Nested Sampling algorithm, which provides our default fitting strategy~\cite{Ethier:2021bye,Giani:2023gfq}.
Table~\ref{table:fit_settings} provides an overview of the input settings adopted for each of the fits discussed in this section.

\begin{table}[htbp]
    \centering
    \footnotesize
     \renewcommand{\arraystretch}{1.5}
     \begin{tabular}
     {l|C{3cm}|C{1.8cm}|C{1cm}|C{3.4cm}}
       \toprule
    Dataset   & EFT coefficients &  Pseudodata?  & EWPOs  & Figs. \& Tables      \\
         \midrule
   {\sc\small  SMEFiT3.0}  & Global (Marginalised)   &  No  &  Exact   & Figs.~\ref{fig:chi2_lin_vs_quad},~\ref{fig:smefit30_marginalised_bounds_bar},~\ref{fig:smefit30_marginalised_bounds_central},~\ref{fig:pull_lin_vs_quad}, \ref{fig:updated-global-fit-quad},~\ref{fig:new-vs-old-quad}, Tables~\ref{eq:chi2-baseline-grouped}, \ref{tab:smefit30_all_bounds_p1}, \ref{tab:smefit30_all_bounds_p2}   \\
   {\sc\small  SMEFiT3.0}  & Individual (one-param)   &  No  &  Exact   &  Tables~\ref{tab:smefit30_all_bounds_p1} \ref{tab:smefit30_all_bounds_p2}  \\
   {\sc\small  SMEFiT2.0}  & Global (Marginalised)   &  No  &  Exact   & Fig.~\ref{fig:new-vs-old-quad}  \\
    {\sc\small  SMEFiT3.0}  & Global (Marginalised)   &  No  &  Approx. & 
  Fig.~\ref{fig:updated-global-fit-quad}\\
  \bottomrule
  \end{tabular}
  \caption{\small Overview of the input settings used in the fits presented in this chapter.
  For each fit, we indicate its dataset, whether the EFT coefficients are extracted from a global fit and then marginalised or instead from one parameter individual fits, whether we use actual data or instead SM pseudo-data is generated, and the treatment of EWPOs (exact versus approximate). 
  In the last column we list the figures and tables in Sects.~\ref{sec:updated_global_fit} where the corresponding results are displayed. 
  For all fits, variants at both $\mathcal{O}\lp \Lambda^{-2}\rp$ and
  $\mathcal{O}\lp \Lambda^{-4}\rp$ in the EFT expansion have been produced. 
  %
  \label{table:fit_settings}
  }
  \end{table}

\begin{figure}[htbp]
    \center
    \includegraphics[height=.9\textheight]{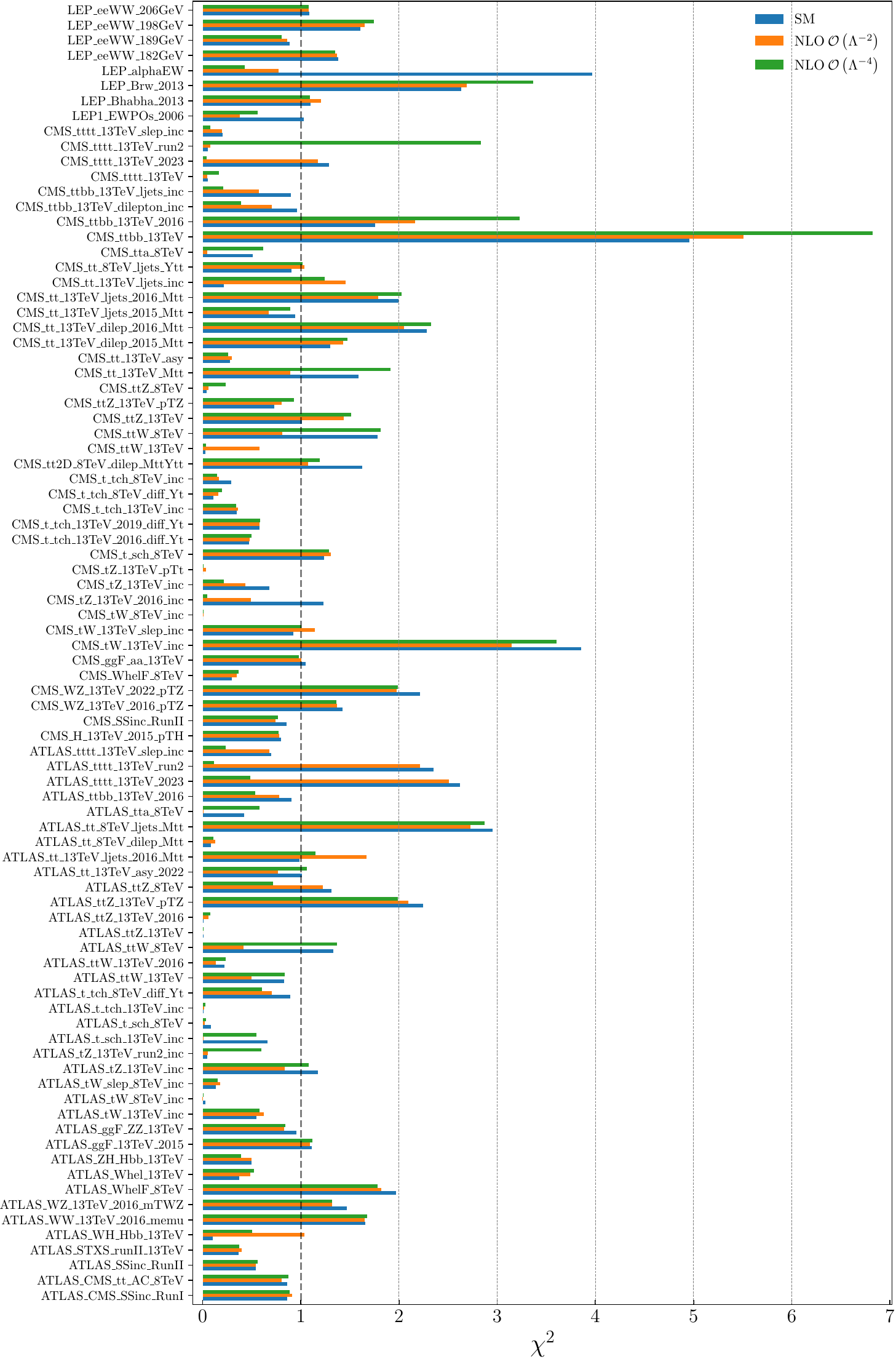}
    \caption{The values of the $\chi^2/n_{\rm dat}$ for the datasets entering the  {\sc\small SMEFiT3.0} analysis. 
    We compare the results based on the SM prediction with the outcome of the SMEFT fits, both at linear and quadratic order in the EFT expansion. 
  See also Table~\ref{eq:chi2-baseline-grouped} for the corresponding results grouped in terms of physical processes.
    }
\label{fig:chi2_lin_vs_quad}
\end{figure}

\myparagraph{Fit quality}
Fig.~\ref{fig:chi2_lin_vs_quad} indicates the values of the $\chi^2/n_{\rm dat}$ for all datasets entering {\sc\small SMEFiT3.0}. 
We compare the values based on the SM predictions with the outcome of the EFT fits, both at linear and quadratic order.
Whenever available, theoretical uncertainties are also included.
The dashed vertical line corresponds to the $\chi^2/n_{\rm dat}=1$ reference.
Note that most of the datasets included in Fig.~\ref{fig:chi2_lin_vs_quad} are composed by just one or a few cross-sections, explaining some of the large fluctuations shown.
The results of Fig.~\ref{fig:chi2_lin_vs_quad} are then tabulated in Table~\ref{eq:chi2-baseline-grouped} at the level of the groups of processes entering the fit. 

\begin{table}[htbp]
  \centering
  \small
   \renewcommand{\arraystretch}{1.40}
   \begin{tabular}
   {l|C{1.0cm}|C{1.4cm}|C{2cm}|C{2cm}}
     \toprule
        \multirow{2}{*}{ Dataset}   & \multirow{2}{*}{$ n_{\rm dat}$} & \multirow{2}{*}{ $\chi^2_{\rm SM}/n_{\rm dat}$} &  $\chi^2_{\rm EFT}/n_{\rm dat}$   & $\chi^2_{\rm EFT}/n_{\rm dat}$     \\
      &   &   & $\mathcal{O}\lp \Lambda^{-2}\rp$ &  $\mathcal{O}\lp \Lambda^{-4}\rp$  \\
        \toprule
 $t\bar{t}$ inclusive  & 115& 1.365 & 1.193   & 1.386  \\
  $t\bar{t}+\gamma$  & 2  & 0.465  & 0.027   & 0.598\\
 $t\bar{t}+V$  & 21  & 1.200  & 1.100   & 1.165   \\
 single-top inclusive & 28  & 0.439   & 0.393   &0.407   \\
 single-top $+V$ &  13 & 0.663  & 0.540   & 0.562  \\
 $t\bar{t}b\bar{b}$ \& $t\bar{t}t\bar{t}$        &  12 & 1.396  & 1.386   & 1.261 \\
 \midrule
 Higgs production \& decay & 129  & 0.687   & 0.692   & 0.676  \\
 \midrule
 Diboson (LEP+LHC)  & 81  & 1.481  & 1.429   & 1.436  \\
 LEP + SLD & 44  & 1.237  &  0.942  & 1.002  \\
 \midrule
 {\bf Total}  & {\bf 445}  &  {\bf 1.087  }   & {\bf 0.992 }  & {\bf 1.048 } \\
\bottomrule
\end{tabular}
\caption{\small Summary of the $\chi^2/n
_{\rm dat}$ results displayed in Fig.~\ref{fig:chi2_lin_vs_quad} in terms of the groups of processes entering the fit. 
\label{eq:chi2-baseline-grouped}
}
\end{table}

The $\chi^2$ values collected in Fig.~\ref{fig:chi2_lin_vs_quad} and Table~\ref{eq:chi2-baseline-grouped} indicate that, for most of the datasets considered here, the SM predictions are in good agreement with the experimental data.
This agreement remains the same, or it is further improved, at the level of the (linear or quadratic) EFT fits.
However, for some datasets, the SM $\chi^2$ turns out to be poor.
In most cases, this happens for datasets containing one or a few cross-section points.
Datasets with a poor $\chi^2$ to the SM include 
{\tt CMS\_ttbb\_13TeV}, {\tt CMS\_tW\_13TeV\_inc}, {\tt ATLAS\_tttt 13TeV\_2023}, {\tt ATLAS\_tt\_8TeV\_ljets\_Mtt}, and \linebreak[4]{\tt ATLAS\_ttZ\_13TeV\_pTZ.}
For these datasets, a counterpart from the complementary experiment is also part of the fit and agreement with the SM is found there, suggesting some tension between the ATLAS and CMS measurements.
See also the discussions in~\cite{Kassabov:2023hbm} for the top quark datasets in light of the covariance matrix decorrelation method~\cite{Kassabov:2022pps}.
This interpretation is supported by the fact that, for these datasets with a poor $\chi^2$ to the SM prediction, accounting for EFT effects does not improve the agreement with the data.
In such cases, the poor $\chi^2$ values may be explained by either internal inconsistencies~\cite{Kassabov:2022pps} or originates from tensions between different measurements of the same process.

Concerning the LEP measurements, good agreement with the SM is observed with the only exception of the electroweak coupling constant $\alpha_{\rm EW}$ and the $W$ branching ratios.
While the $\chi^2$ to the former observable improves markedly once EFT corrections are accounted for, the opposite appears to be true for the LEP $W$-boson branching ratios.
We recall here that, for the $W$ branching fractions in Eq.~(\ref{eq:Wmass_BR}), possible invisible decay channels are not accounted for. 

In Table~\ref{eq:chi2-baseline-grouped}, we present the $\chi^2$ values grouped by physical process, comparing the SM with the best fit parameters found in the SMEFT fits. Notably, the SM demonstrates a commendable per data point $\chi^2=1.087$, a value that further refines to 0.992 and 1.048 for the linear and quadratic EFT fits, respectively, with $n_{\text{eft}}=45$ and 50 parameters in each case. 
While the $\chi^2$ values of the Higgs production and decay dataset are similar in the SM and in the linear and quadratic EFT fits, and likewise for diboson production, more variation is found for the top-quark production datasets, specially for inclusive $t\bar{t}$ production.
In the case of the EWPOs, there is a clear reduction of the $\chi^2$ per data point in the EFT fit as compared to the baseline SM predictions.
It is worth emphasising that these $\chi^2$ values do not  imply that the EFT model offers a superior description to the data. 
Indeed, a thorough hypothesis test mandates normalising the $\chi^2$ by the number of degrees of freedom $n_{\text{dof}} = n_{\text{dat}} - n_{\text{eft}}$. 
In this sense, the SM, having a good $\chi^2$ with less parametric freedom, remains the preferred model to describe the available data.

\begin{figure}[htbp]
  \center
  \includegraphics[height=.9\textheight]{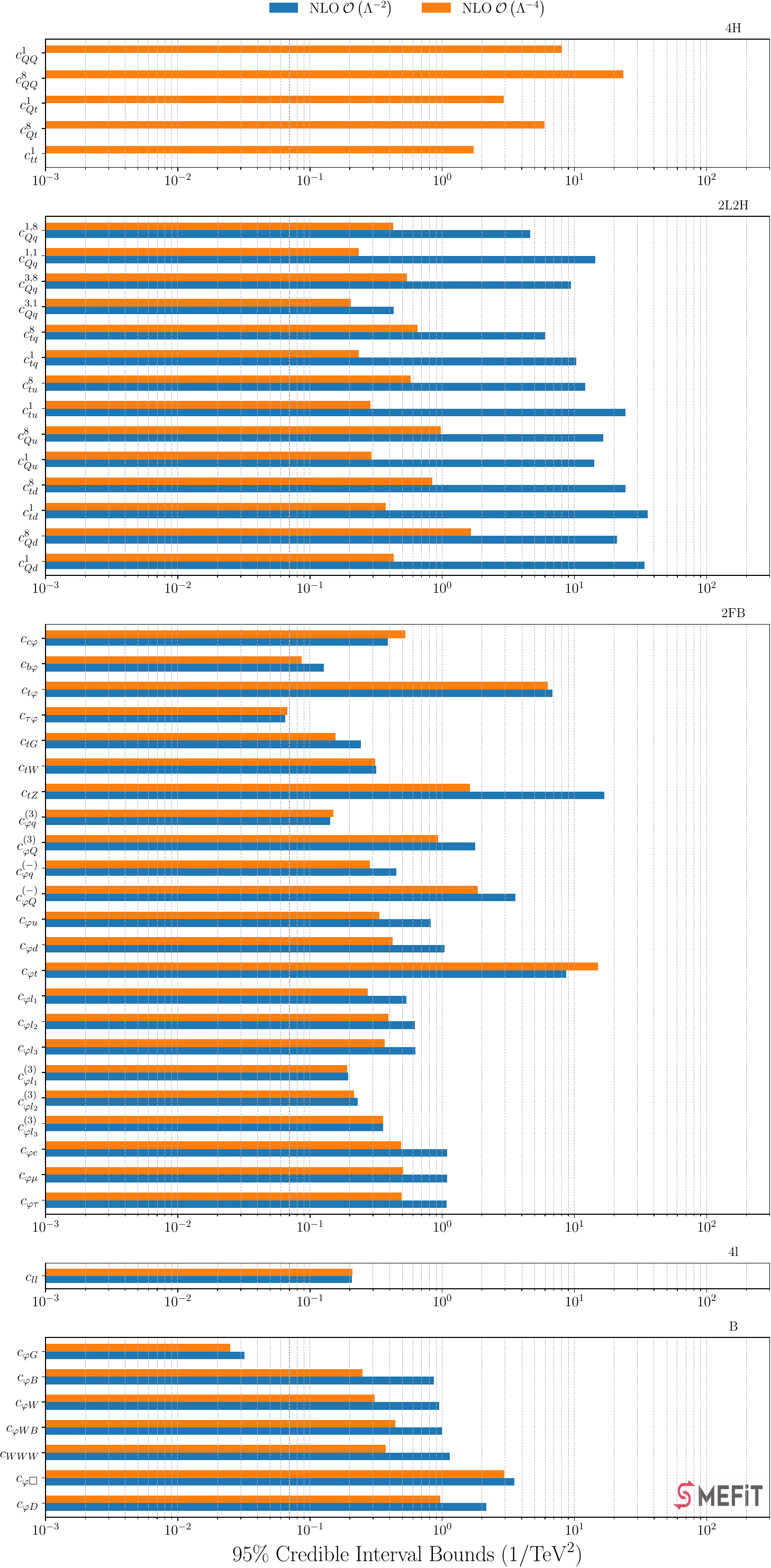}
  
  \caption{The length of the 95\% CI, expressed in units of 1/TeV$^{2}$, for the $n_{\rm eft}=50$ coefficients entering the fit, both for linear and for quadratic (marginalised) analyses. 
  From top to bottom we display the four-heavy quark (except for the linear fit), two-light-two-heavy quark, two-fermion-bosonic, four-lepton, and the purely bosonic coefficients.
  }
\label{fig:smefit30_marginalised_bounds_bar}
\end{figure}

\begin{figure}[htbp]
    \center
    \includegraphics[height=.9\textheight]{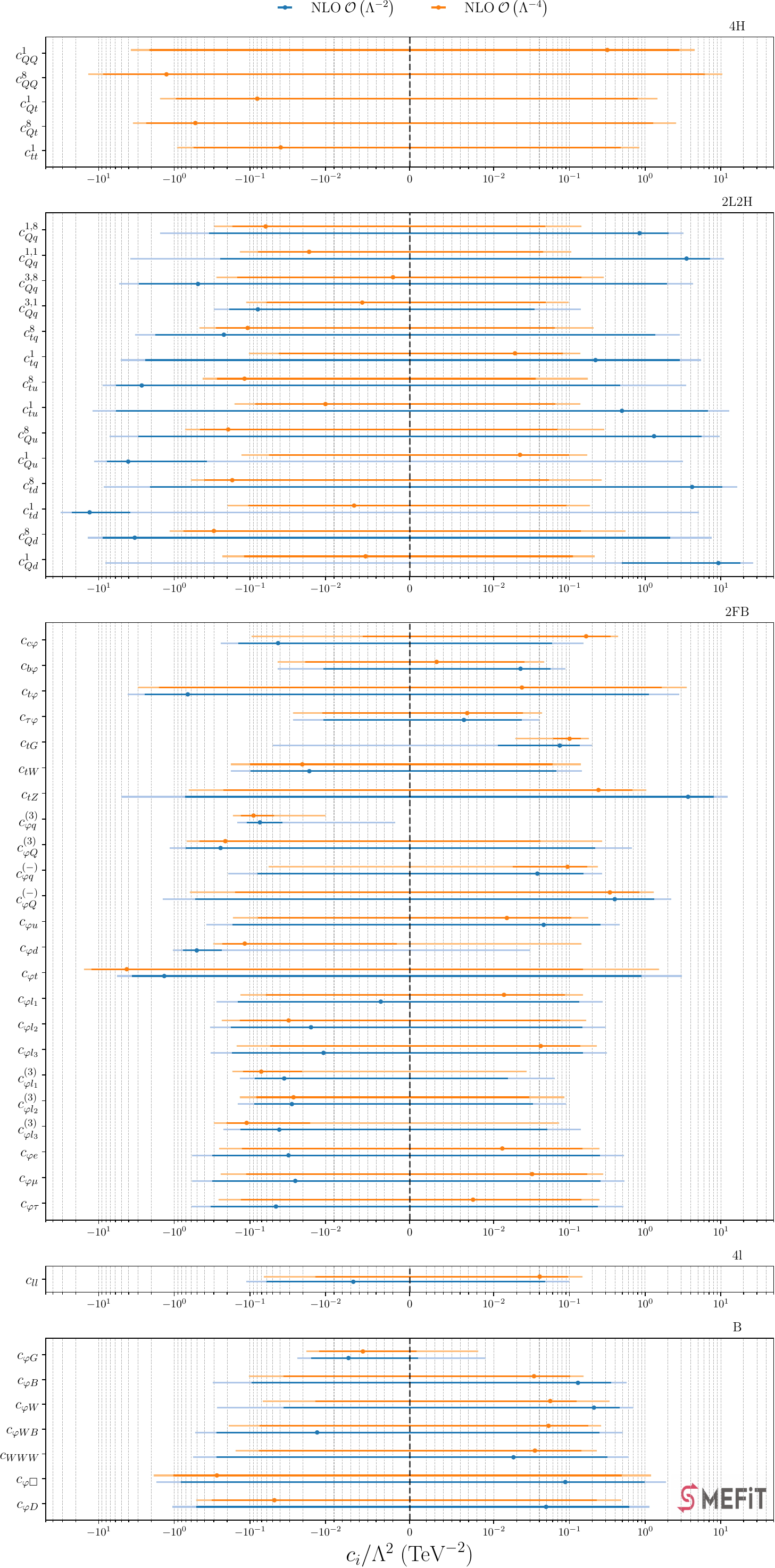}
    \caption{The coefficients $c_i/\Lambda^{2}$ for the same fits as shown in Fig.~\ref{fig:smefit30_marginalised_bounds_bar}, where the thicker (thinner) line indicates the 68\% CI (95\% CI) uncertainties.
    The vertical dashed line indicates the SM expectation for the values of these coefficients.
    }
\label{fig:smefit30_marginalised_bounds_central}
\end{figure}

\myparagraph{Constraints on the EFT operators}
Figs.~\ref{fig:smefit30_marginalised_bounds_bar} and~\ref{fig:smefit30_marginalised_bounds_central} display the results of {\sc\small SMEFiT3.0} at the level of the $n_{\rm eft}=45~(50)$ operators entering the analysis at the linear (quadratic) EFT level.
Fig.~\ref{fig:smefit30_marginalised_bounds_central} shows the best-fit values and the  68\% and 95\% credible intervals (CI), both for the linear and for the quadratic baseline fits.
The reported bounds are extracted from a global fit with all coefficients being varied simultaneously, and then the resultant posterior distributions are marginalised down to individual coefficients.
From top to bottom, we display
the four-heavy quark, two-light-two-heavy quark, two-fermion, four-lepton, and purely bosonic
coefficients.
The corresponding information on the magnitude of the 95\% credible interval is provided in Fig.~\ref{fig:smefit30_marginalised_bounds_bar}.

The bounds displayed in Figs.~\ref{fig:smefit30_marginalised_bounds_bar} and~\ref{fig:smefit30_marginalised_bounds_central} are also collected in Tables~\ref{tab:smefit30_all_bounds_p1} and \ref{tab:smefit30_all_bounds_p2}, where  for completeness we also include the individual bounds obtained from one parameter fits to the data (with all other coefficients set to zero).
It is worth noting that for some operators at the quadratic EFT level the 95\% CI bounds are disjoint, indicating the presence of degenerate solutions. 
For the four-heavy operators, in the linear fit one can only display the individual bounds, since in this sector the SMEFT displays flat directions (for the available data) unless quadratic corrections are included. 

\begin{table}[htb]
 \centering
  \scriptsize
   \renewcommand{\arraystretch}{1.24}
   \begin{tabular}{l|C{0.8cm}|C{2.2cm}|C{2.2cm}|C{2.2cm}|C{2.2cm}}
     \multirow{2}{*}{Class}   &  \multirow{2}{*}{DoF}
     &  \multicolumn{2}{c|}{ 95\% CI bounds, $\mathcal{O}\lp \Lambda^{-2}\rp$} &
     \multicolumn{2}{c}{95\% CI bounds, $\mathcal{O}\lp \Lambda^{-4}\rp$,} \\ 
 &  & Individual & Marginalised &  Individual & Marginalised  \\ \toprule
 \multirow{5}{*}{4H}
 & $c_{QQ}^{1}$            & [1.648, 24.513]     & \textemdash       & [-2.403, 2.153]                       & [-3.765, 4.487]  \\ \cline{2-6}
 & $c_{QQ}^{8}$            & [3.343, 63.182]     & \textemdash       & [-7.196, 6.533]                       & [-13.586, 10.491]  \\ \cline{2-6}
 & $c_{Qt}^{1}$            & [-509.511, 211.968] & \textemdash       & [-1.945, 1.958]                       & [-1.546, 1.455]  \\ \cline{2-6}
 & $c_{Qt}^{8}$            & [1.632, 21.393]     & \textemdash       & [-4.415, 3.607]                       & [-3.500, 2.549]  \\ \cline{2-6}
 & $c_{tt}^{1}$            & [0.768, 12.075]     & \textemdash       & [-1.201, 1.070]                       & [-0.919, 0.836]  \\ \cline{2-6}
 \hline
 \multirow{14}{*}{2L2H}
 & $c_{Qq}^{1,8}$          & [-0.363, 0.201]     & [-1.547, 3.207]   & [-0.292, 0.141]                       & [-0.296, 0.144]    \\ \cline{2-6}
 & $c_{Qq}^{1,1}$          & [-1.154, 0.096]     & [-3.820, 11.011]  & [-0.150, 0.096]                       & [-0.136, 0.105]    \\ \cline{2-6}
 & $c_{Qq}^{3,8}$          & [-1.285, 0.417]     & [-5.313, 4.288]   & [-0.355, 0.229]                       & [-0.278, 0.282]    \\ \cline{2-6}
 & $c_{Qq}^{3,1}$          & [-0.128, 0.106]     & [-0.301, 0.141]   & [-0.092, 0.080]                       & [-0.112, 0.097]    \\ \cline{2-6}
 & $c_{tq}^{8}$            & [-0.639, 0.236]     & [-3.270, 2.885]   & [-0.459, 0.180]                       & [-0.467, 0.208]    \\ \cline{2-6}
 & $c_{tq}^{1}$            & [0.176, 1.188]      & [-5.092, 5.481]   & [-0.073, 0.160]                       & [-0.104, 0.139]    \\ \cline{2-6}
 & $c_{tu}^{8}$            & [-0.675, 0.247]     & [-8.866, 3.490]   & [-0.439, 0.179]                       & [-0.422, 0.175]    \\ \cline{2-6}
 & $c_{tu}^{1}$            & [-1.622, 0.214]     & [-12.084, 12.836] & [-0.178, 0.126]                       & [-0.159, 0.139]    \\ \cline{2-6}
 & $c_{Qu}^{8}$            & [-1.567, 0.076]     & [-7.200, 9.684]   & [-0.702, 0.211]                       & [-0.715, 0.289]    \\ \cline{2-6}
 & $c_{Qu}^{1}$            & [0.210, 1.596]      & [-11.379, 3.183]  & [-0.101, 0.193]                       & [-0.129, 0.171]    \\ \cline{2-6}
 & $c_{td}^{8}$            & [-1.677, 0.206]     & [-8.511, 16.583]  & [-0.685, 0.244]                       & [-0.603, 0.266]    \\ \cline{2-6}
 & $c_{td}^{1}$            & [-3.955, -0.251]    & [-31.597, 5.147]  & [-0.234, 0.172]                       & [-0.198, 0.186]    \\ \cline{2-6}
 & $c_{Qd}^{8}$            & [-3.147, -0.091]    & [-13.997, 7.530]  & [-1.108, 0.326]                       & [-1.158, 0.549]    \\ \cline{2-6}
 & $c_{Qd}^{1}$            & [0.840, 3.755]      & [-8.140, 26.827]  & [-0.149, 0.242]                       & [-0.230, 0.216]    \\ \cline{2-6}
 \hline
\end{tabular}
\caption{ The 95\% CI bounds on the four-fermion
     EFT coefficients considered in this analysis. In the first column, we split the coefficients into classes depending on whether either all fermion fields correspond to the third generation (4H), or only two with the remaining ones corresponding to the first two generations (2L2H). 
     The reported bounds correspond to $\Lambda=1$ TeV and can thus be rescaled for any other value of $\Lambda$.
     We present results both for linear EFT fits, $\mathcal{O}\lp \Lambda^{-2}\rp$, and for quadratic EFT fits, $\mathcal{O}\lp \Lambda^{-4}\rp$.
     In each case, we indicate both individual bounds (from one-parameter fits) and marginalised bounds (from a simultaneous determination of the full set of $n_{\rm eft}$ coefficients).
     }
\label{tab:smefit30_all_bounds_p1}
\end{table}

\begin{table}[htb]
     \centering
      \scriptsize
       \renewcommand{\arraystretch}{1.24}
       \begin{tabular}{l|C{0.8cm}|C{2.2cm}|C{2.2cm}|C{2.2cm}|C{2.2cm}}
         \multirow{2}{*}{Class}   &  \multirow{2}{*}{DoF}
         &  \multicolumn{2}{c|}{ 95\% CI bounds, $\mathcal{O}\lp \Lambda^{-2}\rp$} &
         \multicolumn{2}{c}{95\% CI bounds, $\mathcal{O}\lp \Lambda^{-4}\rp$,} \\ 
     &  & Individual & Marginalised &  Individual & Marginalised  \\ \toprule
     \multirow{23}{*}{2FB}
     & $c_{c \varphi}$         & [-0.022, 0.120]     & [-0.243, 0.154]   & [-0.000, 0.373]                       & [-0.094, 0.442]   \\ \cline{2-6}
     & $c_{b \varphi}$         & [-0.007, 0.040]     & [-0.043, 0.088]   & [-0.403, -0.360] $\cup$ [-0.008, 0.036]                       & [-0.043, 0.046]   \\ \cline{2-6}
     & $c_{t \varphi}$         & [-1.199, 0.327]     & [-4.142, 2.831]   & [-1.168, 0.333]                       & [-3.035, 3.527]   \\ \cline{2-6}
     & $c_{\tau \varphi}$      & [-0.027, 0.036]     & [-0.027, 0.040]   & [-0.024, 0.041]  $\cup$ [0.398, 0.462]                      & [-0.027, 0.043]   \\ \cline{2-6}
     & $c_{tG}$                & [0.004, 0.084]      & [-0.050, 0.199]   & [0.003, 0.080]                  & [0.019, 0.180]    \\ \cline{2-6}
     & $c_{tW}$                & [-0.087, 0.029]     & [-0.180, 0.147]   & [-0.082, 0.029]                 & [-0.177, 0.141]   \\ \cline{2-6}
     & $c_{tZ}$                & [-0.034, 0.102]     & [-4.999, 12.276]  & [-0.038, 0.094]                 & [-0.645, 1.027]   \\ \cline{2-6}
     & $c_{\varphi q}^{(3)}$   & [-0.015, 0.012]     & [-0.147, -0.002]  & [-0.015, 0.012]                       & [-0.166, -0.010]  \\ \cline{2-6}
     & $c_{\varphi Q}^{(3)}$   & [-0.016, 0.023]     & [-1.155, 0.665]   & [-0.016, 0.023]                       & [-0.685, 0.271]   \\ \cline{2-6}
     & $c_{\varphi q}^{(-)}$   & [-0.121, 0.119]     & [-0.193, 0.269]   & [-0.118, 0.119]                       & [-0.056, 0.239]   \\ \cline{2-6}
     & $c_{\varphi Q}^{(-)}$   & [-0.031, 0.046]     & [-1.427, 2.224]   & [-0.031, 0.047]                       & [-0.620, 1.292]   \\ \cline{2-6}
     & $c_{\varphi u}$         & [-0.071, 0.081]     & [-0.375, 0.461]   & [-0.077, 0.079]                       & [-0.168, 0.177]   \\ \cline{2-6}
     & $c_{\varphi d}$         & [-0.140, 0.071]     & [-1.038, 0.030]   & [-0.137, 0.072]                       & [-0.303, 0.143]   \\ \cline{2-6}
     & $c_{\varphi t}$         & [-2.855, 1.036]     & [-5.750, 3.084]   & [-4.000, 0.872]                       & [-15.638, 1.532]  \\ \cline{2-6}
     & $c_{\varphi l_1}$       & [-0.009, 0.012]     & [-0.276, 0.273]   & [-0.008, 0.012]                       & [-0.133, 0.150]   \\ \cline{2-6}
     & $c_{\varphi l_2}$       & [-0.031, 0.017]     & [-0.334, 0.302]   & [-0.030, 0.017]                       & [-0.237, 0.166]   \\ \cline{2-6}
     & $c_{\varphi l_3}$       & [-0.035, 0.025]     & [-0.329, 0.311]   & [-0.034, 0.025]                       & [-0.150, 0.231]   \\ \cline{2-6}
     & $c_{\varphi l_1}^{(3)}$ & [-0.015, 0.009]     & [-0.136, 0.064]   & [-0.015, 0.009]                       & [-0.170, 0.027]   \\ \cline{2-6}
     & $c_{\varphi l_2}^{(3)}$ & [-0.031, 0.002]     & [-0.146, 0.089]   & [-0.031, 0.002]                       & [-0.137, 0.085]   \\ \cline{2-6}
     & $c_{\varphi l_3}^{(3)}$ & [-0.039, 0.017]     & [-0.225, 0.141]   & [-0.039, 0.017]                       & [-0.298, 0.073]   \\ \cline{2-6}
     & $c_{\varphi e}$         & [-0.025, 0.001]     & [-0.583, 0.527]   & [-0.025, 0.001]                       & [-0.254, 0.248]   \\ \cline{2-6}
     & $c_{\varphi \mu}$       & [-0.021, 0.039]     & [-0.582, 0.533]   & [-0.021, 0.038]                         & [-0.245, 0.277]   \\ \cline{2-6}
     & $c_{\varphi \tau}$      & [-0.045, 0.024]     & [-0.597, 0.512]   & [-0.045, 0.024]                          & [-0.261, 0.248]   \\ \cline{2-6}
     \hline
     \multirow{1}{*}{4l}
     & $c_{ll}$                & [-0.008, 0.037]     & [-0.112, 0.100]   & [-0.008, 0.037]                       & [-0.066, 0.149]   \\ \cline{2-6}
     \hline
     \multirow{7}{*}{B}
     & $c_{\varphi G}$         & [-0.001, 0.005]     & [-0.024, 0.009]   & [-0.002, 0.005]                       & [-0.018, 0.008]   \\ \cline{2-6}
     & $c_{\varphi B}$         & [-0.005, 0.002]     & [-0.310, 0.573]   & [-0.005, 0.002] $\cup$ [0.085, 0.092]& [-0.103, 0.152]  \\ \cline{2-6}
     & $c_{\varphi W}$         & [-0.018, 0.006]     & [-0.273, 0.707]   & [-0.017, 0.006] $\cup$ [0.282, 0.305] & [-0.067, 0.338] \\ \cline{2-6}
     & $c_{\varphi WB}$        & [-0.007, 0.003]     & [-0.525, 0.504]   & [-0.007, 0.003]                       & [-0.190, 0.263]   \\ \cline{2-6}
     & $c_{WWW}$               & [-0.479, 0.607]     & [-0.565, 0.609]   & [-0.155, 0.197]                       & [-0.156, 0.230]   \\ \cline{2-6}
     & $c_{\varphi \Box}$      & [-0.416, 1.193]     & [-1.715, 1.879]   & [-0.429, 1.141]                       & [-1.856, 1.199]   \\ \cline{2-6}
     & $c_{\varphi D}$         & [-0.027, -0.003]    & [-1.063, 1.149]   & [-0.027, -0.003]                      & [-0.513, 0.483]   \\ \cline{2-6}
     \hline
\end{tabular}
\caption{Same as Table~\ref{tab:smefit30_all_bounds_p1}, now displaying the 95\% CI bounds on the EFT coefficients in the operator classes two-fermion-bosonic (2FB), four-lepton (4l) and purely bosonic (B).
}
\label{tab:smefit30_all_bounds_p2}
\end{table}

\begin{figure}[htbp]
    \center
    \includegraphics[height=.9\textheight]{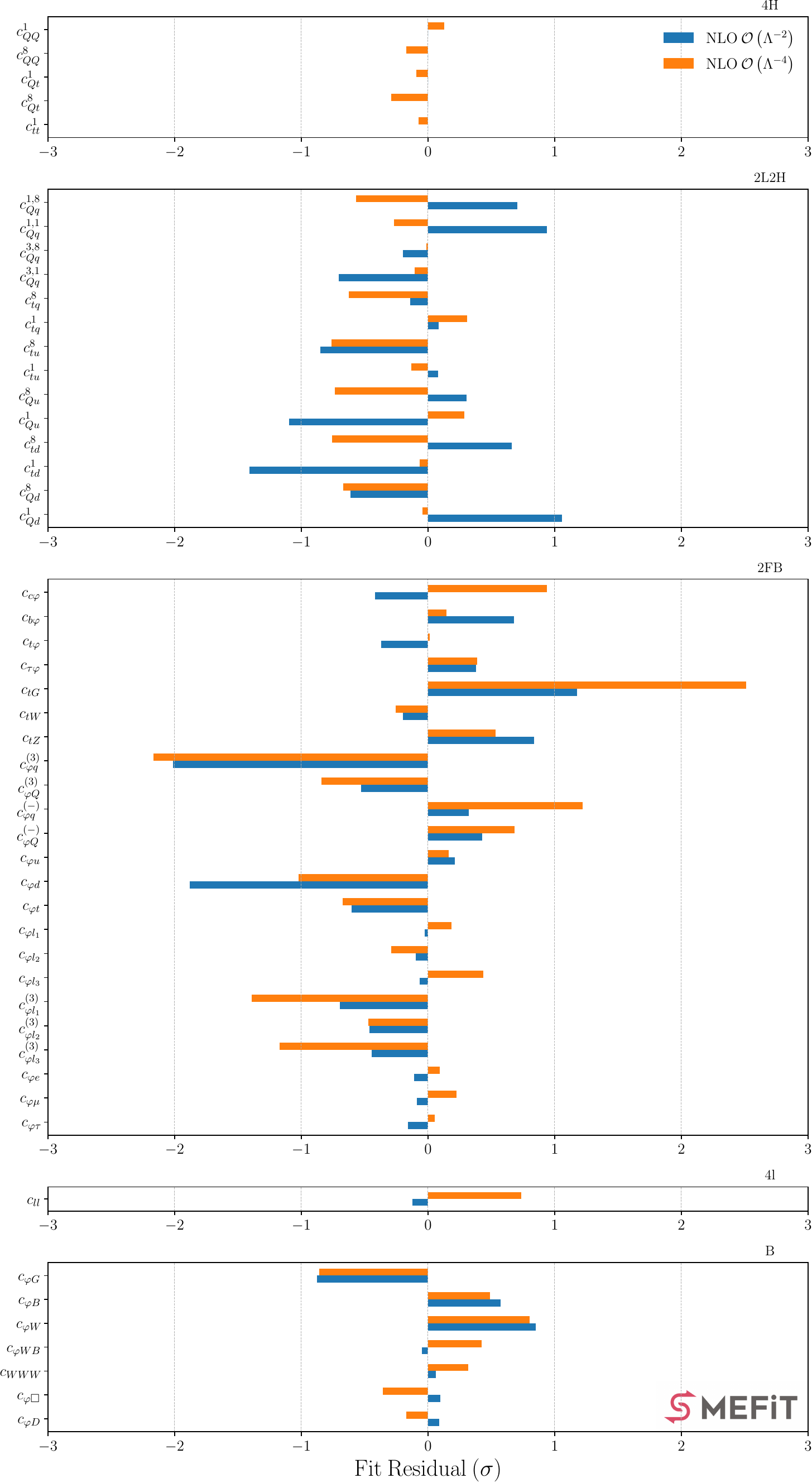}
    \caption{The residuals between the fit results and the SM expectations defined as in Eq.~(\ref{eq:fit_residuals}), for the operators entering Fig.~\ref{fig:smefit30_marginalised_bounds_central} and for both linear and quadratic fits.
    These residuals are normalised to
    the 68\% CI, hence a residual larger than 2 (in absolute value) indicates a coefficient that disagrees with the SM at the 95\% CI.
    }
\label{fig:pull_lin_vs_quad}
\end{figure}

To further quantify the agreement between the SMEFT fit results and the corresponding SM expectations, Fig.~\ref{fig:pull_lin_vs_quad} displays the fit residuals defined as
\begin{equation}
\label{eq:fit_residuals}
P_i \equiv 2\left(\frac{  \left\langle c_i\right\rangle  - c_i^{(\rm SM)}}{
 \lc c_i^{\rm min}, c_i^{\rm max} \rc^{68\%~{\rm CI}}
}\right) \, , \qquad i=1,\ldots, n_{\rm eft} \, ,
\end{equation}
in the same format as that of Fig.~\ref{fig:smefit30_marginalised_bounds_central} for both linear and quadratic fits.
Given that $c_i^{(\rm SM)}=0$ and that Eq.~(\ref{eq:fit_residuals}) is normalised to
    the 68\% CI (which in linear fits correspond to the standard deviation $\sigma$), 
a residual larger than 2 (in absolute value) indicates a coefficient that does not agree with the SM at the 95\% CI.

Several observations can be derived from the inspection of Figs.~\ref{fig:smefit30_marginalised_bounds_bar}~--~\ref{fig:pull_lin_vs_quad} as well as Tables~\ref{tab:smefit30_all_bounds_p1} and \ref{tab:smefit30_all_bounds_p2}.
First, the fit residuals evaluated in Fig.~\ref{fig:pull_lin_vs_quad}, consistently with Fig.~\ref{fig:smefit30_marginalised_bounds_central}, confirm that in general there is a good agreement between the EFT fit results and the experimental data.
For the purely bosonic, four-lepton, four-heavy, and two-light-two-heavy operators, the fit residuals satisfy $|P_i|\lsim 1$,  the only exception being $c_{td}^1$ in the linear fit for which $P_i\sim 1.5$.
Somewhat larger residuals are found for a subset of the two-fermion operators, in particular for the chromomagnetic operator coefficient $c_{tG}$ (only in the quadratic fit), for $c_{\varphi q}^{(3)}$, and for $c_{\varphi d}$ (only in the linear fit).
For these coefficients, the values of $|P_i|$ range between 1.9 and 2.5. 
Below we investigate the origin of these large fit residuals. 

One also finds that quadratic EFT corrections improve the bounds on most operators entering the fit, with a particularly marked impact for the two-light-two-heavy operators.
The only exceptions of this trend are $c_{\varphi t}$, which is poorly constrained to begin with, and the charm Yukawa $c_{c\varphi}$.
For both coefficients, the worse bounds arising in the quadratic fit are explained by the appearance of a second, degenerate solution, as demonstrated by the corresponding posterior distributions displayed in Fig.~\ref{fig:new-vs-old-quad}.
Such degenerate solutions may arise~\cite{Ethier:2021bye} when quadratic corrections become comparable in magnitude with opposite sign to the linear EFT cross-section, a configuration formally equivalent to setting $c_i=0$ and hence reproducing the SM.

As is well known~\cite{Hartland:2019bjb}, quadratic EFT corrections also allow one to bound the four-heavy operator coefficients  $c_{tt}^{1}$, $c_{Qt}^{1}$, $c_{Qt}^{8}$, $c_{QQ}^{1}$, and $c_{QQ}^{8}$.
Within a $\mathcal{O}\lp \Lambda^{-2}\rp$ fit, only two linear combinations of these four-heavy operators can be instead constrained, leaving three flat directions.
These considerations do not hold for one-parameter fits, where the four-heavy operators can be separately constrained. 
From Fig.~\ref{fig:smefit30_marginalised_bounds_bar}, one can also see that for some operators the effects of the quadratic EFT corrections are essentially negligible, indicating that the linear (interference) cross-section dominates the sensitivity.
Specifically, operators for which quadratic corrections are small are the four-lepton operator $c_{\ell\ell}$, the two-fermion operators  $c_{\varphi q}^{3}$, $c_{\varphi \ell_i}^{3}$ (with $i=1,\,2,\,3$), $c_{tW}$, and the tau and top Yukawa couplings, $c_{\tau\varphi}$ and $c_{t\varphi}$ respectively.

The comparison between global (marginalised) and individual (one-parameter) fit results reported in Tables~\ref{tab:smefit30_all_bounds_p1} and \ref{tab:smefit30_all_bounds_p2} indicates that, for the linear EFT fits, one-parameter bounds are always tighter than the marginalised ones.
The differences between individual and marginalised bounds span a wide range of variation, from $c_{WWW}$, which essentially shows no difference, to $c_{\varphi D}$, with individual bounds tighter by two orders of magnitude as compared to the marginalised counterparts.
Concerning the quadratic EFT fits, for the purely bosonic and two-fermion operators the situation is similar as in the linear case, with individual bounds either (much) tighter than the marginalised ones or essentially unchanged (as is the case for $c_{WWW}$
and $c_{\tau \varphi}$, for example).
The situation is somewhat different for the four-heavy and two-light-two-heavy operators. 
For the latter, the marginalised and individual bounds are now similar to each other, as opposed to the linear fit case.
For the four-heavy operators, the marginalised bounds are either similar or a bit broader than the individual ones, except in the case of $c_{Qt}^{(8)}$, whose 95\% CI bounds $[-4.4,3.6]$ (individual) improve to $[-3.5,2.5]$ (marginalised), hence by a factor of approximately 30\%.
In such cases the correlations with other parameters entering the global fit improve the overall sensitivity compared to the one-parameter fits.

\begin{figure}[t]
    \center
    \includegraphics[width=.9\linewidth]{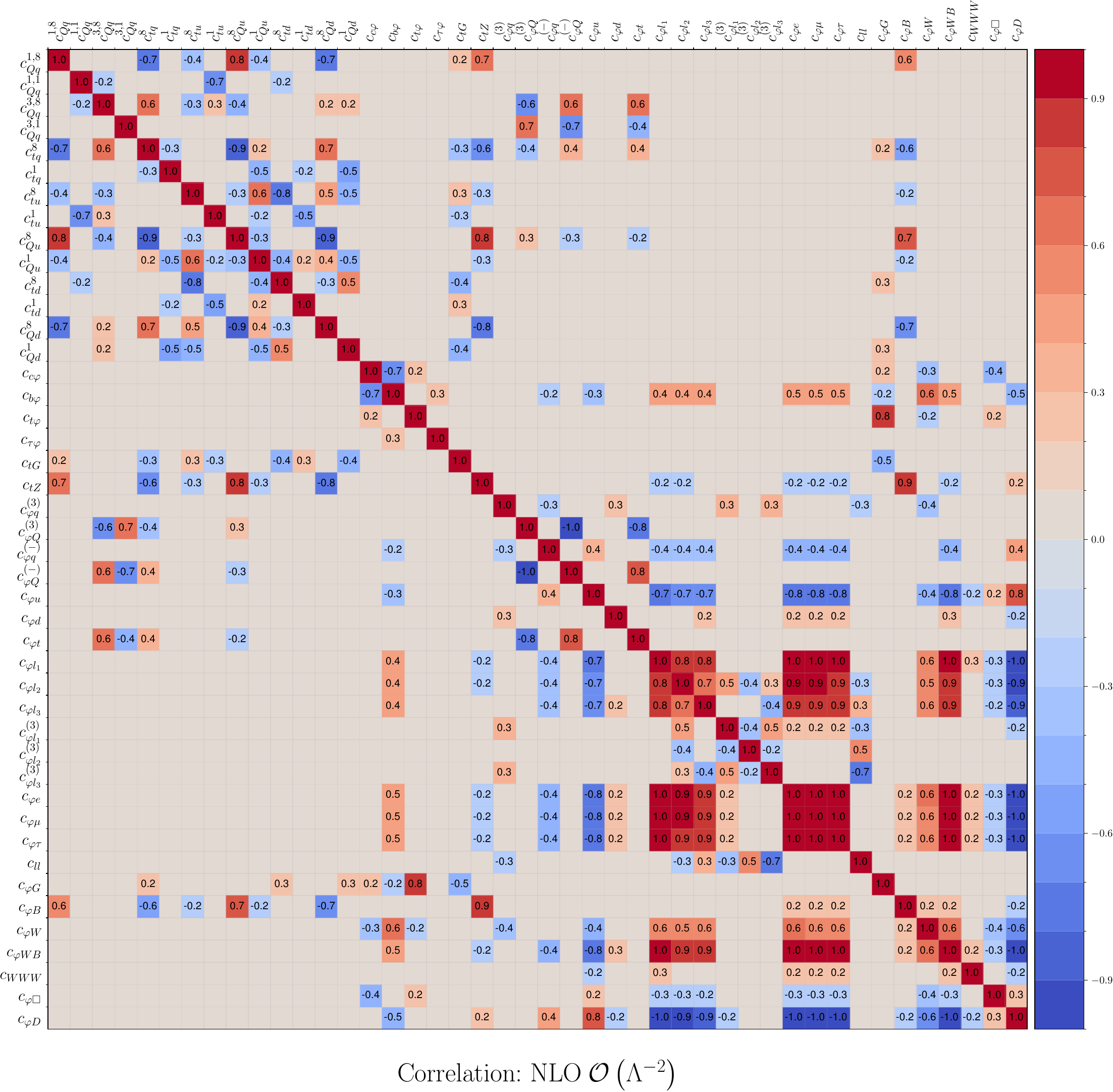}
    \caption{The correlation matrix for the $n_{\rm eff}=45$ coefficients associated to the linear {\sc\small SMEFiT3.0} baseline analysis.
    To facilitate visualisation, the EFT coefficients whose correlation with all other coefficients is $<0.2$ are removed from the plot. 
    }
\label{fig:corr_smefit3_lin}
\end{figure}

When interpreting the bounds on the EFT coefficients and the associated residuals displayed in Figs.~\ref{fig:smefit30_marginalised_bounds_central} and~\ref{fig:pull_lin_vs_quad}, one should recall that in general there are potentially large correlations between them.
To illustrate these, Fig.~\ref{fig:corr_smefit3_lin} displays the entries of the correlation matrix, $\rho_{ij}$, for the $n_{\rm eff}=45$ coefficients associated to the linear {\sc\small SMEFiT3.0} baseline analysis.
To facilitate visualisation, entries with $|\rho_{ij}|<0.2$ (negligible correlations) are not shown in the plot. 
We found significantly weaker correlations in the quadratic fit, especially for the two-light-two-heavy top quark operators as already noticed in~\cite{Hartland:2019bjb}, but also for some purely bosonic and two-fermion operators.
We recall that the correlation patterns in Figs.~\ref{fig:corr_smefit3_lin} depend on the specific fitted dataset, and in particular these patterns change qualitatively once we include the FCC-ee projections in Chapter \ref{chap:smefit_fcc}.

\myparagraph{Coefficients with large residuals}
As mentioned above, the fit residual analysis of Fig.~\ref{fig:pull_lin_vs_quad} indicates that three Wilson coefficients, namely $c_{tG}$ (in the quadratic fit), $c_{\varphi d}$ (in the linear fit), and $c_{\varphi q}^{3}$ (in both cases) do not agree with the SM expectation at the 95\% CI, with pulls of $P_i \simeq +2.5,-1.9$, and $-2.1$ respectively. 
The corresponding individual (one-parameter) analysis of Tables~\ref{tab:smefit30_all_bounds_p1} and \ref{tab:smefit30_all_bounds_p2} indicates that for these coefficients the pulls are $P_i \simeq +2.0,-0.3$, and $-0.2$ respectively, when fitted setting all other operators to zero.
Therefore, the pull on $c_{tG}$ in the quadratic case is somewhat reduced in individual fits but does not go away, while the large pulls on $c_{\varphi q}^{3}$ and $c_{\varphi d}$ completely disappear in the one-parameter fits.
The latter result indicates that the pulls of $c_{\varphi q}^{3}$ and $c_{\varphi d}$ found in the global fit arise as a consequence of the correlations with other fit parameters.

In the case of the chromomagnetic operator coefficient $c_{tG}$, the tension with the SM which arises in the quadratic fit was already present in previous versions of our analysis~\cite{Ethier:2021bye,Giani:2023gfq} and is known to be driven by the CMS top-quark double-differential distributions in $(y_{t\bar{t}},m_{t\bar{t}})$ at $8$~TeV from~\cite{CMS:2017iqf}.
In the context of a linear EFT fit, the obtained residual is consistent with the SM, a finding also in agreement with the independent analysis carried out in~\cite{Kassabov:2023hbm}.
Indeed, if this CMS double-differential $t\bar{t}$ 8 TeV measurement is excluded from the quadratic fit, $c_{tG}$ becomes fully consistent with the SM expectation.
We also note that this dataset, with $\chi^2_{\rm SM}/n_{\rm dat}\simeq 1.7$ for $n_{\rm dat}=16$ points, improves down to $\chi^2_{\rm EFT}/n_{\rm dat}\simeq 1$ after the fit.
Given that $c_{tG}$ modifies the overall normalisation of top-quark pair production, rather than the shape of the distributions, this result may imply that the normalisation of this 2D CMS measurement is in tension with that of other $t\bar{t}$ measurements included in the fit.
All in all, it appears unlikely that this large pull on $c_{tG}$ obtained in the quadratic fit is related to a genuine BSM signal.

Regarding the $c_{\varphi d}$ and $c_{\varphi q}^{(3)}$ Wilson coefficients, we note pulls of approximately $-1.9$ and $-2.1$, respectively, in the linear fit. However, in the quadratic fit, the pull for $c_{\varphi d}$ decreases to around $-1.0$. Notably, the individual constraints are instead entirely consistent with the SM.
This pattern arises from the predominance of LEP data in individual fits, where no deviations from the SM are apparent. However, in a comprehensive global fit, the LEP data exhibit strong inter-coefficient correlations, leading to a notable reduction in their constraining effectiveness.
As is well-established, the complementary nature of LHC diboson measurements to EWPO is crucial to break several of these correlations. For this reason, the LHC diboson data, despite being less precise, can have a surprising impact on the bounds of the EFT coefficients affecting LEP observables. We have confirmed that these measurements are indeed responsible for the observed pulls in the global fit.

\begin{figure}[t]
    \centering
    \includegraphics[width=\linewidth]{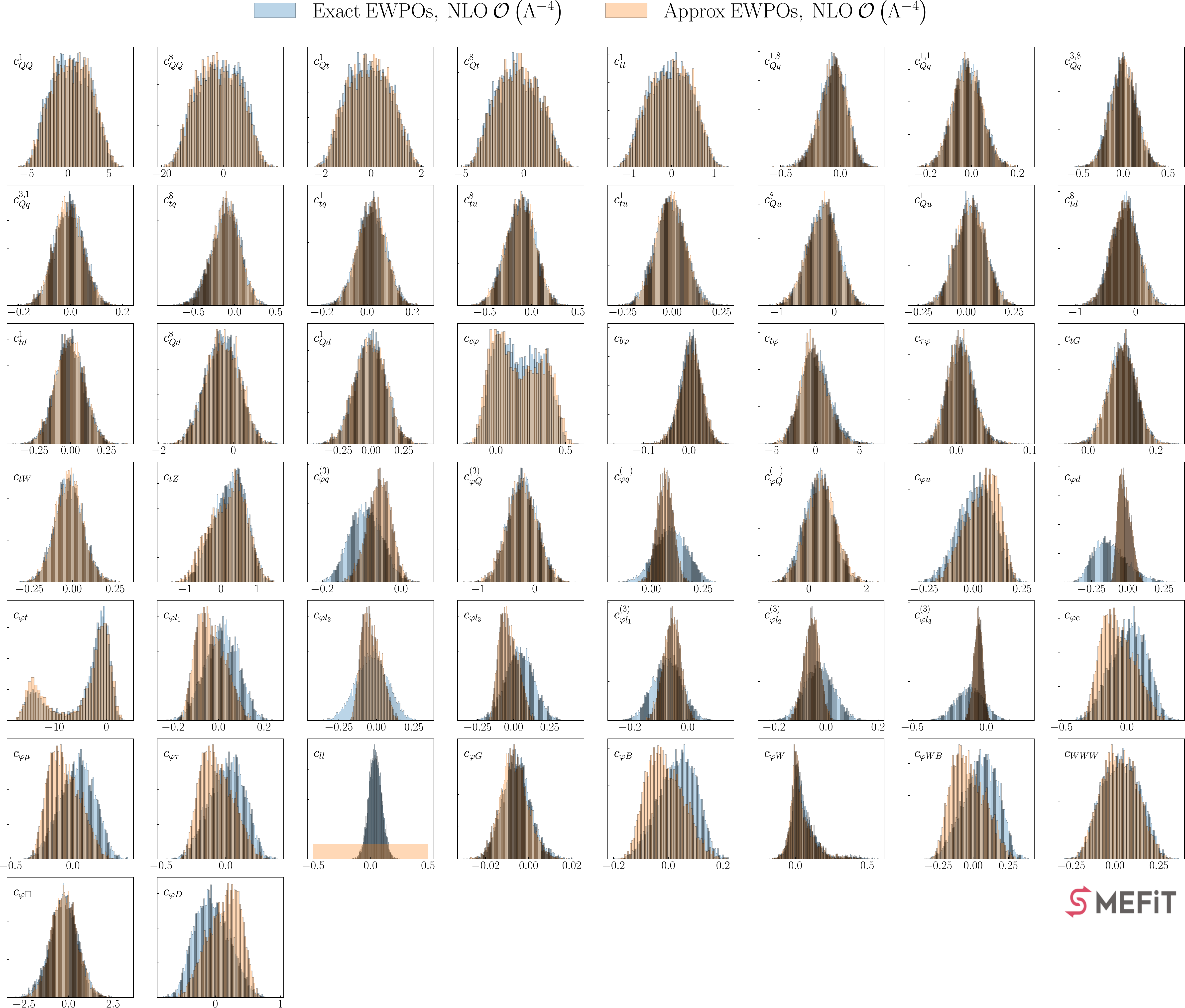}
    \caption{Posterior distributions associated to the $n_{\rm eft}=50$ Wilson coefficients constrained in the 
 {\sc\small SMEFiT3.0} global analysis, carried out at $\mathcal{O}\lp \Lambda^{-4}\rp$ in the EFT expansion.
 A simultaneous determination of all coefficients is performed and then one marginalises for individual degrees of freedom. 
 The baseline results, performed with the exact implementation of the EWPOs, are compared with the approximated implementation used in~\cite{Ethier:2021bye}.
    }
    \label{fig:updated-global-fit-quad}
\end{figure}

\begin{figure}[t]
    \centering
    \includegraphics[width=\linewidth]{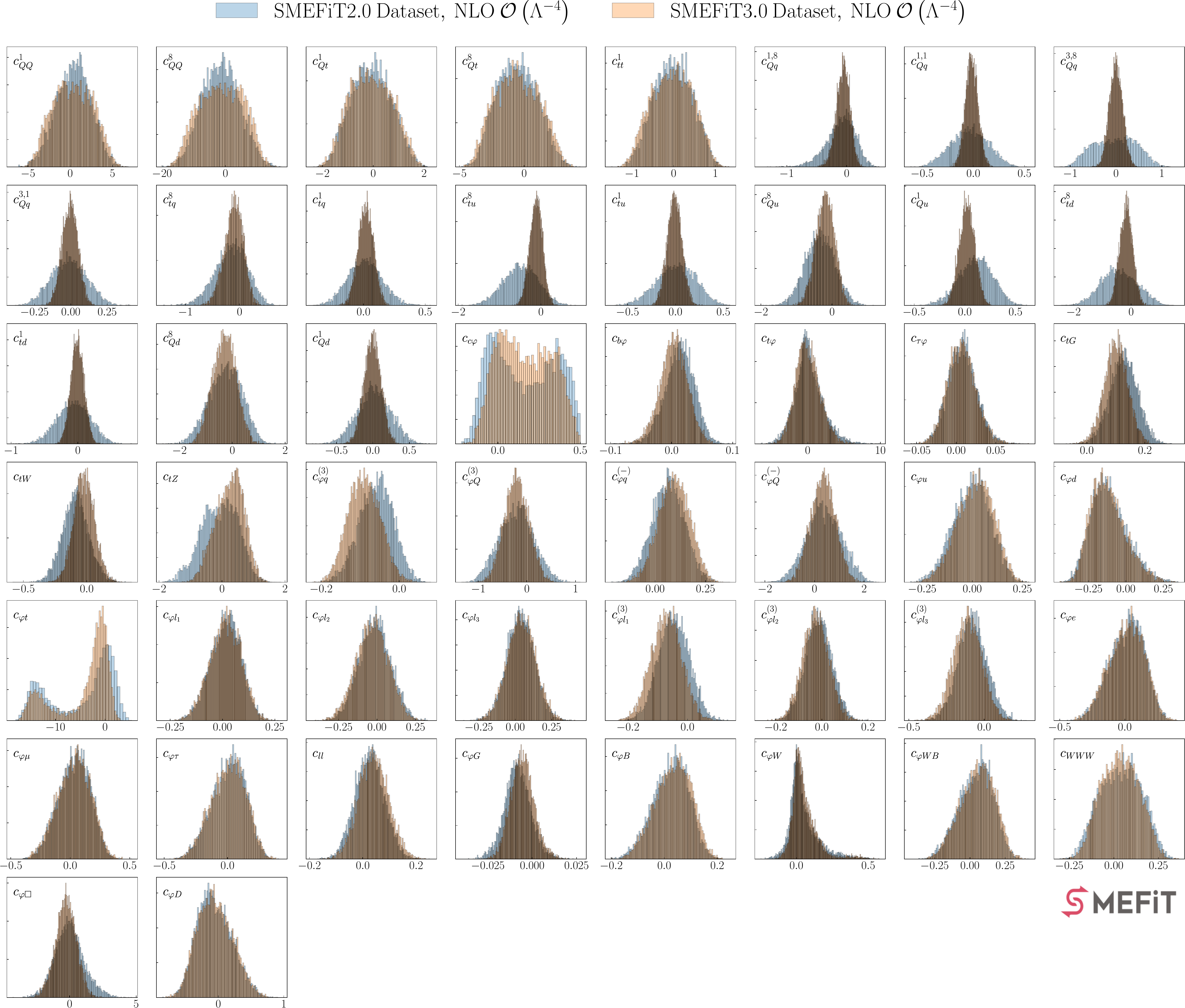}
    \caption{Same as Fig.~\ref{fig:updated-global-fit-quad} now comparing the results of the global analysis based
    on the {\sc\small SMEFiT2.0} and {\sc\small SMEFiT3.0} datasets, all other settings and theory calculations kept the same.
    In particular, in both cases the baseline settings includes
   the exact implementation of the EWPOs.
    }
    \label{fig:new-vs-old-quad}
\end{figure}

\myparagraph{Exact versus approximate implementation of the EWPOs}
Fig.~\ref{fig:updated-global-fit-quad} displays a comparison at the level of the posterior distributions on the Wilson coefficients between the new implementation of the EWPOs presented in Sect.~\ref{sec:ewpos} and used in {\sc\small SMEFiT3.0} and the previous, approximate implementation entering  {\sc\small SMEFiT2.0} and based on imposing the restrictions in Eq.~\eqref{eq:2independents}.
In both cases, these posteriors correspond to global marginalised fits carried out at $\mathcal{O}\lp \Lambda^{-4}\rp$ in the EFT expansion, see also Table~\ref{table:fit_settings}.

From this comparison one observes that the exact implementation of the EWPOs does not lead to major qualitative differences in the posterior distributions. 
Nevertheless, the approximate implementation of the EWPOs was in some cases too aggressive, and when replaced by the exact implementation one observes how the associated posterior distributions may display a broadening, as is the case for instance for the $c_{\varphi \ell_3}^{(3)}$ and $c_{\varphi d}$ coefficients.
Other EFT degrees of freedom for which the posterior distributions are modified following the exact implementation of the EWPOs are $c_{\varphi \ell_3},\,c_{\varphi \ell_2}^{(3)},\,c_{\varphi q}^{(3)},\, c_{\varphi q}^{(-)},\,$ and $c_{\ell\ell}$.  
In particular, we note that the four-lepton coefficient $c_{\ell\ell}$ was set to zero in the approximate implementation, while now it enters as an independent degree of freedom.
 
Two-light-two-heavy and four-heavy operators are constrained mostly by a set of processes not sensitive to the operators entering the EWPOs, i.e. top pair production and four-heavy quark production. 
There is limited cross-talk between the four-heavy operators and those entering the EWPOs, and hence the posteriors of the former remain unchanged comparing the two fits.
Furthermore, for other operators which are not directly sensitive to the EWPOs, we have verified that the residual observed differences arise from their correlations within the global fit with coefficients modifying the electroweak sector of the SMEFT (and they are hence absent in one-parameter individual fits), see also Fig.~\ref{fig:corr_smefit3_lin}.

We conclude from this analysis that, at the level of sensitivity that global SMEFT fits such as the one presented in this work are achieving, it is crucial to properly account for the constraints provided by the precise EWPOs from electron-positron colliders. 

\myparagraph{Impact of new LHC Run II data}
Next we quantify the impact of the new LHC Run II measurements included in the analysis, in comparison with {\sc\small SMEFiT2.0}, and listed in Sect.~\ref{subsec:new_datasets}.
To this end, we compare the baseline global SMEFT fit with a variant in which the input dataset is reduced to match that used in our previous analyses~\cite{Ethier:2021bye,Giani:2023gfq}.
In both cases, methodological settings and theory calculations are kept identical, and in particular both fits include the exact implementation of the EWPOs, quadratic EFT effects, and NLO QCD corrections to the EFT
cross-sections, see also Table~\ref{table:fit_settings}.
Hence the only difference between the two results concerns the LHC Run II data being fitted.

Fig.~\ref{fig:new-vs-old-quad} displays the same comparison of the results of the global analysis based on the {\sc\small SMEFiT2.0} and {\sc\small SMEFiT3.0} datasets.
The most marked impact of the new data is observed for the two-light-two-heavy four-fermion operators, where the narrower posterior distributions reflect improved bounds by a factor between 2 and 3 compared to the fit to the {\sc\small SMEFiT2.0} dataset, depending on the specific operator.
In all cases, the posterior distributions for the two-light-two-heavy operators remain consistent with the SM expectation at the 68\% CI, see also Fig.~\ref{fig:pull_lin_vs_quad}.

Other coefficients for which the new data brings in moderate improvements include the charm Yukawa $c_{c\varphi}$ (thanks to the latest Run II measurements which constrain the Higgs branching ratios and hence the total Higgs width), $c_{tZ}$  and $c_{\varphi t}$ (from the new $t\bar{t}Z$ dataset).
For the other coefficients, the impact of the new datasets is minor.
In particular, the latest measurements on $t\bar{t}t\bar{t}$ and $t\bar{t}b\bar{b}$  leave the posterior distributions of the four-heavy-fermion operators essentially unchanged.
Furthermore, one notes that the bound on the triple-gauge coupling operator $\mathcal{O}_{WWW}$ does not improve upon the inclusion of diboson production measurements based on the full Run II luminosity.
However, we should emphasise that including diboson production in proton-proton collisions in the fit plays an important role in breaking flat directions from EWPOs, for example in the $\lp c_{\varphi q}^{(3)}, c_{\varphi q}^{(-)}\rp$ plane. 
As we will also see in Chapter \ref{chap:smefit_fcc} when we discuss the impact of HL-LHC projections, (HL)-LHC diboson measurements are crucial in improving the bounds on various two-light-fermion coefficients.

\section{Summary and outlook}
\label{sec:summary_smefit3}
In this chapter, we have presented a global SMEFT analyses, {\sc\small SMEFiT3.0}, combining the constraints from the EWPOs from LEP and SLD with those provided by the recent LHC Run II measurements on Higgs boson, top quark, and diboson production, in many cases based on the full integrated luminosity.

Our analysis provides bounds on $n_{\rm eft}=50$ independent Wilson coefficients (45 in the linear fits) associated to dim-6 operators, with EFT cross-sections being evaluated either at $\mathcal{O}\lp \Lambda^{-2}\rp$ or $\mathcal{O}\lp \Lambda^{-4}\rp$ and including NLO QCD corrections whenever possible. We have included 445 data points in the baseline dataset, an increase of 40\% with respect to the previous {\sc\small SMEFiT2.0} analysis. The analysis is based on an independent implementation of the EWPOs, where all operators are treated on the same footing as those entering LHC measurements. This required not only implementing the EWPOs in the SMEFT itself, but also extending and recomputing existing LHC measurements to properly account for the operators sensitive to the EWPOs. 

We observed no major qualitative difference while comparing the posterior distributions associated to the exact and the approximate implementation of the EWPOs. However, we did observe a broadening in the shape of the posterior in some cases when the exact implementation was used. This demonstrates the importance of including a full independent treatment of the EWPOs in a global SMEFT analysis when combined with top, Higgs and diboson data. Furthermore, we found a significant tightening of the posteriors in the two-light-two-heavy class due to the increased {\sc\small SMEFiT3.0} dataset in comparison to the {\sc\small SMEFiT2.0} dataset. 

The {\sc\small SMEFiT3.0} results will serve as a baseline scenario for dedicated projection studies at future colliders that will be presented in Chapter \ref{chap:smefit_fcc}. We will first
consider the HL-LHC, and then the FCC-ee and the CEPC in order to quantify the impact of its respective measurements on the parameter space of the SMEFT in a global interpretation. Furthermore, we will also demonstrate how this impact at the EFT coefficient level translates into the parameter space of representative one-particle and multi-particle extensions of the SM matched to the SMEFT.

   \chapter{Automatised SMEFT-assisted constraints on UV-complete models}
\label{chap:smefit_uv}
\vspace{-1cm}
\begin{center}
\begin{minipage}{1.\textwidth}
    \begin{center}
        \textit{This chapter is based on my results that are presented in Ref.~\mycite{terHoeve:2023pvs}} 
    \end{center}
\end{minipage}
\end{center}

\myparagraph{Introduction} One of the main motivations for the ongoing SMEFT program in Particle
Physics is to streamline the connection between experimental data and UV-complete scenarios of BSM physics that contain new particles which are too heavy to be directly produced at available facilities.
In this paradigm, rather than comparing the predictions of specific UV-complete models
directly with data to derive information on its parameters (masses and couplings),
UV-models are first matched to the SMEFT and subsequently the resulting Wilson coefficients are constrained by means
of a global EFT analysis including a broad range of observables.

The main advantage of this approach is to bypass the need to recompute predictions for physical observables with different UV-complete models. 
The global SMEFT analysis essentially encapsulates, for a well-defined
set of assumptions, the information provided by available experimental observables, while
 the matching relations (see Sect.~\ref{sec:matching_intro} for an introduction) determine how this information relates
to the masses and couplings of the UV-complete model.
This feature becomes especially relevant whenever new BSM models are introduced:
one can then quantify to which extent their parameter space is constrained
by current data from a pre-existing global SMEFT analysis,
rather than having first to provide predictions for a large number of observables and then compare
those with data.

\myparagraph{Motivation} In recent years, several groups~\cite{Fuentes-Martin:2016uol,deBlas:2017xtg,DasBakshi:2018vni,Kramer:2019fwz,Gherardi:2020det,Angelescu:2020yzf,Ellis:2020ivx,Cohen:2020qvb,Fuentes-Martin:2020udw,Summ:2020dda,Carmona:2021xtq,Fuentes-Martin:2022vvu,Fuentes-Martin:2022jrf,DeAngelis:2023bmd,Banerjee:2023iiv,Chakrabortty:2023yke} have systematically studied the matching
between UV-complete models and the Wilson coefficients of the SMEFT,
with various degrees of automation and in many cases accompanied by the release of the corresponding open-source codes.
In order to realise the full potential of such EFT matching studies, it is however necessary to interface these results with global SMEFT analyses parameterising the constraints provided by the experimental data.
Such an interface must be constructed in a manner that benefits from the automation of EFT matching tools
and that does not impose restrictions in the type of UV-models to be matched.
In particular, it must admit non-linear matching relations with 
additional constraints such as parameter positivity. 
Several groups have reinterpreted global SMEFT fits in terms of matched UV models~\cite{Ellis:2020unq,Brivio:2021alv,Anisha:2021hgc,Greljo:2023adz}, but their focus is limited to pre-determined,
relatively simple models with few parameters.
No framework has been released to date that enables performing such fits with generic, user-specified, multi-particle UV-complete models.

In this chapter, we bridge this gap in the SMEFT literature by developing a framework automating the limit-setting procedure  on the parameter space
of generic UV-models which can be matched to dim-6 SMEFT operators.
This is achieved by extending {\sc\small SMEFiT}~\cite{Hartland:2019bjb,vanBeek:2019evb,Ethier:2021ydt,Ethier:2021bye,Giani:2023gfq} with the capabilities of working directly on
 the parameter space of UV-models, given arbitrary matching relations between UV couplings and EFT coefficients as an input.
To this end, we have designed an interface to the {\sc\small MatchMakerEFT} code~\cite{Carmona:2021xtq} such that for any
of the available UV-models it outputs a run card with the relevant Wilson coefficients entering
the {\sc\small SMEFiT} analysis.
This  interface, consisting of
the Mathematica package \matchTOfit~(available on \href{https://github.com/arossia94/match2fit}{ Github~\faicon{github}}),
also provides a list of the UV variables (denoted
as UV invariants) 
that can be inferred from the data
and corresponding to specific combinations
of UV couplings and masses. 
The adopted procedure removes any limitations on the type of matching relations involved.

We exploit this new pipeline to derive bounds on a broad range
of UV-complete scenarios both at linear and quadratic order in the SMEFT expansion
from a global dataset composed by LHC and LEP measurements, using either tree-level or
one-loop matching relations.
We consider both relatively simple single-particle
extensions of the SM as well as more
complex multi-particle extensions,
in particular with a benchmark model
composed by two heavy vector-like fermions
and one heavy vector boson.
We study the stability of the fits results with respect to the order in the EFT expansion
and  the perturbative QCD accuracy for the EFT cross-sections.
We carry out fits directly at the level of UV couplings and hence the constraints between different coefficients are provided implicitly by the matching relations, rather than directly as explicit restrictions in a fit of EFT coefficients. 
  

\myparagraph{Outline}
The structure of this chapter is as follows.
  First, continuing from our introduction in Sect.~\ref{sec:matching_intro}, we discuss in Sect.~\ref{sec:uv_invariants} the notion of UV invariants, which provide a one-to-one mapping between the EFT parameter space and the space of UV masses and couplings. 
  Sect.~\ref{sec:smefit} describes how the {\sc\small SMEFiT} framework is extended
  to enable constraint-setting
 directly in the parameter space of UV-complete models.
  The main results are presented in Sect.~\ref{sec:results_uv}, where we derive
  bounds on single- and multi-particle BSM models from a global dataset.
  %
  

\section{UV invariants}
\label{sec:uv_invariants}
Matching relations define a mapping $f$ from $U$, the parameter space spanned by the UV couplings $\mathbf{g}$,  to $W$, the space spanned by the Wilson coefficients $\mathbf{c}$,
\begin{equation}
    f: U \rightarrow W \, .
\end{equation}
The mapping $f$
associated to UV-models such as the one defined by Eq.~\eqref{eq:Lag_UV_full_Varphi_model} is in general non-injective
and hence non-invertible. 
Therefore, even choosing a particular UV model does not lift completely these degeneracies.
They might be partially or fully removed by either matching at higher loop orders or considering higher-dimensional operators.
In particular, dim-8 operators offer advantages to disentangle UV models~\cite{Zhang:2021eeo}, but their study is beyond the scope of this work.

Since the fit is, at best, only sensitive to $f(\mathbf{g})$, one can only meaningfully discriminate UV parameters $\mathbf{g}$ that map to different points
in the EFT parameter space $W$ under 
the matching relation $f$. 
Thus, we define ``UV invariants'' as those combinations of UV parameters such that $f$ remains invariant under a mapping $h$, defined as
\begin{equation}
    h: U \rightarrow I,
\end{equation}
such that $f(h(\mathbf{g})) = f(h(\mathbf{g}'))$. We denote by $I$ the space of UV invariants that is now bijective with $W$ under $f$ and that contains all the information that we can extract about the UV couplings by measuring $f$ from experimental data in the global EFT fit.

To illustrate the role of UV invariants, we consider again the tree-level matching relations 
for our benchmark model Eq.~(\ref{eq:Lag_UV_full_Varphi_model}) in Sect.~\ref{sec:matching_intro} given by
Eq.~\eqref{eq:matching_result_example_tree}. 
Expressing the UV couplings in terms of the 
EFT coefficients leads to two different solutions,
\begin{align}
\label{eq:uvinvariants_1}
    \nonumber
	\big(y_{\phi}^{d}\big)_{33} &= -  \sqrt{- \lp c_{qd}^{(8)}\rp_{3333} } \frac{m_\phi}{\Lambda}, \\
    \nonumber
    \big(y_{\phi}^{u} \big)_{33} &= -\frac{c_{t\varphi}}{c_{b\varphi}}\sqrt{- \lp c_{qd}^{(8)}\rp_{3333} } \frac{m_\phi}{\Lambda} \\
    \lambda_{\phi} &= \frac{c_{b\varphi}}{\sqrt{-\lp  c_{qd}^{(8)}\rp_{3333}}}\frac{m_\phi}{\Lambda} \, ,
\end{align}
and
\begin{align}
\label{eq:uvinvariants_2}
    \nonumber   
	\big(y_{\phi}^{d}\big)_{33} &=  \sqrt{- \lp c_{qd}^{(8)}\rp_{3333} } \frac{m_\phi}{\Lambda}, \\
    \nonumber
    \big(y_{\phi}^{u}\big)_{33} &= \frac{c_{t\varphi}}{c_{b\varphi}}\sqrt{- \lp c_{qd}^{(8)}\rp_{3333} } \frac{m_\phi}{\Lambda}, \\
    \lambda_{\phi} &= - \frac{c_{b\varphi}}{\sqrt{ \lp  c_{qd}^{(8)}\rp_{3333} }}\frac{m_\phi}{\Lambda},
\end{align}
where we have omitted the flavour indices of the coefficients for clarity.
The resulting sign ambiguity stems from the sensitivity to the sign of only two products of the three UV couplings. 
Other non-vanishing EFT coefficients do not enter these solutions since they are related to the present ones via linear or non-linear relations.

Eqns.~(\ref{eq:uvinvariants_1})
and~(\ref{eq:uvinvariants_2}) hence indicate
that the EFT fit is not sensitive to
the sign of these UV couplings for this specific
model.
The sought-for mapping $h$ between the original UV couplings
$(\big(y_{\phi}^{d}\big)_{33}, \big(y_{\phi}^{u}\big)_{33}, \lambda_\phi)$
and the UV invariants which can be meaningfully
constrained by the global EFT fit is given by
\begin{equation}
\label{eq:example_UVinvariants}
    h: (\big(y_{\phi}^{d}\big)_{33}, \left(y_{\phi}^{u}\right)_{33}, \lambda_\phi) \mapsto \left( |\left(y_{\phi}^{u}\right)_{33}|,\, \lambda_\phi \,\mathrm{sgn}\lp \left(y_{\phi}^{u}\right)_{33}\rp,\, \left(y_{\phi}^{d}\right)_{33}\,\mathrm{sgn}\lp \lambda_\phi\rp \right),
\end{equation}
with $\mathrm{sgn}(x)$ being the sign function. Notice the degree of arbitrariness present in this construction, since for example one could have  also chosen $\left( |\lambda_\phi|,\, \left(y_{\phi}^{u}\right)_{33} \,\mathrm{sgn}\lp \lambda_\phi\rp\right)$ instead of the first two invariants
of Eq.~(\ref{eq:example_UVinvariants}).
This simple example displays only a sign ambiguity, but one could be unable to distinguish two UV couplings altogether since, e.g. they always appear multiplying each other.
The  \matchTOfit~package automates the computation of the transformations $h$ defining the
UV invariants for the models at the tree-level matching level, 
see App.~B from Ref.~\mycite{terHoeve:2023pvs} for more details.

\begin{figure}[htbp]
    \centering
    \includegraphics[width=\linewidth]{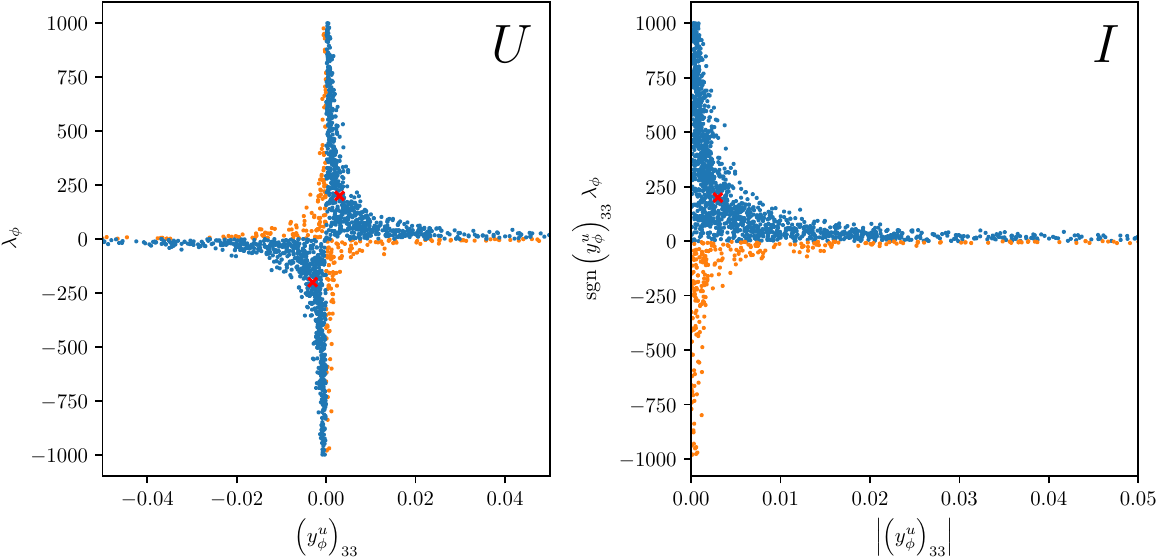}
    \caption{Left: marginalised posterior distributions in the space $U$ of UV parameters
$\lp \big(y_{\phi}^{u}\big)_{33}, \lambda_\phi\rp$
in the heavy scalar doublet model given by Eq.~\eqref{eq:Lag_UV_full_Varphi_model} from Chapter \ref{chap:smeft}
fitted to the data according to the
procedure of Sect.~\ref{sec:smefit}.
Right: the same results represented in the space $I$ of UV invariants
$
\left( |\big(y_{\phi}^{u}\big)_{33}|,\, \lambda_\phi \,\mathrm{sgn}( \big(y_{\phi}^{u}\big)_{33} ) \right).
$
The red points indicate two different
sets of UV couplings in $U$ that
are mapped to the same point in $I$
upon the transformation $h$.
Blue (orange) points indicate positive
(negative) values of the
UV-invariant $\lambda_\phi \,\mathrm{sgn}( \big(y_{\phi}^{u}\big)_{33} )$.
}
    \label{fig:uv_invariant_explained}
\end{figure}

Furthermore, one can also illustrate the concept
of UV invariants by fitting
 the heavy scalar doublet model previously introduced in Sect.~\ref{sec:matching_intro}, Eq.~\eqref{eq:Lag_UV_full_Varphi_model},
 to the experimental data
by following the procedure which will be outlined
in Sect.~\ref{sec:smefit}.
Fig.~\ref{fig:uv_invariant_explained} shows the 
resulting 
 marginalised posterior distributions in the space $U$ of UV parameters
$( \big(y_{\phi}^{u}\big)_{33}, \lambda_\phi)$
and in the space $I$ of UV invariants
$
( |\big(y_{\phi}^{u}\big)_{33}|,\, \lambda_\phi \,\mathrm{sgn}\left( \big( y_{\phi}^{u} \big)_{33} \right) 
$.
The red points indicate two different
sets of UV couplings in $U$,
$\mathbf{g}\ne\mathbf{g}'$, that
are mapped to the same point in $I$
upon the transformation $h$.
Presenting results at the level of UV invariants
has the benefit of making explicit the symmetries
and relations between UV-couplings that
may be hidden otherwise if one presents
results in the UV parameters space $U$. It is worth stressing that UV invariants do not necessarily correspond to combinations of UV parameters that one can constrain in a fit. Rather, they represent what can be said about the UV couplings from given values for the Wilson coefficients (WCs) and hence serve to map out the UV parameter space such that no redundant information is shown.

\section{Implementation in SMEFiT}
\label{sec:smefit}

Here we describe how the {\sc\small SMEFiT} global analysis
framework~\cite{Hartland:2019bjb,vanBeek:2019evb,Ethier:2021bye,Ethier:2021ydt,Giani:2023gfq} has been extended to operate, in addition to at the level
of Wilson coefficients, directly at the level of the parameters
of UV-complete models.

Assume a UV-complete model defined by the Lagrangian $\mathcal{L}_{\rm UV}(\boldsymbol{g})$
which contains $n_{\rm uv}$ free parameters $\boldsymbol{g}$.
Provided that this model has the SM as its low-energy limit with linearly realised electroweak symmetry breaking, one can derive matching relations between the SMEFT coefficients and the UV couplings of the form $\boldsymbol{c} = \boldsymbol{f}(\boldsymbol{g},\mu)$ for a given choice of the matching scale $\mu$.
Once these matching conditions are evaluated, 
the EFT cross-sections $\sigma(\boldsymbol{c})$
entering the fit can be expressed in terms of the UV couplings and masses $\sigma(\boldsymbol{f}(\boldsymbol{g},\mu))$.
By doing so, one ends up with the likelihood function $L$
now expressed in terms of UV couplings, $L(\boldsymbol{g})$.
Bayesian sampling techniques can now be applied directly to $L(\boldsymbol{g})$, assuming
a given prior distribution of the UV coupling space, in  the same manner
as for the fit of EFT coefficients.

The current release of {\sc\small SMEFiT} enables the user to impose these matching conditions $\boldsymbol{c} = \boldsymbol{f}(\boldsymbol{g},\mu)$ via run cards thanks to its support for a wide class of different constraints on the fit parameters, see App.~A of Ref.~\mycite{terHoeve:2023pvs}.
The code applies the required substitutions on the 
theoretical predictions for the observables
entering the fit automatically.
Additionally, the availability of Bayesian sampling means that the functional relationship between the likelihood function $L$ and the fitted parameters 
$\boldsymbol{g}$ is unrestricted.

In order to carry out parameter inference directly at the level of UV couplings within {\sc\small SMEFiT}, three main ingredients are required:

\begin{itemize}

\item First, the matching relations $\boldsymbol{f}$ between the parameters of
the UV Lagrangian $\mathcal{L}_{\rm UV}(\boldsymbol{g})$ and the EFT Wilson coefficients, $\boldsymbol{c} = \boldsymbol{f}(\boldsymbol{g})$ in the
Warsaw basis used in the fit.
As discussed in Sect.~\ref{sec:matching_intro},
this step can be achieved automatically 
both for tree-level and for one-loop matching
relations
by using {\sc\small MatchMakerEFT}~\cite{Carmona:2021xtq}.
Other matching frameworks may also be used
in this step.

\item Second, the conversion between the output of the matching code, {\sc\small MatchMakerEFT} in our case,
and the input required for the {\sc\small SMEFiT} run cards specifying the
dependence of the Wilson coefficients $\boldsymbol{c}$
on the UV couplings $\boldsymbol{g}$ such that
the replacement $\sigma(\boldsymbol{c}) \to \sigma (\boldsymbol{f}(\boldsymbol{g}))$ on the physical observables entering the fit can be implemented.
This step has been automated by modifying accordingly \smefit~and by the release of the new Mathematica package \matchTOfit. The latter translates the output of the matching code, {\sc\small MatchMakerEFT} in our case, to an expression that can be understood by \smefit~and evaluates numerically all those parameters that are not to be fitted. More details on the use of \matchTOfit~can be found in App.~B of Ref.~\mycite{terHoeve:2023pvs}.
%
The automation supports one-loop matching results, reaching the same level as state-of-the-art matching tools.
{\sc\small SMEFiT} then performs the replacements $\sigma(\boldsymbol{c}) \to \sigma (\boldsymbol{f}(\boldsymbol{g}))$ specified by the runcards.

\item Third, a choice of prior volume
in the space $U$ spanned by the UV
couplings $\boldsymbol{g}$ entering the fit.
In this work, we assume a flat prior on the UV parameters $\boldsymbol{g}$ and verify
that results are stable with respect to variations of this prior volume.
We note that, for the typical (polynomial) matching relations between UV couplings and EFT coefficients, this choice of prior implies non-trivial forms for the priors on the space of the latter.
This observation is an important motivation
to support the choice of fitting directly at the level of UV couplings.

\end{itemize}

Once these ingredients are provided,  {\sc\small SMEFiT} performs the global fit by comparing
EFT theory predictions with experimental data
and returning the posterior probability distributions
on the space of UV couplings $\boldsymbol{g}$
or any combination thereof. 
Fig.~\ref{fig:pipeline} displays a schematic representation of the pipeline adopted in this work
to map in an automated manner the parameter space of UV-complete models using the SMEFT as a bridge  to the data and based on the combination
of three tools: {\sc\small MatchMakerEFT}
to derive the matching relations,
\matchTOfit~to transform its output
into the {\sc\small SMEFiT}-compliant format,
and {\sc\small SMEFiT} to infer from
the data bounds on the UV coupling space.

\tikzstyle{io} = [trapezium, trapezium left angle=70, trapezium right angle=110, minimum width=3cm, minimum height=1cm, text centered, draw=black, fill=blue!30]
\tikzstyle{decision} = [diamond, minimum width=3cm, minimum height=1cm, text centered, draw=black, fill=green!30]
\tikzstyle{arrow} = [thick,->,>=stealth]
\tikzstyle{dottedline} = [thick, dotted]
\tikzstyle{process} = [rectangle, minimum width=1.7cm, minimum height=1cm, text centered, draw=black, fill=orange!30, rounded corners, text width=2.5cm]
\tikzstyle{smallbox} = [rectangle, minimum width=1cm, minimum height=1cm, text centered, draw=black, fill=orange!30, rounded corners, text width=1cm]
\tikzstyle{startstop} = [rectangle, minimum width=2cm, minimum height=1cm, text centered, draw=black, fill=blue!30, rounded corners, text width=6cm]
\tikzstyle{textbox} = [rectangle, minimum width=2cm, minimum height=1cm, text centered, rounded corners, text width=3cm]

\begin{figure}[t]
\small
    \centering
    \begin{tikzpicture}[node distance=2cm]

\node (UV1) [process] {$\mathcal{L}_{\rm UV}(\boldsymbol{g}_1)$};
\node (UV2) [process, right of=UV1, xshift=1.5cm] {$\mathcal{L}_{\rm UV}(\boldsymbol{g}_2)$};
\node (UVn) [process, right of=UV2, xshift=4cm] {$\mathcal{L}_{\rm UV}(\boldsymbol{g}_n)$};

\node[fit=(UV1) (UVn),draw,minimum height=2cm, minimum width = 13cm] (Input) {};

\node (match) [process, below of=UV1, yshift=-1.5cm] {$\mathbf{c} = f(\mathbf{g}_1, \mu)$};

\node (xsec) [process, right of=match, xshift=3.0cm] {$\sigma(\mathbf{c})=\sigma(\mathbf{g}_1)$};

\node (chi2) [process, right of=xsec,  xshift=2.5cm] {$\chi^2(\mathbf{g}_1)$};

\node[fit=(match) (xsec),draw,minimum height=2cm, minimum width = 8.5cm] (match2fit) {};

\draw [dottedline] (UV2.east) --(UVn.west);
\draw [arrow] (UV1) -- (match) node[midway, right, xshift=0.3cm] {\sc\footnotesize MatchMakerEFT};
\draw [arrow] (match) -- (xsec);
\draw [arrow] (match) -- (xsec) node[midway, above,yshift=0.2cm] {\sc\footnotesize match2fit};
\draw [arrow] (match2fit) -- (chi2)  node[midway, above, yshift=0.2cm]{\sc\footnotesize SMEFiT} ;

\end{tikzpicture}
    \caption{Schematic representation of the pipeline adopted in this work
    to map the parameter space of UV-complete models using the SMEFT as a bridge
    to the data.
    The starting point is a UV-Lagrangian containing a number of free parameters
    $\boldsymbol{g}$ such as its masses and coupling constants,
    for which a flat prior is assumed.
    We then determine the matching relations between the UV parameters
     $\boldsymbol{g}$ and the corresponding EFT coefficients  $\boldsymbol{c}$ at
     the matching scale $\mu$
     using {\sc\small MatchMakerEFT}.
     Then the \matchTOfit~interface
     enables expressing
     cross-sections for
     processes entering the EFT
     in terms of the  UV-parameters $\boldsymbol{g}$.
     Finally, these UV-parameters
     are constrained from the data using the sampling methods of {\sc\small SMEFiT}
     applied to the figure of merit $\chi^2(\boldsymbol{g})$ evaluated
     on a global dataset.}
    \label{fig:pipeline}
\end{figure}
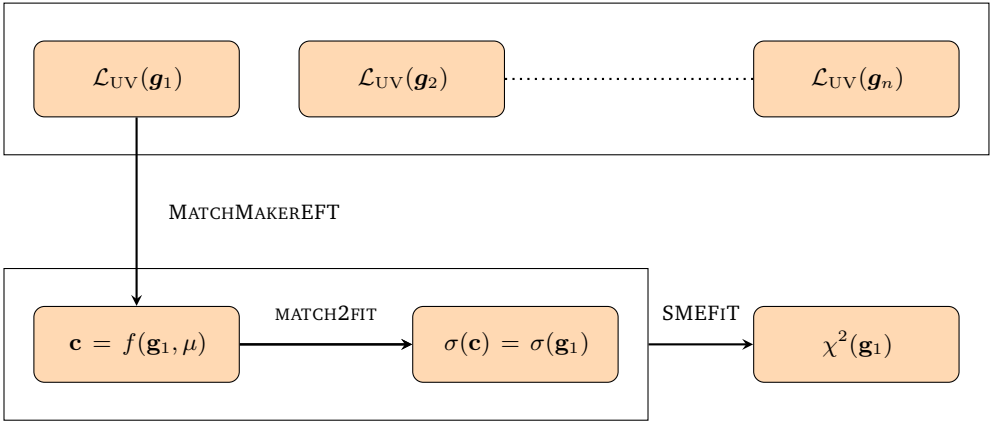


The UV invariants introduced in Sect.~\ref{sec:uv_invariants} are used only after the fit for plotting purposes, even if their functional form can be computed beforehand with \matchTOfit. 
Plotting the posteriors of UV invariants from the results of the fit and the output of \matchTOfit~is also automated by \smefit.
Additionally, we note here that
depending on the specific UV-complete model, one might be able to constrain only the absolute value of certain UV parameters or of their product.
For this reason, here we will display results mostly at the level of UV invariants $\boldsymbol{I}_{\rm UV}(\boldsymbol{g})$  
 determined from the matching relations.
We will find that  certain UV invariants  may nevertheless remain unconstrained, due to the lack of sensitivity to specific EFT
coefficients in the fitted data. 
In any case, the user can easily plot and study
arbitrary combinations
of the fitted UV parameters.

The baseline global {\sc\small SMEFiT} analysis
adopted in this work to constrain
the parameter space of UV-complete models
is based on the analysis presented in \mycite{Celada:2024mcf}, 
which in turn updated the global SMEFT analysis
of Higgs, top quark, and diboson data from \cite{Ethier:2021bye}.
\section{Results}
\label{sec:results_uv}

We now present the main results of this work, namely
the constraints on the parameter space of
a  broad range  of UV-complete models obtained using the {\sc\small SMEFiT} global analysis
integrated with the pipeline described in Sect.~\ref{sec:smefit}.
We discuss the following results in turn:
one-particle models matched at tree-level,
multi-particle models also matched at tree-level,
and one-particle models matched at 
the one-loop level.
For each category of models, we study both the impact of linear and quadratic corrections in the SMEFT expansion as well as of the QCD accuracy. Whenever possible, we provide comparisons with related studies to validate our results. 

The UV models considered in this work are composed of one or more of the heavy BSM particles listed in Table~\ref{tab:UV_particles} and
classified in terms of their spin
as scalar, fermion, and vector
particles.
 For each particle, we indicate the irreducible representation of the SM gauge group under which they transform.
These particles can couple linearly to the SM fields and hence generate dim-6 SMEFT operators after being integrated out at tree-level. 
The complete tree-level matching results for these particles were computed in~\cite{deBlas:2017xtg}, from where we adopt the notation.
The only exception is the heavy scalar field $\varphi$, which we rename $\phi$ to be consistent with the convention used in Sect.~\ref{sec:matching_intro}. 

\begin{landscape}
    \begin{table}[htbp]
    \footnotesize
    \renewcommand{\arraystretch}{1.35}
    \begin{center}
    \rotatebox{-90}{
    \begin{tabular}{|c|C{2.5cm}|c|}
    \toprule
    {\bf Model Label} & {\bf SM irreps} & {\bf UV couplings} \\
    \hline
    \multicolumn{3}{|c|}{\bf Heavy Scalar Models} \\
    \hline
    $\mathcal{S}$ & $(1,1)_{0\phantom{/-3}}$ & $\kappa_{\mathcal{S}}$\\
    $\mathcal{S}_1$ & $(1,1)_{1\phantom{/-3}}$ & $\left(y_{\mathcal{S}_1}\right)_{12}, \left(y_{\mathcal{S}_1}\right)_{21}$\\
    $\phi$ & $(1,2)_{1/2\phantom{-}}$ & $\lambda_{\phi},\,\, (y_{\phi}^u)_{33}$    \\
    $\Xi$  & $(1,3)_{0\phantom{/-3}}$ & $\kappa_{\Xi}$ \\
    $\Xi$  & $(1,3)_{1\phantom{/-3}}$ & $\kappa_{\Xi_1}$ \\
    $\omega_1$  & $(3,1)_{-1/3}$ & $\left(y_{\omega_1}^{qq}\right)_{33}$ \\
    $\omega_4$  & $(3,1)_{-4/3}$ & $\left(y_{\omega_4}^{uu}\right)_{33}$ \\
    $\zeta$  & $(3,3)_{-1/3}$ & $\left(y_{\zeta}^{qq}\right)_{33}$ \\
    $\Omega_1$  & $(6,1)_{1/3\phantom{-}}$ & $\left(y_{\Omega_1}^{qq}\right)_{33}$ \\
    $\Omega_4$  & $(6,1)_{4/3\phantom{-}}$ & $\left(y_{\omega_4}\right)_{33}$ \\
    $\Upsilon$  & $(6,3)_{1/3\phantom{-}}$ & $\left(y_{\Upsilon} \right)_{33}$\\
    $\Phi$  & $(8,2)_{1/2\phantom{-}}$ &  $\left(y_{\Phi}^{qu}\right)_{33}$ \\[1mm]
    \hline
    \multicolumn{3}{|c|}{\bf Heavy Vector Models} \\
    \hline
    $\mathcal{B}$ & $(1,1)_{0\phantom{/-3}}$ & $\left(g_B^u\right)_{33}$, $\left(g_B^q\right)_{33}$, $g_{\mathcal{B}}^{\varphi}$\\
    & & $\left(g_B^e\right)_{11}$, $\left(g_B^e\right)_{22}$, $\left(g_B^e\right)_{33}$, \\
    & & $\left(g_B^\ell\right)_{22}$, $\left(g_B^\ell \right)_{33}$ \\
    $\mathcal{B}_1$ & $(1,1)_{1\phantom{/-3}}$ & $g_{B_1}^{\varphi}$ \\
    $\mathcal{W}$ & $(1,3)_{0\phantom{/-3}}$ & $\left(g_{\mathcal{W}}^\ell\right)_{11},\,\left(g_{\mathcal{W}}^\ell\right)_{22}$,  $g_{\mathcal{W}}^\varphi$, $\left(g_{\mathcal{W}}^q\right)_{33}$ \\
    $\mathcal{W}_1$ & $(1,3)_{1\phantom{/-3}}$ & $g_{\mathcal{W}_1}^{\varphi}$ \\
    $\mathcal{G}_1$ & $(8,1)_{0\phantom{/-3}}$ & $\left(g_{\mathcal{G}}^q \right)_{33}$, $\left(g_{\mathcal{G}}^u\right)_{33}$ \\
    $\mathcal{H}$ & $(8,3)_{0\phantom{/-3}}$ & $\left(g_{\mathcal{H}}\right)_{33}$ \\
    $\mathcal{Q}_5$ & $(3,2)_{-5/6}$ & $\left( g_{\mathcal{Q}_5}^{uq} \right)_{33}$ \\
    $\mathcal{Y}_5$ & $(\bar{6},2)_{-5/6}$ & $\left( g_{\mathcal{Y}_5} \right)_{33}$ \\
    \hline
    \multicolumn{3}{|c|}{\bf Heavy Fermion Models} \\
    \hline
     $N$ & $(1,1)_{0\phantom{/-3}}$ & $\left(\lambda_N^e\right)_3$ \\
    $E$ & $(1,1)_{-1\phantom{/3}}$ & $\left(\lambda_E\right)_3$ \\
    $\Delta_1$ & $(1,2)_{-1/2}$ & $\left(\lambda_{\Delta_1}\right)_3$\\
    $\Delta_3$ & $(1,2)_{-3/2}$ & $\left(\lambda_{\Delta_3}\right)_3$ \\
    $\Sigma$ & $(1,3)_{0\phantom{/-3}}$ & $\left(\lambda_{\Sigma}\right)_3$ \\
    $\Sigma_1$ & $(1,3)_{-1\phantom{/3}}$ & $\left(\lambda_{\Sigma_1}\right)_3$ \\
    $U$ & $(3,1)_{2/3\phantom{-}}$ & $\left(\lambda_{U}\right)_3$ \\
    $D$ & $(3,1)_{-1/3}$ & $\left(\lambda_{D}\right)_3$\\
    $Q_1$ & $(3,2)_{1/6\phantom{-}}$ & $\left(\lambda_{Q_1}^{u} \right)_3$ \\
    $Q_7$ & $(3,2)_{7/6\phantom{-}}$ & $\left(\lambda_{Q_7}\right)_3$ \\
    $T_1$ & $(3,3)_{-1/3}$ & $\left(\lambda_{T_1}\right)_3$  \\
     $T_2$ & $(3,3)_{2/3\phantom{-}}$ & $\left(\lambda_{T_2}\right)_3$  \\
    \bottomrule
    \end{tabular}
    }
    \caption{
    Heavy BSM particles entering the UV-complete models considered in this thesis. 
    For each particle, we indicate the irreducible representation  of the SM gauge group under which they transform, with notation $(\text{SU}(3)_{\text{c}},\text{SU}(2)_{\text{L}})_{\text{U}(1)_{\text{Y}}}$, and the relevant UV couplings entering the associated Lagrangian.
    The model couplings are restricted to respect the {\sc\small SMEFiT} flavour assumption after tree-level matching.
    Multi-particle models can be constructed from combining subsets of the one-particle models.
    \label{tab:UV_particles}
    }
    \end{center}
    \end{table}
\end{landscape}

Concerning one-particle models, we include each of the particles listed
in Table~\ref{tab:UV_particles} one at the time, and then impose restrictions on the couplings between the UV heavy fields  and the SM particles such that these models satisfy the {\sc\small SMEFiT} flavour assumption after matching at tree-level. 
A discussion on the UV couplings allowed for each of the considered models under such restriction can be found in App.~D of Ref.~\mycite{terHoeve:2023pvs}. 
We have discarded from our analysis those heavy particles considered in~\cite{deBlas:2017xtg} for which we could not find a set of restrictions on their UV couplings such that they obeyed the baseline EFT fit flavour assumptions. 

With regard to multi-particle models, we consider  combinations of the one-particle models
mentioned above without additional assumptions on their UV couplings unless specified. 
For illustration purposes,
we will present results for the model that results from adding the custodially-symmetric models with vector-like fermions and an SU$(2)_L$ triplet vector boson, presented in~\cite{Marzocca:2020jze}. 
We analyse the cases of both degenerate and different heavy particle masses.

An overview of which Wilson coefficients are generated by each UV particle at tree-level is provided in Tables~\ref{tab:heavy_scalar_sens}, \ref{tab:heavy_vboson_sens}, \ref{tab:heavy_vfermion_sens} for heavy scalars, vector bosons and heavy vector-like fermions respectively.
Notice that some operators are not generated by any of the models considered in this work, e.g.~four-fermion operators with 
two-light and two-heavy quarks
or with purely light-quark ones.
This is due, in part, to the correlations among operators imposed by UV models and the restrictions imposed by the assumed flavour symmetry.  
We recall that, in general, one-loop
matching will introduce more coefficients
as compared to the tree-level ones
listed in Tables~\ref{tab:heavy_scalar_sens}, \ref{tab:heavy_vboson_sens} and~\ref{tab:heavy_vfermion_sens}

Finally, we will consider the model composed of a heavy scalar doublet $\phi$ matched to the SMEFT at the one-loop level. 
In this case, we impose the {\sc\small SMEFiT} flavour restriction only at tree-level. 
The operators generated by this model at one-loop were already discussed in Sect.~\ref{sec:matching_intro}.

In the following, credible intervals for the UV model parameters are evaluated as highest density intervals (HDI or HPDI)~\cite{Box1993}. The HDI is defined as the interval such that all values inside the HDI have higher probability than any value outside. In contrast to equal-tailed intervals (ETI) which are based on quantiles, HDI does not suffer from the property that some value inside the interval might have lower probability than the ones outside the interval. 
Whenever the lower bound is two orders of magnitude smaller than the width of the CI, we round it to zero.

\renewcommand{\arraystretch}{1.4}
\begin{table}[t]
    \centering
    \footnotesize
    \begin{tabular}
    {c|c|c|c|c|c|c|c|c|c|c|c|c}
    \toprule
& \multicolumn{12}{c}{\bf Heavy Scalars}\\
    \midrule
        &$\mathcal{S}$ 
        & $\mathcal{S}_1$ 
        & $\phi$ & $\Xi$ & $\Xi_1$ & $\omega_1$ & $\omega_4$ & $\zeta$ & $\Omega_1$&
    $\Omega_4$&
    $\Upsilon$&
    $\Phi$ \\
    \hline
        $c_{\varphi \Box}$           & \chm      & &       & \chm & \chm &      &      &      &      &      &      &      \\
        $c_{\varphi D}$              &           & &       & \chm & \chm &      &      &      &      &      &      &      \\
        $c_{\tau \varphi}$           &           & &  & \chm & \chm &      &      &      &      &      &      &      \\
        $c_{b \varphi}$              &           & &  & \chm & \chm &      &      &      &      &      &      &      \\
        $c_{t \varphi}$              &           & & \chm  & \chm & \chm &      &      &      &      &      &      &      \\
        $c_{\ell\ell}$               &           & \chm &       &      &      &      &      &      &      &      &      & \\
        $c_{Qt}^{1}$                 &           & & \chm  &      &      &      &      &      &      &      &      & \chm \\
        $c_{Qt}^{8}$                 &           & & \chm  &      &      &      &      &      &      &      &      & \chm \\
        $c_{QQ}^{1}$                 &           & &       &      &      & \chm &      & \chm & \chm &      & \chm &      \\
        $c_{QQ}^{8}$                 &           & &       &      &      & \chm &      & \chm & \chm &      & \chm &      \\
        $c_{tt}^{1}$                 &           & &       &      &      &      & \chm &      &      & \chm &      &      \\
        \bottomrule
    \end{tabular}
    \caption{The Wilson coefficients, in the 
    {\sc\small SMEFiT} fitting basis, generated at tree-level
    by the heavy scalar 
    whose quantum numbers are listed in Table~\ref{tab:UV_particles}.
   %
    }
    \label{tab:heavy_scalar_sens}
\end{table}

\renewcommand{\arraystretch}{1.4}
\begin{table}[t]
    \centering
    \footnotesize
    \begin{tabular}
    {c|c|c|c|c|c|c|c|c}
    \toprule
& \multicolumn{8}{c}{\bf Heavy Vector Bosons} \\
    \midrule
       &$\mathcal{B}$&
    $\mathcal{B}_1$&
    $\mathcal{W}$&
    $\mathcal{W}_1$&
    $\mathcal{G}$&
    $\mathcal{H}$&
    $\mathcal{Q}_5$&
    $\mathcal{Y}_5$ \\
    \hline
        $c_{\varphi \Box}$            & \chm & \chm & \chm & \chm &      &      &      &    \\
        $c_{\varphi D}$               & \chm & \chm &      & \chm &      &      &      &                \\
        $c_{\tau \varphi}$            &      & \chm & \chm & \chm &      &      &      &          \\
        $c_{b \varphi}$               &      & \chm & \chm & \chm &      &      &      &         \\
        $c_{t \varphi}$               &      & \chm & \chm & \chm &      &      &      &         \\
        $c_{\varphi l_{1,2,3}}^{(3)}$ &      &      & \chm &      &      &      &      &            \\
        $c_{\varphi l_{1,2,3}}^{(1)}$ & \chm &      &      &      &      &      &      &            \\
        $c_{\varphi (e, \mu, \tau)}$  & \chm &      &      &      &      &      &      &            \\
        $c_{\varphi Q}^{(3)}$         &      &      & \chm &      &      &      &      &            \\
        $c_{\varphi Q}^{(-)}$         & \chm &      & \chm &      &      &      &      &            \\
        $c_{\varphi t}$               & \chm &      &      &      &      &      &      &            \\
        $c_{\ell\ell}$                &      &      & \chm &      &      &      &      &            \\
        $c_{Qt}^{1}$                  & \chm &      &      &      &      &      & \chm & \chm       \\
        $c_{Qt}^{8}$                  &      &      &      &      & \chm &      & \chm & \chm       \\
        $c_{QQ}^{1}$                  & \chm &      & \chm &      &      & \chm &      &            \\
        $c_{QQ}^{8}$                  &      &      & \chm &      & \chm & \chm &      &            \\
        $c_{tt}^{1}$                  & \chm &      &      &      & \chm &      &      &            \\
        $c_{{\ell \ell}_{1111}}$      & \chm &      & \chm &      &      &      &      &            \\
        \bottomrule
    \end{tabular}
    \caption{Same as Table~\ref{tab:heavy_scalar_sens}, now for the heavy vector boson particles
    whose quantum numbers are listed in Table~\ref{tab:UV_particles}.}
    \label{tab:heavy_vboson_sens}
\end{table}

\renewcommand{\arraystretch}{1.3}
\begin{table}[t]
    \centering
    \begin{tabular}{c|c|c|c|c|c|c|c|c|c|c}
    \toprule
& \multicolumn{10}{c}{\bf Heavy Fermions} \\
    \midrule
&$N$&
    $E$&
    $\Delta_{1}$,
    $\Delta_{3}$&
    $\Sigma$, $\Sigma_{1}$&
    $U$&
    $D$&
    $Q_1$&
    $Q_5$&
    $Q_7$&
    $T_1$,
    $T_2$ \\
    \hline
    $c_{\varphi \ell_3}^{(3)}$      &\chm    &\chm &     &\chm   &       &       &       &       &       & \\
    $c_{\varphi \ell_3}$           &\chm    &\chm &     &\chm   &       &       &       &       &       & \\
    $c_{\varphi \tau}$             &        &     &\chm &       &       &       &       &       &       & \\
    $c_{\tau \varphi}$                                     &        &\chm &\chm &\chm   &       &       &       &       &       & \\
    $c_{\varphi Q}^{(3)}$                                 &        &     &     &       &\chm   &\chm   &       &       & &\chm  \\
    $c_{\varphi Q}^{(-)}$                                  &        &     &     &       &\chm   &       &       &       & &\chm   \\
    $c_{\varphi t}$                                        &        &     &     &       &       &       &\chm   &   &  \chm     & \\
    $c_{t \varphi}$                                        &        &     &     &       &\chm   &       &\chm   &   &\chm   &\chm  \\
    $c_{b \varphi}$                                        &        &     &     &       &      &\chm   &       & \chm      & & \chm \\
    \bottomrule
    \end{tabular}
    \caption{
    Same as Tables~\ref{tab:heavy_scalar_sens} and \ref{tab:heavy_vboson_sens}, now
for the heavy fermions.
    }
    \label{tab:heavy_vfermion_sens}
\end{table}

\subsection{One-particle models matched at tree-level}
\label{subsec:heavy-scalar}

Here we present results obtained from the tree-level matching of one-particle extensions of
the SM.
Motivated by the discussion in Sect.~\ref{sec:uv_invariants},
we present results at the level of UV invariants
since these are the combinations of UV couplings
that are one-to-one with the Wilson coefficients under the matching relations.
We study in turn the one-particle models
containing heavy scalars,
fermions, and vector bosons
listed in Tables~\ref{tab:heavy_scalar_sens}, \ref{tab:heavy_vboson_sens}
and~\ref{tab:heavy_vfermion_sens}.

\myparagraph{Heavy scalars}
The upper part of Table~\ref{tab:cl-heavy-scalar-fermion} shows the $95\%$ CI obtained for the UV invariants  associated to the one-particle
heavy scalar models considered
in this work and listed in Table~\ref{tab:UV_particles}.
We compare results obtained with
different settings of the global SMEFT fit:
linear and quadratic level in the EFT expansion
and either LO or NLO QCD accuracy
for the EFT cross-sections.
We exclude from Table~\ref{tab:cl-heavy-scalar-fermion} the results of heavy scalar model $\phi$ that is characterised by two UV invariants.
In all cases, we assume that the mass of the heavy particle is $m_{\rm UV}=1$~TeV, and hence
all the UV invariants shown are dimensionless.
The resulting bounds for these heavy scalar models are well below the naive perturbative limit $g\lesssim 4\pi$ in most cases, and only for a few models in linear EFT fits the bounds are close to saturate the perturbative unitary condition, $g \lesssim \sqrt{8\pi}$~\cite{DiLuzio:2016sur}.

\begin{table}[htbp]
\begin{center}
\scriptsize
\renewcommand{\arraystretch}{1.6}\begin{tabular}{c|c|c|c|c|c}
Model & UV invariants & LO $\mathcal{O}\left(\Lambda^{-2}\right)$ &LO $\mathcal{O}\left(\Lambda^{-4}\right)$ & NLO $\mathcal{O}\left(\Lambda^{-2}\right)$ & NLO $\mathcal{O}\left(\Lambda^{-4}\right)$ \\
\toprule
$\mathcal{S}$ & $\left|\kappa_{\mathcal{S}}\right|$ & [0, 1.4] & [0, 1.4] & [0, 1.5] & [0, 1.4] \\
$\mathcal{S}_1$ & $\left(y_{\mathcal{S}_1}\right)_{12} \left(y_{\mathcal{S}_1}\right)_{21}$ & [-0.041, 0.0018] & [-0.040, 0.0042] & [-0.042, 0.0027] & [-0.042, 0.0030] \\
$\Xi$ & $\left|\kappa_{\Xi}\right|$ & [0, 0.067] & [0, 0.069] & [0, 0.069] & [0, 0.069] \\
$\Xi_1$ & $\left|\kappa_{\Xi1}\right|$ & [0, 0.049] & [0, 0.049] & [0, 0.049] & [0, 0.048] \\
$\omega_1$ & $\left|\left(y_{\omega_1}^{qq}\right)_{33}\right|$ & [0, 5.0] & [0, 1.6] & [0, 5.2] & [0, 1.7] \\
$\omega_4$ & $\left|\left(y_{\omega_4}^{uu}\right)_{33}\right|$ & [0.027, 3.6] & [0.021, 1.1] & [0, 3.1] & [0.043, 1.0] \\
$\zeta$ & $\left|\left(y_{\zeta}^{qq}\right)_{33}\right|$ & [0.11, 3.7] & [0.011, 1.0] & [0.14, 3.3] & [0.034, 0.99] \\
$\Omega_1$ & $\left|\left(y_{\Omega_1}^{qq}\right)_{33}\right|$ & [0.021, 4.4] & [0, 1.5] & [0, 4.0] & [0.031, 1.4] \\
$\Omega_4$ & $\left|\left(y_{\Omega_4}\right)_{33}\right|$ & [0.099, 5.165] & [0.059, 1.553] & [0, 4.4] & [0.037, 1.4] \\
$\Upsilon$ & $\left|\left(y_{\Upsilon}\right)_{33}\right|$ & [0, 3.4] & [0, 1.1] & [0, 3.0] & [0.027, 1.0] \\
$\Phi$ & $\left|\left(y_{\Phi}^{qu}\right)_{33}\right|$ & [0.14, 11] & [0, 2.9] & [0.018, 9.8] & [0.014, 2.6] \\
\bottomrule
\toprule
$N$ & $\left|\left(\lambda_N^e\right)_3\right|$ & [0, 0.47] & [0, 0.47] & [0, 0.47] & [0, 0.48] \\
$E$ & $\left|\left(\lambda_E\right)_3\right|$ & [0, 0.24] & [0, 0.25] & [0, 0.25] & [0, 0.25] \\
$\Delta_1$ & $\left|\left(\lambda_{\Delta_1}\right)_3\right|$ & [0, 0.21] & [0, 0.20] & [0, 0.21] & [0, 0.20] \\
$\Delta_3$ & $\left|\left(\lambda_{\Delta_3}\right)_3\right|$ & [0, 0.26] & [0, 0.27] & [0.0015, 0.26] & [0, 0.27] \\
$\Sigma$ & $\left|\left(\lambda_{\Sigma}\right)_3\right|$ & [0, 0.29] & [0, 0.28] & [0, 0.28] & [0, 0.29] \\
$\Sigma_1$ & $\left|\left(\lambda_{\Sigma_1}\right)_3\right|$ & [0, 0.42] & [0, 0.42] & [0, 0.43] & [0, 0.42] \\
$U$ & $\left|\left(\lambda_{U}\right)_3\right|$ & [0, 0.84] & [0, 0.85] & [0, 0.82] & [0, 0.84] \\
$D$ & $\left|\left(\lambda_{D}\right)_3\right|$ & [0, 0.23] & [0, 0.24] & [0, 0.24] & [0, 0.23] \\
$Q_1$ & $\left|\left(\lambda_{\mathcal{Q}_1}^u\right)_3\right|$ & [0, 0.94] & [0, 0.95] & [0, 0.93] & [0, 0.92] \\
$Q_7$ & $\left|\left(\lambda_{\mathcal{Q}_7}\right)_3\right|$ & [0, 0.95] & [0, 0.93] & [0, 0.91] & [0 0.91] \\
$T_1$ & $\left|\left(\lambda_{T_1}\right)_3\right|$ & [0, 0.46] & [0, 0.46] & [0, 0.45] & [0, 0.47] \\
$T_2$ & $\left|\left(\lambda_{T_2}\right)_3\right|$ & [0, 0.39] & [0, 0.38] & [0, 0.38] & [0, 0.38] \\
\bottomrule
\end{tabular}
\end{center}
\caption{The 95\% CI for the UV invariants 
relevant for the heavy scalar (upper part) and heavy fermion (lower part of the table) 
one-particle models matched at tree-level.
We quote
the 95\% CI upper limit 
and the lower limit 
is rounded to 0 whenever it is two orders of magnitudes smaller than the total CI width. For each model we compare results
obtained at the linear and quadratic level
in the EFT expansion and using either LO
or NLO perturbative QCD corrections to
the EFT cross-sections.
In all cases, we set the mass of the heavy particle to $m_{\rm UV}=1$ TeV.
Note that the model with heavy scalar  $\phi$ is considered separately in Table~\ref{tab:cl-phi}, given that it is parameterised in terms of multiple UV invariants.}
\label{tab:cl-heavy-scalar-fermion}
\end{table}


Comparing the impact of linear versus quadratic corrections, we notice a significant improvement in sensitivity for the heavy scalar models $\omega_1$, $\omega_4$, $\zeta$, $\Omega_1$, $\Omega_4$, $\Upsilon$ and $\Phi$. 
This can be explained by the fact that they generate four-fermion operators, as these are characterised by large quadratic corrections.
Indeed, four-heavy operators, constrained
in the EFT fit by $t\bar{t}t\bar{t}$
and $t\bar{t}b\bar{b}$ cross-section data, have limited
sensitivity in the linear EFT fit.
In the remaining models, the impact of
quadratic EFT corrections
on the UV-invariant bounds is small.
Considering the impact of the QCD perturbative accuracy
used for the EFT cross-sections, one notices a moderate improvement of $\sim 10\%$ for models that are sensitive to four-fermion operators with the exception of $\omega_1$. 

\begin{figure}[t]
    \centering
    \includegraphics[width=\linewidth]{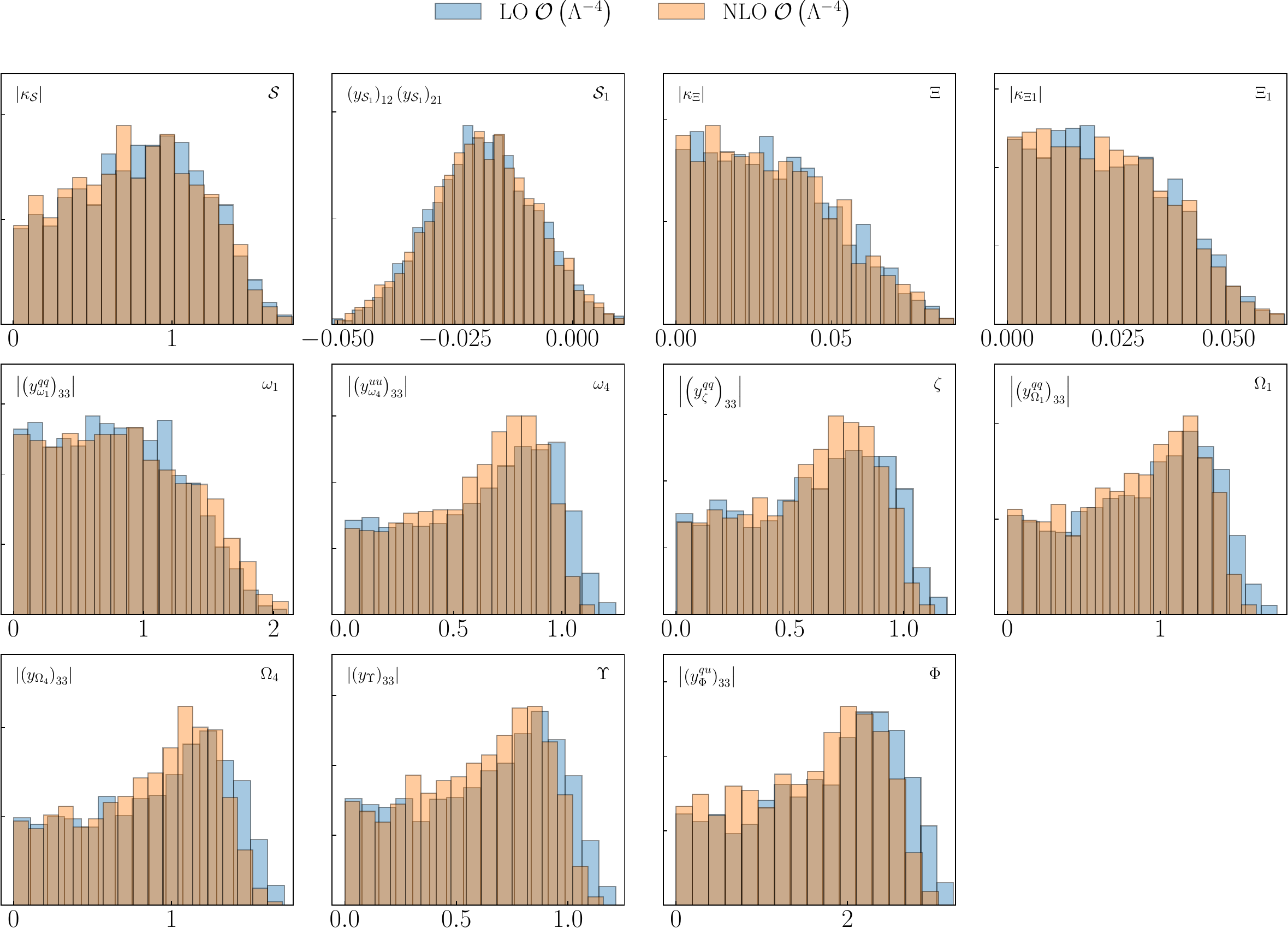}
    \caption{Same as Fig.~\ref{fig:heavy_scalar_nlo}, now comparing the baseline results,
    based on NLO QCD cross-sections
    for the EFT cross-sections, with
    the fit variant restricted to LO
    accuracy. }
    \label{fig:heavy_scalar_nlo}
\end{figure}

The posterior distributions 
associated to the heavy scalar models
listed in Table~\ref{tab:cl-heavy-scalar-fermion} are shown in Figs.~\ref{fig:heavy_scalar_quad}
and~\ref{fig:heavy_scalar_nlo}, comparing
results with different EFT expansion order
and QCD accuracy respectively.
To facilitate visualisation of the bulk region,
the distributions are
cut after the distribution has dropped to $5\%$ of its maximum, though
this choice is independent from the calculation
of the CI bounds.
One can see how
using quadratic EFT corrections (NLO QCD cross-sections)
improve significantly (moderately) the bounds on models that are sensitive to four-fermion operators for the reasons mentioned above.
The bounds most affected by quadratic EFT corrections are also the most susceptible to be modified when including dim-8 effects.
In a few cases, one can observe a worsening of the bound when including quadratic corrections, which are due to numerical accuracy limitations.
For all models considered, the posterior
distributions indicate agreement
with the SM, namely vanishing UV model couplings.

Along the same lines,
from Figs.~\ref{fig:heavy_scalar_quad}
and~\ref{fig:heavy_scalar_nlo} one
also observes that the posterior distributions of the absolute values of UV couplings tend to exhibit a most likely value (mode) away from zero. 
This feature is not incompatible with a posterior distribution on the EFT coefficient space that favours the SM case.
The reason is that when transforming the probability density function from one space to the other, one must consider the Jacobian factor that depends on the functional relation between EFT coefficients and UV couplings. 
For most matching relations, this Jacobian factor can generate these peaks away from zero in the UV coupling space even when the EFT fit favours the SM solution.

To illustrate this somewhat counter-intuitive result, consider a toy model for the probability density of a (positive-definite) Wilson coefficient $c$,
\begin{equation}
P(c) = \frac{2}{\sqrt{\pi}}e^{-c^2} \, ,\qquad \int_{0}^\infty dc\,P(c) = 1 \, ,
\end{equation}
where the underlying law is the SM.
For a typical matching condition of the form $c=g^2$, (see i.e. the heavy
scalar model of Eq.~(\ref{eq:matching_result_example_tree})),
the transformed probability distribution for
the ``UV-invariant'' $|g|$ is
\begin{equation}
P(|g|) = \frac{4}{\sqrt{\pi}}|g|e^{-|g|^4} \, , \qquad \int_{0}^\infty d|g| \,P(|g|) = 1 \, ,
\end{equation}
which is maximised by $g=1/\sqrt{2}\ne 0$.
Hence posterior distributions in the UV
coupling space favouring non-zero values
do not (necessarily) indicate preference
for BSM solutions in the fit. On the other hand, posteriors on the UV that are approximately constant near zero correspond to posteriors in the WC space that diverge towards zero.

One also observes from Table~\ref{tab:cl-heavy-scalar-fermion} that two of the considered 
heavy scalar models, specifically those containing the
 $\Xi$ and $\Xi_1$ particles, lead to bounds which are at least two orders of magnitude more stringent than for the rest.
 These two
 models generate the same operators with slightly different matching coefficients,
 and as indicated in Table~\ref{tab:heavy_scalar_sens} they are the only scalar particles that generate the Wilson
coefficient $c_{\varphi D}$, which is  
strongly constrained by the EWPOs. 
Therefore, one concludes that the heavy scalar models
with the best constraints are those
whose generated operators are sensitive
to the high-precision electron-positron collider data.

\begin{table}[t]
\footnotesize
\begin{center}
\renewcommand{\arraystretch}{1.5}\begin{tabular}{c|c|c|c|c|c}
\toprule
Model & UV invariants & LO $\mathcal{O}\left(\Lambda^{-2}\right)$ &LO $\mathcal{O}\left(\Lambda^{-4}\right)$ & NLO $\mathcal{O}\left(\Lambda^{-2}\right)$ & NLO $\mathcal{O}\left(\Lambda^{-4}\right)$ \\
\midrule
\multirow{2}{*}{$\phi$ (tree-level)} & $\left|\lambda_{\phi}\right|$ & [0, $8.2\cdot 10^{2}$] & [0, $7.4\cdot 10^{2}$] & [0, $8.0\cdot 10^{2}$] & [0, $7.9\cdot 10^{2}$] \\
 & $\mathrm{sgn}(\lambda_{\phi})\left(y_{\phi}^u\right)_{33}$ & [-0.11, 1.0] & [-0.20, 2.1] & [-0.19, 0.62] & [-0.18, 1.7]  \\
 \midrule
\multirow{2}{*}{$\phi$ (one-loop)} & $\left|\lambda_{\phi}\right|$ & [0, 7.6] & [0, 7.6] & [0, 7.6] & [0, 7.1] \\
 & $\mathrm{sgn}(\lambda_{\phi})\left(y_{\phi}^u\right)_{33}$ & [-0.81, 2.8] & [-1.2, 2.3] & [-0.80, 2.2] & [-0.87, 2.1]  \\
\bottomrule
\end{tabular}
\end{center}
\caption{Same as Table~\ref{tab:cl-heavy-scalar-fermion} for the heavy scalar $\phi$ model,
which has associated two independent
UV invariants.
See Fig.~\ref{fig:tree_vs_loop} for
the associated posterior distributions.
}
\label{tab:cl-phi}
\end{table}

\begin{figure}[t]
    \centering
    \includegraphics[width=\linewidth]{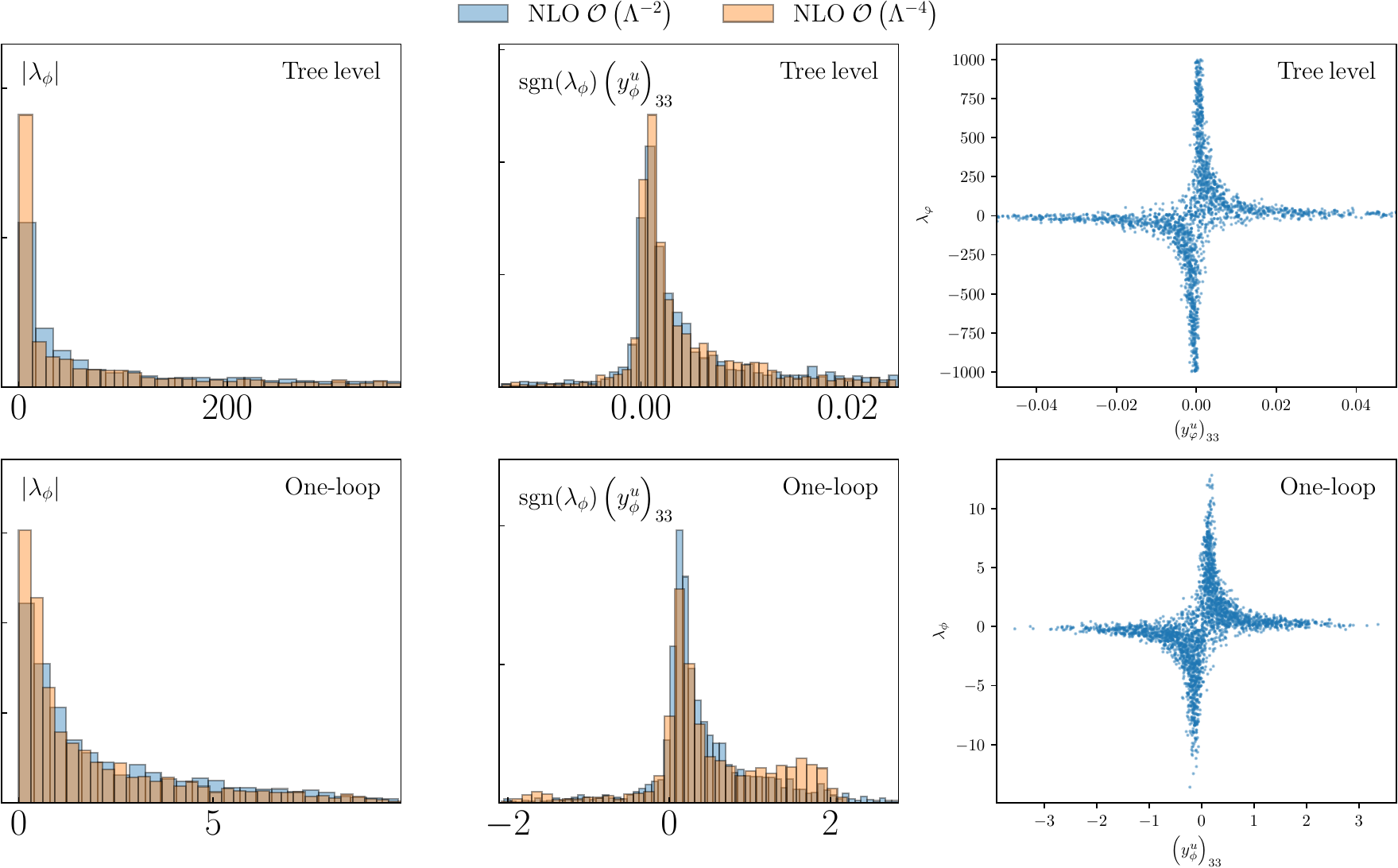}
    \caption{Marginalised posterior distributions of the two UV invariants associated
    to  the heavy scalar model $\phi$, comparing the impact of linear and quadratic EFT corrections after matching at tree-level (upper panel) and at one-loop level (lower panel). 
    The rightmost panels illustrate how 
    the individual UV couplings $\lambda_\phi$ and $(y_{\phi}^u)_{33}$ are correlated, as expected
    given that only their product
    can be constrained from the data. 
    See Table~\ref{tab:cl-phi} for
    the corresponding 95\% CI.
    }
    \label{fig:tree_vs_loop}
\end{figure}

The heavy scalar models that we have discussed so far 
and listed in Table~\ref{tab:cl-heavy-scalar-fermion}
depend on a single UV invariant.
On the other hand, as discussed
in Sect.~\ref{sec:uv_invariants}, the heavy scalar $\phi$ model depends on two
different UV couplings, $\lambda_\phi$ and $(y_{\phi}^u)_{33}$, resulting in two independent invariants.
We present the corresponding $95\%$ CI from tree-level 
matching in the upper part of Table~\ref{tab:cl-phi} and their distributions
in Fig.~\ref{fig:tree_vs_loop}. 
One finds that this model exhibits a degeneracy along 
the $\big(y_{\phi}^{u}\big)_{33} = 0$ direction, meaning that $\lambda_\phi$ can only be constrained whenever $\big(y_{\phi}^u\big)_{33} \ne 0$. This feature can be traced back to the tree-level matching relations in Eq.~\eqref{eq:matching_result_example_tree} with $\big(y_{\phi}^{d,e}\big)_{33}=0$ and the fact that there is no observable sensitive to  the $c_{\varphi}$
operator in the {\sc\small SMEFiT} dataset as well as the fact that the data does not prefer a non-zero $\big(y_{\phi}^{u}\big)_{33}$.
As we discuss below, this flat direction
is lifted once one-loop corrections to the matching
relations are accounted for.
As opposed to the UV invariants in Table~\ref{tab:cl-heavy-scalar-fermion}, for this model the constrained UV-invariant is not positive-definite.

\begin{figure}[t]
    \centering
    \includegraphics[width=\linewidth]{_inputs/smefit_uv/figures/Granada_posteriors_HO_vs_NHO_NLO_scalars.pdf}
    \caption{Posterior distributions
    associated to the UV invariants in the one-particle heavy scalar
 models listed in the upper part of Table~\ref{tab:cl-heavy-scalar-fermion}
 and
   obtained from tree-level matching.
   Note that all UV invariants
   for the considered models
   are positive-definite. 
   We
   compare the
   results
   based on linear
   and quadratic corrections 
   in the EFT expansion, in both cases
   with NLO QCD accuracy.
   To facilitate visualisation, the posterior distributions are cut after they have dropped to $5\%$ of its maximum, though
   this does not affect the calculation
   of the bounds listed in Table~\ref{tab:cl-heavy-scalar-fermion}.
   }\label{fig:heavy_scalar_quad}
\end{figure}

\begin{figure}[t]
    \centering
    \includegraphics[width=\linewidth]{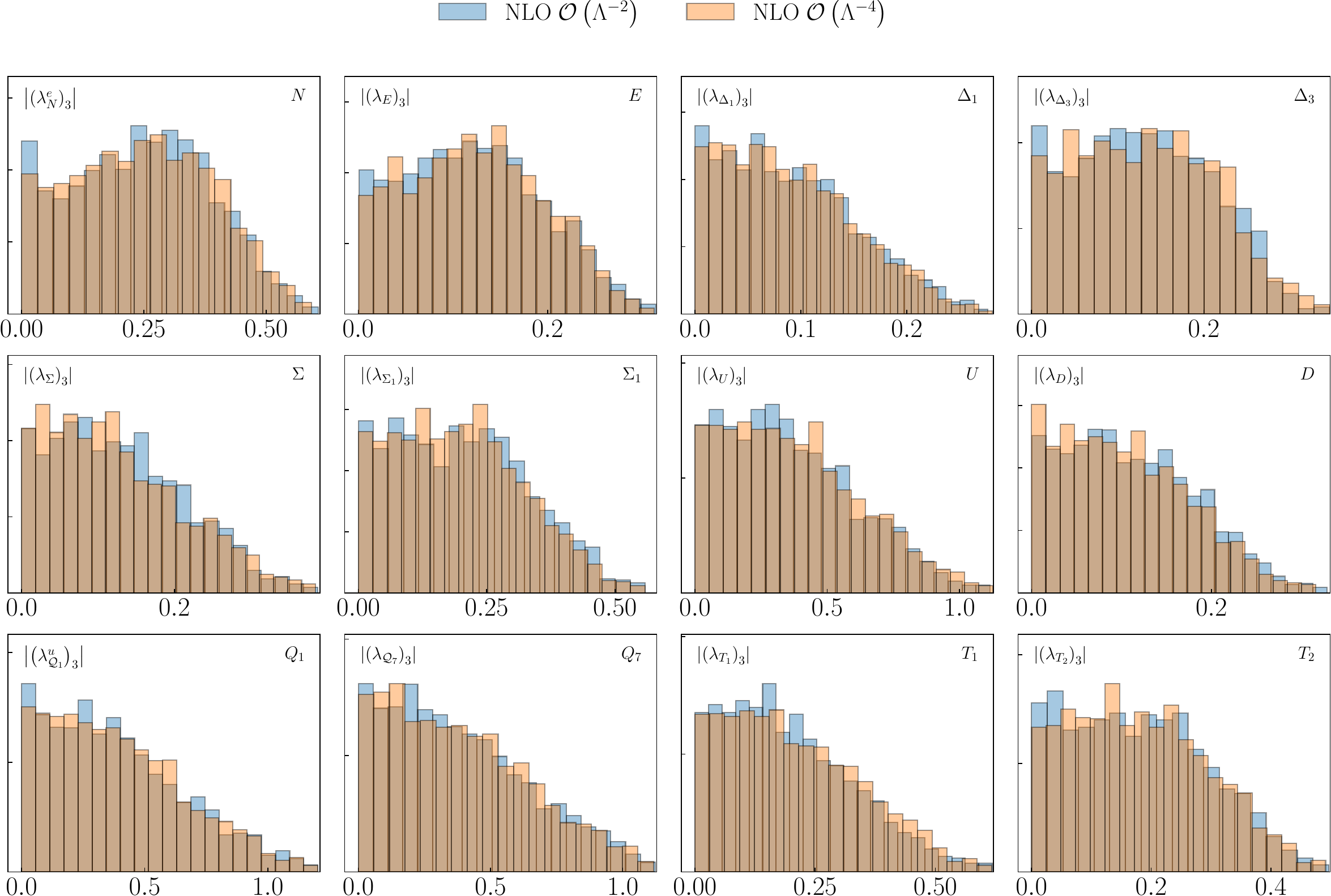}
    \caption{Same as
    Fig.~\ref{fig:heavy_scalar_quad}
    for the one-particle models
    composed of heavy vector-like
    fermions}
    \label{fig:heavy_vfermion_nlo}
\end{figure}

\myparagraph{Heavy fermions}
Following the discussion
of the results for the single-particle
BSM extensions with heavy scalars,
we move to the corresponding
analysis involving the heavy vector-like fermions listed in Table~\ref{tab:UV_particles}.
We provide their 95\% CI bounds
in the lower part of Table~\ref{tab:cl-heavy-scalar-fermion}, with the corresponding posterior distributions shown in Fig.~\ref{fig:heavy_vfermion_nlo}. 
All the (positive-definite) UV invariants
can be constrained from the fit, leaving no flat
directions, and are consistent
with the SM hypothesis.
One observes how the constraints
achieved in all the heavy fermion models are in general
similar, with differences no larger
than a factor 4
depending on the specific gauge representation. Additionally, all the bounds are $\mcO(0.1)$, indicating that current data probes weakly-coupled fermions with masses around $1$~TeV.

For these heavy fermion models, 
differences between fits carried
out at the linear and quadratic levels
in the EFT expansion are minimal. 
These reason is that these UV scenarios
are largely constrained by the precisely measured EWPOs from LEP and SLD,
composed by processes for which quadratic EFT corrections
are very small~\cite{ewpos}.
The same considerations apply
to the stability with respect to higher-order
QCD corrections, which are negligible
for the EWPOs.

\begin{figure}[t]
    \centering
    \includegraphics[width=\linewidth]{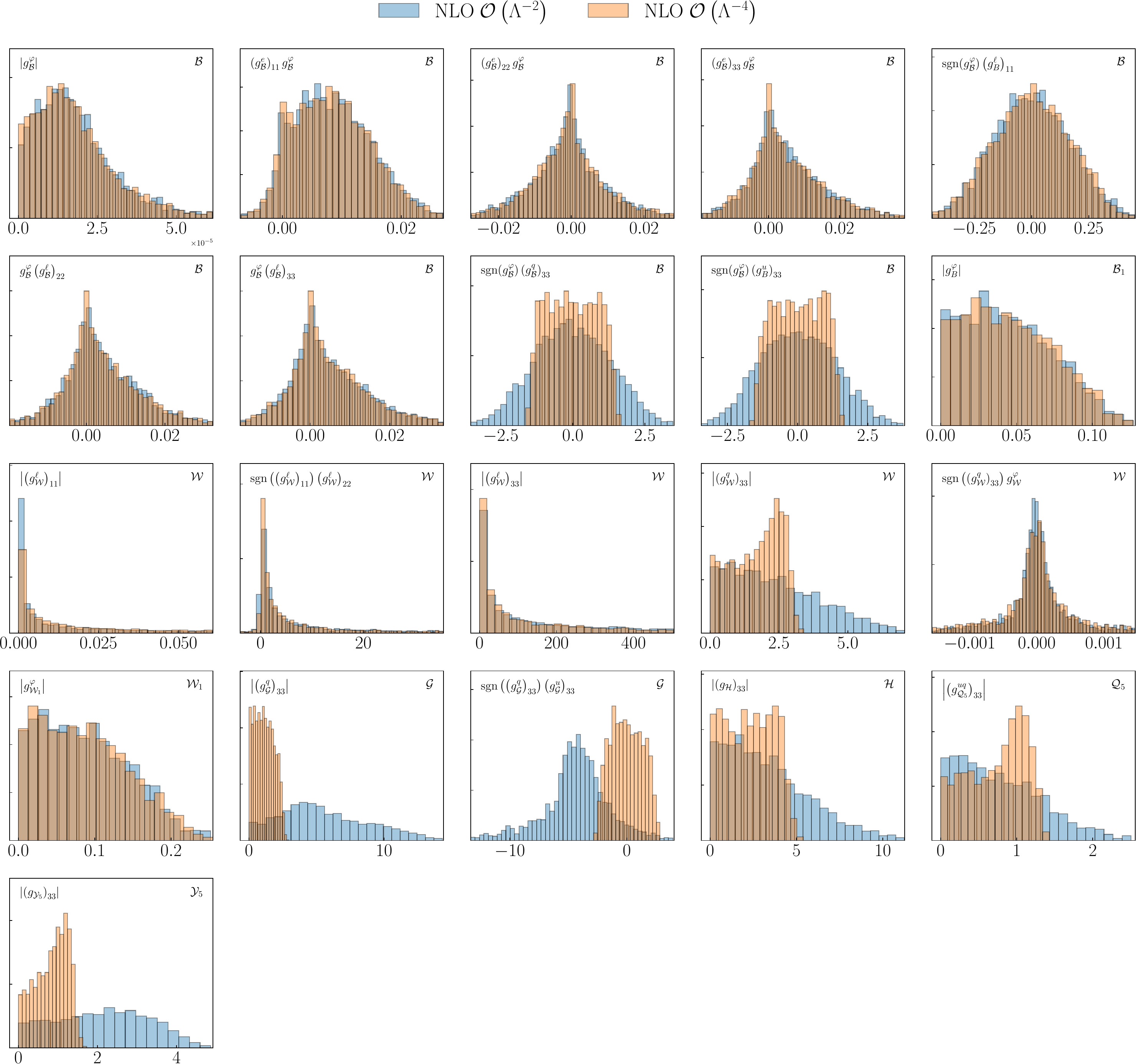}
    \caption{Same as
    Fig.~\ref{fig:heavy_scalar_quad}
    for the one-particle models
    composed of heavy vector-like
    bosons}
    \label{fig:heavy_vbosons_nlo}
\end{figure}

\myparagraph{Heavy vector bosons}
The last category of single-particle models
listed in Table~\ref{tab:UV_particles}
is the one composed of heavy vector bosons.
We provide their
$95\%$ marginalised CI in Table~\ref{tab:cl-heavy-boson}, with the corresponding posterior distributions in Fig.~\ref{fig:heavy_vbosons_nlo} in which we compare linear and quadratic EFT corrections at NLO QCD.

 A first observation is that most of the heavy vector models depend on multiple UV invariants. To be specific, the $\mathcal{B}$, $\mathcal{W}$ and $\mathcal{G}$ vector models have associated 9, 5 and 2 UV invariants respectively, while the other vector models are characterised by a single UV-invariant. For all models considered, results are consistent with the SM scenario, corresponding to vanishing UV parameters. 
 
Regarding model $\mathcal{B}$, its results can be understood as follows. First of all, we observe a strong increase in sensitivity at the quadratic level in the last two invariants, $\mathrm{sgn}\lp g_{\mathcal{B}}^\varphi\rp \lp g_{\mathcal{B}}^q\rp_{33}$ and $\mathrm{sgn}\lp g_{\mathcal{B}}^\varphi\rp \lp g_{\mathcal{B}}^u\rp_{33}$. Indeed, they generate the four-heavy operators $c_{tt}^{(1)}$ and $c_{QQ}^{(1)}$, respectively, which are characterised by large quadratic corrections. 
The other UV invariants in the $\mathcal{B}$ model generate operators sensitive to the EWPOs which are strongly constrained from LEP data leading to strong bounds in all cases. For example, the operators  $c_{\varphi\Box}$ and $c_{\varphi D}$ are generated by $g_{\mathcal{B}}^\varphi$,
\begin{equation}
\frac{c_{\varphi\Box}}{\Lambda^2} = \frac{1}{2}\frac{\lp g_{\mathcal{B}}^\varphi\rp^2}{m_{\mathcal{B}}^2} \, ,\qquad  \frac{ c_{\varphi D} }{\Lambda^2} = -2 \frac{\lp g_{\mathcal{B}}^\varphi\rp^2}{m_{\mathcal{B}}^2}\, ,
\end{equation}
and thus provide strong bounds on the invariant $\left|g_{\mathcal{B}}^\varphi\right|$. Furthermore, the invariant $\lp g_{\mathcal{B}}^e\rp_{ii} g_{\mathcal{B}}^\varphi$ is sensitive to the leptonic Yukawa operators $c_{\varphi (e, \mu, \tau)}$ and therefore gets well constrained by LEP data as well. Yet another example related to the EWPOs is the invariant $g_{\mathcal{B}}^\varphi \lp g_{\mathcal{B}}^\ell\rp_{ii}$ that generates the operators $c_{\varphi\ell_2}$ and $c_{\varphi\ell_3}$ for $i=1,2$, respectively. Finally, $\lp g_{\mathcal{B}}^\ell\rp_{11}$ generates $\lp c_{\ell\ell}\rp_{1111}$ and thus gets constrained by Bhabha scattering. For this model, we have chosen the UV invariants such that they agree with the constrained directions. Had we built them in the same way that for other models, it would be explicit that there are 5 poorly constrained directions along $|\big(g_{\mathcal{B}}^{e}\big)_{11}|$, $|\big(g_{\mathcal{B}}^{e}\big)_{22}|$, $|\big(g_{\mathcal{B}}^{e}\big)_{33}|$, $|\big(g_{\mathcal{B}}^{\ell}\big)_{22}|$, and $|\big(g_{\mathcal{B}}^{\ell}\big)_{33}|$. The model $\mathcal{B}_1$ is only sensitive to operators that can be constrained via EWPOs, hence no improved sensitivity is observed after adding quadratic EFT effects.

Moving on to model $\mathcal{W}$, we observe two flat directions along $\lp g_{\mathcal{W}}^{\ell}\rp_{33}$ and \\$\mathrm{sgn}\left(\left(g_{\mathcal{W}}^\ell\right)_{11}\right)\left(g_{\mathcal{W}}^\ell\right)_{22}$. In the matching relations, the UV coupling $\lp g_{\mathcal{W}}^{\ell}\rp_{33}$ enters exclusively in a product together with $g_{\mathcal{W}}^\varphi$. Now, $g_{\mathcal{W}}^\varphi$ already gets strongly constrained via other independent relations to $c_{\varphi\Box}, c_{b\varphi}, c_{t \varphi}$ and $c_{\tau, \varphi}$. As a result, the UV invariant $\left|\lp g_{\mathcal{W}}^{\ell}\rp_{33}\right|$ is left essentially unconstrained as no other matching relation exists to disentangle $ \lp g_{\mathcal{W}}^{\ell}\rp_{33}$ from $g_{\mathcal{W}}^\varphi$.
A similar argument holds for the second flat direction we observe along $\mathrm{sgn}\left(\left(g_{\mathcal{W}}^\ell\right)_{11}\right)\left(g_{\mathcal{W}}^\ell\right)_{22}$. The UV parameter $\lp g_{\mathcal{W}}\rp_{22}$ only enters as a product with either $\lp g_W^\ell\rp_{11}$ or $g_W^\varphi$, both of which already get constrained via other independent matching relations, e.g. $\lp c_{\ell\ell}\rp_{1111} \sim \lp g_{\mathcal{W}}^\ell\rp^2$ and the aforementioned reason in case of $g_{\mathcal{W}}^\varphi$. In fact, these bounds can be considered meaningless since they are of the order or above the perturbative limit $4\pi$ and well in excess of more refined perturbative unitarity bounds~\cite{Barducci:2023lqx}. 
The four-heavy operators $c_{QQ}^{(1, 8)}$ are responsible for the increased sensitivity we observe in $\lp g_{\mathcal{W}}^q\rp_{33}$ after including quadratic EFT corrections.

Finally, concerning the models $\mathcal{G}$, $\mathcal{H}$, $\mathcal{Q}_5$ and $\mathcal{Y}_5$, we observe a significant tightening of the bounds after quadratic EFT corrections are accounted for. This is entirely due to their sensitivity to the four-heavy operators, as can be seen in Table~\ref{tab:heavy_vboson_sens}.

\begin{table}[t]
\begin{center}
\scriptsize

\renewcommand{\arraystretch}{1.5}\begin{tabular}{C{0.68cm}|C{2.65cm}|C{1.76cm}|C{1.76cm}|C{1.76cm}|C{1.76cm}}
Model & UV invariants & LO $\mathcal{O}\left(\Lambda^{-2}\right)$ &LO $\mathcal{O}\left(\Lambda^{-4}\right)$ & NLO $\mathcal{O}\left(\Lambda^{-2}\right)$ & NLO $\mathcal{O}\left(\Lambda^{-4}\right)$ \\
\toprule
\multirow{9}{*}{$\mathcal{B}$} & $\left|g_{\mathcal{B}}^\varphi\right|$ & [0, $4.4\cdot 10^{-5}$] & [0, $4.4\cdot 10^{-5}$] & [0, $4.6\cdot 10^{-5}$] & [0, $4.6\cdot 10^{-5}$] \\
 & $\left(g_{\mathcal{B}}^e\right)_{11} g_{\mathcal{B}}^\varphi$ & [-0.0033, 0.021] & [-0.0028, 0.021] & [-0.0022, 0.022] & [-0.0032, 0.021]  \\
 & $\left(g_{\mathcal{B}}^e\right)_{22} g_{\mathcal{B}}^\varphi$ & [-0.026, 0.022] & [-0.026, 0.022] & [-0.025, 0.025] & [-0.025, 0.025]  \\
 & $\left(g_{\mathcal{B}}^e\right)_{33} g_{\mathcal{B}}^\varphi$ & [-0.015, 0.030] & [-0.015, 0.029] & [-0.016, 0.029] & [-0.014, 0.030]  \\
 & $\mathrm{sgn}(g_{\mathcal{B}}^\varphi) \left(g_B^\ell\right)_{11}$ & [-0.32, 0.33] & [-0.33, 0.32] & [-0.32, 0.33] & [-0.32, 0.30]  \\
 & $g_{\mathcal{B}}^\varphi \left(g_{\mathcal{B}}^\ell \right)_{22}$ & [-0.016, 0.027] & [-0.016, 0.027] & [-0.016, 0.028] & [-0.016, 0.029]  \\
 & $g_{\mathcal{B}}^\varphi \left(g_{\mathcal{B}}^\ell \right)_{33}$ & [-0.014, 0.028] & [-0.015, 0.027] & [-0.014, 0.029] & [-0.015, 0.028]  \\
 & $\mathrm{sgn}(g_\mathcal{B}^\varphi)\left(g_\mathcal{B}^q\right)_{33}$ & [-3.1, 3.0] & [-1.4, 1.5] & [-2.4, 2.6] & [-1.4, 1.4]  \\
 & $\mathrm{sgn}(g_\mathcal{B}^\varphi)\left(g_B^u\right)_{33}$ & [-3.2, 2.9] & [-1.5, 1.4] & [-2.5, 2.6] & [-1.4, 1.4]  \\
 \midrule
$\mathcal{B}_1$ & $\left|g_B^\varphi\right|$ & [0, 0.099] & [0, 0.098] & [0, 0.098] & [0, 0.099] \\
\midrule
\multirow{5}{*}{$\mathcal{W}$}& $\left|\left(g_{\mathcal{W}}^\ell\right)_{11}\right|$ & [0, 0.33] & [0, 0.32] & [0, 0.28] & [0, 0.34] \\
 & $\mathrm{sgn}\left(\left(g_{\mathcal{W}}^\ell\right)_{11}\right)\left(g_{\mathcal{W}}^\ell\right)_{22}$ & [-34, $4.6\cdot 10^{2}$] & [-20, $5.3 \cdot 10^{2}$] & [-17, $7.0\cdot 10^{2}$] & [-21, $3.6\cdot 10^{1}$]  \\
 & $\left|\left(g_{\mathcal{W}}^\ell\right)_{33}\right|$ & [0, $7.9\cdot 10^{2}$] & [0, $7.8\cdot 10^{2}$] & [0, $8.0\cdot 10^{2}$] & [0, $7.7\cdot 10^{2}$]  \\
 & $\left|\left(g_{\mathcal{W}}^q\right)_{33}\right|$ & [0, 6.4] & [0, 3.1] & [0, 5.6] & [0, 2.9]  \\
 & $\mathrm{sgn}\left(\left(g_{\mathcal{W}}^q\right)_{33}\right)g_{\mathcal{W}}^\varphi$ & [-0.022, 0.019] & [-0.024, 0.017] & [-0.026, 0.020] & [-0.024, 0.020]  \\
\midrule
\multirow{2}{*}{$\mathcal{G}$}& $\left|\left(g_{\mathcal{G}}^q\right)_{33}\right|$ & [0, 12] & [0, 2.5] & [0, 12] & [0, 2.3] \\
 & $\mathrm{sgn}\left(\left(g_{\mathcal{G}}^q\right)_{33}\right)\left(g_{\mathcal{G}}^u\right)_{33}$ & [-11, 1.8] & [-2.3, 2.6] & [-11, 2.2] & [-2.2, 2.4]  \\
 \midrule
$\mathcal{H}$ & $\left|\left(g_{\mathcal{H}}\right)_{33}\right|$ & [0, 8.0] & [0, 4.3] & [0, 8.0] & [0, 4.4] \\
\midrule
$\mathcal{Q}_5$ & $\left|\left(g_{\mathcal{Q}_5}^{uq}\right)_{33}\right|$ & [0, 2.1] & [0.026, 1.4] & [0, 1.8] & [0, 1.3] \\
\midrule
$\mathcal{Y}_5$ & $\left|\left(g_{\mathcal{Y}_5}\right)_{33}\right|$ & [0.046, 4.5] & [0.031, 1.6] & [0, 4.0] & [0.053, 1.4] \\
\bottomrule
\end{tabular}
\end{center}
\caption{Same as Table~\ref{tab:cl-heavy-scalar-fermion}
for the UV invariants associated
to the one-particle heavy vector boson extensions
}
\label{tab:cl-heavy-boson}
\end{table}

\subsection{Multi-particle models matched at tree-level}

Moving beyond one-particle models,
we now study  UV-completions
of the SM which include multiple heavy
particles.
Specifically, we consider a UV model 
which includes two heavy vector-like
fermions, $Q_1$ and $Q_7$,
and a heavy vector-boson, $\mathcal{W}$, see Table~\ref{tab:UV_particles} for their
quantum numbers and gauge group
representation.
In  the case of equal masses and couplings
for the two heavy fermions, this model
satisfies
custodial symmetry~\cite{Marzocca:2020jze}.
The two heavy fermions
$Q_1$ and $Q_7$ generate the same two operators, namely $c_{t\varphi}$ and $c_{\varphi t}$. 
A contribution to the top Yukawa operator $c_{t\varphi}$ is also generated by 
the heavy vector-boson $\mathcal{W}$, introducing an interesting interplay between the quark bidoublets on the one hand and the neutral vector triplet on the other hand. 
As indicated in Table~\ref{tab:heavy_vboson_sens},
several other operators in addition
to $c_{t\varphi}$ are generated
when integrating out
the heavy vector boson $\mathcal{W}$.

It should be emphasised that
we make this choice of 
multi-particle model for illustrative
purposes as well as to compare
with the benchmark studies of~\cite{Marzocca:2020jze}, and that
results for any other combination
of the heavy BSM particles listed
in Table~\ref{tab:UV_particles} 
can easily be obtained within our approach.
The only limitations are 
that the number of UV couplings
must remain smaller than the number
of EFT coefficients entering the analysis,
and that the input data used in the global
fit exhibits sensitivity to the matched
UV model.

\begin{figure}[t]
    \centering
    \includegraphics[width=\linewidth]{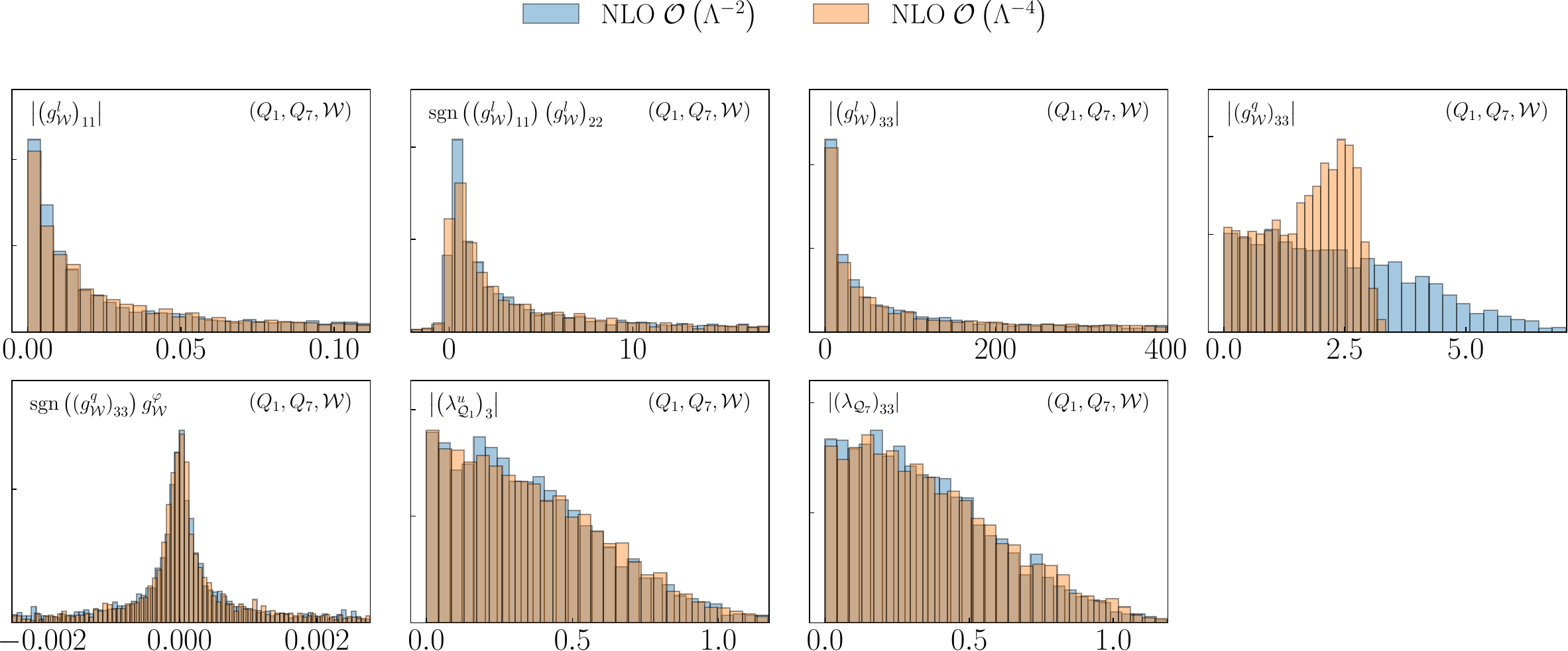}
    \caption{The posterior distributions of the UV invariants in the
three-particle model consisting
of two heavy fermions $Q_1,Q_7$ and a heavy
vector boson $\mathcal{W}$, see
also  see Table~\ref{tab:cl-mp}, 
comparing the impact of linear (blue) versus quadratic (orange) corrections in the EFT expansion.
}
    \label{fig:mp-posteriors}
\end{figure}

We provide in Table~\ref{tab:cl-mp} the $95\%$ CI for the UV invariants associated to this three-particle model. A common
value of the heavy mass, 
 $m_{Q_1}=m_{Q_7}=
m_\mathcal{W} = 1$ TeV, is assumed.
As in the case of the one-particle models,
we compare results at the linear and quadratic EFT level and at LO and NLO in the QCD expansion.
The associated posterior distributions
for the UV invariants
comparing the impact of linear  versus quadratic EFT corrections are shown in Fig.~\ref{fig:mp-posteriors}.
On the one hand, one notices an  increase in sensitivity at the quadratic level in case of $(g_{\mathcal{W}}^{q})_{33}$, which is
consistent with the results of 
the one-particle model
analysis shown in Table~\ref{tab:cl-heavy-boson}. 
On the other hand, for some of the UV invariants in the
model, such as $|(g_{\mathcal{W}}^l)_{11}|$,
the bounds become looser once quadratic EFT
corrections are accounted for, presumably due to the appearance of a second minimum
in the marginalised $\chi^2$ profiles.
For this specific model, it turns out that one has a quasi-flat direction
in the UV-invariant $\left|\left(g_{\mathcal{W}}^l\right)_{33}\right|$.

\begin{table}[t]
\begin{center}
\scriptsize
\renewcommand{\arraystretch}{1.5}\begin{tabular}{C{0.68cm}|C{2.65cm}|C{1.76cm}|C{1.76cm}|C{1.76cm}|C{1.76cm}}
\toprule
Model & UV invariants & LO $\mathcal{O}\left(\Lambda^{-2}\right)$ &LO $\mathcal{O}\left(\Lambda^{-4}\right)$ & NLO $\mathcal{O}\left(\Lambda^{-2}\right)$ & NLO $\mathcal{O}\left(\Lambda^{-4}\right)$ \\
\midrule
\multirow{7}{*}{\makecell{$Q_1$ \\ + \\ $Q_7$ \\ + \\ $\mathcal{W}$}} & $\left|\left(g_{\mathcal{W}}^l\right)_{11}\right|$ & [0, 0.43] & [0, 0.31] & [0, 0.36] & [0, 0.28] \\
 & $\mathrm{sgn}\left(\left(g_{\mathcal{W}}^l\right)_{11}\right)\left(g_{\mathcal{W}}^l\right)_{22}$ & [-3.3, 40] & [-3.9, 49] & [-4.7, 53] & [-4.8, 49]  \\
 & $\left|\left(g_{\mathcal{W}}^l\right)_{33}\right|$ & [0, $7.7\cdot 10^{2}$] & [0, $7.9\cdot 10^{2}$] & [0, $7.4\cdot 10^{2}$] & [0, $7.6\cdot 10^{2}$]  \\
 & $\left|\left(g_{\mathcal{W}}^q\right)_{33}\right|$ & [0, 6.5] & [0.011, 3.1] & [0, 5.4] & [0.014, 2.9]  \\
 & $\mathrm{sgn}\left(\left(g_{\mathcal{W}}^q\right)_{33}\right)g_{\mathcal{W}}^\varphi$ & [-0.013, 0.012] & [-0.014, 0.019] & [-0.013, 0.0099] & [-0.014, 0.011]  \\
 & $\left|\left(\lambda_{\mathcal{Q}_1}^u\right)_3\right|$ & [0, 0.87] & [0, 0.88] & [0, 0.86] & [0, 0.86]  \\
 & $\left|\left(\lambda_{\mathcal{Q}_7}\right)_{33}\right|$ & [0, 0.88] & [0, 0.87] & [0, 0.84] & [0, 0.87]  \\
\bottomrule
\end{tabular}
\end{center}
\caption{Same as Table~\ref{tab:cl-heavy-scalar-fermion}
for the UV invariants associated
to the three-particle model consisting
of two heavy fermions $Q_1,Q_7$ and a heavy
vector boson $\mathcal{W}$ obtained from
tree-level matching.
A common value of the heavy mass is assumed for
the three particles, $m_{Q_1}=m_{Q_7}=
m_\mathcal{W} = 1$ TeV.
See Fig.~\ref{fig:mp-different-masses} for the corresponding
results in a scenario with $m_{Q_1} \ne m_{Q_7} \ne
m_\mathcal{W}$. 
For this model we have a flat direction
in the UV-invariant $\left|\left(g_{\mathcal{W}}^l\right)_{33}\right|$.
}
\label{tab:cl-mp}
\end{table}

\begin{figure}[tbp]
    \centering
    \includegraphics[width=\linewidth]{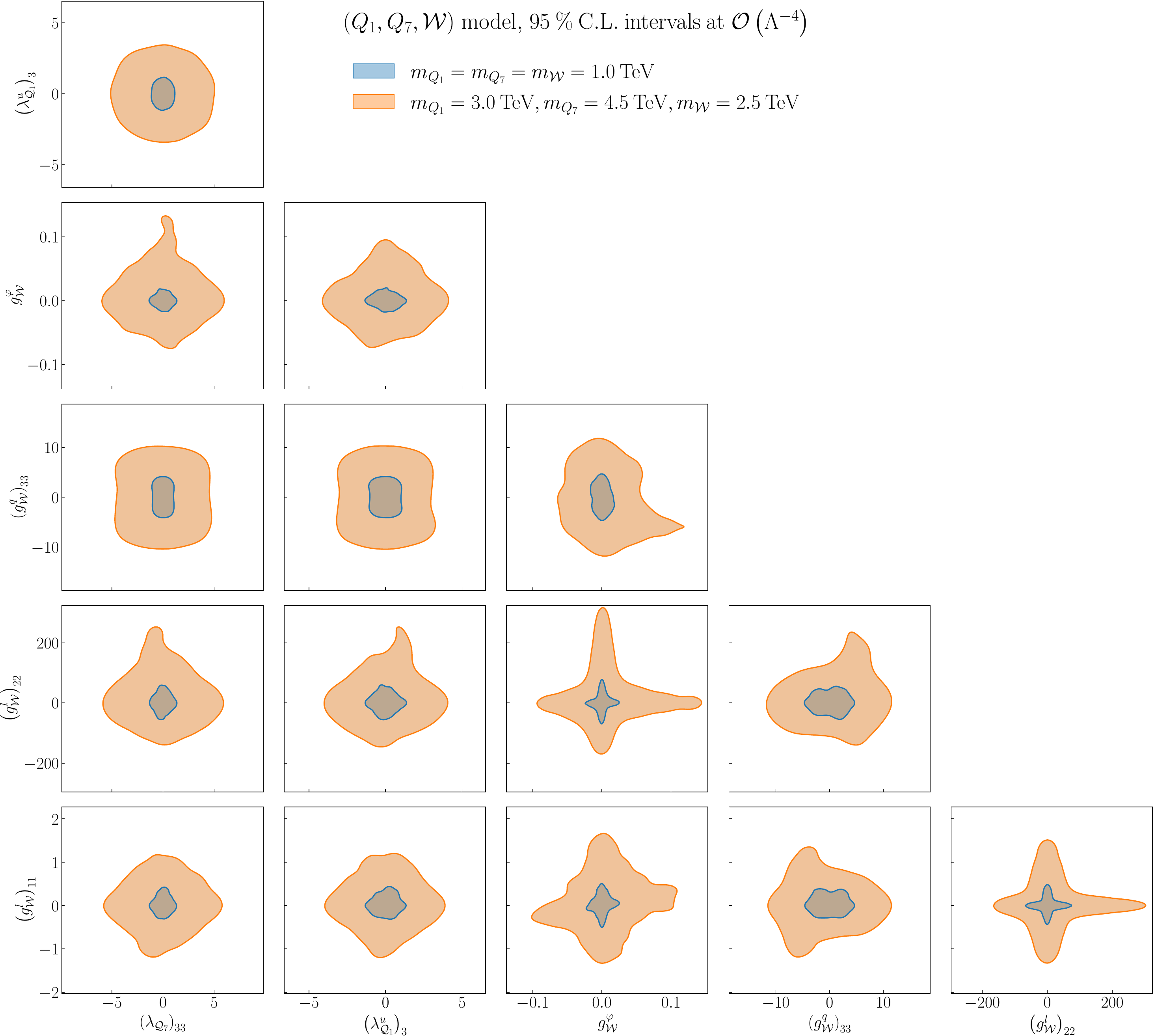}
    \caption{Pair-wise marginalised 
95\% contours for the UV couplings
associated
to the three-particle model consisting
of two heavy fermions, $Q_1$ and $Q_7$, and the heavy
vector boson $\mathcal{W}$, see also Fig.~\ref{fig:mp-posteriors} and Table~\ref{tab:cl-mp}
for the corresponding single-invariant posterior distributions and 95\% CI bounds.
We compare results for a common
value of the heavy mass, $m_{Q_1}=m_{Q_7}=
m_\mathcal{W} = 1$ TeV, with another
scenario
with different heavy masses,
$m_{Q_1}=3$ TeV, $m_{Q_7}=4.5$ TeV,
and $m_\mathcal{W}=2.5$ TeV. 
The EFT fit is carried out at NLO QCD
with $\mathcal{O}\lp \Lambda^{-4}\rp$
corrections included.
We exclude
$\left(g_{\mathcal{W}}^l\right)_{33}$ given that it is essentially
unconstrained from the fitted data.
}
    \label{fig:mp-different-masses}
\end{figure}

While the results
of Table~\ref{tab:cl-heavy-boson} assume
a common value of the heavy particle masses,
it is trivial to extend them
to different masses for each
of the different particles in the model.
This way, one can assess the dependence
of the UV-coupling fit results
on the assumptions of the heavy particle
masses.
With this motivation, Fig.~\ref{fig:mp-different-masses} displays pair-wise marginalised 95\% contours for the original Lagrangian parameters in the model.
The baseline results with a common mass of
1 TeV are compared to 
a scenario
with three different heavy masses,
$m_{Q_1}=3$ TeV, $m_{Q_7}=4.5$ TeV,
and $m_\mathcal{W}=2.5$ TeV. 
We exclude
$\left(g_{\mathcal{W}}^\ell\right)_{33}$ from this comparison, given that it is essentially
unconstrained from the fitted data.

From the comparison in  Fig.~\ref{fig:mp-different-masses} one observes, as expected, that assuming heavier BSM particles
leads to looser constraints on the
UV couplings.
Taking into account that different terms in the matching relations scale with the heavy particle
masses in a different manner, it
is not possible in general to rescale
the bounds obtained from the equal mass
scenario to another with different heavy
masses.
Nevertheless, given that in a single-particle extension we know that bounds worsen by
a factor $\lp m_{\rm UV}^{*}/m_{\rm UV}\rp^2$
if the heavy particle mass is increased
from $m_{\rm UV}$ up to $m_{\rm UV}^*$,
one can expect that in this three-particle scenario
the bounds are degraded by an amount between $\sim 5$ and $\sim 20$ depending on the specific UV coupling, a estimate which is consistent with the results in Fig.~\ref{fig:mp-different-masses}.
This comparison also highlights how
for very heavy particles we lose all
sensitivity and the global EFT fit cannot provide competitive constraints 
on the UV model parameters.

\subsection{Single-particle models matched at one-loop}
\label{subsec:results_oneloop}
All results presented so far relied on tree-level matching relations. 
We now present results for UV-coupling fits
in the case of matching relations obtained at the one-loop level, and study their effect on the UV parameter space as compared
to results based on tree-level matching.
For this purpose we can also deploy {\sc\small MatchMakerEFT}~interfaced to {\sc\small SMEFiT} via \matchTOfit~in order to obtain one-loop matching relations suitable for their use in the fit in an automated way.
In its current incarnation, this pipeline
enables using any
of the single-particle heavy scalar
and heavy fermion models (and combinations
thereof) listed
in Table~\ref{tab:UV_particles} 
when matched at the one-loop level.
We note here that the automation of the one-loop matching for  heavy vector bosons has not been 
achieved yet.
For concreteness, here
we present results based on the one-loop matching of the heavy scalar $\phi$ defined
by Eq.~(\ref{eq:Lag_UV_full_Varphi_model}), and also discussed in Sect.~\ref{subsec:heavy-scalar}, and the heavy fermions $T_1$ and $T_2$. We compare its bounds to those previously obtained from the tree-level matching analysis. 

Table~\ref{tab:cl-phi} compares
the 95\% CI bounds obtained for the two UV invariants in the $\phi$ model following either
tree-level or one-loop matching relations,
with the associated marginalised posterior
distributions shown in Fig.~\ref{fig:tree_vs_loop}.
In contrast to the bounds obtained at tree-level, one-loop corrections enable lifting the flat direction along the $|\lambda_\phi|$ invariant.
This effect is a consequence of the operators $\mathcal{O}_{\varphi \Box}, \mathcal{O}_{b\varphi}$ and $\mathcal{O}_{t\varphi}$, which receive additional one-loop matching contributions resulting in independent constraints on $\lambda_\phi$.
Specifically, the one-loop matching relations
for the Wilson coefficients
associated to $\mathcal{O}_{\varphi \Box}$ and $\mathcal{O}_{t\varphi}$ read

\begin{align}
    \label{eq:matching-1loop-lambda}
    \frac{c_{\varphi\Box}}{\Lambda^2} = & -\frac{g_1^4}{7680\pi^2}\frac{1}{m_\phi^2}-\frac{g_2^4}{2560\pi^2}\frac{1}{m_\phi^2}-\frac{3}{32\pi^2} \frac{\lambda_{\phi}^2}{m_\phi^2}, \nonumber \\
    \frac{c_{t\varphi}}{\Lambda^2} = & -\frac{\lambda_\phi\, \big(y_{\phi}^{u}\big)_{33}}{m_\phi^2} - \frac{g_{2}^4 y_{t}^{\text {SM}}}{3840 \pi^2}\frac{1}{m_\phi^2} + \frac{y_{t}^{\text {SM}}}{16\pi^2 }\frac{\lambda_\phi^2 }{m_\phi^2} +\frac{(4 \left(y_{b}^{\text{SM}} \right)^2 - 13 \left(y_{t}^{\text{SM}}\right)^2 )}{64\pi^2}\frac{\lambda_\phi\, \big(y_{\phi}^{u}\big)_{33}}{m_\phi^2} \nonumber\\
    & - \lp 12 \lambda_{\varphi}^{\text {SM}} + \left(y_{b}^{\text{SM}} \right)^2 - 11 \left(y_{t}^{\text{SM}} \right)^2 \rp \frac{y_{t}^{\text {SM}} }{64 \pi^2} \frac{\big(y_{\phi}^{u}\big)_{33}^2}{m_\phi^2} + \frac{3 }{128\pi^2}\frac{\lambda_\phi\, \big(y_{\phi}^{u}\big)_{33}^3}{m_\phi^2}\, ,
\end{align}

\noindent
where $\lambda_{\varphi}^{\text{SM}}$ is the quartic Higgs self-coupling in the SM.
In Eq.~\eqref{eq:matching-1loop-lambda}, 
all terms with a factor $1/\pi^2$ 
arise from one-loop corrections, indicating that
$ c_{\varphi\Box}$ is entirely loop-generated
while for $ c_{t\varphi}$
the tree-level matching relation is simply
$c_{t\varphi}=-\lambda_\phi\, \big(y_{\phi}^{u}\big)_{33}/m_\phi^2$.
This additional  dependence on $\lambda_\phi$
arising from one-loop corrections
is hence responsible for closing the tree-level
flat direction.
%

On the other hand, as a consequence of one-loop corrections to the matching relations
one also observes a degradation
in the sensitivity 
to the UV-invariant
$\mathrm{sgn}(\lambda_{\phi})\big(y_{\phi}^u\big)_{33}$.
The reason is that the parameter region
around  arbitrarily small values of
the coupling $\big(y_{\phi}^{u}\big)_{33}$ is disfavoured now, translating into a flattening of the distribution in $\big(y_{\phi}^{u}\big)_{33}$ as observed in the rightmost panels of Fig.~\ref{fig:tree_vs_loop}. 
These results showcase that one-loop corrections
bring in not only precision but also
accuracy, in that the stronger bound
on this UV-invariant at tree-level was a consequence of a flat direction which
is lifted as soon as perturbative
corrections are accounted for.

Interestingly, the middle panels of Fig.~\ref{fig:tree_vs_loop} also indicate the appearance of a double-peak structure in the distribution of $\mathrm{sgn}(\lambda_\phi) \big(y_{\phi}^{u}\big)_{33}$ at quadratic order in the EFT expansion which is absent in case of tree-level matching.
Such structure is associated to a second
minimum in the $\chi^2$ favouring
non-zero UV couplings, and may
be related to cancellations between
different terms in the matching relations.\footnote{Such cancellations may arise in EFT fits whenever linear and quadratic corrections become comparable.}
For one-loop matching relations
such as those displayed in Fig.~\ref{eq:matching-1loop-lambda},
the minimised figure of merit
will in general be a complicated higher-order
polynomial in the UV couplings, and
in particular 
in the presence of quadratic EFT
corrections, the $\chi^2(\mathbf{c})$
will be a quartic form of the Wilson
coefficients.
Therefore, for the specific case of 
the heavy scalar model, the $\chi^2(\mathbf{g})$ expressed in terms
of the UV couplings will include 
terms of up to $\mathcal{O}\lp \lambda_\phi^8\rp$
and $\mathcal{O}\lp \big( y_{\phi}^{u}\big)^{16}_{33}\rp$,
see Eqns~(\ref{eq:matching-1loop-lambda}) and~(\ref{eq:oneloop_logs_example}),
as well as the various cross-terms.
The minima structure 
of $\chi^2(\mathbf{g})$ can then only be resolved
by means of a numerical analysis
such as the one enabled by our strategy.

\begin{figure}
    \centering
    \includegraphics[width=.9\linewidth]{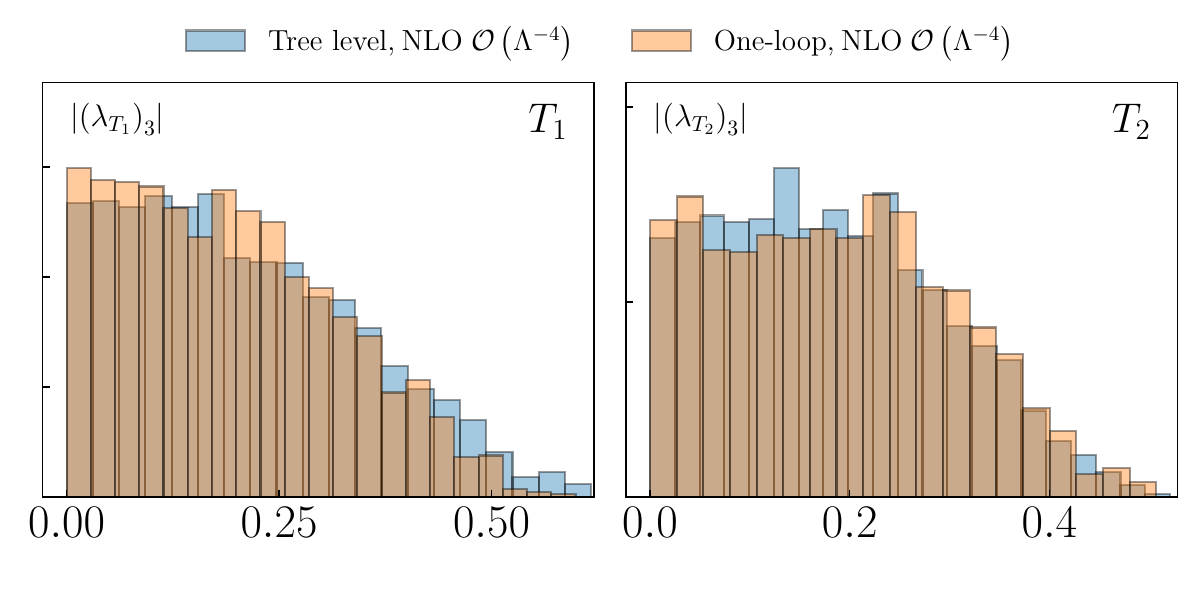}
    \caption{Posterior distributions associated to the UV invariants in the one-particle heavy fermion models $T_1$ and $T_2$, comparing the effect of tree and one-loop level matching. 
    In both cases, we display results based on quadratic corrections in the EFT expansion and at NLO QCD accuracy.}
    \label{fig:tree_vs_loop_T12}
\end{figure}

Regarding heavy fermions matched at one-loop, we include in Fig.~\ref{fig:tree_vs_loop_T12} a comparison of the effect of one-loop matching as compared to only tree-level for the models $T_1$ and $T_2$. We notice a slightly increased sensitivity thanks to the one-loop matching effects, with this additional sensitivity entering through the extra operators generated at one-loop. Indeed, these two models generated only four operators at tree-level, while at one-loop they generate all but two of the operators in our fitting basis. However, since the main constraint on these models come from EWPOs, the improvement on the bounds is relatively small. We stress that this limited impact of one-loop matching is not a general result, but limited to these vector-like fermion models.

The above discussion with the models $\phi$, $T_1$ and $T_2$ as examples demonstrates that one-loop corrections to the SMEFT matching relations provide non-trivial information on the parameter space of the considered
UV models. 
 Loop corrections not only  modify the sensitivity to UV couplings already constrained by tree-level,  but 
can also lift flat or quasi-flat  directions.
This result further motivates
ongoing efforts to increase
the perturbative accuracy
of EFT matching relations. 
\section{Summary and outlook}
\label{sec:summary}
In this chapter, we have presented and validated
a general strategy to scrutinise the parameters of BSM models  by using the SMEFT
as an intermediate bridge to the experimental
data.
By combining {\sc\small MatchMakerEFT} (for the matching between UV models and the EFT) with {\sc\small match2fit} (for the conversion of the matching results to fit run card) and {\sc\small SMEFiT} (for the comparison with data
and derivation of posterior distributions), this
pipeline enables to constrain the masses and
couplings of any UV model that can be matched
to the SMEFT either at tree-level or at one-loop.
While in this work we adopted {\sc\small MatchMakerEFT}, our approach is fully
general and any other matching framework could be adopted.

This flexible pipeline resulted
in the automation of the bound-setting
procedure on UV-complete models that can be matched to the SMEFT.
To illustrate its reach,
we applied it to derive constraints on a wide
range of single-particle
models extending the SM with heavy scalars,
fermions, or vector bosons with different
quantum numbers.
We also considered a three-particle
model which combines two heavy vector-like
fermions with a vector boson.
While most of the results presented arise
from tree-level matching relations, 
our approach applies equally well to one-loop matching results as shown by our results with one heavy scalar doublet model and two vector-like fermion models.


Several directions could be considered for future work generalising the results presented here. An obvious one is the automation of the UV invariants computation at one-loop and testing our techniques with one-loop matched vector boson models, once the matching results for the latter class of models become available. The generalisation of the flavour symmetry assumed for the fit would also allow to fit more general models consistently.

As we have shown in the case
of the heavy scalar model,  one-loop matching
allows us to bound directions in the UV parameter space which are
flat after tree-level matching. This can be crucial in resolving degeneracies among models, which could be further helped by including  dim-8 operators and their positivity bounds in the fit~\cite{Zhang:2021eeo}.
While a bottom-up determination of dim-8 operators from data is hampered by their large number
in the absence of extra assumptions, specific UV models only generate a relatively small number
of dim-8 operators, facilitating
their integration in the SMEFT fit. 

One could also consider accounting for RGE effects between the cut-off scale given by the UV model and the scale at which the global fit is performed. 
The latter could be combined with the inclusion of RGE effects in the cross sections entering the fit~\cite{Aoude:2022aro}.
%
%
%
Finally, it would be beneficial to consider
more flexible flavour symmetries which
are automatically consistent with the fitting
code, something which
would anyway be required for a full extension
of multi-particle models to one-loop matching.

All in all, the results presented in this chapter provide valuable
insights on ongoing model-building efforts by using the 
SMEFT to connect them with the experimental data,
hence providing complementary information to that derived
from direct searches for new heavy particles at the
LHC and elsewhere.
The pipeline adopted
in this chapter brings closer one of the original
goals of the EFT program at the LHC, namely
bypassing the need to directly compare
the predictions of UV-complete models
with the experimental measurements
and instead using the SMEFT
as the bridge connecting UV physics with data.

  \chapter{Optimal observables from Machine Learning}
\label{chap:ml4eft}
\vspace{-1cm}
\begin{center}
\begin{minipage}{1.\textwidth}
    \begin{center}
        \textit{This chapter is based on my results that are presented in Ref.~\mycite{GomezAmbrosio:2022mpm}} 
    \end{center}
\end{minipage}
\end{center}

\myparagraph{Introduction} The combined SMEFT interpretation of LHC data from the Higgs, top-quark, diboson and electroweak sectors presented in Chapter \ref{chap:smefit3}, as well as efforts from other groups \cite{DeBlas:2019ehy,Ethier:2021bye,Ethier:2021ydt, Ellis:2020unq,Bissmann:2020mfi,Bruggisser:2021duo,deBlas:2021wap}, rely on unfolded binned distributions provided by the experiments. That is,
they are based on reinterpreting ``SM measurements'' within the SMEFT framework.
In addition to such a combination of multiple datasets and processes, another avenue towards
improved SMEFT analyses is provided by the design
of tailored observables characterised by
enhanced, or even maximal, sensitivity to the underlying Wilson coefficients
for a given process. In this chapter, we introduce and develop \textit{optimal observables} based on Machine Learning techniques in the context of global SMEFT interpretations and show how they compare against approaches that rely on unfolded binned distributions. 

\myparagraph{Motivation} Optimal observables are able to maximally exploit the kinematic
information contained within a given measurement, event by event,
to carry out parameter inference by comparing with the corresponding theoretical predictions.
The low-multiplicity final states present in electron-positron collisions make them particularly
amenable to this strategy, and optimal observables have been used  in the
context of parameter fitting at
LEP, e.g.~\cite{DELPHI:2010ykq,Diehl:1993br}, and for future lepton collider studies~\cite{Durieux:2018tev}.
Constructing optimal observables is instead more difficult
in hadron collisions, where the higher complexity and
multiplicity of the final state, the significant
QCD shower and non-perturbative effects, and the need to
account for detector simulation make difficult the evaluation of the event-by-event likelihood.
This is one of the reasons why
most LHC measurements are presented as unfolded binned cross-sections,
with the exact statistical model~\cite{Cranmer:2021urp} replaced by a multi-Gaussian approximation.

Despite technical challenges associated to their definition and their
presentation~\cite{Arratia:2021otl},
there is growing evidence that at the LHC unbinned multivariate observables accounting for
the full event-by-event
kinematic information are advantageous to constrain the SMEFT parameters.
As an illustration, the most stringent limits on top quark EFT operators from CMS
data are those arising from unbinned detector-level observables~\cite{CMS:2022hjj,CMS:2021aly}.
As compared to traditional measurements, unbinned observables
enhance the sensitivity to EFT coefficients
by preventing the information loss incurred when adopting a specific binning or
when restricting the analysis to a subset of the possible final-state kinematic variables.
Constructing such observables for hadronic collisions can be achieved
with the analytical evaluation of the event likelihood using e.g. the Matrix Element
Method (MEM)~\cite{Campbell:2012cz,Artoisenet:2010cn,Gainer:2013iya,Fiedler:2010sg,Martini:2015fsa}
or numerically by means of Monte Carlo (MC) simulations.
In the latter case, Machine Learning techniques
provide a powerful toolbox to efficiently construct high-sensitivity
observables for EFT studies~\cite{Chen:2020mev,DAgnolo:2019vbw,DAgnolo:2018cun,Brehmer:2019xox,Brehmer:2018kdj,Brehmer:2018eca,Letizia:2022xbe,Chatterjee:2022oco,Chatterjee:2021nms,Brehmer:2019gmn,Bortolato:2020zcg,Butter:2021rvz,Arganda:2022qzy,Arganda:2022cl,Arganda:2022zbs,Gritsan:2020pib}, see
also~\cite{DeCastro:2018psv,Brehmer:2018hga,Wunsch:2020iuh,dAgnolo:2021aun,Coccaro:2019lgs} for related work.
Such optimal observables are relevant in other contexts beyond
EFTs such
as PDF fits~\cite{Gao:2017yyd,Kovarik:2019xvh}, see~\cite{Aggarwal:2022cki}
for a recent example.

In this chapter, we present a general framework
enabling the integration of tailored unbinned multivariate observables
from LHC processes within global SMEFT fits.
Our strategy, implemented in the {\sc\small python} open source package
 {\sc\small ML4EFT}, combines Machine Learning regression and classification techniques to
parameterise high-dimensional likelihood ratios for an arbitrary number of kinematic inputs
and EFT coefficients.
Once the likelihood ratio is parametrised in terms of
neural networks trained on MC simulations,
the posterior probability distributions
in the EFT coefficients  can be inferred by means of Nested Sampling.
The Monte Carlo replica method is used
to estimate methodological uncertainties, such
as those associated to the finite number of training events, and to propagate
them to the inferred confidence level intervals.
A key feature of {\sc\small ML4EFT}
is that the number of networks to be trained, which scales quadratically with the number of
EFT parameters, can be fully parallelised.
While previous studies of ML-assisted optimised observables for EFT fits
consider relatively small operator bases, our
framework is hence well-suited to construct general unbinned multivariate observables which
depend on up to tens of EFT coefficients  as required in global fits.

As a proof of concept of the  {\sc\small ML4EFT} framework,
we construct unbinned multivariate observables for
inclusive top-quark pair production
in the $ b \ell^+  \nu_{\ell} \bar{b}  \ell^-  \bar{\nu}_{\ell}$ (dilepton)
final state. We refer to our work in Ref.~\mycite{GomezAmbrosio:2022mpm} for additional results for 
Higgs boson production in association with a $Z$ boson in the $b\ell^+\nu_\ell\bar{b}\ell^-\bar{\nu}_\ell$ (dilepton) final state. 
We consider fiducial regions where these
measurements are statistically-limited and therefore systematic errors can be neglected.
Whenever possible, we compare the results based on the ML parametrisation
with those provided by the analytical evaluation of the exact event-by-event likelihood.
We demonstrate the improved constraints that these unbinned multivariate observables
provide on the SMEFT parameter space as compared to their binned counterparts,
and study the information gain associated to the inclusion of multiple kinematic inputs.
Our analysis motivates and defines a possible roadmap towards the measurement (and delivery)
of unbinned observables tailored to SMEFT parameters at the LHC.

\myparagraph{Outline} The outline of this chapter is as follows.
%
%
First of all, Sect.~\ref{sec:trainingMethodology} introduces unbinned multivariate observables in the context of the SMEFT and how
Machine Learning is deployed to parametrise high-dimensional likelihood functions.
Sect.~\ref{sec:pseudodata} describes our pipeline for
the MC simulation of LHC events in the SMEFT and the settings
of the pseudo-data generation.
Our results are presented in Sect.~\ref{sec:results_ml4eft}, which
quantifies the constraints on the EFT parameter space
provided by unbinned observables.
Finally, in Sect.~\ref{sec:summary_ml4eft} we summarise and discuss possible
future avenues.

\section{Unbinned observables from Machine Learning}
\label{sec:trainingMethodology}

Here we describe our approach
 to construct unbinned multivariate
observables tailored for global EFT analyses by means
of supervised Machine Learning.
We discuss how neural networks are deployed as universal unbiased interpolants in order
to parametrise likelihood ratios associated to the theoretical models
of the SM and EFT differential cross-sections, making possible the efficient
 evaluation of the likelihood functions for arbitrary values
of the Wilson coefficients as required for parameter inference.
We emphasise the scalability and robustness of our approach with respect
to the number of coefficients and to the dimensionality of the final state
kinematics, and validate the performance of the neural network training.

\subsection{Differential cross-sections}
\label{sec:param_eft_xsec}
Let us consider a given process
whose associated measurement $\mathcal{D}$ consists of
$ N_{\rm ev}$ events, each of them characterised by $n_{k}$ final state
variables,
\begin{equation} 
\mathcal{D}=\{ \boldsymbol{x}_i\}\, \qquad
\boldsymbol{x}_i= \lp x_{i,1}, x_{i,2},\ldots, x_{i,n_{k}} \rp \, ,\qquad  i=1, \ldots, N_{\rm ev} \, .
\end{equation} 
The kinematic variables (features) $\boldsymbol{x}$ under consideration depend on the type
of measurement that is being carried out.
For instance, for a top quark measurement at the parton level, one would have
that the $\boldsymbol{x}_i$ are the transverse momenta and rapidities of the top quark,
while for the corresponding particle-level measurement, one would use instead
$b$-jet and leptonic kinematic variables.
Likewise,  $\boldsymbol{x}_i$ could also correspond
to detector-level kinematic variables for measurements
carried out without unfolding.
The inclusive cross-section case corresponds to $n_k=0$
when one integrates over all final state kinematics subject to fiducial cuts.
The probability distribution associated to the events constituting $\mathcal{D}$
is given by the differential cross-section
\begin{equation}
  \label{eq:prob_density_sect3}
f_\sigma\lp \boldsymbol{x}, \boldsymbol{c} \rp= \frac{1}{\sigma_{\rm fid}(\boldsymbol{c}) }\frac{d\sigma(\boldsymbol{x}, \boldsymbol{c})}{d\boldsymbol{x}} \, ,
\end{equation}
in terms of the model parameters $\boldsymbol{c}$.

In general not all $n_k$ kinematic variables that one can consider for a given process
will be independent.
For example, $2\to 2$ processes with on-shell
particles (like $pp\to t\bar{t}$
before decay) are fully described by three independent final-state variables.
For more exclusive measurements, $n_k$ grows rapidly yet the final-state variables
remain partially correlated to each other.
The best choice of $\boldsymbol{x}$ and $n_k$ should in this respect
be studied separately from the impact associated
to the use of unbinned observables as compared to their binned
counterparts.
Furthermore, in the same manner that one expects that the constraints provided by a binned observable
tend to those from unbinned ones in the narrow bin limit,
these constraints will also saturate once $n_k$ becomes large enough
that adding more variables does not provide independent information.

In the specific case of the SMEFT,
the parameters of the theory framework $\mathcal{T}\lp \boldsymbol{c}\rp$
are the Wilson coefficients associated
to the $n_{\rm eft}$
higher dimensional operators that enter the description of the processes
under consideration for a given set of flavour assumptions.
Given that a differential cross-section $d\sigma(\boldsymbol{x}, \boldsymbol{c})$ in the dim-6 SMEFT
exhibits at most a quadratic dependence with the Wilson coefficients,
one can write the differential probability density in
Eq.~(\ref{eq:prob_density_sect3}) as
\begin{equation} 
\label{eq:EFT_structure}
d\sigma(\boldsymbol{x}, \boldsymbol{c})  =
d\sigma(\boldsymbol{x},\boldsymbol{0}) + \sum_{j=1}^{n_{\rm{eft}}}
d\sigma^{(j)}(\boldsymbol{x})c_j
+\sum_{j=1}^{n_{\rm{eft}}}\sum_{k \ge j}^{n_{\rm{eft}}}
d\sigma^{(j,k)}(\boldsymbol{x})c_j c_k \, ,
\end{equation} 
where the cut-off scale $\Lambda$ is absorbed into the definition of the Wilson coefficients, $d\sigma(\boldsymbol{x},\boldsymbol{0}) $ corresponds to the SM
cross-section, $d\sigma^{(j)}(\boldsymbol{x})$ indicates the linear EFT corrections
arising from the interference with the SM amplitude,
and $d\sigma^{(j,k)}(\boldsymbol{x})$ corresponds to the quadratic corrections
associated to the square of the EFT amplitude.
We note that while $d\sigma(\boldsymbol{x},\boldsymbol{0}) $
and $d\sigma^{(j,k)}(\boldsymbol{x})$ arise from squared amplitudes
and hence are positive-definite, this is not necessarily the case
for the interference cross-section $d\sigma^{(j)}(\boldsymbol{x})$.

The SM and EFT differential
cross-sections $d\sigma(\boldsymbol{x},\boldsymbol{0}), d\sigma^{(j)}(\boldsymbol{x})$,
and $d\sigma^{(j,k)}(\boldsymbol{x})$ can be evaluated in perturbation theory, and one can account
for different types of effects such as parton shower, hadronisation, or detector simulation,
depending on the observable under consideration.
The SM differential cross-sections $d\sigma(\boldsymbol{x},\boldsymbol{0})$ can be
computed at NNLO QCD (eventually matched
to parton showers) for most of the LHC processes relevant for global EFT fits, while
for the EFT linear and quadratic corrections the accuracy frontier is NLO QCD~\cite{Degrande:2020evl}.
The settings of the calculation should be chosen to reproduce
as close as possible those of the corresponding
experimental measurement, while aiming to minimise
the associated theoretical uncertainties.
In this chapter, we evaluate the differential cross-sections $d\sigma(\boldsymbol{x},\boldsymbol{c})$
numerically, cross-checking with analytic calculations  whenever possible.

In order to construct unbinned observables in an efficient manner it
is advantageous to work in terms of the ratio between EFT and SM
cross-sections, which,
accounting for the quadratic structure of the EFT cross-sections in Eq.~(\ref{eq:EFT_structure}),
can be expressed as
\begin{equation} 
\label{eq:EFT_structure_v2}
r_{\sigma}(\boldsymbol{x}, \boldsymbol{c}) \equiv \frac{d\sigma(\boldsymbol{x}, \boldsymbol{c})}{d\sigma(\boldsymbol{x}, \boldsymbol{0})} =
1 + \sum_{j=1}^{n_{\rm eft}}
r^{(j)}_\sigma(\boldsymbol{x})c_j
+\sum_{j=1}^{n_{\rm eft}}\sum_{k \ge j}^{n_{\rm eft}}
r^{(j,k)}_\sigma(\boldsymbol{x})c_j c_k \, ,
\end{equation} 
where we have defined the linear and quadratic ratios to the SM cross-section as
\begin{equation} 
r^{(j)}_\sigma(\boldsymbol{x}) = \frac{d\sigma^{(j)}(\boldsymbol{x})}{
  d\sigma(\boldsymbol{x},\boldsymbol{0})} \, ,\qquad
r^{(j,k)}_\sigma(\boldsymbol{x}) = \frac{d\sigma^{(j,k)}(\boldsymbol{x})}{
d\sigma(\boldsymbol{x},\boldsymbol{0})} \, .
\end{equation} 
Parameterising the ratios between the EFT and SM cross-sections, Eq.~(\ref{eq:EFT_structure_v2}), is
beneficial as compared to directly
parameterising the absolute cross-sections since in general EFT effects
represent a moderate distortion of the SM baseline prediction.

The profile likelihood ratio, first defined in Eq.~\eqref{eq:plr_4}, is used to derive limits on the EFT coefficients and
can be expressed in terms of the ratio Eq.~(\ref{eq:EFT_structure_v2}).
Indeed, in the case of the dim-6 SMEFT
the profile likelihood ratio reads
\begin{equation} 
  q_{\boldsymbol{c}}&=& 2\left[\nu_{\rm tot}(\boldsymbol{c})-\sum_{i=1}^{N_{\rm ev}} \log\lp
    1 + \sum_{j=1}^{n_{\rm eft}}
r^{(j)}_\sigma(\boldsymbol{x}_i)c_j
+\sum_{j=1}^{n_{\rm eft}}\sum_{k \ge j}^{n_{\rm eft}}
r^{(j,k)}_\sigma(\boldsymbol{x}_i)c_j c_k\rp
\right] \nonumber \\
 &-& 2\left[\nu_{\rm tot}(\boldsymbol{\hat{c}})-\sum_{i=1}^{N_{\rm ev}} \log
\lp 1 + \sum_{j=1}^{n_{\rm eft}}
r^{(j)}_\sigma(\boldsymbol{x}_i)\hat{c}_j
+\sum_{j=1}^{n_{\rm eft}}\sum_{k \ge j}^{n_{\rm eft}}
r^{(j,k)}_\sigma(\boldsymbol{x}_i)\hat{c}_j \hat{c}_k\rp
    \right] \, .
\label{eq:plr_6}
\end{equation} 
where the  $\boldsymbol{\hat{c}}$
denotes the maximum likelihood estimator of the Wilson coefficients.
We emphasise that in this derivation the SM serves as a natural reference hypothesis in the EFT parameter space - ratios expressed with respect to another reference point, say $\boldsymbol{c'}$, are trivially equivalent according to the following identity
\begin{equation} 
\label{eq:EFT_structure_v3}
r_{\sigma}(\boldsymbol{x}, \boldsymbol{c}) = \frac{d\sigma(\boldsymbol{x},\boldsymbol{c})}{d\sigma(\boldsymbol{x},\boldsymbol{0})} =
\frac{d\sigma(\boldsymbol{x},\boldsymbol{c})}{d\sigma(\boldsymbol{x},\boldsymbol{c'})} \frac{d\sigma(\boldsymbol{x},\boldsymbol{c'})}{d\sigma(\boldsymbol{x},\boldsymbol{0})} \, .
\end{equation} 

The main challenge in applying limit setting to unbinned observables
by means of the profile likelihood ratio of Eq.~(\ref{eq:plr_6}) is that the evaluation
of the EFT cross-section ratios $r^{(j)}_\sigma(\boldsymbol{x})$ and $r^{(j,k)}_\sigma(\boldsymbol{x})$ is
computationally intensive, and in many cases intractable, specifically for high-multiplicity
observables and  when the number of events considered $N_{\rm ev}$ is large.
As we explain next, in this work we bypass this challenge by parameterising the EFT cross-section ratios in terms
of feed-forward neural networks, with the kinematic variables $\boldsymbol{x}$ as inputs,
trained on the outcome of Monte Carlo simulations.

\subsection{Cross-section parametrisation}
\label{sec:methodology_xsec_param}
The profile likelihood ratio provides an optimal test statistic
in the sense that no statistical power is lost in the process of mapping the high-dimensional feature vector $\boldsymbol{x}$
onto the scalar ratio $r_\sigma(\boldsymbol{x}, \boldsymbol{c})$.
Performing inference on the Wilson coefficients
using the profile likelihood ratio from Eq.~(\ref{eq:plr_6})
requires a precise knowledge about the differential cross section ratio $r_\sigma(\boldsymbol{x}, \boldsymbol{c})$ for arbitrary values
of $\boldsymbol{c}$.
However, in general one does not have direct access to $r_\sigma(\boldsymbol{x}, \boldsymbol{c})$
whenever MC event generators can only be run in the forward mode, i.e. used to generate samples.
The inverse problem, namely statistical inference, is often rendered intractable due to the many paths in parameter space that lead from the theory parameters $\boldsymbol{c}$ to the final measurement in the detector.
In the Machine Learning literature this intermediate (hidden) space is known as the latent space.

Feed-forward neural networks are suitable in this context
as model-independent unbiased interpolants to construct a surrogate of the true profile likelihood ratio.
Consider two balanced
datasets $\mathcal{D}_{\rm eft}(\boldsymbol{c})$ and $\mathcal{D}_{\rm sm}$ generated
based on the theory hypotheses $\mathcal{T}(\boldsymbol{c})$ and $\mathcal{T}(\boldsymbol{0})$ respectively,
where by balanced we mean that the same number of unweighted Monte Carlo events are generated in both cases.
We would like to determine the decision boundary
function $g(\boldsymbol{x,c})$ which can be used to classify an event
$\boldsymbol{x}$ into either $\mathcal{T}(\boldsymbol{0})$, the Standard Model,
or $\mathcal{T}(\boldsymbol{c})$, the SMEFT hypothesis for point $\boldsymbol{c}$
in parameter space.
We can determine this decision boundary
by using the balanced datasets $\mathcal{D}_{\rm eft}(\boldsymbol{c})$ and $\mathcal{D}_{\rm sm}$
to train a binary classifier by means of the cross-entropy loss-functional, defined as
\begin{equation} 
L[g(\boldsymbol{x},\boldsymbol{c})] = -\int d\boldsymbol{x}\frac{d\sigma(\boldsymbol{x},\boldsymbol{c})}{d\boldsymbol{x}}\log(1-g(\boldsymbol{x},\boldsymbol{c})) - \int d\boldsymbol{x}\frac{d\sigma(\boldsymbol{x},\boldsymbol{0})}{d\boldsymbol{x}}\log g(\boldsymbol{x},\boldsymbol{c}) \, .
\label{eq:loss_CE}
\end{equation} 
In practice, the integrations required in the evaluation of the cross-entropy loss Eq.~(\ref{eq:loss_CE}) are carried out numerically from the generated Monte Carlo events, such that
\begin{equation} 
L[g(\boldsymbol{x},\boldsymbol{c})] = - \sigma_{\rm fid}(\boldsymbol{c})\sum_{i=1}^{N_{\rm ev}}\log(1-g(\boldsymbol{x}_i,\boldsymbol{c})) - \sigma_{\rm fid}(\boldsymbol{0}) \sum_{j=1}^{N_{\rm ev}} \log g(\boldsymbol{x}_j,\boldsymbol{c}) \, ,
\label{eq:loss_CE_numeric}
\end{equation} 
where $\sigma_{\rm fid}(\boldsymbol{c})$ and $\sigma_{\rm fid}(\boldsymbol{0})$
represent the integrated fiducial cross-sections in the SMEFT and the SM
respectively.
Recall that we have two independent sets of $N_{\rm ev}$ events each generated under
$\mathcal{T}(\boldsymbol{0})$ and $\mathcal{T}(\boldsymbol{c})$ respectively,
and hence in Eq.~(\ref{eq:loss_CE_numeric}) the first (second) term
in the RHS involves the sum over the $N_{\rm ev}$ events generated according to $\mathcal{T}(\boldsymbol{c})$
($\mathcal{T}(\boldsymbol{0})$).

It is also possible to adopt other loss functions for the binary classifier Eq.~(\ref{eq:loss_CE}),
such as the quadratic loss used in~\cite{Chen:2020mev}.
The outcome of the classification
should be stable with respect to alternative choices of the loss function, and indeed
we find that both methods lead to consistent results, while the cross entropy formulation
benefits from a faster convergence due to presence of stronger gradients as compared
to the quadratic loss.

In the limit of an infinitely large training dataset
and sufficiently flexible parametrisation, one can take the functional derivative of $L$ with respect to the
decision boundary function $g(\boldsymbol{x}, \boldsymbol{c})$ to determine that it is given by
\begin{equation} 
\frac{\delta L}{\delta g} = 0 \implies g(\boldsymbol{x}, \boldsymbol{c}) = \lp 1 +\frac{d\sigma(\boldsymbol{x},\boldsymbol{c})}{d\boldsymbol{x}}
\Bigg/ \frac{d\sigma(\boldsymbol{x},\boldsymbol{0})}{d\boldsymbol{x}} \rp^{-1} = \frac{1}{1 + r_\sigma(\boldsymbol{x}, \boldsymbol{c})} \, ,
\label{eq:optimal_f}
\end{equation} 
and hence in this limit
the solution of the classification problem defined by the cross-entropy
function Eq.~(\ref{eq:loss_CE}) is given by the EFT ratios $r_\sigma(\boldsymbol{x}, \boldsymbol{c})$
that need to be evaluated
in order to determine the associated profile likelihood ratio.
Hence our strategy will be to parametrise $r_\sigma(\boldsymbol{x}, \boldsymbol{c})$ with neural networks,
benefiting from the characteristic quadratic structure of the EFT cross-sections, and then
training these Machine Learning classifiers by minimising the loss function Eq.~(\ref{eq:loss_CE}).

In practice, one can only expect to obtain a reasonably good estimator $\hat{g}$ of the true result due to finite size effects
in the Monte Carlo training data $\mathcal{D}_{\rm eft}$ and $\mathcal{D}_{\rm sm}$ and in the neural network
architecture.
Since EFT and SM predictions largely overlap in a significant region of the phase space,
it is crucial to obtain a decision boundary trained with as much precision as possible
in order to have a reliable test statistic to carry out inference.
The situation is in this respect different from  usual classification problems, for which an imperfect decision boundary
parameterised by $g$ can still achieve high performances whenever most features are disjoint, and hence
a slight modification of  $g$ does not lead to a significant performance drop.
In order to estimate the uncertainties associated to the fact that
the actual estimator $\hat{g}$ differs from the true result $g(\boldsymbol{x}, \boldsymbol{c})$,
in this work we use the Monte Carlo replica method
described in Sect.~\ref{sec:nntraining}.

Given the quadratic structure of the EFT cross-sections and their ratios to the SM prediction, Eqns.~(\ref{eq:EFT_structure})
and~(\ref{eq:EFT_structure_v2}) respectively, once the  linear and quadratic ratios $r_\sigma^{(j)}({\boldsymbol{x}})$ and
$r_\sigma^{(j, k)}({\boldsymbol{x}})$ are determined throughout the entire phase space one can
straightforwardly evaluate the EFT
differential cross sections (and their ratios to the SM) for
any point in the EFT parameter space.
Here we exploit this property during the neural network
training by decoupling the learning problem of the linear cross section ratios from that
of the quadratic ones.
This allows one to extract $r_\sigma^{(j)}$ and $r_\sigma^{(j, k)}$ independently from each other, meaning that the neural network classifiers can be trained in parallel and also that the
training scales at most quadratically with the number of EFT operators considered
$n_{\rm eft}$. 

To be specific, at the linear level we determine the EFT cross-section ratios
$r_\sigma^{(j)}({\boldsymbol{x}})$ by training the binary classifier from the cross-entropy loss
Eq.~(\ref{eq:loss_CE}) on a reference dataset $\mathcal{D}_{\rm sm}$ and an EFT dataset defined by
\begin{equation} 
\label{eq:eft_dataset_onecoeff}
\mathcal{D}_{\rm eft}(\boldsymbol{c}=(0,\ldots,0,c^{\rm (tr)}_j,0,\ldots,0)) \, ,
\end{equation} 
and generated at linear order,  $\mathcal{O}\left(\Lambda^{-2}\right)$, in the EFT expansion with all Wilson coefficients
set to zero except for the $j$-th one, which we denote by $c^{\rm (tr)}_j$.
For such model configuration, the EFT cross-section ratio can be parametrised as
\begin{equation} 
r_\sigma(\boldsymbol{x}, c^{\rm (tr)}_j) = 1 + c^{\rm (tr)}_j \mathrm{NN}^{(j)}(\boldsymbol{x}) \, ,
\label{eq:nn_param_lin}
\end{equation} 
where only the individual coefficient $c^{\rm (tr)}_j$ has survived
the sum in Eq.~(\ref{eq:EFT_structure_v2}) since all other EFT parameters
are switched off by construction.
Comparing Eq.~(\ref{eq:nn_param_lin}) and Eq.~(\ref{eq:EFT_structure_v2}) we see that in the large sample limit
\begin{equation} 
\mathrm{NN}^{(j)}(\boldsymbol{x}) \rightarrow r_{\sigma}^{(j)}(\boldsymbol{x}) \, .
\label{eq:trained_lin_coeff}
\end{equation} 
In practice, this relation will only be met with a certain finite accuracy due to statistical fluctuations in the finite training sets.
This limitation is especially relevant in phase space regions where the cross-section is suppressed, such as in the tails of invariant mass distributions,
and indicates that it is important to account for these methodological uncertainties associated to the training procedure.
By means of the Monte Carlo replica method one can estimate and propagate these uncertainties
first to the parametrisation of the EFT ratio $r_{\sigma}$ and then to the associated limits on the
Wilson coefficients.

Concerning the training of the EFT quadratic cross-section ratios $r_\sigma^{(j, k)}$, we follow the same strategy as in the linear case, except that now we construct the EFT dataset at quadratic order without any linear contributions.
By omitting the linear term, we reduce the learning problem at the quadratic level to a linear one.
Specifically, we generate events at pure $\mathcal{O}\left(\Lambda^{-4}\right)$ level, without the
interference (linear) contributions,
in the EFT by switching off all Wilson coefficients except two of them,
denoted by $c^{\rm (tr)}_j$ and $c^{\rm (tr)}_k$, 
\begin{equation} 
\label{eq:eft_dataset_twocoeff}
\mathcal{D}_{\rm eft}(\boldsymbol{c}=(0,\ldots,0,c^{\rm (tr)}_j,0,\ldots,0, c^{\rm (tr)}_k, 0, \ldots )) \, ,
\end{equation} 
and parametrise the cross-section ratio as
\begin{equation} 
r_\sigma(\boldsymbol{x}, c^{\rm (tr)}_j, c^{\rm (tr)}_k) = 1 + c^{\rm (tr)}_j c^{\rm (tr)}_k \mathrm{NN}^{(j, k)}(\boldsymbol{x}) \, ,
\label{eq:nn_param_quad}
\end{equation} 
where only purely quadratic terms with both $c^{\rm (tr)}_j$ and $c^{\rm (tr)}_k$ have survived the sum.
Note that when $j \neq k$, this parametrisation of the cross-section ratio $r_\sigma(\boldsymbol{x}, c^{\rm (tr)}_j, c^{\rm (tr)}_k)$ depends only on the product $c_{j} c_{k}$,
whereas when $j=k$ it depends only on terms proportional to $c_{j}^{2}$.
The cross-section ratio is parametrised in this way
to facilitate separate training of the $c_{j}^{2}$, $c_{k}^{2}$ and $c_{j} c_{k}$ terms, and we make use of training data
in which the contributions from each of these terms has been separately generated, as discussed in more detail in Sect.~\ref{sec:pipeline}. 
By the same reasoning as above, in the large sample limit we will have that
\begin{equation} 
\mathrm{NN}^{(j, k)}(\boldsymbol{x}) \rightarrow r_{\sigma}^{(j, k)}(\boldsymbol{x}) \, .
\label{eq:trained_quad_coeff}
\end{equation} 
We note that in the case that the Monte Carlo generator used to evaluate
the theory predictions $\mathcal{T}(\boldsymbol{c})$ does not allow
the separate evaluation of the EFT quadratic terms, one can always subtract the linear
contribution numerically by means of the outcome of Eq.~(\ref{eq:trained_lin_coeff}).

By repeating this procedure $n_{\rm eft}$ times for
the linear terms and $n_{\rm eft}(n_{\rm eft}+1)/2$ times for the quadratic terms, one ends up with the
set of functions that parametrise the EFT cross-section ratio Eq.~(\ref{eq:EFT_structure_v2}),
\begin{equation} 
\label{eq:set_of_trained_NNs}
\{ \mathrm{NN}^{(j)}(\boldsymbol{x}) \} \quad {\rm and}\quad \{ \mathrm{NN}^{(j,k)}(\boldsymbol{x}) \} \, ,
\quad  j, k=1,\ldots,n_{\rm eft}\,, \quad k\ge j \, .
\end{equation} 
The similar structure that is shared between Eq.~(\ref{eq:nn_param_lin}) and Eq.~(\ref{eq:nn_param_quad}) implies that parameterising the quadratic EFT contributions
in this manner is ultimately a linear problem, i.e. redefining the product $c^{\rm (tr)}_j c^{\rm (tr)}_k$ as  $\tilde{c}^{\rm (tr)}_{j,k}$ maps the quadratic learning problem back to a linear one:
\begin{equation} 
r_\sigma(\boldsymbol{x}, \tilde{c}^{\rm (tr)}_{j,k}) = 1 + \tilde{c}^{\rm (tr)}_{j,k}\mathrm{NN}^{(j, k)}(\boldsymbol{x}) \, .
\label{eq:nn_param_quad_redef}
\end{equation} 
Eq.~(\ref{eq:set_of_trained_NNs}) represents the final outcome of the training procedure,
namely an approximate parametrisation  $\hat{r}_{\sigma}(\boldsymbol{x}, \boldsymbol{c})$ of the true
EFT cross-section ratio  $r_\sigma(\boldsymbol{x}, \boldsymbol{c})$,
\begin{equation} 
\label{eq:EFT_structure_v4}
\hat{r}_{\sigma}(\boldsymbol{x}, \boldsymbol{c}) =
1 + \sum_{j=1}^{n_{\rm eft}}
 \mathrm{NN}^{(j)}(\boldsymbol{x}) c_j
+\sum_{j=1}^{n_{\rm eft}}\sum_{k \ge j}^{n_{\rm eft}}
\mathrm{NN}^{(j, k)}(\boldsymbol{x}) c_j c_k \, ,
\end{equation} 
valid for any point in the model parameter $\boldsymbol{c}$,
as 
required to evaluate the profile likelihood ratio, Eq.~(\ref{eq:plr_4}),
and to
perform inference on the Wilson coefficients.
Below we provide technical details about how the neural network training
is carried out and how uncertainties are estimated by means of the replica method.

\myparagraph{Cross-section positivity during training}
While the differential cross-section $f_\sigma\lp \boldsymbol{x}, \boldsymbol{c}\rp$ (and its ratio
to the SM) is positive-definite, this is not necessarily the case for the linear
(interference) EFT term, and hence in principle Eq.~(\ref{eq:nn_param_lin}) is unbounded from below.

At the level of the training pseudo-data, we avoid the issue of negative cross-sections by generating our
pseudo-data at fixed values of the Wilson coefficients, specifically chosen such that the differential cross sections 
are always positive.  For example, in the case of negative interference between the EFT and the SM, we generate our training pseudo-data
assuming a negative Wilson coefficient such that the net effect of the EFT is an enhancement relative to the SM.
The choices of Wilson coefficients used in our study will be further discussed in Sect.~\ref{sec:inputs_traiing} and in Table~\ref{tab:settings_nntrain}.

It only then remains to ensure  
that the physical requirement of cross-section positivity
is satisfied at the level
of neural network training, and hence that the parameter space region leading
to negative cross-sections is avoided.
Cross-section positivity can be implemented at the training level
by means
of adding a penalty term to the loss function
whenever the likelihood ratio becomes negative through a Lagrange multiplier.
That is, the loss function is extended as
\begin{equation} 
\label{eq:cross-section-pos}
L[g] \rightarrow L[g] + \lambda \, \mathrm{ReLU}\left(\frac{g(\boldsymbol{x},\boldsymbol{c})-1}{g(\boldsymbol{x},\boldsymbol{c})}\right) = L[g] + \lambda \, \mathrm{ReLU}\left(
- r_{\sigma}(\boldsymbol{x},\boldsymbol{c})\right)  \, ,
\end{equation} 
where $\mathrm{ReLU}$ stands for the Rectified Linear Unit activation function.
Such a Lagrange multiplier penalises configurations where the likelihood ratio
becomes negative,  with the penalty increasing the more negative
$r_{\sigma}$ becomes. The value of the hyperparameter $\lambda$
should be chosen such that the training in the physically
allowed region is not distorted.
This is the same method used in the NNPDF4.0 
analysis to implement PDF positivity and integrability~\cite{NNPDF:2021uiq,NNPDF:2021njg}
at the training level without having to impose these constraints in at the parametrisation level.

However, the Lagrange multiplier method defined by Eq.~(\ref{eq:cross-section-pos}) is not compatible with the cross-entropy loss function of Eq.~(\ref{eq:loss_CE}), given that this loss function is only well defined for $0 < g(\boldsymbol{x}, \boldsymbol{c}) < 1$ corresponding to positive likelihood ratios.
We note that this is not the case for other loss-functions for which configurations
with $r_\sigma(\boldsymbol{x}, \boldsymbol{c}) < 0$ are allowed, such as the quadratic loss-function used by~\cite{Chen:2020mev}, making them in principle compatible with the Lagrange multiplier method
to ensure cross-section positivity.

Instead of using the Lagrange multiplier method, in this work we introduce an alternative parameterisation of the cross section ratio $r_\sigma$ such that cross-section positivity is guaranteed by construction.
Specifically, we modify  Eq.~(\ref{eq:nn_param_lin}) to enforce positivity, namely
the condition
\begin{equation} 
r_{\sigma}(\boldsymbol{x}, \boldsymbol{c}) = \lp 1 + c^{(\rm tr)}_{j}\cdot \mathrm{NN}^{(j)}(\boldsymbol{x}) \rp > 0 \, ,
\end{equation} 
for any value of $\boldsymbol{x}$ and $ \boldsymbol{c}$,
by transforming the outcome of the neural network $\mathrm{NN}^{(j)}(\boldsymbol{x})$
as follows
\begin{equation}
  \mathrm{NN}^{(j)}(\boldsymbol{x}) \rightarrow \widetilde{\mathrm{NN}}^{(j)}(\boldsymbol{x}; c^{(\rm tr)}_{j}) =
  \begin{cases}
        \mathrm{ReLU}(\mathrm{NN}^{(j)}(\boldsymbol{x})) - 1/c^{(\rm tr)}_{j} + \epsilon,  & \text{if}\ c^{(\rm tr)}_{j}>0 \\
        - \mathrm{ReLU}(\mathrm{NN}^{(j)}(\boldsymbol{x})) - 1/c^{(\rm tr)}_{j} - \epsilon, & \text{if}\ c^{(\rm tr)}_{j}<0
      \end{cases},
  \label{eq:xsec_pve_2}
  \end{equation}
where $\epsilon$ is an infinitesimal positive constant to ensure $r_\sigma(\boldsymbol{x}, \boldsymbol{c}) > 0 $ when the linear contribution becomes negative, $\mathrm{NN}^{(j)}(\boldsymbol{x})<0$.
The transformation of Eq.~(\ref{eq:xsec_pve_2}) can be thought of as adding a custom activation function at the end of the network such that the cross-entropy loss is well-defined throughout the entire training procedure.
We stress that it is the transformed neural network $\widetilde{\mathrm{NN}}^{(j)}$ which is subject to training and not the original $\mathrm{NN}^{(j)}$.
Regarding imposing cross-section positivity at the quadratic level, we note that the transformation of Eq.~(\ref{eq:xsec_pve_2}) applies just as well in the quadratic case by virtue of Eq.~(\ref{eq:nn_param_quad_redef}),
and therefore the same approach can be taken there.
The main advantage of Eq.~(\ref{eq:xsec_pve_2}) as compared to the Lagrange multiplier
method is that we always work with a positive-definite likelihood ratio
as required by the cross-entropy loss function.

\subsection{Neural network training}
\label{sec:nntraining}

Here we describe the settings of the neural network training leading to
the parametrisation of Eq.~(\ref{eq:EFT_structure_v3}).
We consider in turn the choice of neural network architecture, minimiser, and other
related hyperparameters; how the input data is preprocessed;
the settings of the stopping criterion used to avoid overfitting;  the estimate
of methodological uncertainties by means of the Monte Carlo replica method;
the scaling
of the ML training with respect to the number of EFT parameters;
and finally the validation procedure where the Machine Learning model
is compared to the analytic calculation of the likelihood ratio.

\myparagraph{Architecture, optimizer, and hyperparameters}
Table~\ref{tab:training_settings} specifies the training settings that are
adopted for each process, e.g. the features that were trained on, the architecture of the hidden layers, the
learning rate $\eta$ and the number of mini-batches.
Given a process for which the SMEFT parameter space is spanned by $n_{\rm eft}$ Wilson coefficients,
there are a maximum of $N_{\rm nn}=(n_{\rm eft}^2 + 3n_{\rm eft})/2$
independent neural networks to be trained.
In practice, this number can be smaller due to vanishing contributions,
in which case we will mention this explicitly. 
We have verified that we select redundant architectures, meaning that
training results are stable in the event that a somewhat less flexible architecture were
to be adopted.
For every choice of  $n_k$ kinematic features, these $N_{\rm nn}$ neural
networks share the same hyperparameters listed there.
The last column of Table~\ref{tab:training_settings}
indicates the average training time per replica and the corresponding standard deviation,
evaluated over the $N_{\rm nn}\times N_{\rm rep}$ networks to be trained
for a given process.
In future work one can consider an automated process to optimise
the choice of the hyperparameters listed in Table~\ref{tab:training_settings} along the lines of 
the strategy adopted for the NNPDF4.0 analysis~\cite{NNPDF:2021njg,Carrazza:2019mzf}.

\begin{table}[t]
  \centering
  \small
  \renewcommand{\arraystretch}{1.5}
    \begin{tabular}{c|C{1.3cm}|c|C{0.8cm}|c|c}
      \toprule
        process &  feat. & hidden layers & $\eta$ & $n_{\mathrm{batch}}$ & time (min) \\
         \midrule
         \multirow{3}{*}{ $pp\to t\bar{t}\to b\bar{b}\ell^+\ell^- \nu_\ell \bar{\nu}_{\ell}$} & $p_T^{\ell\bar{\ell}}$ & $25\times 25\times 25 $& $10^{-3}$&1&$46.8\pm35.0$\\
          & $p_T^{\ell\bar{\ell}}, \eta_\ell$ & $ 25\times 25\times 25 $& $10^{-3}$&1&$53.7\pm 29.9$\\
     & $18$ & $100\times 100 \times 100 $& $10^{-4}$&$50$&$5.4 \pm 2.7$\\
         \bottomrule
    \end{tabular}
    \vspace{0.3cm}
    \caption{Overview of the settings for the neural network trainings.
      For the top-pair production process to be described in Sect.~\ref{sec:pseudodata}, we specify the $n_k$ kinematic features $\boldsymbol{x}$ used
      for the likelihood ratio parametrisation (feat.), the architecture, the learning rate $\eta$, the number of mini-batches, and the training time per network averaged over all replicas.
      As indicated by Eq.~(\ref{eq:EFT_structure_v4}),
      given a process for which the parameter space is spanned by $n_{\rm eft}$ Wilson coefficients,
      there are  $(n_{\rm eft}^2 + 3n_{\rm eft})/2$
      independent neural networks to be trained.
      For each choice of  $n_k$ kinematic features, these neural
      networks share the settings listed here.
    }
    \label{tab:training_settings}
\end{table}

We train these neural networks by performing (mini)-batch gradient descent on the cross-entropy loss function Eq.~(\ref{eq:loss_CE_numeric}) using the \verb|AdamW|~\cite{Loshchilov2017} optimiser. Training was implemented in \verb|PyTorch|~\cite{Paszke2019} and run on \verb|AMD Rome| with $19.17$ HS06 per CPU core. We point the interested reader
to Sect.~\ref{subsec:ml4eft_code} and the corresponding online
documentation where the main features of the {\sc\small ML4EFT} software
framework are highlighted. 

\myparagraph{Data preprocessing}
The kinematic features $\boldsymbol{x}$ that enter
the evaluation of the likelihood function $f_\sigma(\boldsymbol{x}, \boldsymbol{c})$ and
its ratio $r_\sigma(\boldsymbol{x}, \boldsymbol{c})$
cannot be used directly as an input to the neural network training algorithm and should be preprocessed
first to ensure that the input information is provided to the neural nets
in their region of maximal sensitivity.
For instance, considering parton-level top quark pair production at the LHC,
the typical invariant masses $m_{t\bar{t}}$ to be used for the training
cover the range between 350 GeV and 3000 GeV,
while the rapidities are dimensionless and restricted to the range $y_{t\bar{t}}\in [-2.5,2.5]$.
To ensure a homogeneous and well-balanced training, especially for high-multiplicity
observables, all features should be transformed 
to a common range and their distribution in this range should be reasonably
similar.

A common data preprocessing
method for Gaussianly distributed variables is to standardise all features to zero mean and unit variance.
However, for typical LHC process the kinematic distributions are highly non-Gaussian,
in particular invariant mass and $p_T$ distributions are very skewed.
In such cases,  one instead should perform a rescaling based on a suitable interquartile range,
such as the $68\%$ CL interval.
This method is particularly interesting for our application because of its robustness to outliers at high invariant masses and transverse momenta, in the same way that the median is less sensitive to them than the sample mean.
In our approach we use a robust feature scaler which
subtracts the median and scales to an inter-quantile range,
resulting into input feature distributions peaked around zero with their bulk well contained within the
$\lc -1,1\rc$ region, which is not necessarily the case for the standardised Gaussian scaler.
Further justification of this choice will be provided in Sect.~\ref{sec:pseudodata}.
See also~\cite{Carrazza:2021yrg} for a recent application of feature scaling to the training
of neural networks in the context of PDF fits.

\myparagraph{Stopping and regularisation}
The high degree of flexibility of neural networks in supervised learning applications
has an associated risk of overlearning, whereby the model ends up learning the statistical
fluctuations present in the data rather than the actual underlying law.
This implies that for a sufficiently flexible architecture a training with a fixed number of epochs will result
in either underlearning or overfitting, and hence that the optimal
number of epochs should be determined separately for each individual training by means
of a stopping criterion.

Here the optimal stopping point is determined separately
for each trained neural network by means of a variant
of the cross-validation dynamical stopping algorithm introduced in~\cite{NNPDF:2014otw}.
Within this approach, one splits up randomly each of the input datasets
$\mathcal{D}_{\rm sm}$ and $\mathcal{D}_{\rm eft}$ into two
disjoint sets known as the training set and the validation set, in a $80\%/20\%$ ratio.
The points in the validation subset are excluded from the optimisation procedure,
and the loss function evaluated on them, denoted by $L_{\rm val}$, is used as a diagnosis tool
to prevent overfitting.
The minimisation of the training loss function $L_{\rm tr}$ is carried out while
monitoring the value of $L_{\rm val}$.
One continues training until $L_{\rm val}$ has stopped decreasing following $n_{\rm p}$ (patience) epochs with respect to its last recorded local minimum.
The optimal network parameters, those with the smallest generalisation error,
then correspond to those at which $L_{\rm val}$
exhibits this global minimum within the patience threshold.

\begin{figure}[t]
  \centering
  \includegraphics[width=\textwidth]{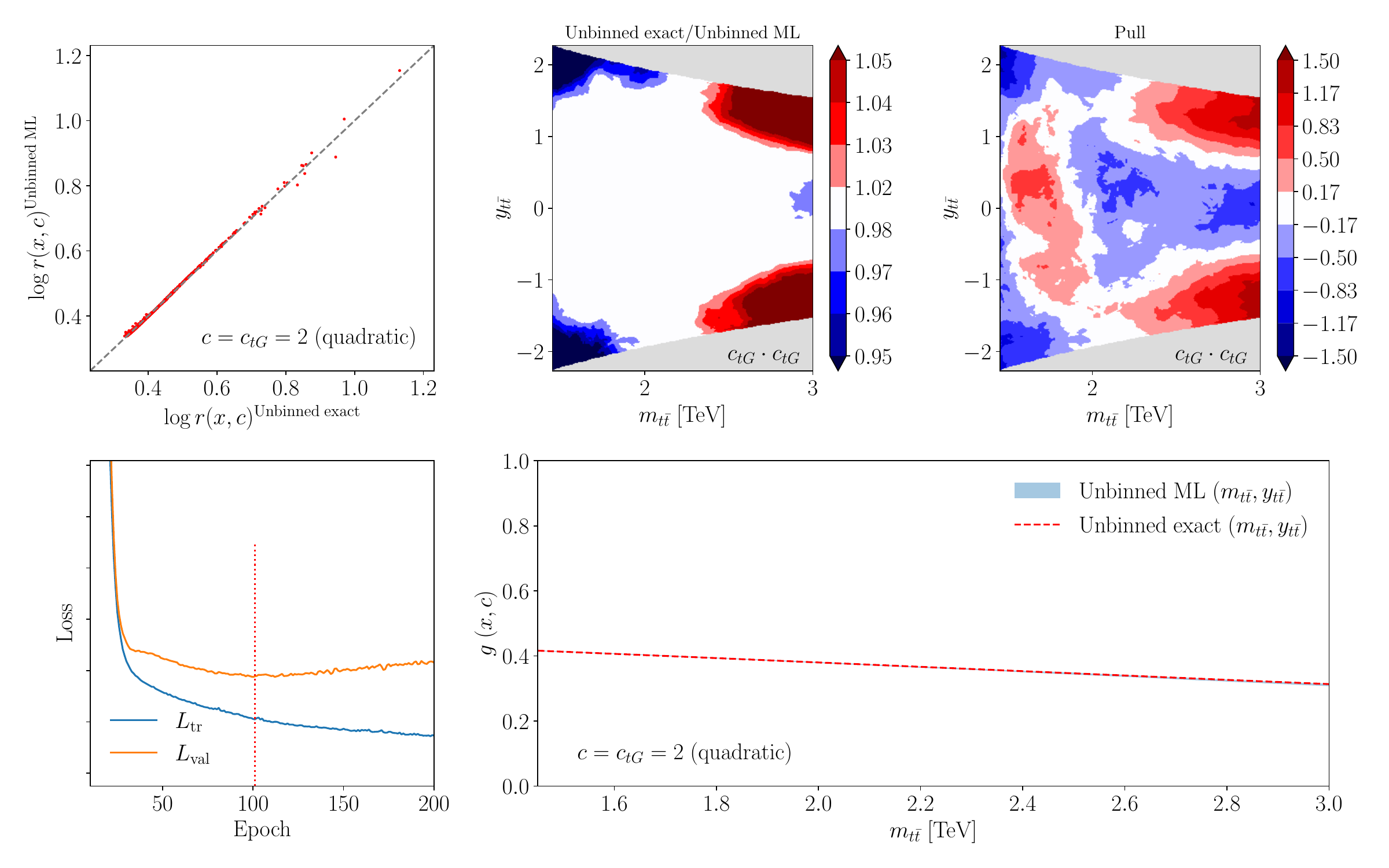}
  \caption{Validation of the Machine Learning parametrisation
    of the EFT cross-section ratios
    when applied to  the case of parton-level top quark pair production.
    The results shown here correspond to the training of
    the quadratic neural network
    $\mathrm{NN}^{(j, j)}(m_{t\bar{t}},y_{t\bar{t}})$ in Eq.(\ref{eq:EFT_structure_v4})
    with $j$ indicating the chromomagnetic operator $c_{tG}$.
    From left to right and top to bottom we display a point-by-point comparison of the log-likelihood ratio in the ML model and the corresponding analytical calculation; the median
    of the ratio  between the ML model
    and the analytical calculation and the associated pull in the
    $(m_{t\bar{t}},y_{t\bar{t}})$ feature space; the evolution of the loss function split in training and validation sets for a representative replica as a function of the number of training epochs; and the resultant decision boundary $g(\boldsymbol{x,c})$ for $c_{tG}=2$ including MC replica uncertainties at the end of the training procedure.}
  \label{fig:nn_overview_perf}
\end{figure}

The bottom-left plot of Fig.~\ref{fig:nn_overview_perf} illustrates the dependence of
the training and validation loss functions in a representative training.
While $L_{\rm tr}$ continues to decrease as the number of epochs increases,
at some point  $L_{\rm val}$ exhibits a global minimum
and does not decrease further during $n_{\mathrm{p}}$ epochs.
The position of this global minimum is indicated with a vertical dashed line,
corresponding to the optimal stopping point.
The parameters of the trained network are stored for each iteration,
and once the optimal stopping point has been identified the final parameters
are assigned to be those of the epoch where $L_{\rm val}$ has its global minimum.

\myparagraph{Uncertainty estimate from the replica method}
In general the ML parametrisation  $\hat{r}_{\sigma}(\boldsymbol{x}, \boldsymbol{c})$ will
differ from the true
EFT cross-section ratio $r_\sigma(\boldsymbol{x}, \boldsymbol{c})$ for two main reasons:
first, because of the finite statistics of the MC event samples used for the
neural network training, leading to a functional uncertainty in the ML model,
and second, due to residual inefficiencies of the optimisation and stopping algorithms.
In order to quantify these sources of methodological uncertainty and their impact
on the subsequent EFT parameter inference procedure, we adopt the
neural network replica method developed in the
context of PDF determinations~\cite{Giele:2001mr,Giele:1998gw,Ball:2008by,DelDebbio:2004xtd}.

The basic idea is to generate $N_{\mathrm{rep}}$
replicas of the MC training dataset,
each of them statistically independent, and then train separate sets of neural networks
on each of these replicas.
As explained in Sect.~\ref{sec:methodology_xsec_param}, we train
the decision boundary $g({\boldsymbol{x}},{\boldsymbol{c}})$
from a balanced sample of SM and EFT events.
If we aim to carry out
the training of $\hat{r}_{\sigma}$ on a sample of $N_{\rm ev}$ events
(balanced between the EFT and SM hypotheses), one generates
a total of  $N_{\rm ev}\times N_{\rm rep}$ events and divides them into $N_{\mathrm{rep}}$
replicas, each of them containing the same amount of information on the
underlying EFT cross-section $r_\sigma$.
Subsequently, one trains the full set of neural networks required
to parametrise $\hat{r}_{\sigma}$ separately for each of these replicas,
using in each case different random seeds for the initialisation of the network
parameters and other settings of the optimisation algorithm.

In this manner, at the end of the training procedure, one ends up 
instead of Eq.~(\ref{eq:EFT_structure_v4}) with  an ensemble of  $N_{\rm rep}$
 replicas of the cross-section ratio parametrisation,
\begin{equation} 
\label{eq:EFT_structure_v5}
\hat{r}^{(i)}_{\sigma}(\boldsymbol{x}, \boldsymbol{c}) \equiv
1 + \sum_{j=1}^{n_{\rm eft}}
 \mathrm{NN}^{(j)}_i(\boldsymbol{x}) c_j
+\sum_{j=1}^{n_{\rm eft}}\sum_{k \ge j}^{n_{\rm eft}}
\mathrm{NN}^{(j, k)}_i(\boldsymbol{x}) c_j c_k \, , \qquad i=1,\ldots,N_{\rm rep}
\end{equation} 
which estimates the methodological uncertainties associated to the parametrisation
and training.
Confidence level intervals associated with these uncertainties can then be determined in the usual way, for instance by taking suitable lower and upper quantiles.
In other words, the replica ensemble given by Eq.~(\ref{eq:EFT_structure_v5}) provides
a suitable representation of the probability density in the space of NN models,
which can be used to quantify the impact of methodological
uncertainties at the level of EFT parameter inference.
For the processes considered in this work we find that values of  $N_{\rm rep}$ between 25 and 50
are sufficient to estimate the impact of these procedural uncertainties at the level
of EFT parameter inference.

\myparagraph{Scaling with number of EFT parameters}
If unbinned observables such as those constructed here are to be integrated
into global SMEFT fits, their scaling with the number of EFT operators $n_{\rm eft}$ considered
should be not too computationally costly, given that typical fits involve up
to $n_{\rm eft}\sim 50$ independent degrees of freedom.
In this respect, exploiting the polynomial structure of  EFT  cross-sections as done in this work
allows for an efficient scaling of the neural network training and makes complete parallelisation possible.
We note that most related approaches in the literature, such as e.g.~\cite{Cowan:2010js}, are limited to a small
number of EFT parameters and hence not amenable to global fits.
In other approaches, e.g.~~\cite{Chen:2020mev}, the proposed ML parametrisation
is such that the coefficients of the linear and the quadratic terms mix, and in such case  no separation between linear and quadratic terms
and between different Wilson coefficients is possible.
This implies that in such approaches all
 neural networks
 parameterising the likelihood functions
 need to be trained simultaneously and hence that parallelisation is not possible.

 Within our framework,
 assembling the parametrisation of the cross-section ratio Eq.~(\ref{eq:EFT_structure_v3}) involves
 $n_{\rm eft}$ independent trainings for the linear contributions followed by
$n_{\rm eft}(n_{\rm eft}+1)/2$ ones for the quadratic terms.
Hence the total number of independent neural network trainings required will be given by
\begin{equation} 
 N_{\rm nn} = \frac{n_{\rm eft}^2 + 3n_{\rm eft}}{2} \, ,
\end{equation} 
which scales polynomially ($n_{\rm eft}^2$) for a large number of EFT parameters.
Furthermore, since each neural net is trained independently,
the procedure is fully parallelizable and the total computing time
required scales rather as $n_{\rm eft}^2/n_{\rm proc}$
with $n_{\rm proc}$ being the number of available processors.
Thanks to this property, even for the case in which $n_{\rm eft}\sim 40$ in a typical cluster with $\sim 10^3$
 nodes the computational effort required to construct Eq.~(\ref{eq:EFT_structure_v3}) is only 50\%
larger as compared to the case with $n_{\rm eft}=1$.
This means that our method is well suited
for the large parameter spaces considered in global EFT analyses.

Furthermore, for each unbinned multivariate
observable that is constructed we  repeat the training of the
neural networks $N_{\rm rep}$ times  to estimate methodological
uncertainties.
Hence the maximal number of neural network trainings involved
will be given by
\begin{equation} 
\#\,{\rm trainings} = N_{\rm rep}\times N_{\rm nn} = \frac{N_{\rm rep}\lp n_{\rm eft}^2 + 3n_{\rm eft}\rp}{2} \, .
\end{equation} 
For example, for $hZ$ production with quadratic EFT corrections
we will have $n_{\rm eft}=7$ coefficients and $N_{\rm rep}=50$ replicas,
resulting into a maximum of 1750 neural networks to be trained.\footnote{In this
  case the actual number of trainings is smaller, $\#\,{\rm trainings}=1500$, given
that some quadratic cross-terms vanish.}
While this number may appear daunting, these trainings are parallelizable
and the total integrated computing requirements end up being not too different
from those of the single-network training.

\myparagraph{Validation with analytical likelihood}
As will be explained in Sect.~\ref{sec:pseudodata},
for relatively simple processes one can evaluate the cross-section ratios
Eq.~(\ref{eq:EFT_structure_v3}) also in a purely analytic manner.
In such cases, the profile likelihood ratio and the associated parameter inference can be evaluated exactly
without the need to resort to numerical simulations.
The availability of such analytical calculations offers the possibility
to independently validate its Machine Learning counterpart, Eq.~(\ref{eq:EFT_structure_v4}),
at various levels during the training process.

Fig.~\ref{fig:nn_overview_perf} presents an overview of representative validation checks
of our procedure
that we carry out whenever the analytical cross-sections are available.
In this case the process under consideration is parton-level top quark pair production where the kinematic features
are the top quark pair invariant mass $m_{t\bar{t}}$  and rapidity $y_{t\bar{t}}$, that is,
the feature array is given by
$\boldsymbol{x} = \lp m_{t\bar{t}},y_{t\bar{t}} \rp$.
The neural
network training shown corresponds to the quadratic term
$\mathrm{NN}^{(j, j)}$ with $j$ being the chromomagnetic operator $c_{tG}$.

First, we display a point-by-point comparison of the log-likelihood ratio in the ML model and the
corresponding analytical calculation, namely comparing Eqns.~(\ref{eq:EFT_structure_v4})
and~(\ref{eq:EFT_structure_v3}) evaluated on the kinematics of the Monte Carlo
events generated for the training in the specific case of  $c_{tG}=2$.
One obtains excellent agreement within the full phase space considered.
Then we show the median value (over replicas)
of the ratio between the analytical and Machine Learning calculations of $\mathrm{NN}^{(j, j)}$ 
evaluated in the $\lp m_{t\bar{t}},y_{t\bar{t}} \rp$ kinematic feature space, with $j$ again being the chromomagnetic operator $c_{tG}$.
We also show the pull between the analytical and numerical calculations in units
of the Monte Carlo replica uncertainty.
From the median plot we see that the parametrised ratio $\hat{r}_{\sigma}$ reproduces
the exact result within a few percent except for low-statistics regions (large  $|y_{t\bar{t}}|$
and $ m_{t\bar{t}},y_{t\bar{t}}$ tails), and that these differences are in general well contained
within the one-sigma MC replica uncertainty band. 

The bottom right plot of Fig.~\ref{fig:nn_overview_perf}
displays the resultant decision boundary $g(\boldsymbol{x,c})$
for $y_{t\bar{t}}=0$ as a function of the invariant mass $m_{t\bar{t}}$
in the training of
the quadratic cross-section ratio proportional to $c_{tG}^2$ in
the specific case of also for $c_{tG}=2$.
The band in the ML model is evaluated as the 68\% CL interval
over the trained MC replicas, and is the largest at high $m_{t\bar{t}}$ values
where statistics are the smallest.
Again we find that the ML parametrisation is in agreement within uncertainties
when compared to the exact analytical calculation, further validating the procedure.
Similar good agreement is observed for other EFT operators
both for the linear and for the quadratic cross-sections.

\section{Theoretical modelling}
\label{sec:pseudodata}

We describe here the settings adopted
for the theoretical modelling and simulation of unbinned observables at the LHC
and their subsequent SMEFT interpretation.
We consider a representative process relevant
for global EFT fits, namely top-quark pair production. We also considered Higgs boson production in association with a $Z$-boson, see
Ref.~\cite{GomezAmbrosio:2022mpm} for its original discussion. 
We describe the calculational setups used for the SM and EFT cross-sections
at both the parton and the particle level, justify the 
choice of EFT operator basis, motivate the selection
and acceptance cuts applied to final-state particles, present
the validation
of our numerical simulations with  analytical calculations whenever possible,
and summarise the inputs to the neural network training.

\subsection{Benchmark processes and simulation pipeline}
\label{sec:pipeline}

We apply the methodology developed in
Sect.~\ref{sec:trainingMethodology} to construct
unbinned observables for inclusive top-quark pair production.
We evaluate theoretical predictions in the SM and in the SMEFT
for both processes at LO, which suffices
in this context given that we are considering pseudo-data.
For particle-level event generation we consider the fully leptonic decay channel of top quark
pair production,
\begin{equation} 
\label{eq:tt_particlelevel}
{\rm p}+{\rm p} \to t + \bar{t} \to b + \ell^+ + \nu_{\ell} + \bar{b} + \ell^- + \bar{\nu}_{\ell} \, .
\end{equation} 
The evaluation of the SM and SMEFT cross-sections at LO is carried out
with \linebreak[4]{\tt MadGraph5\_aMC@NLO}~\cite{Alwall:2014hca} interfaced to {\sc\small SMEFTsim}~\cite{Brivio:2017btx,Brivio:2020onw}
with \linebreak[4]{\tt NNPDF31\_nnlo\_as\_0118} as the input PDF set~\cite{Ball:2012cx}. 
\VerbatimFootnotes
As discussed in Sect.~\ref{sec:methodology_xsec_param}, at the quadratic level in the EFT we parametrise 
the cross-section ratio such that we have separate neural networks for terms proportional to $c_{j}^{2}$ and for the quadratic mixed terms $c_{j} c_{k}$ \:\footnote{
To generate the corresponding training sample we make use of the \verb|MadGraph5_aMC@NLO| syntax which allows for the evaluation
of cross sections dependent only on the product $c_{j} c_{k}$, for example
$$
\verb|p p > t t~  NP<=1 NP^2==2 NPc[j]^2==1 NPc[k]^2==1| \, .
$$
}.
In addition to Eq.~(\ref{eq:tt_particlelevel}),
we also carry out $t\bar{t}$ simulations at the undecayed
parton level with the goal of comparing with the corresponding exact analytical calculation of
the likelihood ratio for benchmarking purposes.
Such analytical evaluation becomes more difficult (or impossible) for 
realistic unbinned multivariate measurements  presented
in terms of  particle-level or detector-level observables.
These analytical
calculations also make possible validating the accuracy of the neural
network training, as exemplified in
Fig.~\ref{fig:nn_overview_perf}, and indeed the agreement persists at the level
of the training of the decision boundary  $g(\boldsymbol{x},\boldsymbol{c})$.

\myparagraph{Dominance of statistical uncertainties}
As discussed in Sect.~\ref{subsec:unbinned_likelihood} in Chapter \ref{chap:methdologies}, we restrict our analysis
to measurements dominated by statistical uncertainties for which correlated systematic uncertainties
can be neglected.
This condition can be enforced by
restricting the fiducial phase space
such that the number of events per bin satisfies
\begin{equation} 
\label{eq:condition_luminosity}
\frac{\delta \sigma_i^{(\rm stat)}}{\sigma_i(\boldsymbol{0})} = \frac{1}{\sqrt{\nu_i(\boldsymbol{0})}} \ge \delta_{\rm min}^{(\rm stat)} \, ,
\qquad i=1,\ldots,N_{b} \, ,
\end{equation} 
where $\nu_i(\boldsymbol{0})$ is the number of expected events in bin $i$ according to the
SM hypothesis  after applying selection, acceptance, and efficiency cuts.
The threshold parameter $\delta_{\rm min}^{(\rm stat)}$
is set to $\delta_{\rm min}^{(\rm stat)}=0.02$ for our baseline analysis.
We have verified that our qualitative findings are not modified upon moderate variations of its value.
Since Eq.~(\ref{eq:condition_luminosity}) must apply for all possible binning choices, it should also hold
for $N_{b}=1$, namely for the total fiducial cross-section.
Therefore, we require that the  selection and acceptance cuts applied lead
to a fiducial region satisfying
$(\delta \sigma_{\rm fid}^{(\rm stat)}/\sigma_{\rm fid}) \ge \delta_{\rm min}^{(\rm stat)}$.
This condition implies that the requirement of Eq.~(\ref{eq:condition_luminosity}) will also be satisfied
for any particular choice of binning, including the narrow bin limit, i.e.~the unbinned case.

Within our approach there are two options by which the condition Eq.~(\ref{eq:condition_luminosity}) can be enforced
when applied to the fiducial cross-section, given by
\begin{equation} 
\label{eq:nu_tot_lumi_discussion}
\nu_{\rm tot}(\boldsymbol{0}) = \mathcal{L}_{\rm int} \times \sigma_{\rm fid}(\boldsymbol{0}) \, .
\end{equation} 
The first option is adjusting the integrated luminosity $\mathcal{L}_{\rm int} $
corresponding to this measurement.
In this work we will take a fixed baseline luminosity $\mathcal{L}_{\rm int}=300$ fb$^{-1}$, corresponding to the
integrated luminosity accumulated at the end of Run III.
The second option is to adjust the fiducial region
such that Eq.~(\ref{eq:condition_luminosity}) is satisfied.
Taking into account Eqns.~(\ref{eq:condition_luminosity}) and~(\ref{eq:nu_tot_lumi_discussion}),
for a given luminosity $ \mathcal{L}_{\rm int}$ the fiducial (SM) cross-section should satisfy
\begin{equation} 
\sigma_{\rm fid}(\boldsymbol{0}) \ge \lc \lp  \delta_{\rm min}^{(\rm stat)}  \rp^2\mathcal{L}_{\rm int} \rc^{-1} \, .
\end{equation} 
In this work we take the second option,
imposing kinematic cuts restricting the events
to the high-energy, low-yield tails of distributions, such as by means of a strong $m_{t\bar{t}}$ cut
in the case of top quark pair production, see Table~\ref{tab:tt_cuts}.
It is then possible to generalise the
results presented in this work for  $\mathcal{L}_{\rm int}=300$ fb$^{-1}$
to higher integrated luminosities by  making
the cuts that define the fiducial region more stringent.

\subsection{Top-quark pair production}
\label{subsec:ttbar_theorysim}


In the particle-level case, where the top-quark events are decayed into the
$ b \ell^+  \nu_{\ell} \bar{b}  \ell^-  \bar{\nu}_{\ell}$ final state,
one considers a broader set of kinematic features.
As in the parton level case, SM and EFT events are simulated
with {\sc\small MadGraph5\_aMC@NLO} at LO in the QCD expansion,
though now the analytical calculation
is not available as a cross-check.
In order to select the relevant EFT operators,
we adopt the following strategy.
Since we consider a single process, it
is only possible to constrain a subset of operators,
which are taken to be the  $n_{\rm eft}$ Wilson coefficients with the highest Fisher information value,
namely those that can be better determined from the fit.
%
Constraining additional Wilson coefficients would require extending
the analysis to consider unbinned observables for processes such as $t\bar{t}V$
which span complementary directions in the parameter space.

In Table~\ref{tab:op_defn_tt} we indicate the SMEFT operators 
entering inclusive top-quark pair production considered in this analysis.
For each operator we provide its definition in terms of the SM fields
and the notation used to refer to the corresponding
Wilson coefficients in {\sc\small SMEFTsim} (in the {\tt topU3l} flavour scheme),
{\sc\small SMEFiT}, and {\sc\small SMEFT@NLO}~\cite{Degrande:2020evl}.
These operator definitions are consistent with those used in the  {\sc\small SMEFiT}
global analyses~\cite{Hartland:2019bjb,Ethier:2021bye} as required
for the eventual integration of the unbinned observables there.

\begin{table}[t]
  \centering
  \small
  \renewcommand{\arraystretch}{1.5}
    \begin{tabular}{C{1.5cm}|C{1.5cm}|c|c|c}
    \toprule
        $\qquad$ Operator $\qquad$& $\qquad$\verb|SMEFiT| $\qquad$ & $\qquad$\verb|SMEFTsim|$\qquad$&\verb|SMEFT@NLO|&Definition  \\
         \midrule
         $\mathcal{O}_{tG}$ & \verb|ctG|& $-\verb|ctGRe|/\verb|gs|$ &\verb|ctG|&$ig_s(\bar{Q}\tau^{\mu\nu}T_A t)\Tilde{\varphi}G_{\mu\nu}^A + \mathrm{h.c.}$ \\
         
         $\mathcal{O}_{Qq}^{1,8}$&\verb|c81qq|& \verb|cQj18| &\verb|cQq18|&$\sum\limits_{i=1,2}c_{qq}^{1(i33i)} + 3c_{qq}^{3(i33i)}$\\
         
         $\mathcal{O}_{Qq}^{3,8}$&\verb|c83qq|& \verb|cQj38| &\verb|cQq38|&$\sum\limits_{i=1,2}c_{qq}^{1(i33i)}-c_{qq}^{3(i33i)}$ \\
         
         $\mathcal{O}_{tq}^{8}$& \verb|c8qt|& \verb|ctj8| &\verb|ctq8|&$\sum\limits_{i=1,2}c_{qu}^{8(ii33)}$ \\
         
         $\mathcal{O}_{tu}^{8}$ &\verb|c8ut|&
         \verb|ctu8|&\verb|ctu8|&$\sum\limits_{i=1,2}2c_{uu}^{(i33i)}$\\
         
         $\mathcal{O}_{Qu}^8$ &\verb|c8qu|&
         \verb|cQu8|&\verb|cQu8|&$\sum\limits_{i=1,2}c_{qu}^{8(33ii)}$\\
         
         $\mathcal{O}_{td}^8$ &\verb|c8dt|&
         \verb|ctd8|\:\verb|=|\:\verb|ctb8|&\verb|ctd8|&$\sum\limits_{j=1,2,3}c_{ud}^{8(33jj)}$\\
         
         $\mathcal{O}_{Qd}^8$ &\verb|c8qd|&
         \verb|cQd8|\:\verb|=|\:\verb|cQb8|&\verb|cQd8|&$\sum\limits_{j=1,2,3}c_{qd}^{8(33jj)}$\\
         
         \bottomrule
    \end{tabular}
    \vspace{0.3cm}
    \caption{The SMEFT operators entering inclusive top-quark pair production.
    We indicate their definition in terms of the SM fields
    and the notation used for corresponding
    Wilson coefficients in {\sc\small SMEFTsim} (in the {\tt topU3l} flavour scheme),
    {\sc\small SMEFiT}, and {\sc\small SMEFT@NLO}.
The {\sc\small =} sign indicates that two coefficients
are fixed to  the same value~\cite{Brivio:2020onw}.
}
    \label{tab:op_defn_tt}
\end{table}

\begin{table}[t]
    \centering
    \small
    \begin{tabular}{C{4cm}|c}
        \toprule
         kinematic feature   & $\qquad$ cut$\qquad$ \\
         \midrule
         $m_{t\bar{t}}$&$>1.45$ TeV\\
         $p_T^\ell\:\rm{~(leading)}$&$>25$ GeV\\
         $p_T^\ell\:\rm{~(trailing)}$&$>20$ GeV\\
         $p_T^j$&$>30$ GeV\\
         $|\eta^j|$&$< 2.5$\\
         $m_{\ell\bar{\ell}}$&$m_{\ell\bar{\ell}} > 106$ GeV, or $m_{\ell\bar{\ell}} < 76$ GeV and $m_{\ell\bar{\ell}}>20$ GeV\\
         $\Delta R (j, \ell)$&$>0.4$\\
         $p_T^{\rm{miss}}$&$>40$ GeV\\
              \bottomrule
                  \end{tabular}
    \vspace{0.3cm}
    \caption{Selection and acceptance cuts
    imposed on the final-state particles of the $t\bar{t}\to b\bar{b}\ell^+\ell^-\nu_\ell\bar{\nu}_\ell$ process.}
    \label{tab:tt_cuts}
\end{table}

The selection and acceptance cuts
imposed on the final-state particles of
the $t\bar{t}\to b\bar{b}\ell^+\ell^-\nu_\ell\bar{\nu}_\ell$ process
are adapted from the Run II dilepton CMS analysis~\cite{CMS:2018adi}
and listed in Table~\ref{tab:tt_cuts}.
Concerning the  array of kinematic features ${\boldsymbol{x}}$,
it is composed of $n_k=18$ features:
$p_T$ of the lepton $p_T^\ell$,
$p_T$ of the antilepton $p_T^{\bar{\ell}}$,
leading $p_T^\ell$,
trailing $p_T^\ell$,
lepton pseudorapidity  $\eta_\ell$,
antilepton pseudorapidity  $\eta_{\bar{\ell}}$,
leading $\eta_\ell$,
trailing $\eta_\ell$,
$p_T$ of the dilepton system $p_T^{\ell\bar{\ell}}$,
invariant mass of the dilepton system $m_{\ell\bar{\ell}}$,
absolute difference in azimuthal angle $|\Delta \phi(\ell, \bar{\ell})|$,
difference in absolute rapidity $\Delta \eta(\ell, \bar{\ell})$,
leading $p_T$ of the $b$-jet, 
trailing $p_T$ of the $b$-jet,
pseudorapidity of the leading $b$-jet $\eta_b$,
pseudorapidity of the trailing $b$-jet $\eta_b$,
$p_T$ of the $b\bar{b}$ system $p_T^{b\bar{b}}$, and
invariant mass of the $b\bar{b}$ system $m_{b\bar{b}}$.
These features are partially correlated among them, and hence
maximal sensitivity of the unbinned
observables to constrain the EFT coefficients will be achieved
for $n_k<18$.

Since no parton shower  or hadronisation effects are included, the $b$-quarks can be reconstructed
without the need of jet clustering and assuming  perfect tagging efficiency.
These simulation settings are not suited to describe
actual data but suffice for the present analysis based on pseudo-data,
whose goal is the consistent comparison of the impact
on the EFT parameter space of unbinned multivariate ML observables
with
their binned counterparts.

\subsection{Inputs to the neural network training}
\label{sec:inputs_traiing}

\begin{figure}[htbp]
    \centering
    \includegraphics[width=\textwidth]{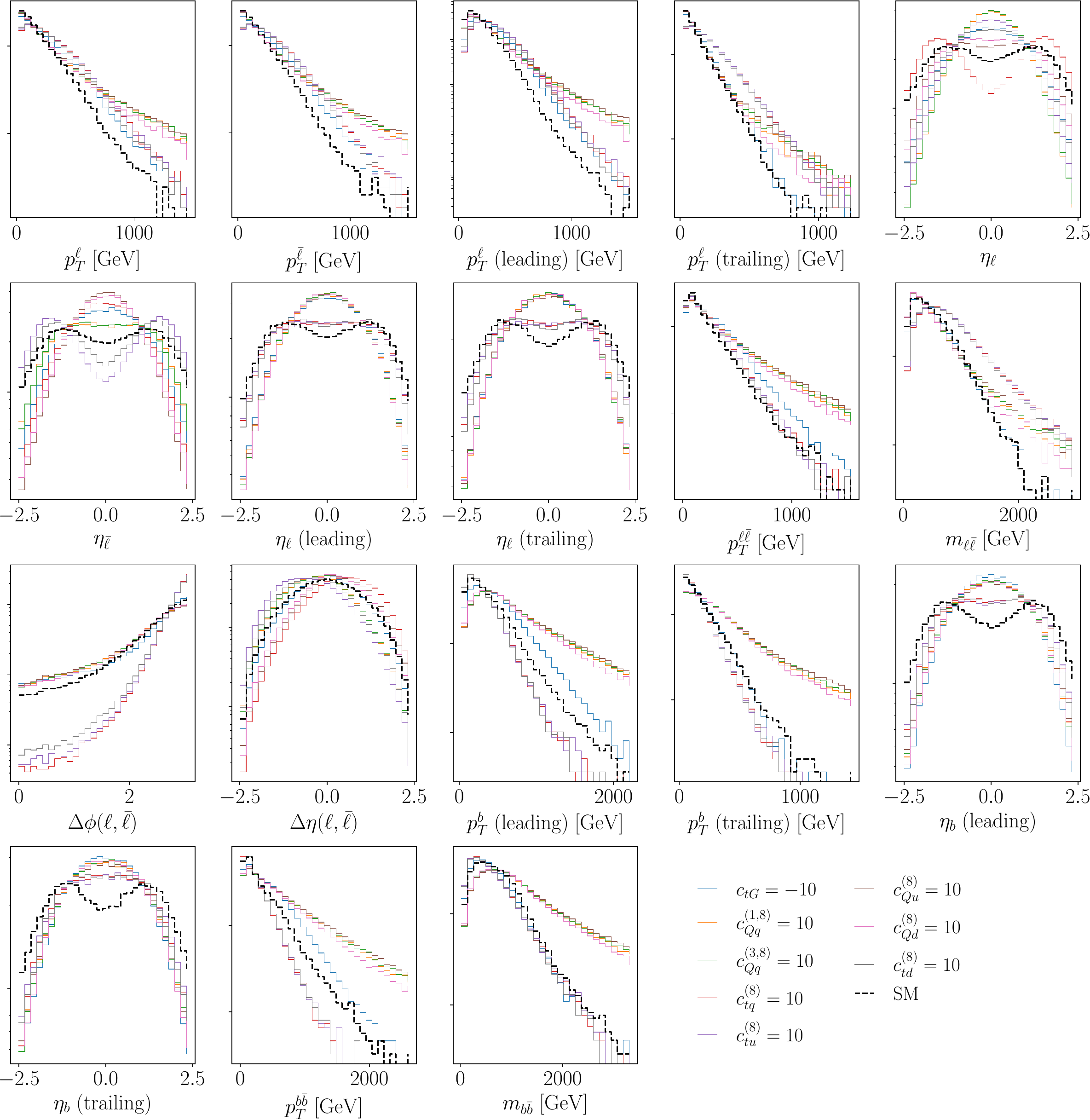}
    \caption{Differential distributions
          in the $n_k=18$ kinematic features used
          to parametrise the likelihood ratio in the $t\bar{t}\to b\bar{b}\ell^+\ell^- \nu_\ell \bar{\nu}_{\ell}$ process.
          We compare the SM predictions with those obtained in the SMEFT
          when individual operators are activated for coefficient values
          used for the neural network training.
          Results are shown
          at the quadratic-only level
          and normalised to their fiducial
          cross-sections.
    }
    \label{fig:tt_quad_dist}
\end{figure}

Fig.~\ref{fig:tt_quad_dist} displays the differential distributions
in the kinematic features used
to parametrise the likelihood ratio Eq.~(\ref{eq:EFT_structure_v4}) in the 
$t\bar{t}\to b\bar{b}\ell^+\ell^- \nu_\ell \bar{\nu}_{\ell}$ process.
We compare the SM predictions with those obtained in the SMEFT
when individual operators are activated for the values of
the Wilson coefficients used for the neural network training.
Results are shown at the quadratic-only level,
to highlight our approach separates the training of the linear
from the quadratic cross-section ratios, see also Sect.~\ref{sec:methodology_xsec_param}.
In order to illustrate shape (rather than normalisation)
differences of the NN inputs,
all distributions are normalised by their fiducial
cross-sections.
%

From Figs.~\ref{fig:tt_quad_dist}
one can observe how each operator modifies the qualitative shape of the various kinematic
features in different ways.
Furthermore, in general the EFT quadratic-only corrections enhance the shift
with respect to the SM distributions as compared to the linear ones.
The complementarity of the information provided by each kinematic feature motivates
the inclusion of as many final-state variables as possible when constructing unbinned
observables, though as mentioned above the limiting sensitivity will typically be saturated
before reaching the total number of kinematic features used for the training.

As mentioned in Sect.~\ref{sec:nntraining}, an efficient neural network
training strategy demands that the kinematic features $\boldsymbol{x}$ entering
the evaluation of
the $r_\sigma(\boldsymbol{x}, \boldsymbol{c})$
cross-section ratios are preprocessed
to ensure that the input information is provided to the neural networks
in the region of maximal sensitivity.
That is, all features should be transformed 
to a common range and their distribution within this range should be reasonably similar.
Here we use a robust scaler to ensure that this condition is satisfied, which subtracts the median and scales to the $95\%$ inter-quantile range.


Table~\ref{tab:settings_nntrain} summarises
the settings adopted for the neural network training of the
likelihood ratio function Eq.~(\ref{eq:EFT_structure_v4}) for the processes
considered.
For each process, we indicate the number of replicas $N_{\rm rep}$ generated,
the values of the EFT coefficients that enter the training as specified in Eqns.~(\ref{eq:nn_param_lin})
and~(\ref{eq:nn_param_quad}),
the number of Monte Carlo events generated $\widetilde{N}_{\rm ev}$ for each replica, and
the number of neural networks to be trained per replica $N_{\rm nn}$.
The values of the Wilson coefficients are chosen to be sufficiently large so as to mitigate 
the effect of MC errors that might otherwise dominate the SM-EFT discrepancy.  Furthermore, 
the sign of each Wilson coefficient is chosen such that the effect of the EFT is an enhancement relative to the SM,
and therefore the differential cross sections are consistently positive.  For example, in the case of negative EFT-SM 
interference, we select negative values of Wilson coefficients.
Cross-section positivity must also be maintained during training of the neural networks, and this 
is further discussed in Sect.~\ref{sec:methodology_xsec_param}.
The last column indicates the total number of trainings required to assemble the full parametrisation including
the $N_{\rm rep}$ replicas, namely $\#{\rm trainings}=N_{\rm rep}\times N_{\rm nn}$.
For particle-level top-quark pair production
our procedure requires the training of 1000 networks respectively
in the case of the quadratic EFT analysis.\footnote{We note that the actual value of $\#{\rm trainings}$
  can differ from the maximum value $N_{\rm rep}\times N_{\rm nn}$ since some quadratic cross-terms vanish. }
As discussed in Sect.~\ref{sec:trainingMethodology} these trainings are parallelisable
and the overall computational overhead remains moderate.

The values listed in
the last two columns of Table~\ref{tab:settings_nntrain} correspond to the case of quadratic EFT fits,
since as will be explained in Sect.~\ref{sec:results_ml4eft} at the linear level the presence of degenerate
directions requires restricting the subset of operators for which inference can be performed.
For each process, the total number of Monte Carlo events in the SMEFT that need to be generated is
therefore $\widetilde{N}_{\rm ev}\times N_{\rm rep}\times N_{\rm nn}$, and in addition the training needs
a balanced SM sample composed by $\widetilde{N}_{\rm ev}\times N_{\rm rep}$ events.
For example, in the case of $hZ$ production the total number of SMEFT events to be generated
is $10^5\times 50 \times 30 = 1.5\times 10^8$ events.

\begin{table}[htbp]
  \centering
  \scriptsize
  \renewcommand{\arraystretch}{1.5}
    \begin{tabular}{c|c|c|c|c|c}
      \toprule
      \small
        $\qquad $ Process  $\qquad $ & $\quad N_{\rm rep}\quad $ &  $c_j^{(\rm tr)}$ & $\widetilde{N}_{\rm ev}$~(per replica) & $\quad N_{\rm nn}\quad $  & \#trainings\\
         \midrule
         $pp\to t\bar{t}\to b\bar{b}\ell^+\ell^- \nu_\ell \bar{\nu}_{\ell}$ &25&
         \begin{tabular}{l r}
           $c_{td}^{(8)}=10$ &\\
           $c_{Qd}^{(8)}=10$&\\
           $c_{Qq}^{(1,8)}=10$&\\
           $c_{Qq}^{(3,8)}=10$&\\
           $c_{Qu}^{(8)}=10$&\\
           $c_{tG}=-10$&\\
           $c_{qt}^{(8)}=10$&\\
           $c_{tu}^{(8)}=10$& \\
         \end{tabular}& $10^5$ &$40$ & 1000\\
         \bottomrule
    \end{tabular}
    \vspace{0.3cm}
    \caption{Overview of the  settings for the neural network training of the
      likelihood ratio Eq.~(\ref{eq:EFT_structure_v4}).
      We indicate the number of replicas $N_{\rm rep}$,
      the values of the EFT coefficients that enter the training as specified in Eqns.~(\ref{eq:nn_param_lin})
      and~(\ref{eq:nn_param_quad}),
      the number of Monte Carlo events generated $\widetilde{N}_{\rm ev}$ for each replica, and
      the number of neural networks to be trained per replica $N_{\rm nn}$
      taking into account that  some EFT cross-sections vanish at the linear level.
      The values of the Wilson coefficients are chosen such that the EFT has a large effect relative to the SM, mitigating the 
	effect of MC errors that could otherwise dominate the SM-EFT discrepancy.  The signs of the Wilson coefficients are chosen such that 
	the EFT always leads to an enhancement relative to the SM; for example, negative Wilson coefficients are chosen
	to compensate for negative SM-EFT interference.  This ensures positive differential cross-sections throughout the training samples.
      The last column indicates the total number of trainings
      required, $\#{\rm trainings}=N_{\rm rep}\times N_{\rm nn}$.
      The last two columns correspond to the case of quadratic EFT fits;
      at the linear EFT level the presence of quasi-flat directions
      restricts the subset of operators for which inference can be performed.
      The total number of Monte Carlo EFT events generated is
      $\widetilde{N}_{\rm ev}\times N_{\rm rep}\times N_{\rm nn}$, and in addition we need
      balanced SM samples requiring $\widetilde{N}_{\rm ev}\times N_{\rm rep}$ events.
}
    \label{tab:settings_nntrain}
\end{table}

\section{EFT constraints from unbinned multivariate observables}
\label{sec:results_ml4eft}

We now present the
constraints on the EFT parameter space provided by the unbinned observables
constructed in Sect.~\ref{sec:trainingMethodology}
in comparison with those provided by their binned counterparts.
We study the dependence of these results on the choice of binning and
on the kinematic features.
We also quantify how much the EFT constraints are modified
when restricting the analysis to linear 
$\mathcal{O}(\Lambda^{-2})$ effects as compared to when the quadratic
$\mathcal{O}(\Lambda^{-4})$ contributions are also included.

First, we describe the method adopted to infer the posterior distributions
of the EFT parameters for a given observable, either binned or unbinned.
Second, we consider particle level top quark pair production in the dilepton final
state (Sect.~\ref{subsec:ttbar_theorysim}), and quantify the information gain resulting from
unbinned observables and its dependence on the choice of kinematic features
used in the training.
Finally, we will discuss the impact of methodological uncertainties, discussed in 
Sect.~\ref{sec:nntraining}, on the constraints we obtain on the EFT parameter space.
The results presented here can be reproduced and extended to other processes by means of the {\sc\small ML4EFT}
framework, summarised in Sect.~\ref{subsec:ml4eft_code}.

\subsection{EFT parameter inference}
\label{subsec:EFT_inference}
For each of the LHC processes considered in Sect.~\ref{sec:pseudodata}, Monte Carlo samples
in the SM and the SMEFT are generated in order to train the decision boundary
$g({\boldsymbol x},\boldsymbol{c})$ from the minimisation of the cross-entropy loss
function Eq.~(\ref{eq:loss_CE}). See Table~\ref{tab:tt_cuts} for the pseudodata generation settings.
The outcome of the neural network training is a parametrisation of the cross-section ratio
$\hat{r}_{\sigma}(\boldsymbol{x}, \boldsymbol{c}) $,
Eq.~(\ref{eq:EFT_structure_v4}), which in the limit of large statistics and perfect training
reproduces the true result $r_{\sigma}(\boldsymbol{x}, \boldsymbol{c}) $,
Eq.~(\ref{eq:EFT_structure_v2}).
To account for finite-sample and finite network flexibility effects,
we use the Monte Carlo replica method as described in Sect.~\ref{sec:nntraining} to
estimate the associated methodological uncertainties. 
Therefore, the actual requirement that defines a satisfactory neural
network parametrisation $\hat{r}_{\sigma}$ is that it reproduces the exact result $r_{\sigma}$,
within the 68\% CL replica uncertainties evaluated from the ensemble Eq.~(\ref{eq:EFT_structure_v5}).

From the exact result for the cross-section ratio $r_{\sigma}(\boldsymbol{x}, \boldsymbol{c}) $,
or alternatively its ML representation $\hat{r}_{\sigma}(\boldsymbol{x}, \boldsymbol{c})$,
one can carry out inference on the EFT Wilson coefficients by means of the profile
likelihood ratio, Eq.~(\ref{eq:plr_4}), applied to the $N_{\rm ev}$ generated pseudo-data events.
We emphasise that the $N_{\rm ev}$ events entering
in Eq.~(\ref{eq:plr_6}) are the physical events
expected after acceptance and selection cuts for a given integrated luminosity $\mathcal{L}_{\rm int}$,
see also the discussion in Sect.~\ref{sec:pipeline}.
The Monte Carlo events used
to train the ML classifier are instead  different: in general, the classifier is trained
on a larger event sample than the one expected for the actual measurement.
Once the Machine Learning parametrisation of the cross-section ratio, Eq.~(\ref{eq:EFT_structure_v4}),
has been determined alongside its replica uncertainties,
the profile likelihood ratio can be used to infer the posterior
probability distribution in the EFT parameter space and thus determine confidence level
intervals associated to the unbinned observables.
The same method is applied to binned likelihoods.

In this work, the EFT parameter inference based on the Machine Learning parametrisation of the
profile likelihood ratio is carried out by means of the Nested Sampling algorithm, in particular
via the MultiNest implementation~\cite{Feroz:2013hea}.
We recall that this
is the baseline sampling algorithm used in the {\sc\small SMEFiT} analysis~\cite{Ethier:2021bye}
to determine the posterior distributions in the  EFT parameter space composed of $n_{\rm eft}=36$
independent Wilson coefficients.
The choice of MultiNest is motivated from its previous applications
to EFT inference problems of comparable complexity to those relevant here, and also to
facilitate the integration of unbinned observables into global EFT fits.
The posterior distributions provided by MultiNest
are represented by $N_{s}$ samples in the EFT parameter space,
where $N_{s}$ is determined 
by requiring a given accuracy in the sampling procedure.
From this finite set of samples, contours of the full posterior can be reconstructed via the kernel density
estimator (KDE) method, also used e.g. in the context of Monte Carlo
PDF fits~\cite{Carrazza:2015hva,Butterworth:2015oua}.

Methodological uncertainties associated to the $N_{\rm rep}$ replicas
are propagated to the constraints on the EFT 
parameter space in two complementary ways.
Firstly,
after evaluating the neural networks on the $N_{\rm ev}$ events entering into the inference procedure,
we calculate the median of each $\mathrm{NN}^{(j)}_i(\boldsymbol{x})$ and $\mathrm{NN}^{(j, k)}_i(\boldsymbol{x})$ in
Eq.~(\ref{eq:EFT_structure_v5}).
The corresponding median profile likelihood ratio is then calculated and 
used in the Nested Sampling algorithm, from which contours of the 
posterior distribution are obtained.
The results shown in the following Sect.~\ref{subsec:top_particle} have been produced using this method.
Alternatively, one can assess the impact of the 2-$\sigma$ methodological uncertainties on
these contours.  We do so by defining $N_{\rm rep}$ profile likelihood ratios, one from each of the $N_{\rm rep}$ replicas,
and performing Nested Sampling for each, resulting in $N_{\rm rep}$ sets of $N_{\rm s}$ 
samples from the EFT posterior distribution.  The samples are then combined, and the KDE method used to determine
contours of the posterior distribution.
In Sect.~\ref{subsec:eft_uncert} we will assess the differences observed between these two methodologies.

\subsection{Results}
\label{subsec:top_particle}
We now assess the bounds on the EFT parameter space obtained from unbinned observables in the case
of particle-level top quark pair production, specifically in the dilepton final state
described in Sect.~\ref{subsec:ttbar_theorysim},
$pp \to t\bar{t} \to b \ell^+ \nu_{\ell} \bar{b} \ell^- \bar{\nu}_{\ell}$.
As discussed there, this process is most sensitive to the $n_{\rm eft}=8$ operators
with the highest Fisher information in
inclusive $t\bar{t}$ production, as 
listed in Table~\ref{tab:op_defn_tt}.
While for the quadratic EFT analysis it is possible to derive the posterior
distribution associated to the full set of $n_{\rm eft}=8$ operators,
at the linear level there are quasi-flat directions that destabilise the ML training
of the likelihood ratio and the subsequent parameter inference.
For this reason, in the linear EFT analysis of this process we consider a subset of $n_{\rm eft}=5$ operators
that can be simultaneously constrained from inclusive
top-quark pair production~\cite{Brivio:2019ius}, given by
\begin{equation} 
\label{eq:subset_operator_topparticle_linear}
c_{tG}, c_{qt}^{(8)}, c_{Qu}^{(8)}, c_{Qq}^{(1,8)}, c_{td}^{(8)} \, .
\end{equation} 

\begin{figure}[htbp]
    \centering
    \includegraphics[width=0.85\textwidth]{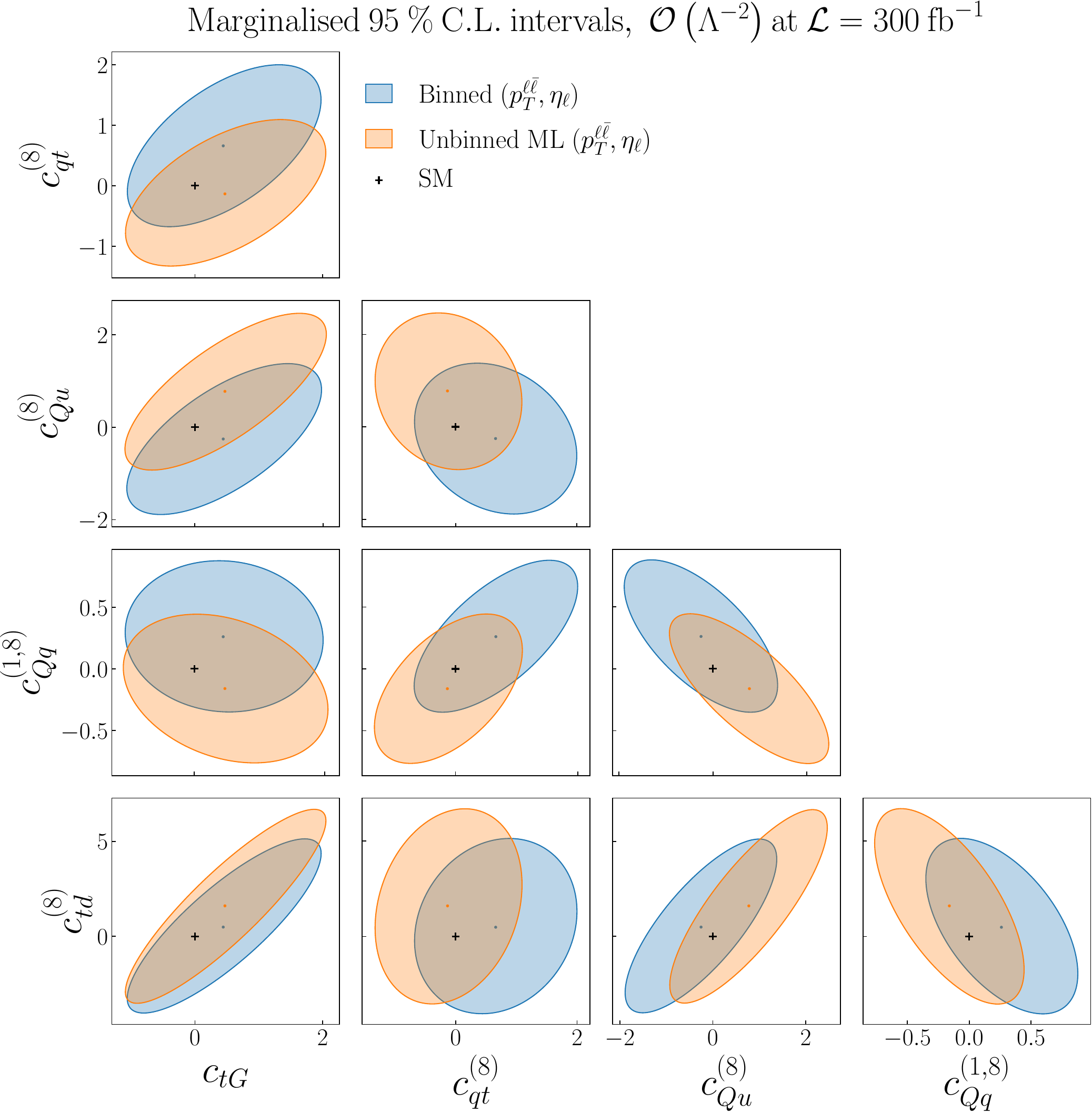}
    \caption{Pair-wise 95\% CL  contours for the 
      Wilson coefficients entering  top quark pair production
      in the dilepton final state, see Sect.~\ref{subsec:ttbar_theorysim} for
      more details.
      These contours are obtained by marginalising over the full posterior
      distribution provided by Nested Sampling.
      We consider here $n_{\rm eft}=5$ Wilson coefficients that
      can be simultaneously constrained from inclusive
      top-quark pair production at the linear level in the EFT expansion.
      We compare the results obtained from both binned
      and unbinned ML  observables constructed
      on the $(p_T^{\ell\bar{\ell}},\eta_\ell)$ kinematic features.
      The black cross indicates the SM values used to generate
      the pseudo-data that enters the inference.
      The comparison of the  unbinned ML
      observable trained on $(p_T^{\ell\bar{\ell}},\eta_\ell)$
      with its counterpart trained on the full set of $n_{k}=18$ kinematic features
      is displayed in Fig.~\ref{fig:tt_glob_lin_nn}.
    }
    \label{fig:tt_glob_lin_nn_binned}
\end{figure}

For this process we construct $n_k=18$  kinematic features
built from the four-vectors
of the final-state leptons and $b-$quarks.
Considering further kinematic variables such as the missing $E_T$ would
be redundant and not provide additional information.
The distribution of these kinematic features in the SM and when
quadratic EFT corrections are accounted for is displayed in Fig.~\ref{fig:tt_quad_dist}, showing
how different features provide complementary information to constrain the EFT and
thus making a fully-fledged multivariate analysis both interesting and necessary in order to
fully capture the EFT effects.
For example, the pseudo-rapidity distributions bend around $\eta=0$ corresponding to the direction transverse to the beam pipe,
while the transverse momenta are sensitive to energy growing effects in the
high-$p_T$ tails of the distributions.

Fig.~\ref{fig:tt_glob_lin_nn_binned} displays
the pair-wise 95\% CL intervals for the $n_{\rm eft}=5$ Wilson coefficients
in Eq.~(\ref{eq:subset_operator_topparticle_linear}) relevant
for the description of top quark pair production at the linear $\mathcal{O}(\Lambda^{-2})$
level.
The black cross indicates the SM scenario used to generate
the pseudo-data that enters the inference.
The  95\% CL contours shown are obtained from the full posterior
distribution provided by Nested Sampling, marginalising over the 
remaining Wilson coefficients for each of the operator pairs, with the ellipses drawn from the $N_s$ posterior samples provided by MultiNest.
These are compared with 
the bounds provided by a binned observable
based on the dilepton transverse momentum
$p_T^{\ell\bar{\ell}}$ and the lepton pseudorapidity  $\eta_\ell$
as kinematic features, where the binning is defined as
\begin{equation} 
p_T^{\ell\bar{\ell}} &\in& \lc 0,10,20,40,60, 100, 150, 400, \infty\rp \, \textrm{GeV}, \nonumber \\
\eta^{\ell} &\in& \lc 0, 0.3, 0.6, 0.9, 1.2, 1.5, 1.8, 2.1, 2.5\rc \, . \nonumber 
\end{equation} 
As for the parton-level case, a cut in the invariant mass $m_{t\bar{t}}$
is applied to ensure that Eq.~(\ref{eq:condition_luminosity})
is satisfied and that statistical uncertainties dominate.
Note that for all observables the pseudo-dataset used for the EFT parameter inference
is the same and was not used during training.

\begin{figure}[htbp]
    \centering
    \includegraphics[width=0.85\textwidth]{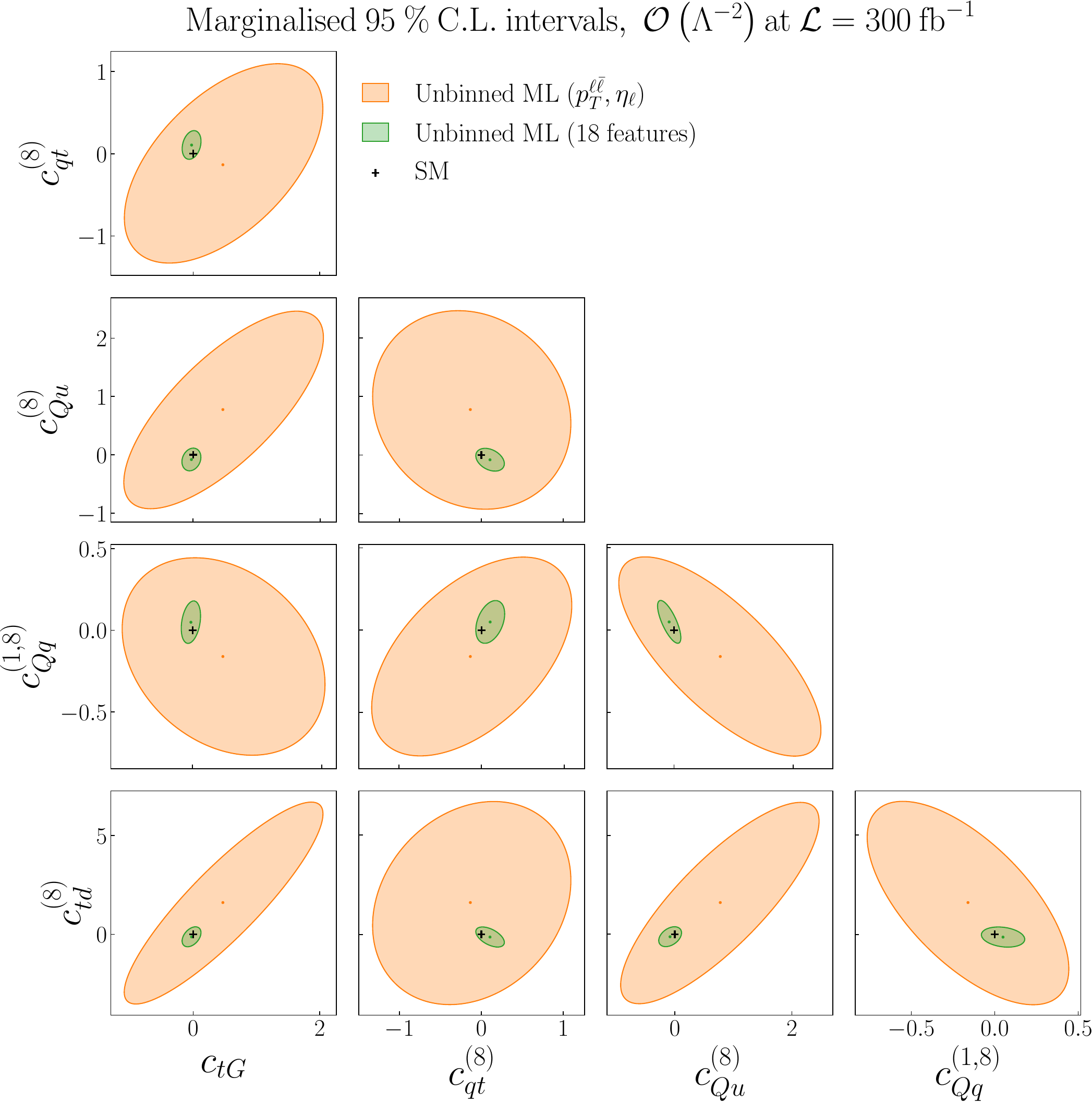}
    \caption{Same as Fig.~\ref{fig:tt_glob_lin_nn_binned} comparing
      the bounds on the EFT coefficients from the ML unbinned observables
trained on $(p_T^{\ell\bar{\ell}},\eta_\ell)$ and on
the full set of $n_k=18$ kinematic features listed in Sect.~\ref{subsec:ttbar_theorysim},
see also Fig.~\ref{fig:tt_quad_dist}.}
    \label{fig:tt_glob_lin_nn}
\end{figure}

From the comparisons in Fig.~\ref{fig:tt_glob_lin_nn_binned}
one can observe how, for all observables and all operator pairings, the  95\% CL intervals
include the SM values used to generate the pseudo-data.
The bounds obtained from the binned and unbinned observables are in general
similar and compatible, and no major improvement is obtained with the adoption of the latter
in this case.
However, here we consider only $n_k=2$ kinematic features and hence ignore the potentially useful
information contained in other variables that can be constructed from the final state kinematics.
In order to assess their impact, Fig.~\ref{fig:tt_glob_lin_nn} displays
the same pair-wise marginalised comparison,
now between  ML unbinned observables with only $p_T^{\ell\bar{\ell}}$ and  $\eta_\ell$
as input features and with the full set of $n_k=18$ kinematic variables displayed
in Fig.~\ref{fig:tt_quad_dist}.
We now find a very significant change in the bounds on the
EFT parameter space, improving by up
to an order of magnitude or better in all cases.
Again, the 95\% CL contours include the SM hypothesis used to generate
the pseudo-data.
Furthermore, the inclusion of the full set of kinematic features
 reduces the correlations between operator pairs that arise
when only $p_T^{\ell\bar{\ell}}$ and $\eta_\ell$ are considered, indicating a breaking
of degeneracies in parameter space.
This result indicates that a multivariate analysis improves the information
on the EFT parameter space that can be extracted from this process
as compared to that obtained from a subset of kinematic features.

\begin{figure}[htbp]
    \centering
    \includegraphics[width=\textwidth]{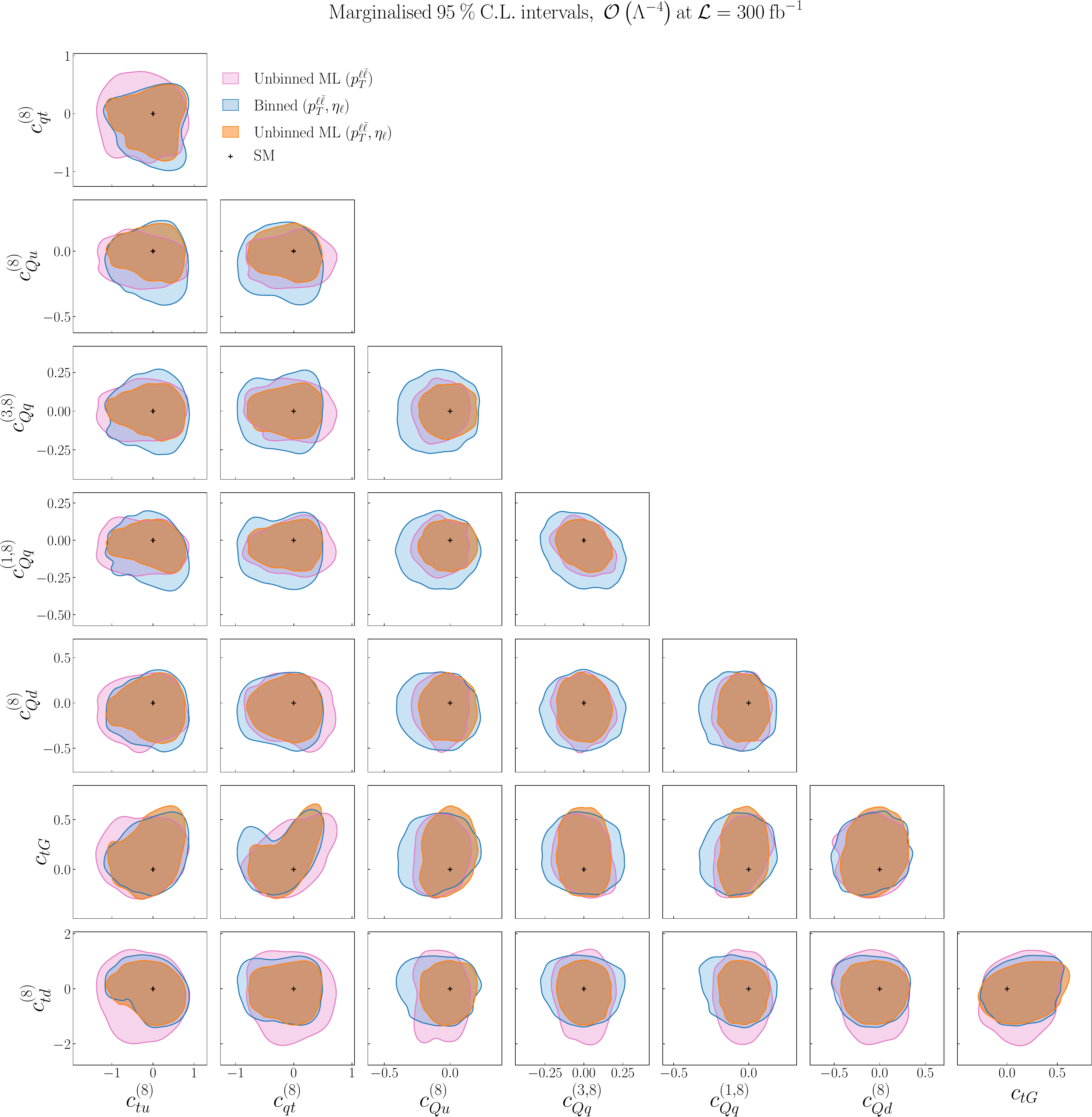}
    \caption{Same as Fig.~\ref{fig:tt_glob_lin_nn_binned} in the case of the EFT analysis
      carried out at the quadratic $\mathcal{O}(\Lambda^{-4})$ level.
      We display the results for pair-wise contours obtained
      from the marginalisation of the posterior distribution
      in the space of $n_{\rm eft}=8$ Wilson coefficients.
      In comparison with the linear EFT analysis, it becomes possible
      to constrain three more coefficients from the same process once quadratic
      corrections are accounted for.
      We also include in this comparison the results obtained
      from the unbinned ML analysis based on $p_T^{\ell\bar{\ell}}$
      as the single input feature.
    }
    \label{fig:tt_glob_quad_nn_binned}
\end{figure}

We note that that the results displayed in Figs.~\ref{fig:tt_glob_lin_nn_binned}
and~\ref{fig:tt_glob_lin_nn} are expected to differ should the starting point be a global
EFT analysis rather than a flat prior as is the case in the present proof-of-concept analysis.
For instance, the two-light-two-heavy operators entering $t\bar{t}$ production
at the linear level are highly correlated,
which means that adopting a multivariate analysis leads to a sizeable effect
partly because it breaks degeneracies in the parameter space.
Hence, the actual impact  of unbinned multivariate observables
for EFT analyses depends on which other datasets and processes are considered
and can only be assessed on a case-by-case basis.
In this respect, beyond the implications for the specific processes considered in this work,
what our framework provides is a robust method to quantify the information on
the EFT parameters provided by different types
of observables constructed on exactly the same dataset: binned versus unbinned,
different choices of binnings, and different numbers and types of kinematic
features.

\begin{figure}[htbp]
    \centering
    \includegraphics[width=\textwidth]{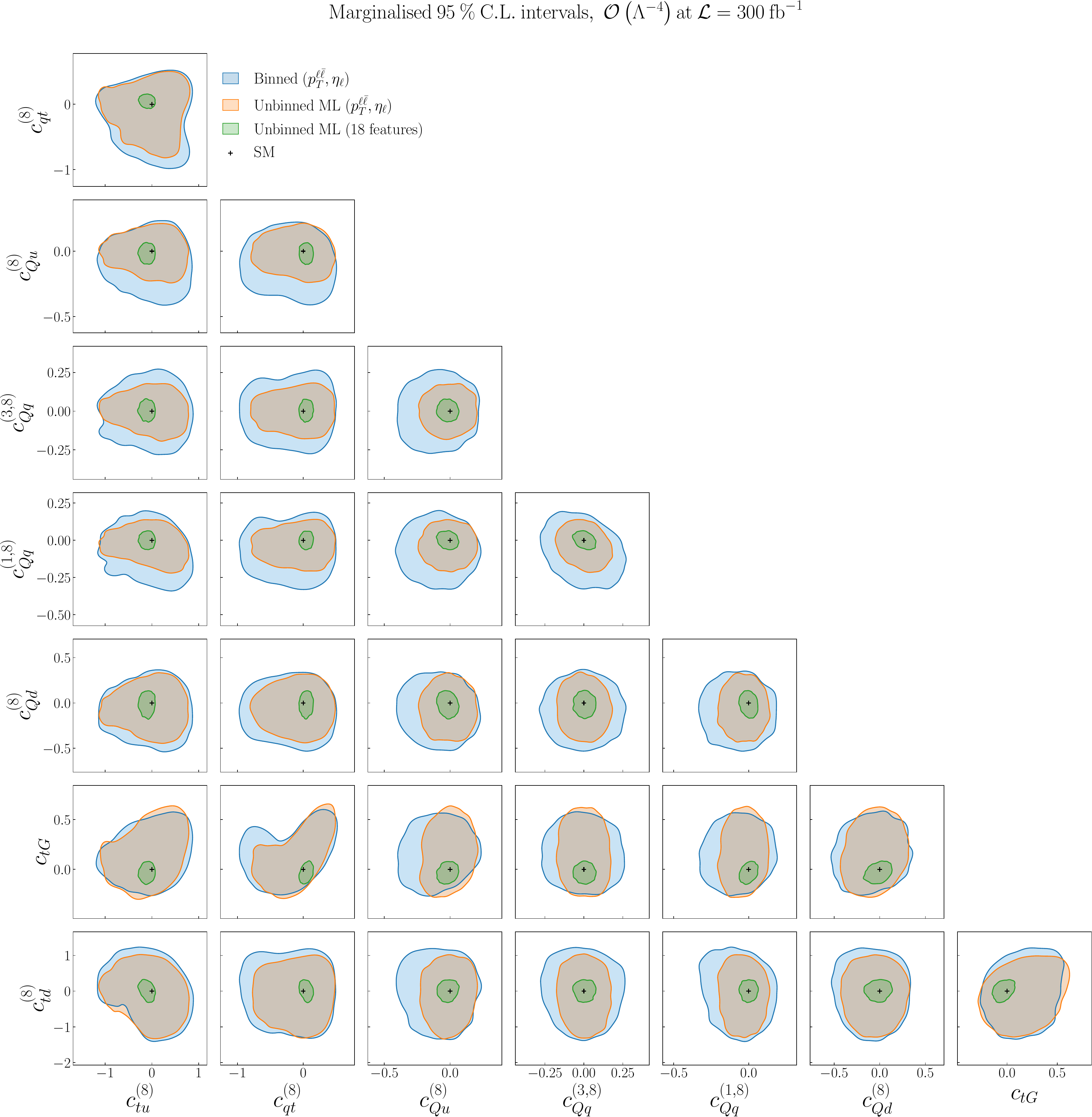}
    \caption{Same as Fig.~\ref{fig:tt_glob_quad_nn_binned}
      comparing the bounds obtained
      from the binned and unbinned observables built upon two
      kinematic features,  $p_T^{\ell\bar{\ell}}$ and $\eta_\ell$, with the corresponding
      results for the unbinned ML observable trained on the full set of
      $n_k=18$ kinematic features.
      \label{fig:tt_glob_quad_nn}}
\end{figure}

As is well known, in top quark pair production quadratic
EFT corrections are important for most operators, in particular due to energy-growing effects
in the tails of distributions.
These sizeable effects are highlighted by the distortions with respect to the SM
distributions induced by quadratic EFT effects, shown in
Fig.~\ref{fig:tt_quad_dist}, in the kinematic features
used to train the ML classifier for this process.
In order to investigate how the results based on linear EFT calculations vary once
quadratic corrections are considered,
in Figs.~\ref{fig:tt_glob_quad_nn_binned} and~\ref{fig:tt_glob_quad_nn}
we present the analogous comparisons to Figs.~\ref{fig:tt_glob_lin_nn_binned}
and~\ref{fig:tt_glob_lin_nn} respectively in the case of EFT calculations that also
include the quadratic corrections.
We display the results for pair-wise contours obtained
from the marginalisation of the posterior distribution
in the space of the full set of $n_{\rm eft}=8$ Wilson coefficients,
given that once quadratic corrections are accounted for it becomes
possible to constrain the full set of relevant operators simultaneously.
We also include in Fig.~\ref{fig:tt_glob_quad_nn_binned}  the results obtained
from the unbinned ML analysis based on $p_T^{\ell\bar{\ell}}$
as the single input feature, while Fig.~\ref{fig:tt_glob_quad_nn}
also displays the two-feature binned contours as reference.
We note that once quadratic effects are accounted for the 95\% CL contours
will in general not be elliptic, and may even be composed of disjoint regions in the case
of degenerate maxima.

Comparing the constraints provided by the binned $(p_T^{\ell\bar{\ell}},\eta_\ell)$ observable
in Fig.~\ref{fig:tt_glob_quad_nn_binned} with those in Figs.~\ref{fig:tt_glob_lin_nn_binned},
one observes how the bounds on the EFT coefficients are improved
when accounting for the quadratic EFT corrections.
This improvement is consistent with the large quadratic corrections
to the kinematic distributions entering the likelihood ratio of this process which lead
to an enhanced sensitivity,
 and with previous studies~\cite{Ethier:2021bye,Hartland:2019bjb} within global SMEFT analyses.
From the results Fig.~\ref{fig:tt_glob_quad_nn_binned} we find that
for all operator pairs considered, the most stringent bounds are provided
by the unbinned observable built upon the $(p_T^{\ell\bar{\ell}},\eta_\ell)$
pair of kinematic features.
Furthermore, one can identify the factor which brings in more information: either
using the full event-by-event kinematics for a given set of features, or increasing
the number of kinematic features being considered.
For some operator combinations, the dominant effect turns out to be
that of adding
a second kinematic feature, for instance in the case of $(c_{tu}^{(8)},c_{td}^{(8)})$
binned and unbinned observables coincide when using the $(p_T^{\ell\bar{\ell}},\eta_\ell)$
features while the unbinned $p_T^{\ell\bar{\ell}}$ observable results in larger
uncertainties.
For other cases, it is instead the information provided by the event-by-event kinematics
which dominates, for example in the $(c_{Qq}^{(3,8)},c_{Qu}^{(8)})$ plane
the constraints from binned $(p_T^{\ell\bar{\ell}},\eta_\ell)$ are clearly looser
than those from their unbinned counterparts.

The comparison between the constraints on the EFT parameter space provided by the unbinned observable
based on two kinematic features, $(p_T^{\ell\bar{\ell}},\eta_\ell)$,
and those provided by its counterpart based on the full set $n_k=18$ kinematic
features is displayed in Fig.~\ref{fig:tt_glob_quad_nn}.
It confirms that at the quadratic level there is also a marked gain in constraining power
obtained from increasing the number of kinematic variables considered in the analysis.
For reference, the plot also displays the bounds obtained with the two-feature binned observable
in the same process.
Interestingly, when considering the full set of kinematic features, one obtains
posterior contours which display a reduced operator correlation,
highlighting how the unbinned multivariate operator is especially effective
at removing (quasi)-degeneracies in the EFT coefficients.
As for the linear case, the improvement in the bounds obtained
with the multivariate can reach up to an order
of magnitude as compared to the two-feature unbinned analysis.

One can compare the bounds obtained in the present study with those from the corresponding
{\sc\small SMEFiT} global analysis of LHC data.s
For the same EFT settings, the
95\% CL marginalised bounds on the chromomagnetic operator
$c_{tg}$ are $\lc 0.021,0.18\rc$ in {\sc\small SMEFiT},
to be compared with a bound of
$\Delta c_{tg} \simeq 0.4$ obtained from the binned and unbinned $(p_T^{\ell\bar{\ell}},\eta_\ell)$ observable
and that of $\Delta c_{tg} \simeq 0.1$ from the multivariate unbinned observable trained using
all features.
While there are too many differences in the input dataset and other settings
to make possible a consistent comparison, this initial estimate suggests that unbinned
multivariate measurements
based on Run III data could provide competitive constraints on the EFT parameter space
as compared to the available binned observables.

All in all, we find that for inclusive top quark pair production,
deploying unbinned multivariate observables makes it possible to tap into a source of information
on the EFT parameter space which is not fully exploited in binned observables,
both at the level of linear and quadratic EFT analyses, and that increasing
the number of  kinematic features considered in the likelihood parametrisation
enhances the constraining power of the measurement.

\subsection{Methodological uncertainties}
\label{subsec:eft_uncert}

The results for the unbinned observables derived so far
are obtained from  EFT parameter inference carried out with Eq.~(\ref{eq:EFT_structure_v4}), the parametrisation of the
likelihood ratio, as described in Sect.~\ref{subsec:EFT_inference}.
As explained in Sect.~\ref{sec:nntraining}, within our approach, rather than a single best model,
we produce a distribution of models, denoted as replicas, each of them trained on a different set
of Monte Carlo events.
Hence the end result of our procedure is Eq.~(\ref{eq:EFT_structure_v5}),
the representation of the probability distribution of the likelihood ratio
$\{\hat{r}^{(i)}_{\sigma}(\boldsymbol{x}, \boldsymbol{c})\}$ composed of $N_{\rm rep}$ equiprobable replicas.
The spread of this distribution provides a measure of methodological and
procedural uncertainties
associated e.g. to finite training datasets and inefficiencies of the optimisation and
stopping algorithms.
Results presented in Sect.~\ref{subsec:top_particle} are based on using the median of the
replica distribution
$\{\hat{r}^{(i)}_{\sigma}(\boldsymbol{x}, \boldsymbol{c})\}$ to determine the bounds in the EFT
parameter space, and here we assess the impact of methodological
uncertainties by displaying results for parameter inference based on the full
distribution of replicas of the likelihood ratio parametrisation.

\begin{figure}[htbp]
    \centering
    \includegraphics[width=\textwidth]{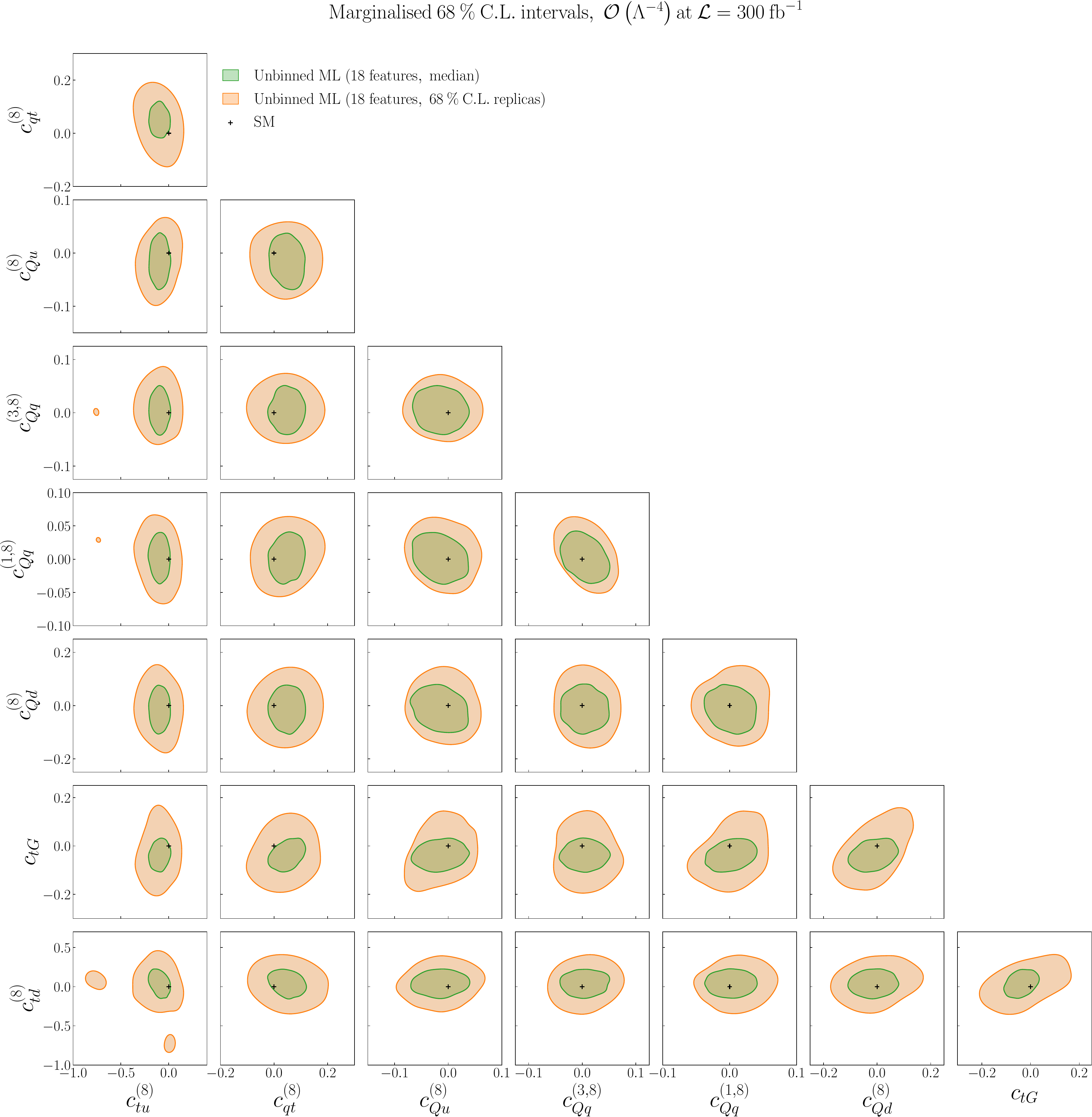}
    \caption{Same as Fig.~\ref{fig:tt_glob_quad_nn} for the 68\% CL contours
       in the EFT parameters obtained with the unbinned multivariate
       observable for particle-level top-quark pair production.
       We compare the bounds obtained from the median of the replica
       distribution of the likelihood ratio
       parametrisation, as  done in Fig.~\ref{fig:tt_glob_quad_nn},
       with the corresponding bounds obtained taking into account the full replica
       distribution.
    }
    \label{fig:tt_quad_envelope}
\end{figure}

Fig.~\ref{fig:tt_quad_envelope} displays
the bounds in the parameter space obtained from the unbinned multivariate
observables based on the complete set of kinematic features and in the case of theory
simulations that include quadratic EFT corrections.
We compare results for parameter inference based on the median of
the replica distribution
$\{\hat{r}^{(i)}_{\sigma}(\boldsymbol{x}, \boldsymbol{c})\}$, which coincide with
those of Fig.~\ref{fig:tt_glob_quad_nn},
with results based on the full distribution of replicas.
Namely, in the latter case  one starts from the $N_{\rm rep}$ 
individual replicas of the profile likelihood ratio functions and performs Nested Sampling inference
on each of them, to subsequently combine the resulting samples and estimate the
posterior distribution by means of the KDE method.
In this manner, the differences between the contours  shown
in Fig.~\ref{fig:tt_quad_envelope} provide an estimate of how
methodological uncertainties associated to the Machine Learning training procedure
impact the derived bounds in the EFT parameter space.

Inspection of Fig.~\ref{fig:tt_quad_envelope} and
comparison with the corresponding bounds displayed in
Fig.~\ref{fig:tt_glob_quad_nn} confirm that
these procedural uncertainties, as estimated with the replica method, do not modify
the qualitative results obtained from the median of the likelihood ratio.
In particular, the observation that bounds obtained using multivariate unbinned
observables are much stronger than those based on unbinned  models based on one or two
kinematic features remains valid once
replica uncertainties are accounted for.
Furthermore, we note that in principle it is possible to reduce the spread of
the replica distribution by training on higher-statistics samples and adopting more
stringent stopping criteria.
In this respect, our approach provides a strategy to quantify under which
conditions the computational overhead required to achieve more accurate
Machine Learning trainings  is justified from the point of view of the
impact on the EFT coefficients.
This said, these results also indicate that in realistic scenarios
based on finite training samples methodological uncertainties cannot
be neglected, and should be accounted for in studies of the
impact of ML-based observables in EFT analyses.

\section{Summary and outlook}
\label{sec:summary_ml4eft}

The general problem of identifying novel types of measurements which, given
a theoretical framework,
provide enhanced or even maximal sensitivity to the parameters of interest
 is ubiquitous in modern particle physics.
 Constructing 
 such optimised observables has a two-fold motivation:
 on the one hand, to achieve
 the most stringent constraints
 on the model parameters from a specific process,
 and on the other hand, to provide bounds
 on the maximum  amount of information that can be extracted
 from the same process.
Optimised observables can hence be used to  design
traditional observables in a way that approaches or saturates
the limiting sensitivity by highlighting the best choices
of binning and kinematic features.

In this chapter, we have presented a new framework
for the design of optimal observables for EFT applications
at the LHC, making use of Machine Learning techniques
to parametrise multivariate likelihoods for an arbitrary number
of higher-dimensional operators.
To illustrate the reach of our method,
we have constructed multivariate
unbinned observables for top-quark pair production.
We have demonstrated how these observables  either lead
to a significant improvement in the constraints
on the EFT parameter space, or indicate the conditions upon which
a traditional binned analysis already saturates the EFT sensitivity. 

The {\sc\small ML4EFT} framework  presented in this chapter and the accompanying simulated
event samples are made available as an open source code which
can be interfaced with existing global EFT fitting tools
such as {\sc\small SMEFiT}, {\sc\small FitMaker}~\cite{Ellis:2020unq},
{\sc\small HepFit}~\cite{DeBlas:2019ehy}, and {\sc\small Sfitter}~\cite{Brivio:2019ius}.
Its scaling behaviour with the number of parameters,
together with its parallelisation capabilities, 
make it suitable for its integration within global EFT fits which involve several tens
of independent Wilson coefficients.
While in its current implementation the parametrised likelihood ratios are provided
in terms of the output of the trained network replicas, work in progress aims
to tabulate this output in terms of  fast interpolation grids, as done
customarily in the case of PDF analyses~\cite{Buckley:2014ana,Bertone:2014zva,Carrazza:2020gss,Carli:2010rw}.
The resulting interpolated unbinned likelihood ratios can then be
combined with Gaussian or Poissonian binned measurements within a global fit.
While currently {\sc\small ML4EFT} can only be used
in combination with simulated Monte Carlo pseudo-data,
all of the ingredients required for the analysis of actual LHC measurements are already in place.

The results presented in this chapter could be extended along
several directions.
First, one could include experimental and theoretical
correlated systematic uncertainties, as required for the interpretation of measurements
which are not dominated by statistics.
Second, other Machine Learning algorithms could be adopted,
which may offer
performance advantages as compared to those used here:
one possibility could be graph neural networks~\cite{ArjonaMartinez:2018eah,Abdughani:2018wrw} which make possible
varying the kinematic features used for the training on an event-by-event basis.
Third, {\sc\small ML4EFT} could be applied to more realistic final states
with higher order corrections, such as by means
of NNLO+PS simulations
of $t\bar{t}$ and $hV$~\cite{Haisch:2022nwz,Mazzitelli:2021mmm,Zanoli:2021iyp}, and  accounting
for detector effects to bridge the gap between
the theory predictions and the experimentally accessible quantities.
It would also be interesting to compare
the performance of different approaches
to construct ML-assisted optimised observables
for EFT applications in specific benchmark scenarios.

Beyond hadron colliders, the framework developed here could also be
relevant to construct optimal observables for EFT analyses in the case of
high-energy lepton colliders, such as electron-positron
collisions at CLIC~\cite{CLICPhysicsWorkingGroup:2004qvu,Linssen:2012hp}
or at a multi-TeV muon collider~\cite{MuonCollider:2022xlm,Chen:2022msz,Buttazzo:2020uzc}.
As mentioned,
statistically optimal observables were actually first designed for electron-positron colliders, where the  simpler final state facilitates the calculation of the exact event
likelihood for parameter inference~\cite{DELPHI:2010ykq,Diehl:1993br,Durieux:2018tev}.
These methods may not be suitable to the high  multiplicity environment
of multi-TeV lepton colliders, in particular due to the complex pattern
of electroweak radiation.
Therefore, the {\sc\small ML4EFT}
method could provide a suitable alternative to construct unbinned multivariate observables
achieving maximal EFT sensitivity at high-energy lepton colliders.

In addition to applications to the SMEFT, our framework
could also be relevant to other types of theory interpretations of collider data such as
global PDF fits.
In particular, PDFs at large values of Bjorken-$x$ are
poorly constrained~\cite{Beenakker:2015rna}
due to the limited amount of experimental data available.
This lack of knowledge degrades the reach of searches for both resonant
and non-resonant new physics in the high-energy tail of
 differential distributions,
as  recently emphasised for the case of the forward-backward asymmetry
in neutral-current Drell-Yan production~\cite{afb}.
Given that measurements of these high-energy tails are often dominated by statistical
uncertainties, the {\sc\small ML4EFT} method could be used
to construct unbinned multivariate observables tailored
to constrain large-$x$ PDFs at the LHC.
Our method could also be applied in the context
of a joint extraction of PDFs and EFT coefficients~\cite{Greljo:2021kvv,Carrazza:2019sec,Iranipour:2022iak, Liu:2022plj, Gao:2022srd},
required to disentangle QCD effects
from BSM ones in kinematic regions where
they potentially overlap, such as large Bjorken-$x$.

  \chapter{The SMEFT at the HL-LHC and future colliders}
\label{chap:smefit_fcc}
\vspace{-1cm}
\begin{center}
\begin{minipage}{1.\textwidth}
    \begin{center}
        \textit{This chapter is based on my results that are presented in Ref.~\mycite{Celada:2024mcf}} 
    \end{center}
\end{minipage}
\end{center}

\myparagraph{Introduction}
The vast amount of data collected by the LHC during Run II has significantly increased our knowledge of fundamental particle physics. 
As we showed in Chapter \ref{chap:smefit3}, a global interpretation of these measurements is required to understand how well the SM describes nature at the TeV scale and how much room (and where) is left for its extensions.
The magnitude of this task, as well as our knowledge, keeps increasing thanks to the ongoing LHC Run III, which will provide exciting new insights.

Following Run III of the LHC, its High-Luminosity upgrade~\cite{Cepeda:2019klc,Azzi:2019yne} is scheduled to start in 2029 and operate for the next decade, accumulating a total integrated luminosity of up to 3 ab$^{-1}$ per experiment.
Beyond the HL-LHC program, several proposals for future particle colliders have been put forward and are being actively discussed by the global community.
These proposals include electron-positron colliders, either circular such as the FCC-ee~\cite{FCC:2018byv,FCC:2018evy} and the CEPC~\cite{CEPCPhysicsStudyGroup:2022uwl}, or linear such as the ILC~\cite{Behnke:2013xla,ILC:2013jhg}, the C3~\cite{Vernieri:2022fae}, and CLIC~\cite{Linssen:2012hp}, high-energy proton-proton colliders such as the FCC-hh~\cite{FCC:2018byv,FCC:2018vvp} and the SppC~\cite{Tang:2015qga}, muon colliders~\cite{Accettura:2023ked,Aime:2022flm}, and high-energy electron-proton/ion colliders such as the LHeC and the FCC-eh~\cite{LHeC:2020van,FCC:2018byv}. 
Furthermore, with a focus on QCD and hadronic physics but also with a rich program of electroweak measurements and BSM searches, the Electron Ion Collider (EIC)~\cite{AbdulKhalek:2021gbh} has already been approved and is expected to see its first collisions in the early 2030s.

These proposed facilities envisage a significant expansion of our knowledge of nature at the smallest accessible scales.
Novel insights would be provided via unprecedented sensitivity to subtle quantum effects, distorting the properties of known particles, and via the direct production of new particles, for instance, heavy (TeV-scale) particles or lighter ones but feebly interacting. 
While opening unique opportunities, the theoretical interpretation of future collider measurements also poses several challenges that must be tackled beforehand.
In the specific case of leptonic colliders, significant theoretical progress both within the SM and beyond would be required to match the expected precision of the experimental measurements. 
This is illustrated with the Tera-$Z$ running program~\cite{TLEPDesignStudyWorkingGroup:2013myl} of circular $e^+e^-$ colliders, which would lead to up to $10^{12}$ $Z$-bosons at the FCC-ee, with hence extremely small statistical uncertainties.

\myparagraph{Motivation} Making an informed decision about which of these future particle colliders should be built demands, in addition to feasibility and cost/effectiveness studies, the quantitative assessment of their scientific reach.
Several dedicated studies comparing the reach of future colliders have been presented in the last years, in particular in the context of the European Strategy for Particle Physics Update (ESPPU)~\cite{EuropeanStrategyforParticlePhysicsPreparatoryGroup:2019qin} and of the Snowmass US-based community process~\cite{Narain:2022qud}. 
The latter has recently culminated in the P5 report, one of whose main recommendations is endorsing an off-shore Higgs factory located
in either Europe or Japan.
As the European community ramps up its activities towards the next update of its long-term strategy, continuing, diversifying, and deepening these quantitative assessments of the reach of future colliders is more timely than ever.

The advantages of using SMEFT (and EFTs in general) for the assessment of future colliders have been leveraged by several groups in the past. 
First, in studies that considered a limited set of measurements and then in fits to a much larger dataset, see e.g.~\cite{Ellis:2015sca,deBlas:2016nqo,Ellis:2017kfi,Durieux:2017rsg,Barklow:2017suo,Barklow:2017awn,DiVita:2017vrr,Chiu:2017yrx,Durieux:2018tev,deBlas:2019rxi,DeBlas:2019qco,LCCPhysicsWorkingGroup:2019fvj,Jung:2020uzh,deBlas:2021jlt,MuonCollider:2022xlm,deBlas:2022ofj,Allwicher:2023shc}. 
The latter group of studies has highlighted the need for the inclusion of LEP and SLD data and HL-LHC projections to avoid overestimating the benefits of future colliders~\cite{deBlas:2019rxi,DeBlas:2019qco,deBlas:2022ofj}. 
SMEFT studies at future lepton colliders could also probe some of the fundamental principles underlying Quantum Field Theory such as unitarity, locality and Lorentz invariance~\cite{Gu:2020thj,Gu:2020ldn}.

The goal of this chapter is to quantify and assess the constraints on the EFT operators of projected measurements from the HL-LHC,
and subsequently from two of the proposed high-energy electron-positron colliders, namely the circular variants FCC-ee and the CEPC.  
This result is achieved by extending the {\sc\small SMEFiT} framework with novel functionalities streamlining the inclusion of projections for future experimental facilities into the global fit.

%

Our analysis emphasises the profound interplay between measurements at leptonic and hadronic colliders to constrain complementary directions in the EFT parameter space. 
It illustrates the potential of future colliders, first the HL-LHC and then the FCC-ee and CEPC, to inform indirect BSM searches via high-precision measurements extending the sensitivity provided by existing data.
This unprecedented reach is quantified both at the level of Wilson coefficients as well as in terms of the parameters (masses and couplings) of representative UV-complete models, in the latter case benefiting from progress in the interfacing with automated matching tools~\cite{Carmona:2021xtq,Fuentes-Martin:2022jrf,terHoeve:2023pvs}, as presented in Chapter \ref{chap:smefit_uv}.

\myparagraph{Outline} Starting from the {\sc\small SMEFiT3.0} baseline scenario as presented in Chapter \ref{chap:smefit3}, we estimate in Sect.~\ref{sec:impact_hl_lhc_data} the constraints on the SMEFT coefficients which may be achieved at the HL-LHC. 
In Sect.~\ref{sec:results}, projections for FCC-ee and CEPC measurements are added to the HL-LHC baseline to determine the ultimate sensitivity of future circular $e^+e^-$ colliders to the SMEFT parameter space.
Sect.~\ref{subsec:uvmodels} presents the results of our global analyses of LHC Run II, HL-LHC and FCC-ee measurements at the level of the couplings and masses of UV-complete models matched onto the SMEFT. 
\section{Projections for the HL-LHC}
\label{sec:impact_hl_lhc_data}

We start by assessing the impact of projected HL-LHC measurements when added on top of the {\sc\small SMEFiT3.0} baseline fit from Chapter \ref{chap:smefit3}.
These projections are constructed following the procedure described in Sect.~\ref{sec:hl_lhc_projections}. We list the processes considered in Table~\ref{tab:proj-datasets} for Higgs and diboson production, and Table~\ref{tab:proj-datasets-top} processes involving 
top-quarks. We only project datasets that are both part of SMEFiT3.0, as described in Chapter \ref{chap:smefit3}, and which are based on the highest integrated luminosity for a given process type. Note that for each process type and final state one can include at most two different projections, given that both ATLAS and CMS will perform independent measurements in the HL-LHC data-taking period. Table~\ref{tab:fit_settings_hllhc_fcc} provides an overview of the input settings used for the fits discussed in this section and the next one.

\begin{table}[htbp]
    \centering
    \footnotesize
     \renewcommand{\arraystretch}{1.70}
     \begin{tabular}
     {l|C{3cm}|C{1cm}|C{1cm}|C{1.5cm}}
       \toprule
    Dataset   & EFT coefficients &  Pseudodata?  & EWPOs  & Figs.      \\
         \midrule
    {\sc\small  SMEFiT3.0}  & Global (Marginalised)   &  Yes  &  Exact   &  \ref{fig:spider_nlo_lin_glob},~\ref{fig:spider_nlo_quad_glob} \\
    {\sc\small  SMEFiT3.0}\,+\,HL-LHC  & Global (Marginalised)   &  Yes  &  Exact   & \ref{fig:spider_nlo_lin_glob},~\ref{fig:spider_nlo_quad_glob} \\
    {\sc\small  SMEFiT3.0}\,+\,HL-LHC  & Individual (one-param)   &  Yes  &  Exact   &  \ref{fig:spider_nlo_lin_glob},~\ref{fig:spider_nlo_quad_glob}\\
    \midrule
     {\sc\small  SMEFiT3.0}\,+\,HL-LHC\,+\,FCC-ee  & Global (Marginalised)   &  Yes  &  Exact  & \makecell{\ref{fig:spider_fcc_nlo_lin_glob},~\ref{fig:fisher_fcc_lin},~\ref{fig:spider_fcc_nlo_quad_glob},\\~\ref{fig:spider_fcc_energy_variations},~\ref{fig:spider_cepc_nlo_lin_glob}} \\ 
      {\sc\small  SMEFiT3.0}\,+\,HL-LHC\,+\,CEPC  & Global (Marginalised)   &  Yes  &  Exact & \ref{fig:spider_cepc_nlo_lin_glob}\\ 
  \bottomrule
  \end{tabular}
  \caption{\small Similar to Table~\ref{table:fit_settings}, now for the fits presented in the current chapter. The HL-LHC dataset is constructed following the projection described in Sect.~\ref{sec:hl_lhc_projections}.
  \label{tab:fit_settings_hllhc_fcc}
  }
  \end{table}

%

\begin{table}[t]
  \centering
  \scriptsize
   \renewcommand{\arraystretch}{1.65}
  \begin{tabular}{l|C{0.7cm}|c|C{2.2cm}|C{0.5cm}|C{0.5cm}}
   \toprule
 Dataset  & $\mathcal{L}$ $\lp\rm{fb}^{-1}\rp$  & Info  &  Observables  & $n_{\rm dat}$ & Ref.    \\
\midrule
     \midrule
     \multirow{3}{*}{ {\tt ATLAS\_STXS\_RunII\_13TeV\_2022} }  &\multirow{3}{*}{139}  &\multirow{3}{*}{
  $gg$F, VBF, $Vh$, $t\bar{t}h$, $th$} & $d\sigma/dp_T^h$    &   \multirow{3}{*}{36}    &  \multirow{3}{*}{ \cite{ATLAS:2022vkf} } \\
     &    &  & $d\sigma/dm_{jj}$&     &    \\
     &    &  & $d\sigma/dp_T^V$&     &    \\
    \midrule  
{\tt CMS\_ggF\_aa\_13TeV}&$77.4$&$gg$F, $h\rightarrow \gamma\gamma$&$\sigma_{gg\rm{F}}(p_T^h, N_{\rm jets})$&6&\cite{CMS:2019xnv}\\   
{\tt ATLAS\_ggF\_ZZ\_13TeV}&$79.8$&$gg$F, $h\rightarrow ZZ$&$\sigma_{gg\rm{F}}(p_T^h, N_{\rm jets})$&6&\cite{ATLAS:2019nkf}\\ 
{\tt ATLAS\_ggF\_13TeV\_2015}&$36.1$&$gg$F, $h\rightarrow ZZ, h\rightarrow \gamma\gamma$&$d\sigma(gg\mathrm{F})/dp_T^h$&9&\cite{ATLAS:2018pgp}\\ 
\multirow{2}{*}{\tt ATLAS\_WH\_Hbb\_13TeV}&\multirow{2}{*}{$79.8$}&\multirow{2}{*}{$Wh, h\rightarrow b\bar{b}$}& \multirow{2}{*}{ $d\sigma^{(\rm fid)}/dp_T^W$ }&\multirow{2}{*}{2}&\multirow{2}{*}{\cite{ATLAS:2019yhn}}\\
&&& \multirow{2}{*}{(stage 1 STXS)}&&\\
\multirow{2}{*}{\tt ATLAS\_ZH\_Hbb\_13TeV}&\multirow{2}{*}{$79.8$}&\multirow{2}{*}{$Zh, h\rightarrow b\bar{b}$}& \multirow{2}{*}{ $d\sigma^{(\rm fid)}/dp_T^Z$ }&\multirow{2}{*}{2}&\multirow{2}{*}{\cite{ATLAS:2019yhn}}\\
&&& \multirow{2}{*}{(stage 1 STXS)}&&\\
\multirow{2}{*}{\tt CMS\_H\_13TeV\_2015\_pTH}&\multirow{2}{*}{$35.9$}&$h\rightarrow b\bar{b}, h\rightarrow \gamma \gamma,$&\multirow{2}{*}{$d\sigma/dp_T^h$}&\multirow{2}{*}{9}&\multirow{2}{*}{\cite{CMS:2018gwt}}\\
&    & $ h\rightarrow ZZ$ & &     &    \\
\midrule
{\tt ATLAS\_WW\_13TeV\_2016\_memu}&$36.1$&fully leptonic&$d\sigma^{(\mathrm{fid})}/dm_{e\mu}$&13&\cite{ATLAS:2019rob}\\
{\tt ATLAS\_WZ\_13TeV\_2016\_mTWZ}&$36.1$&fully leptonic&$d\sigma^{(\mathrm{fid})}/dm_T^{WZ}$&6&\cite{ATLAS:2019bsc}\\
{\tt CMS\_WZ\_13TeV\_2016\_pTZ}&$35.9$&fully leptonic&$d\sigma^{(\mathrm{fid})}/dp_T^Z$&11&\cite{CMS:2019efc}\\
{\tt CMS\_WZ\_13TeV\_2022\_pTZ}&$137$&fully leptonic&$d\sigma/dp_T^Z$&11&\cite{CMS:2021icx}\\
\bottomrule
    \end{tabular}
  \caption{\small LHC Run II datasets for Higgs and diboson production ($\sqrt{s}=13$ TeV) used as input to the HL-LHC extrapolations.
  For each dataset, we indicate its internal {\sc\small SMEFiT} label together with its integrated luminosity $\mathcal{L}$ (in $\rm{fb}^{-1}$), the final state or the specific production mechanism, the physical observable, the number of data points, and the corresponding publication reference.
  We only project datasets that are part of {\sc\small SMEFiT3.0}, which are selected on the basis of the highest integrated luminosity for each process type.
     \label{tab:proj-datasets}}
\end{table}

\begin{table}[htbp]
  \centering
   \scriptsize
   \renewcommand{\arraystretch}{1.65}
  \begin{tabular}{l|C{0.7cm}|c|c|c|c}
   \toprule
 Dataset  & $\mathcal{L}$ $\lp\rm{fb}^{-1}\rp$  & Info  &  Observables  & $n_{\rm dat}$ & Ref.    \\
\midrule
     \midrule
{\tt ATLAS\_tt\_13TeV\_ljets\_2016\_Mtt}&$36.1$&$
\ell$+jets&$d\sigma/dm_{t\bar{t}}$&7&\cite{ATLAS:2019hxz}\\
{\tt CMS\_tt\_13TeV\_dilep\_2016\_Mtt}&$35.9$&dilepton&$d\sigma/dm_{t\bar{t}}$&7&\cite{CMS:2018adi}\\
{\tt CMS\_tt\_13TeV\_Mtt}&$137$&$
\ell$+jets&$1/\sigma d\sigma/dm_{t\bar{t}}$&14&\cite{CMS:2021vhb}\\
{\tt CMS\_tt\_13TeV\_ljets\_inc}&$137$&$
\ell$+jets&$\sigma(t\bar{t})$&1&\cite{CMS:2021vhb}\\
\midrule
{\tt ATLAS\_tt\_13TeV\_asy\_2022}&$139$&$\ell$ + jets&$A_C$&5&\cite{ATLAS:2022waa}\\
{\tt CMS\_tt\_13TeV\_asy}&$138$&$\ell$ + jets&$A_C$&3&\cite{CMS:2022ged}\\
\midrule
{\tt     ATLAS\_Whel\_13TeV} & 139 &
  $W$-helicity fraction & $F_0, F_L$    &    2   &   \cite{ATLAS:2022rms}\\
\midrule
{\tt ATLAS\_ttbb\_13TeV\_2016}&$36.1$&lepton + jets&$\sigma_{\rm tot}(t\bar{t}b\bar{b})$&1&\cite{ATLAS:2018fwl}\\
{\tt CMS\_ttbb\_13TeV\_2016}&$35.9$&all-jets&$\sigma_{\rm tot}(t\bar{t}b\bar{b})$&1&\cite{CMS:2019eih}\\
{\tt CMS\_ttbb\_13TeV\_dilepton\_inc}&$35.9$&dilepton&$\sigma_{\rm tot}(t\bar{t}b\bar{b})$&1&\cite{CMS:2020grm}\\
{\tt CMS\_ttbb\_13TeV\_ljets\_inc}&$35.9$&lepton + jets&$\sigma_{\rm tot}(t\bar{t}b\bar{b})$&1&\cite{CMS:2020grm}\\
\midrule
{\tt ATLAS\_tttt\_13TeV\_run2}&$139$&multi-lepton&$\sigma_{\rm tot}(t\bar{t}t\bar{t})$&1&\cite{ATLAS:2020hpj}\\
{\tt CMS\_tttt\_13TeV\_run2}&$137$&same-sign or multi-lepton&$\sigma_{\rm tot}(t\bar{t}t\bar{t})$&1&\cite{CMS:2019rvj}\\
{\tt ATLAS\_tttt\_13TeV\_slep\_inc}&$139$&single-lepton&$\sigma_{\rm tot}(t\bar{t}t\bar{t})$&1&\cite{ATLAS:2021kqb}\\
{\tt CMS\_tttt\_13TeV\_slep\_inc}&$35.8$&single-lepton&$\sigma_{\rm tot}(t\bar{t}t\bar{t})$&1&\cite{CMS:2019jsc}\\
{\tt ATLAS\_tttt\_13TeV\_2023}&$139$&multi-lepton&$\sigma_{\rm tot}(t\bar{t}t\bar{t})$&1&\cite{ATLAS:2023ajo}\\
{\tt CMS\_tttt\_13TeV\_2023}&$139$&same-sign or multi-lepton&$\sigma_{\rm tot}(t\bar{t}t\bar{t})$&1&\cite{CMS:2023ftu}\\
\midrule
  {\tt CMS\_ttZ\_13TeV\_pTZ}&$77.5$&$t\bar{t}Z$&$d\sigma(t\bar{t}Z)/dp_T^Z$&4&\cite{CMS:2019too}\\
{\tt ATLAS\_ttZ\_13TeV\_pTZ}&$139$&$t\bar{t}Z$&$d\sigma(t\bar{t}Z)/dp_T^Z$&7&\cite{ATLAS:2021fzm}\\
\midrule
{\tt ATLAS\_ttW\_13TeV\_2016}&$36.1$&$t\bar{t}W$&$\sigma_{\mathrm{tot}}(t\bar{t}W)$&1&\cite{ATLAS:2019fwo}\\
{\tt CMS\_ttW\_13TeV}&$35.9$&$t\bar{t}W$&$\sigma_{\mathrm{tot}}(t\bar{t}W)$&1&\cite{CMS:2017ugv}\\
\midrule
{\tt ATLAS\_t\_tch\_13TeV\_inc}&$3.2$&$t$-channel&$\sigma_{\mathrm{tot}}(tq), \sigma_{\mathrm{tot}}(\bar{t}q)$&2&\cite{ATLAS:2016qhd}\\ 
{\tt CMS\_t\_tch\_13TeV\_2019\_diff\_Yt}&$35.9$&$t$-channel&$d\sigma/d|y_t|$&5&\cite{CMS:2019jjp}\\ 
{\tt ATLAS\_t\_sch\_13TeV\_inc}&$139$&$s$-channel&$\sigma(t+\bar{t})$&1&\cite{ATLAS:2022wfk}\\ 
\midrule
{\tt ATLAS\_tW\_13TeV\_inc}&$3.2$&multi-lepton&$\sigma_{\mathrm{tot}}(tW)$&1&\cite{ATLAS:2016ofl}\\
{\tt CMS\_tW\_13TeV\_inc}&$35.9$&multi-lepton&$\sigma_{\mathrm{tot}}(tW)$&1&\cite{CMS:2018amb}\\
{\tt CMS\_tW\_13TeV\_slep\_inc}&$36$&single-lepton&$\sigma_{\mathrm{tot}}(tW)$&1&\cite{CMS:2021vqm}\\
\midrule
{\tt ATLAS\_tZ\_13TeV\_run2\_inc}&$139$&multi-lepton + jets&$\sigma_{\rm fid}(t\ell^+\ell^- q)$&1&\cite{ATLAS:2020bhu}\\
{\tt     CMS\_tZ\_13TeV\_pTt}  & 138 &
  multi-lepton + jets& $d\sigma_{\rm fid}(tZj)/dp_T^t$    &    3   &   \cite{CMS:2021ugv}  \\
\bottomrule
    \end{tabular}
  \caption{\small Same as Table~\ref{tab:proj-datasets} for processes involving top quark production.
     \label{tab:proj-datasets-top}}
\end{table}

%
In a nutshell, we take existing Run II measurements for a given process, focusing on datasets obtained from the highest luminosity, and extrapolate their statistical and systematic uncertainties to the HL-LHC data-taking period. 
Specifically, the statistical uncertainties in the projected pseudo-data are reduced by a factor depending on the ratio of luminosities, while systematic uncertainties are reduced by a fixed factor (taken to be 1/2 in our case) based on the expected performance improvement of the detectors. 

Within the adopted procedure, we maintain the settings and binning of the original Run II analysis unchanged, and assume the SM as the underlying theory.
We note that our projections are not optimised, and in particular with a higher luminosity one could also extend the kinematic coverage of the high-$p_T$ regions~\cite{Durieux:2022cvf}, adopt a finer binning, or attempt multi-differential measurements.
Nevertheless, our approach benefits from being exhaustive and systematic, and is also readily extendable once new Run II and III measurements become available. 

Since the considered HL-LHC projections assume the SM as the underlying theoretical description, and to avoid introducing possible inconsistencies, we generate Level-1 SM pseudo-data for the full {\sc\small SMEFiT3.0} dataset and use it to produce a baseline fit for the subsequent inclusions of the HL-LHC pseudo-data, see also Table~\ref{tab:fit_settings_hllhc_fcc}.
In Level-1, the pseudo-data is fluctuated randomly within uncertainties around the central SM theoretical prediction (see Sect.~\ref{sec:hl_lhc_projections}).
This is the same strategy adopted in the closure tests entering the NNPDF proton structure analyses~\cite{NNPDF:2014otw, NNPDF:2021njg}.
A dataset consistent with the SM as underlying theory throughout enables to cleanly separate the sensitivity of the projected data to the SMEFT parameter space from other possible factors, such as dataset inconsistency, eventual BSM signals, or the interplay with QCD uncertainties such as those associated to the PDFs.
We have verified that, in the SMEFT analyses based on pseudo-data generated this way, the fit quality satisfies $\chi^2/n_{\rm dat}\sim 1$ as expected, both for the baseline fit and once the HL-LHC (and later the FCC-ee and CEPC) projections are included. 
For this reason, in the following we only present results for the relative reduction of the uncertainties associated to the Wilson coefficients, since the central values are by construction consistent with the SM expectations. 

In the rest of this section, we present results in terms of $R_{\delta c_i}$, defined as the ratio between the magnitude of the 95\% CI for a given EFT coefficient $c_i$, to that of the same quantity in the baseline fit:
\begin{equation}
\label{eq:RatioC}
R_{\delta c_i} = \frac{ \lc c_i^{\rm min}, c_i^{\rm max} \rc^{95\%~{\rm CI}}~({\rm baseline+\mathrm{HL}\,\textnormal{-}\,\mathrm{LHC}})}{
 \lc c_i^{\rm min}, c_i^{\rm max} \rc^{95\%~{\rm CI}}~({\rm baseline})
} \, , \qquad i=1,\ldots, n_{\rm eft} \, .
\end{equation}
In the case of a disjoint 95\% CI, we add up the magnitudes of the separate regions.
From the definition of Eq.~(\ref{eq:RatioC}) it is clear that, for a given  coefficient $c_i$, the smaller the value of $R_{\delta c_i}$ the more significant the impact of the new data.

As mentioned above, central values are by construction in agreement with the SM expectation ($c_i^{\rm (SM)} =0$) within uncertainties, and hence it is not necessary to display them in these comparisons.
The ratio Eq.~(\ref{eq:RatioC}) can be evaluated both in one-parameter fits as well as in the global fit followed by marginalisation. 
%

\begin{figure}[t]
    \centering
    \includegraphics[width=0.85\linewidth]{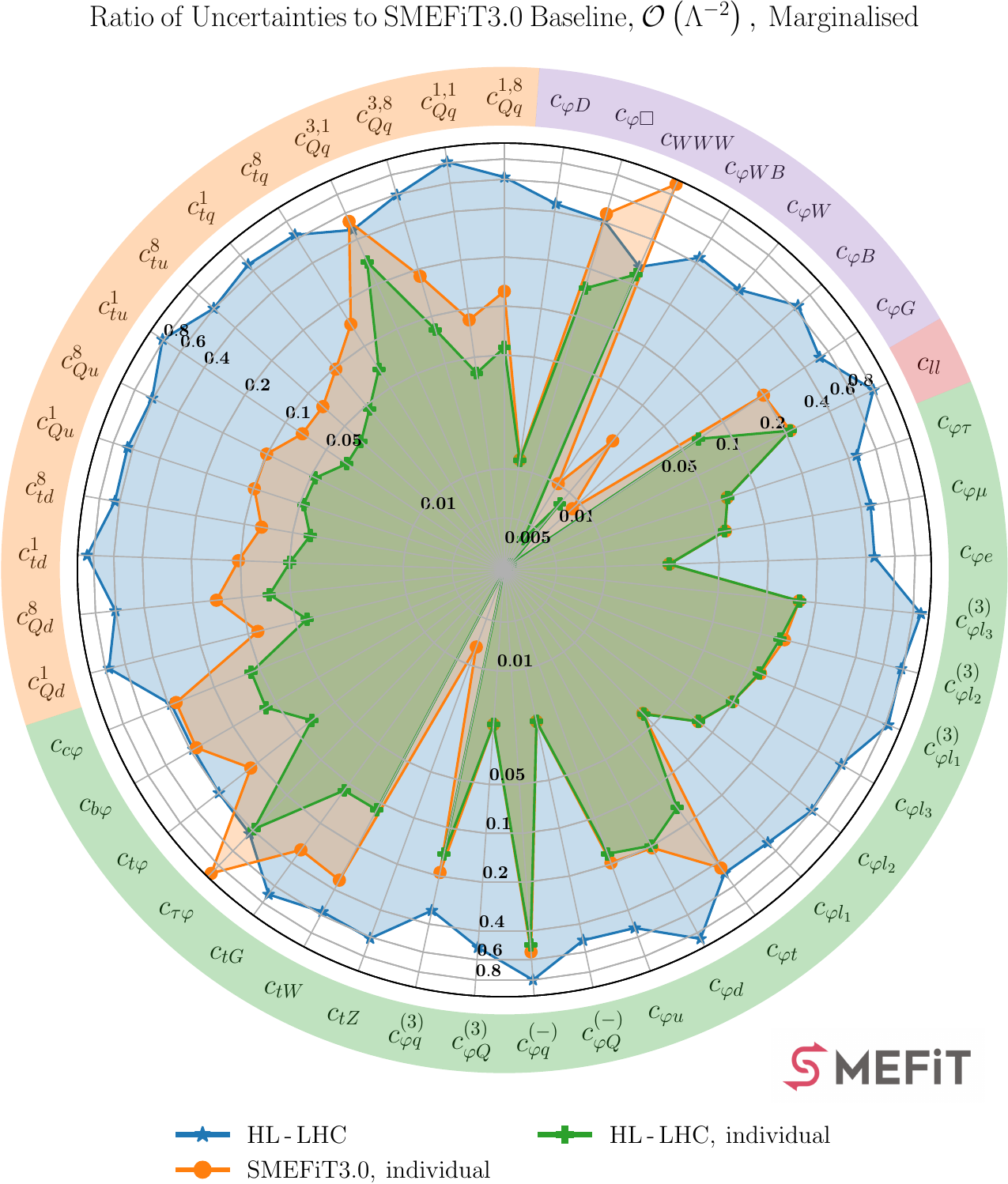}
    \caption{The ratio of uncertainties $R_{\delta c_i}$, defined in Eq.~(\ref{eq:RatioC}), for the $n_{\rm eft}=45$ coefficients entering the linear EFT fit, quantifying the impact of the HL-LHC projections when added on top of the {\sc\small SMEFiT3.0}
    baseline. 
    We display both the results of one-parameter fits and those of the marginalised analysis.
    The different colour codes indicate the relevant groups of SMEFT operators: two-light-two-heavy operators (orange), two-fermion operators (green), purely
    bosonic operators (purple), and the four-lepton operator $c_{\ell \ell}$ (red).
    Note that here  the baseline is a fit to pseudo-data for the {\sc\small SMEFiT3.0} dataset generated assuming the SM, rather than the fit to real data, see also Table~\ref{tab:fit_settings_hllhc_fcc}.
   }
    \label{fig:spider_nlo_lin_glob}
\end{figure}

\begin{figure}[t]
    \centering
    \includegraphics[width=0.85\linewidth]{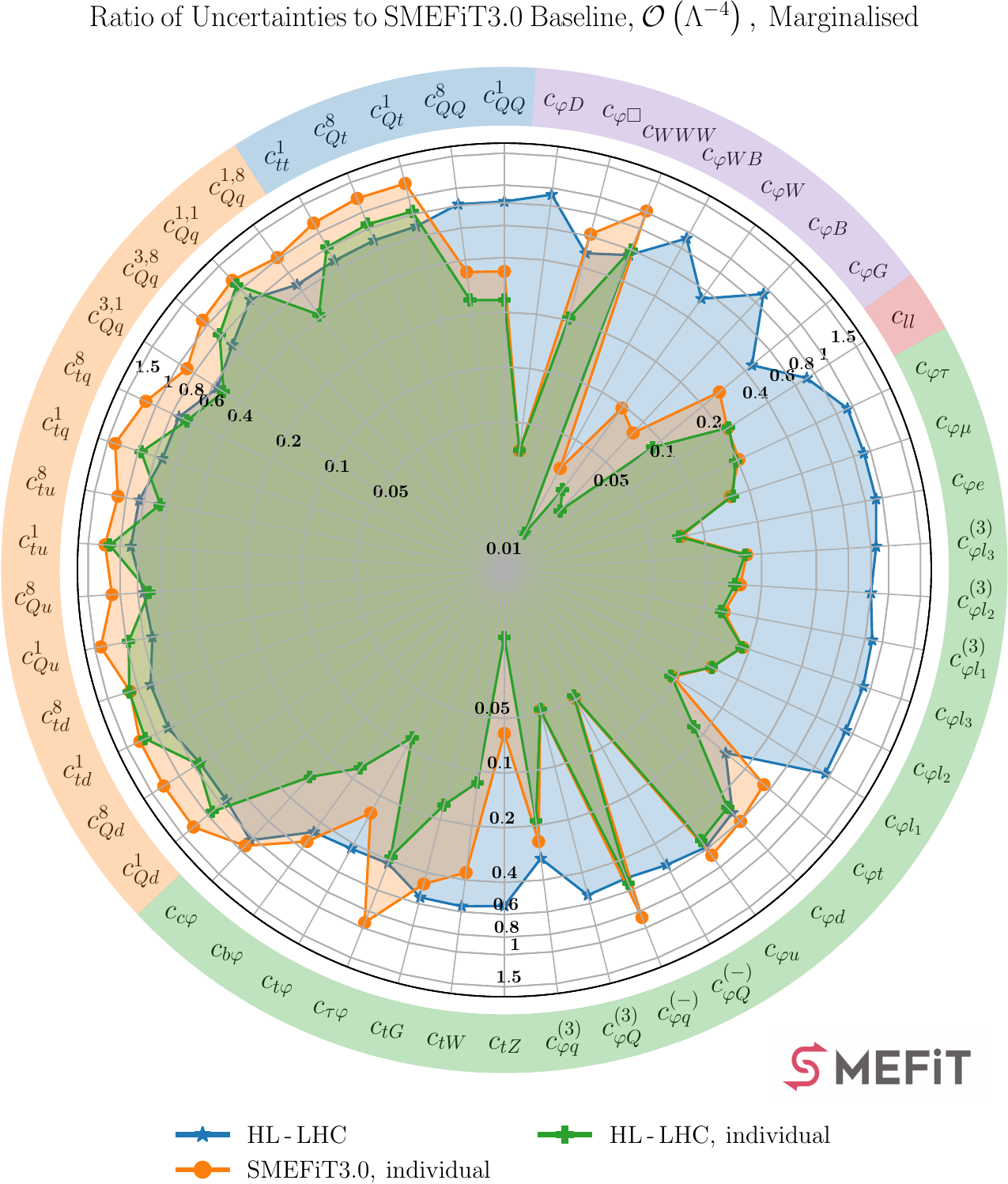}
    \caption{Same as Fig.~\ref{fig:spider_nlo_lin_glob}, now in the case of the analysis with quadratic EFT corrections included. 
    Note that in comparison with the linear fits,
    we now have an extra group of operators (indicated in blue), namely the four-heavy four-fermion operators
    that are constrained by $t\bar{t}t\bar{t}$
    and $t\bar{t}b\bar{b}$ production data.   
 }
    \label{fig:spider_nlo_quad_glob}
\end{figure}

Fig.~\ref{fig:spider_nlo_lin_glob} displays the ratio of uncertainties $R_{\delta c_i}$,  Eq.~(\ref{eq:RatioC}), for the $n_{\rm eft}=45$ Wilson coefficients entering the linear EFT fit, quantifying the impact of the HL-LHC projections when added on top of the {\sc\small SMEFiT3.0}
    baseline. 
We display both the results of the global fit, as well as those of one-parameter fits where all other coefficients are set to zero. 
Whenever available, as in the rest of this work, NLO QCD corrections for the EFT cross-sections are accounted for.
Then in Fig.~\ref{fig:spider_nlo_quad_glob} we
show the same comparison now in the case of the analysis with quadratic EFT corrections included in the theory calculations.

To facilitate visualisation, in Figs.~\ref{fig:spider_nlo_lin_glob}  and~\ref{fig:spider_nlo_quad_glob} results are presented with a ``spider plot'' format, with the different colours on the perimeter indicating the relevant groups of SMEFT operators: two-light-two-heavy four-fermion operators, two-fermion operators, purely bosonic operators, four-heavy four-fermion operators, and the four-lepton operator $c_{\ell \ell}$.
Recall that the four-heavy operators, constrained by $t\bar{t}t\bar{t}$  and $t\bar{t}b\bar{b}$ production data, are excluded from the linear fit due to its lack of sensitivity.
In this plotting format, coefficients whose values for $R_{\delta c_i}$ are closer to the centre of the plot correspond to the operators which are the most constrained by the HL-LHC projections, in the sense of the largest reduction of the corresponding uncertainties.
These plots adopt a logarithmic scale for the radial coordinate, to better highlight the large variations between the $R_{\delta c_i}$  values obtained for the different coefficients.

Several observations are worth drawing from the results of Figs.~\ref{fig:spider_nlo_lin_glob} and~\ref{fig:spider_nlo_quad_glob}.
Considering first the linear EFT fits, 
one observes that the projected HL-LHC observables are expected to improve the precision in the determination of the considered Wilson coefficients by an amount which ranges between around 20\% and a factor 3, depending on the specific operator, in the global marginalised fits. 
For instance, we find values of $R_{\delta c_i}\simeq 0.3$ for the triple gauge coupling $c_{WWW}$ and of $R_{\delta c_i}\simeq 0.4$ for the charm, bottom, and tau Yukawa couplings $c_{b\varphi}$, $c_{c\varphi}$, and $c_{\tau \varphi}$.\footnote{We note that our HL-LHC projections do not include measurements directly sensitive to the $h\to c\bar{c}$ decay.} 
For the two-light-two-heavy operators bounds, driven by $t\bar{t}$ distributions, $R_{\delta c_i}$ ranges between 0.8 and 0.5 hence representing up to a factor two of improvement. 
For the operators driven by top quark production, our estimate of the impact of the HL-LHC data can be compared to the results of~\cite{deBlas:2022ofj,Durieux:2022cvf}. 
Their analysis includes dedicated HL-LHC observables which especially help for the two-light-two-heavy operators, resulting in tighter bounds as compared to our fit by around a factor two for some of these coefficients.
%
\begin{figure}[t]
    \centering
    \includegraphics[width=0.6\linewidth]{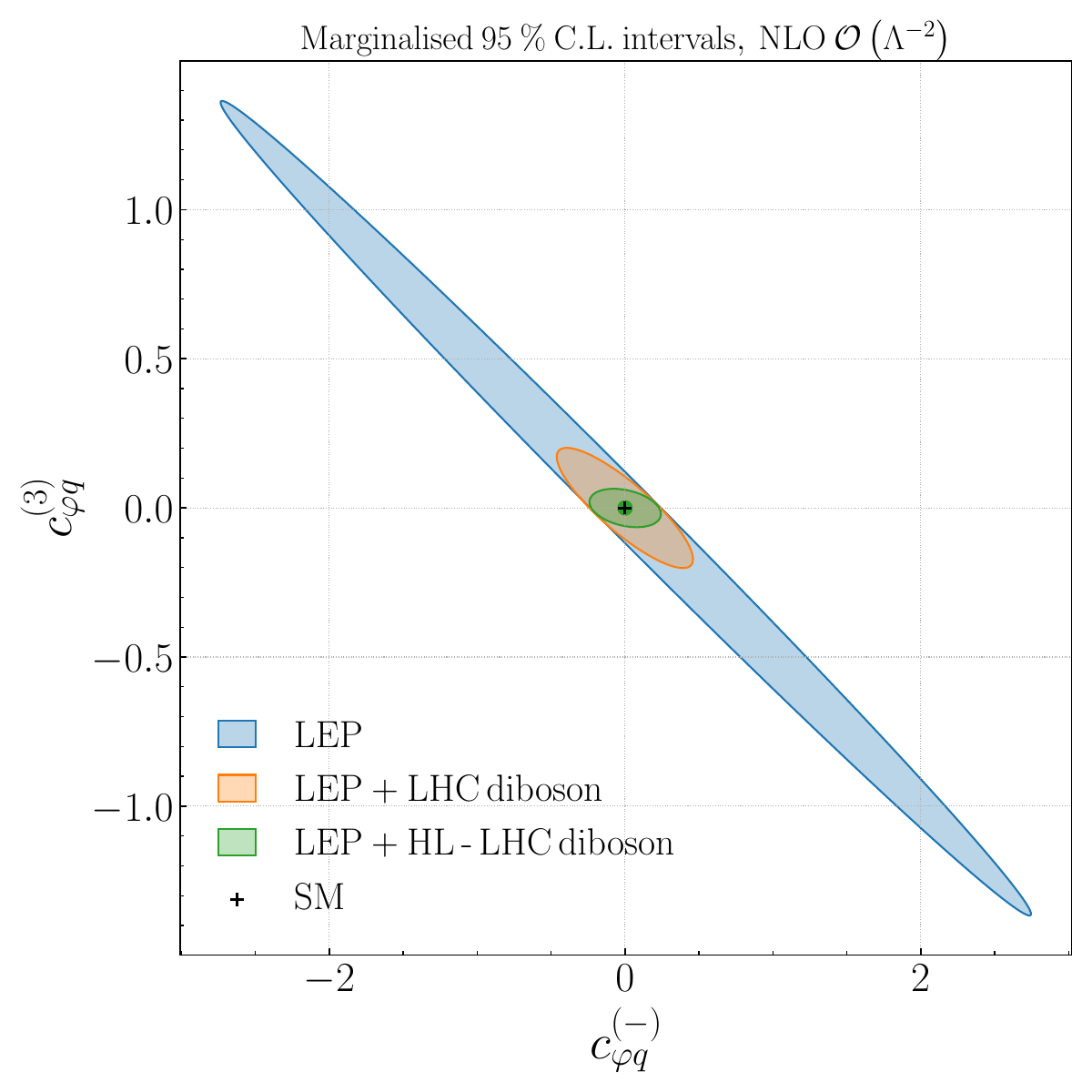}
    \caption{The marginalised $95\%$ CI in the $\lp c_{\varphi q}^{(3)}, c_{\varphi q}^{(-)}\rp$ plane from linear EFT fits to different datasets. 
    We compare the result of a LEP-only fit (blue) with those adding either the LHC Run II diboson data (orange) or the HL-LHC diboson projections (green).
    The three fits are carried out with Level-0 pseudo-data, see Sect.~\ref{sec:hl_lhc_projections}.}
    \label{fig:diboson_effect}
\end{figure}

Additionally, HL-LHC measurements can improve the bounds imposed by EWPOs along directions that are a linear combination of individual coefficients, as illustrated by the $\lp c_{\varphi q}^{(3)}, c_{\varphi q}^{(-)}\rp$ analysis of Fig.~\ref{fig:diboson_effect} where we show the impact of diboson production on  the $95\%$ CI in the  $\lp c_{\varphi q}^{(3)}, c_{\varphi q}^{(-)}\rp$ plane. 
We compare the marginalised bounds from a global linear LEP-only fit with those resulting from combining LEP with either LHC Run II or the HL-LHC diboson data. 
For consistency, the three fits are carried out with Level-0 pseudo-data.
LEP data results in a quasi-flat direction in this plane, which is then well constrained by diboson data at LHC Run II (and subsequently at the HL-LHC), confirming the long-predicted complementarity between LEP and LHC diboson measurements~\cite{Falkowski:2015jaa, Butter:2016cvz, Alioli:2017nzr,Franceschini:2017xkh,Banerjee:2018bio,Grojean:2018dqj}.

The broad reach of the HL-LHC program is illustrated by the fact that essentially all operators considered have associated tighter bounds even in the conservative analysis we perform.  
The comparison between the linear EFT marginalised and individual bounds displayed in Fig.~\ref{fig:spider_nlo_lin_glob} indicates that in the one-parameter fits the sensitivity is typically much better than in the global fit, in some cases by more than an order of magnitude, for example for the $c_{\varphi W B}$, $c_{\varphi B}$,
 $c_{\varphi W}$, and $c_{tZ}$ coefficients.
The exception of this trend are coefficients which are determined by specific subsets of measurements which do not affect other degrees of freedom, such as $c_{WWW}$, constrained from diboson data, and $c_{\tau\varphi}$, constrained only from $h \to \tau\tau$ decays.
This comparison between individual fits to the {\sc\small SMEFiT3.0} and HL-LHC datasets also highlights which operators are constrained mostly by the EWPOs, namely the two-light-fermion operators, the four-lepton operator $c_{\ell\ell}$, and the purely bosonic operator $c_{\varphi \square}$.
For these coefficients, the improvements found in the global marginalised HL-LHC fit arise from indirect improvements in correlated coefficients.

Fig.~\ref{fig:spider_nlo_quad_glob} presents the same comparison as that in Fig.~\ref{fig:spider_nlo_lin_glob} now with the quadratic EFT corrections accounted for, and including also the results for the four-heavy four-fermion operators. 
Quadratic fits break degeneracies and correlations present in the linear fit. 
Hence, some operators not well probed at (HL-)LHC, such as the two-lepton ones, show an $R_{\delta c_i}$ closer to $1$ than in the linear case.
Likewise, for these operators $R_{\delta c_i}$ is unchanged in the individual fits before and after the inclusion of the HL-LHC projections. 
%
%
%
%
Finally, we note that in scenarios relevant to the matching to UV models, which involve a subset of EFT operators, the relevant constraints would be in between the global and the individual bounds shown in Fig.~\ref{fig:spider_nlo_quad_glob}.

Overall, our analysis indicates that, in the context of a global SMEFT fit, the extrapolation of Run II measurements to the HL-LHC results into broadly improved bounds, ranging between 20\% and a factor 3 better depending on the specific coefficient.
Qualitatively similar improvements arising from HL-LHC constraints  will be observed once matching to UV models in Sect.~\ref{subsec:uvmodels}.
We note again that the HL-LHC constraints derived here are conservative, as they may be significantly improved through optimised analyses, exploiting features not accessible with the Run II dataset.

\section{The impact  of future $e^+e^-$ colliders on the SMEFT}
\label{sec:results}

We now present the quantitative assessment of the constraints on the SMEFT coefficients provided by measurements to be carried out at the two proposed high-energy $e^+e^-$ circular colliders, FCC-ee and CEPC.
The baseline for these projections is the global SMEFT analysis augmented with the dedicated HL-LHC projections from Sect.~\ref{sec:impact_hl_lhc_data}.
Here first of all we describe the FCC-ee and CEPC observables and running scenarios considered, for which we follow the recent Snowmass study~\cite{deBlas:2022ofj} with minor modifications.
Then we present results at the level of SMEFT coefficients, highlighting the correlation between LHC- and $e^+e^-$-driven constraints.

\subsection{Observables and running scenarios}
\label{sec:future_projections}

Several recent studies~\cite{deBlas:2019rxi,DeBlas:2019qco,deBlas:2022ofj,Durieux:2022cvf} have assessed the 
physics potential of the various
proposed leptonic colliders, including FCC-ee, ILC, CLIC, CEPC, and a muon collider, 
in terms of global fits to SMEFT coefficients and in some cases also matched to UV-complete models.
Here we describe the projections for FCC-ee and CEPC measurements that will be used to constrain the SMEFT parameter space.
We focus on these two colliders as 
representative examples of 
possible new leptonic colliders, though the same strategy can be straightforwardly applied to any other future facility.

The Future Circular Collider~\cite{Benedikt:2020ejr,FCC:2018byv} in its electron-positron mode (FCC-ee)~\cite{FCC:2018evy,Bernardi:2022hny}, originally known as TLEP~\cite{TLEPDesignStudyWorkingGroup:2013myl},
is a proposed electron-positron collider operating in a tunnel of approximately 90 km of circumference in the CERN site and based on well-established accelerator technologies similar to those of LEP. 
Running at several centre-of-mass energies  is envisaged, starting from the $Z$-pole all the way up to $\sqrt{s}=365$~GeV, above the top-quark pair production threshold. 
Possible additional runs at $\sqrt{s}=125$ GeV (Higgs pole) and for $\sqrt{s}< m_Z$  (for QCD studies) are under consideration for the FCC-ee.
This circular $e^+e^-$ collider would represent the first stage of a decades-long scientific exploitation of the same tunnel, eventually followed by a $\sqrt{s}\sim 100$ TeV proton-proton collider (FCC-hh).
Here we adopt the same scenarios for the FCC-ee running as in the Snowmass study of~\cite{deBlas:2022ofj} but updated to consider 4 interaction points (IPs), along the lines of the recent midterm feasibility report~\cite{FCCfeasibility}.\footnote{Different running scenarios for the FCC-ee are being discussed, including the integrated luminosity at each $\sqrt{s}$, and therefore the contents of Table~\ref{tab:FCCee_runs} may change as the project matures.} 
We summarise the running scenarios in Table~\ref{tab:FCCee_runs}.
As done in previous studies~\cite{deBlas:2019rxi}, we combine the information associated to the data taken at the runs with $\sqrt{s}=350$~GeV and $365$~GeV and denote this combination as ``$\sqrt{s}=365$~GeV'' in the following.

\begin{table}[t]
    \centering
      \renewcommand{\arraystretch}{1.50}
    \begin{tabular}{|l|l|l|c|}
    \toprule
         \multirow{2}*{Energy $(\sqrt{s})$ } & \multicolumn{2}{c|}{$\mathcal{L}_{\rm int}$ (Run time)} & \multirow{2}*{$\mathcal{L}_{\rm FCC-ee}/\mathcal{L}_{\rm CEPC}$ } 
         \\
        & FCC-ee (4 IPs) & CEPC (2 IPs) & \\
    \hline
  $91$~GeV ($Z$-pole) & $300$~ab$^{-1}$ (4 years)  & $100$~ab$^{-1}$ (2 years) & 3\\
          $161$~GeV ($2\,m_W$) & $20$~ab$^{-1}$ (2 years) & $6$~ab$^{-1}$ (1 year) & 3.3\\
          $240$~GeV & $10$~ab$^{-1}$ (3 years) & $20$~ab$^{-1}$ (10 years) & 0.5\\
          $350$~GeV & $0.4$~ab$^{-1}$ (1 year) & $0.2$~ab$^{-1}$  & 2 \\
          $365$~GeV ($2\,m_t$) & $3$~ab$^{-1}$ (4 years) & $1$~ab$^{-1}$ (5 years) & 3\\
         \bottomrule
    \end{tabular}
    \caption{The running scenarios considered in our analysis for the FCC-ee and the CEPC, following~\cite{deBlas:2022ofj,Bernardi:2022hny,CEPCPhysicsStudyGroup:2022uwl} and the mid-term FCC feasibility report~\cite{FCCfeasibility}.
    Our projections assume 4 interaction points (IPs) for the FCC-ee and 2 for the CEPC.
    For each centre of mass energy $\sqrt{s}$,
    we indicate the expected  luminosity as well as the number
    of years in which this luminosity will be collected.
    The last column displays the ratio between the expected integrated luminosities at the FCC-ee and the CEPC.
    When presenting our results, we combine the information associated to the data taken at the runs with $\sqrt{s}=350$~GeV and $365$~GeV and denote this combination as ``$\sqrt{s}=365$~GeV''.
    \label{tab:FCCee_runs}}
\end{table}

The Circular Electron Positron Collider (CEPC)~\cite{An:2018dwb} is a proposed electron-positron collider to be built and operated in China.
The current plan envisages a collider tunnel of around 100 km, and it would operate in stages at different centre of mass energies, with a maximum of $\sqrt{s}=365$ GeV above  the top-quark pair production threshold.
The current baseline design assumes two interaction points for the CEPC.
In the same manner as for the FCC-ee,  Table~\ref{tab:FCCee_runs} indicates the expected integrated luminosity (and number of years required to achieve it) in the current running scenarios for each value of the center-of-mass energy $\sqrt{s}$.
The main differences between the projected statistical uncertainties for the FCC-ee and CEPC arise from the different data-taking plans as well as the different number of IPs. 
For instance, CEPC plans a longer running period at $\sqrt{s}=240$ GeV, which would lead to a reduction of statistical errors as compared to the FCC-ee observables corresponding to the same center-of-mass energy.

In the last column of Table~\ref{tab:FCCee_runs} we display the
ratio between the integrated luminosities at the FCC-ee and the CEPC, $\mathcal{L}_{\rm FCC-ee}/\mathcal{L}_{\rm CEPC}$, for each of the data-taking periods at a common centre of mass energy.
The FCC-ee is expected to accumulate a luminosity 3 times larger than the CEPC for the runs at $\sqrt{s}=91$ GeV, 161 GeV, and 365 GeV, while for $\sqrt{s}=240$ GeV it would accumulate half of the CEPC luminosity, given that the latter is planned to run for 10 years as opposed to the 3 years of the FCC-ee.

In our analysis, we consider five different classes of observables that are accessible at high-energy circular electron-positron colliders such as the FCC-ee and the CEPC.
These are the EWPOs at the $Z$-pole; light fermion (up to $b$ quarks and $\tau$ leptons) pair production; Higgs boson production in both the $hZ$ and $h\nu\nu$ channels; gauge boson pair production;  and top quark pair production.
Diboson  ($W^+W^-$) production becomes available at $\sqrt{s}=161$ GeV ($WW$ threshold), Higgs production opens up at $\sqrt{s}=240$ GeV, and top quark pair production is accessible starting from $\sqrt{s}=350$ GeV, above the $t\bar{t}$ threshold.

Among these processes, the $Z$-pole EWPOs, light fermion-pair, $W^+W^-$, and Higgs production data are included at the level of inclusive cross-sections, accounting also for the corresponding branching fractions.
The complete list of observables considered, together with the projected experimental uncertainties entering the fit, are collected in App.~E of Ref.~\mycite{Celada:2024mcf}.
For diboson and top quark pair production, we consider also unbinned normalised measurements within the optimal observables approach \cite{Diehl:1993br}.
 We  briefly review below these groups of processes.

\myparagraph{EWPOs at the $Z$-pole}
The $Z$-pole electroweak precision observables that would be measured at the FCC-ee and CEPC coincide with those already measured by LEP and SLD, see Table~\ref{tab:ew-datasets}.
The main difference is the greatly improved precision that will be achieved at future electron-positron colliders, due to the increased luminosity and the expected reduction of systematic uncertainties.
Specifically, here we include projections for the QED coupling constant at the $Z$-pole, $\alpha(m_Z)$; the decay widths of the
$W$ and $Z$ bosons, $\Gamma_W$
and $\Gamma_Z$; the asymmetry between vector
and axial couplings $A_f$
for $f=e,\mu,\tau,c,b$;
the total cross-section
for $e^+e^-\to {\rm hadrons}$,
$\sigma_{\rm had}$; 
and the partial decay widths ratio to the total hadronic width $R_f = \Gamma_f/\Gamma_{\mathrm{had}}$
for $f=b,c$ and $R_\ell=\Gamma_\ell/\Gamma_{\mathrm{had}}$ for $\ell=e,\mu,\tau$.

\myparagraph{Light fermion pair production above the $Z$-pole}
The light fermion pair production measurements considered here, $e^+e^- \to f\bar{f}$, consist of both the total cross sections, $\sigma_{\rm tot}(f\bar{f})$, and the corresponding forward-backward asymmetries, $A^f_{\rm FB}$, with $f=e,\mu,\tau,c,b$, defined first in Eq.~(\ref{eq:Rbc}) at $\sqrt{s}=240$ and 365 GeV.
As for the rest of projections, the corresponding central values are taken from the SM predictions.
The production of top quark pairs, available at $\sqrt{s}=350$ GeV and $365$ GeV, is discussed separately below. 

\myparagraph{Higgs production}
We consider here Higgs production in the two dominant mechanisms relevant for electron-positron colliders, namely associated production with a $Z$ boson, also known as Higgsstrahlung,
\begin{equation}
e^+e^- \to Zh \, ,
\end{equation}
and in the vector-fusion mode via $W^+W^-$ fusion,
\begin{equation}
e^+e^- \to \bar{\nu}_eW^+\nu_e W^- \to 
\bar{\nu}_e \nu_e h \, .
\end{equation}
For $\sqrt{s}=240$ GeV, the total Higgs production cross-section is fully dominated by $Zh$ production, while for $\sqrt{s}=365$ GeV the VBF contribution reaches up to 25\% of the total cross-section.
Higgs production via the $ZZ$ fusion channel, $e^+e^- \to e^+e^- ZZ \to e^+e^-h$, is suppressed by a factor 10 in comparison with $WW$ fusion (for both values of $\sqrt{s}$) and is therefore neglected in this analysis.

These two production modes are included in the fit for all the decay modes that become accessible at electron-positron colliders: $c\bar{c}$, $b\bar{b}$, $gg$, $\mu^+\mu^-$, $\tau^+\tau^-$, $ZZ$, $WW$,  $\gamma\gamma$, and $\gamma Z$.
For $\sqrt{s}=240$ GeV, only the dominant decay channel to $b\bar{b}$ is accessible in the vector-boson fusion mode.
We note that possible Higgs decays into invisible final states are also constrained at $e^+e^-$ colliders, a unique feature possible due to the fact that the initial-state energy of the collision $\sqrt{s}$ is precisely known, by means of the direct measurement of the $\sigma_{Zh}$ cross-section via the $Z$-tagged recoil method. 
However, in the present analysis, given that we assume no invisible BSM decays of the Higgs boson, such a direct measurement of the $\sigma_{Zh}$ cross-section does not provide any additional constraints in the EFT parameter space.

Projections for Higgs production measurements are included at the level of inclusive cross-sections times branching ratio, $\sigma\times \text{BR}_X$ (signal strengths), separately for 
$e^+e^- \to Zh$ and $e^+e^- \to 
\bar{\nu}_e \nu_e h$.
Projections for the total inclusive $\sigma_{Zh}$ cross-section are also considered.
The information from differential distributions of the Higgsstrahlung process could in principle also be included, however, its impact on the SMEFT fit is limited~\cite{Durieux:2017rsg,deBlas:2019rxi} and hence we neglect them. 

\myparagraph{Gauge boson pair production}
We consider weak boson pair production, specifically in the $e^+e^-\to W^+W^-$ final state.
Diboson production in $e^+e^-$ collisions, already measured at LEP, enables the study of the electroweak gauge structure via the search for anomalous triple gauge couplings (aTGCs). 
We include the information from this process by means of a two-fold procedure, considering separately the overall signal strengths, namely the fiducial cross-section for $W^+W^-$ production times the $W$ leptonic and hadronic branching ratios, and the information contained in the shape of the differential distributions.
The latter is accounted for via the optimal observables strategy \cite{Celada:2024mcf, Diehl:1993br}.
Projections corresponding to the three center-of-mass energies relevant for diboson production, $\sqrt{s}=161$ GeV, 240 GeV, and 365 GeV, are included.

In our analysis, we assume that the $W$ boson does not have any exotic (invisible) decay modes and thus we impose that the separate leptonic and hadronic branching fractions add up to unity, 
\begin{equation}
    \text{BR}_{W\to e\nu} + \text{BR}_{W\to \mu\nu} + \text{BR}_{W\to \tau\nu} + \text{BR}_{W\to q\bar{q}}=1 \, .
\label{eq:no_exotic_Wdecay}
\end{equation}
This requirement allows one to determine the expected precision for the measurement of the fiducial cross-section and all the branching ratios from the measurement of the relevant $WW$ decay channels, see also Table~6 in~\cite{deBlas:2019rxi}.

\myparagraph{Top quark pair production}
The top quark pair production process, $e^+e^- \to t\bar{t}$, becomes accessible above $\sqrt{s}=2m_t\simeq 350$ GeV, the kinematic production threshold.
Measurements of this process provide information on the electroweak couplings of the top quark, complementing existing LHC measurements, and also enable a determination of the top quark mass with excellent ($\lsim 20$ MeV) precision. 
Similarly to the case of gauge boson pair production, here we include this process in the fit in terms of (now absolute) unbinned measurements within the optimal observables framework for the dominant $e^+e^- \to t\bar{t}\to W^+bW^-\bar{b}$ final state.
We neglect systematic uncertainties and adopt the same settings on the acceptance, identification, and reconstruction efficiencies as in the Snowmass EFT study of~\cite{Durieux:2022cvf}. 
Since here we do not consider normalised observables, it is not necessary to include separately the signal strengths for $e^+e^- \to t\bar{t}$ production and decay as done for the $W^+W^-$ case.

\myparagraph{Implementation}
A new module has been added to {\sc\small SMEFiT} which enables the integration of external, user-provided likelihoods into the figure of merit entering the global fit.
We have verified, in the specific case of $W^+W^-$ production, that using such an external likelihood is equivalent to the baseline implementation of datasets in {\sc\small SMEFiT} and based on separate data tables for the theory calculations and for the experimental measurements.
This new functionality is used here to include the constraints from the optimal observables in $t\bar{t}$ production by reusing the results derived in~\cite{Durieux:2018tev,Durieux:2022cvf,deBlas:2022ofj} with adjustments whenever required.

\subsection{Impact on the SMEFT coefficients}
\label{subsec:coeffs}

\begin{figure}[t]
    \centering
    \includegraphics[width=0.85\linewidth]{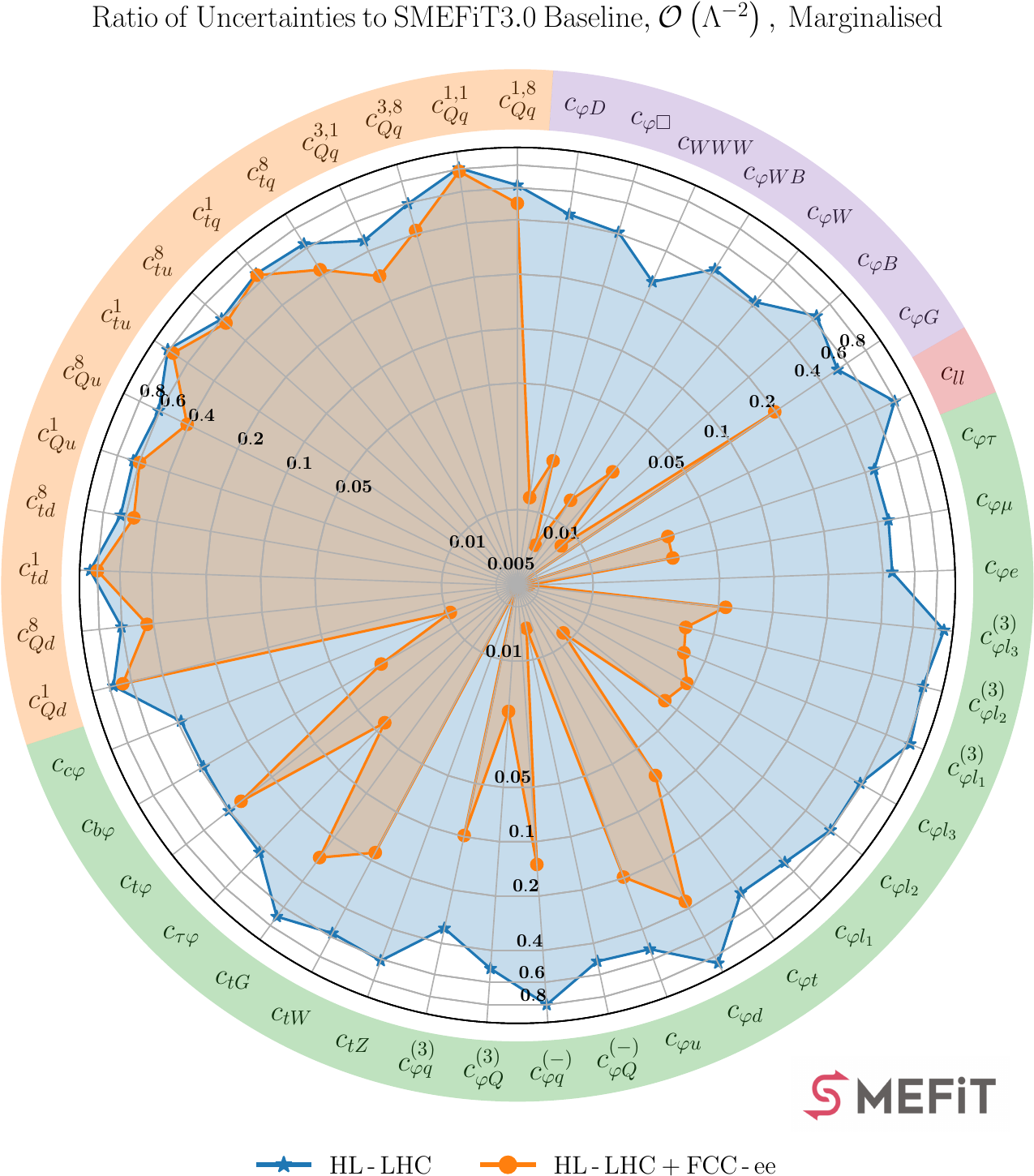}
    \caption{Same as Fig.~\ref{fig:spider_nlo_lin_glob} for the marginalised bounds on the SMEFT operators from a global $\mathcal{O}\lp \Lambda^{-2}\rp$  analysis, displaying the ratio of uncertainties to the 
    {\sc\small SMEFiT3.0} baseline of fits which include first the HL-LHC projections and subsequently both the HL-LHC and the FCC-ee observables. 
    All fits shown here are carried out based on Level-1 pseudo-data.
    }
    \label{fig:spider_fcc_nlo_lin_glob}
\end{figure}

Next we quantify the impact on the EFT coefficients of measurements from the FCC-ee and CEPC when added on top of the baseline fit extended with the HL-LHC projections discussed in Sect.~\ref{sec:impact_hl_lhc_data}.
Since our projections reproduce the assumed SM theory for the fitted observables by construction, one is interested only in the reduction of the length of the 95\% CI for the relevant operators, as quantified by the ratio $R_{\delta c_i}$ in Eq.~(\ref{eq:RatioC}), and hence we do not display the fit central values.
We focus first on the FCC-ee, and then compare with the CEPC results.

\begin{figure}[t]
    \centering
    \includegraphics[width=0.99\linewidth]{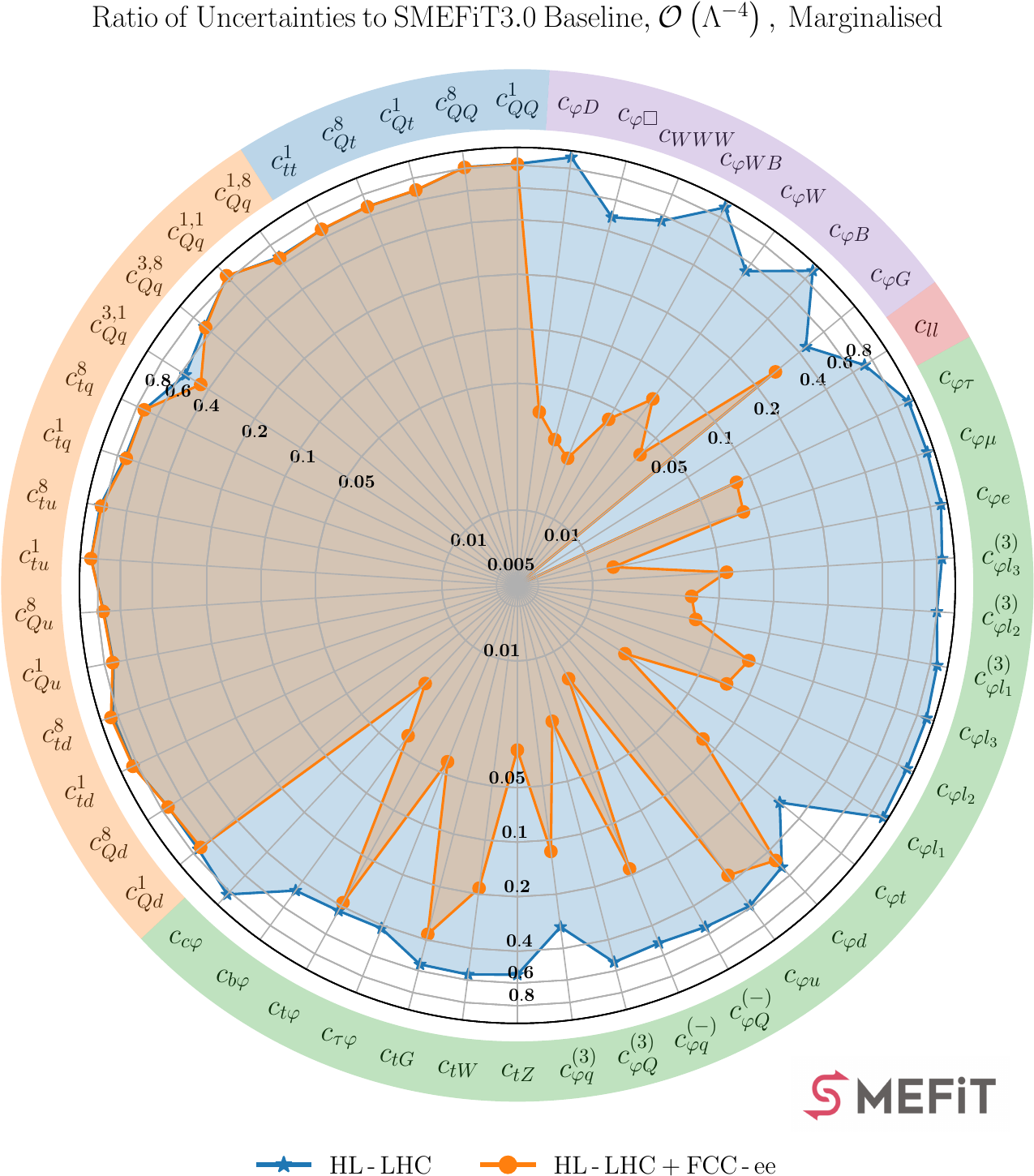}
    \caption{Same as Fig.~\ref{fig:spider_fcc_nlo_lin_glob} in the 
    case of the fits with EFT cross-sections including both the linear, $\mathcal{O}\lp \Lambda^{-2}\rp$, 
    and the quadratic, $\mathcal{O}\lp \Lambda^{-4}\rp$, corrections associated to the dim-6 operators considered in the analysis.}
    \label{fig:spider_fcc_nlo_quad_glob}
\end{figure}

Fig.~\ref{fig:spider_fcc_nlo_lin_glob} (\ref{fig:spider_fcc_nlo_quad_glob}) displays, in the same format as that of Fig.~\ref{fig:spider_nlo_lin_glob}, the marginalised bounds on the SMEFT operators from a global linear (quadratic) analysis, displaying the ratio of uncertainties to the  {\sc\small SMEFiT3.0} baseline of fits which include, first, the HL-LHC projections, and subsequently, both the HL-LHC and the FCC-ee observables, see Table~\ref{tab:fit_settings_hllhc_fcc}.
Inspection of Fig.~\ref{fig:spider_fcc_nlo_lin_glob} demonstrates the substantial impact that FCC-ee measurements would provide on the SMEFT parameter space in comparison with a post-HL-LHC baseline.
Indeed, the FCC-ee would constrain a wide range of directions in the EFT coefficients, except for the four-quark operators.
For the latter only moderate indirect improvements are expected, which go away in the quadratic fits shown in Fig.~\ref{fig:spider_fcc_nlo_quad_glob}. This is due to the fact that no FCC-ee observables included in the fit are sensitive to four-quark operators, and thus any impact of FCC-ee on their constraints arises through marginalisation. 

From Fig.~\ref{fig:spider_fcc_nlo_lin_glob} one also observes how the bounds on some of the purely bosonic, four-lepton, and two-fermion operators achieved at the FCC-ee could improve on the HL-LHC ones by almost two orders of magnitude. 
For instance, the 95\% CI for the coefficient $c_{WWW}$, which modifies the triple gauge boson interactions, is reduced by a factor $R_{\delta c_i} \simeq 0.3$ at the end of HL-LHC and then down to $R_{\delta c_i}=0.008$ at the FCC-ee, corresponding to a relative improvement on the bound by a factor around $40$. 
Likewise, our analysis finds values of $R_{\delta c_i} \simeq 0.4,\,0.4$ and $0.6$ at the end of the HL-LHC for the coefficients $c_{\varphi e}$, $c_{c\varphi}$,
 and  $c_{\varphi B}$
respectively, which subsequently go down to 
$R_{\delta c_i} \simeq 0.005,\, 0.01$ and $0.008$ upon the inclusion of the FCC-ee pseudo-data.
This translates into relative improvements by factors of around $80$, $40$, and $70$ for each EFT coefficient, respectively. 
While these are only representative examples, they highlight how precision measurements at FCC-ee will provide stringent constraints on the SMEFT parameter space, markedly improving on the limits achievable at the HL-LHC.

In Figs.~\ref{fig:spider_fcc_nlo_lin_glob} and~\ref{fig:spider_fcc_nlo_quad_glob}, the impact of the FCC-ee measurements on the EFT coefficients is presented in terms of the relative improvement with respect to either the {\sc\small SMEFiT3.0} baseline or its variant accounting also for the HL-LHC projections.
In order to compare with previous related studies, it is useful to also provide the absolute magnitude of the resulting bounds, in addition to their relative improvement as compared to the baseline.
With this motivation, Table~\ref{tab:fcc_all_bounds} indicates the 95\% CI upper bounds on the EFT coefficients obtained from the fits including both the HL-LHC and the FCC-ee projections, assuming $\Lambda=1$ TeV.
We provide these bounds for linear and quadratic EFT fits and both at the individual and marginalised level.
The impact of the FCC-ee is clearly visible specially for the purely bosonic and
two-fermion operators, with most of them constrained to be $|c|\lsim 0.1$ (for $\Lambda=1$ TeV) in the global marginalised fit, and several of them at the $|c|\lsim 10^{-2}$ level or better.
We demonstrate in Sect.~\ref{subsec:uvmodels} how these constraints on the Wilson coefficients translate into the mass reach for UV-complete models.

\begin{figure}[htbp]
    \centering
    \includegraphics[width=.8\linewidth ]{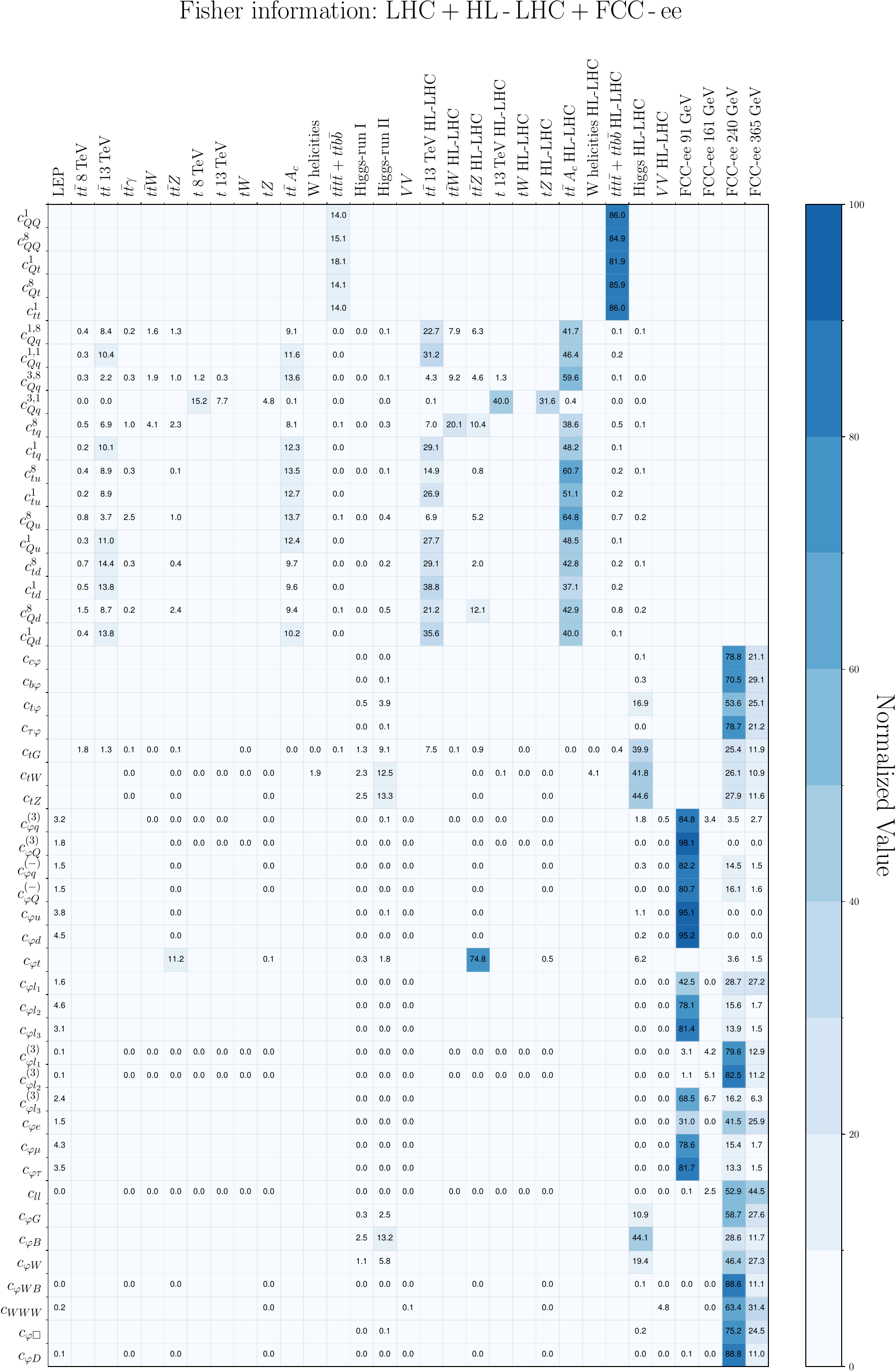}
    \caption{Diagonal entries of the Fisher information matrix evaluated at $\mathcal{O}\lp \Lambda^{-2}\rp$ in the EFT expansion, Eq.~(\ref{eq:fisher_smeft_lin}), for the complete dataset (LEP EWPOs + LHC$_{\rm Run1}$ + LHC$_{\rm Run2}$ + HL-LHC + FCC-ee) considered in this work.
    Each row is normalised to 100. 
    The LHC$_{\rm Run2}$ and HL-LHC datasets are displayed separately.
    For the FCC-ee observables, we evaluate the Fisher matrix individually for each of the relevant $\sqrt{s}$ values. 
    Empty entries indicate lack of direct sensitivity, while ``0.0'' entries indicate a non-zero sensitivity which is less than 0.05\% in magnitude.
    }
    \label{fig:fisher_fcc_lin}
\end{figure}
     

\begin{landscape}
\begin{table}[htbp]
 \centering
  \scriptsize
   \renewcommand{\arraystretch}{1.24}
   \rotatebox{-90}{
   \begin{tabular}{l|C{0.8cm}|C{1.85cm}|C{1.85cm}|C{1.85cm}|C{1.85cm}}
     \multirow{2}{*}{Class}   &  \multirow{2}{*}{DoF}
     &  \multicolumn{2}{c|}{ 95\% CI bounds, $\mathcal{O}\lp \Lambda^{-2}\rp$} &
     \multicolumn{2}{c}{95\% CI bounds, $\mathcal{O}\lp \Lambda^{-4}\rp$,} \\ 
 &  & Individual & Marginalised &  Individual & Marginalised  \\ \toprule
 \multirow{23}{*}{2FB}
 & $c_{c \varphi}$& 2.82$\cdot 10^{-3}$& 3.69$\cdot 10^{-3}$& 2.79$\cdot 10^{-3}$& 3.68$\cdot 10^{-3}$ \\ \cline{2-6}
 & $c_{b \varphi}$& 1.9$\cdot 10^{-3}$& 3.61$\cdot 10^{-3}$& 1.87$\cdot 10^{-3}$& 3.4$\cdot 10^{-3}$ \\ \cline{2-6}
 & $c_{t \varphi}$& 2.96$\cdot 10^{-1}$& 2.18& 2.88$\cdot 10^{-1}$& 2.19 \\ \cline{2-6}
 & $c_{\tau \varphi}$& 1.92$\cdot 10^{-3}$& 2.85$\cdot 10^{-3}$& 1.93$\cdot 10^{-3}$& 2.79$\cdot 10^{-3}$ \\ \cline{2-6}
 & $c_{tG}$& 2.38$\cdot 10^{-2}$& 6.89$\cdot 10^{-2}$& 2.45$\cdot 10^{-2}$& 6.52$\cdot 10^{-2}$ \\ \cline{2-6}
 & $c_{tW}$& 3.83$\cdot 10^{-3}$& 5.82$\cdot 10^{-2}$& 3.74$\cdot 10^{-3}$& 6.02$\cdot 10^{-2}$ \\ \cline{2-6}
 & $c_{tZ}$& 4.79$\cdot 10^{-3}$& 6.92$\cdot 10^{-2}$& 4.7$\cdot 10^{-3}$& 7.19$\cdot 10^{-2}$ \\ \cline{2-6}
 & $c_{\varphi q}^{(3)}$& 4.9$\cdot 10^{-3}$& 1.39$\cdot 10^{-2}$& 4.84$\cdot 10^{-3}$& 1.29$\cdot 10^{-2}$ \\ \cline{2-6}
 & $c_{\varphi Q}^{(3)}$& 5.28$\cdot 10^{-3}$& 3.5$\cdot 10^{-2}$& 5.32$\cdot 10^{-3}$& 1.92$\cdot 10^{-2}$ \\ \cline{2-6}
 & $c_{\varphi q}^{(-)}$& 2.89$\cdot 10^{-2}$& 6.03$\cdot 10^{-2}$& 2.88$\cdot 10^{-2}$& 4.06$\cdot 10^{-2}$ \\ \cline{2-6}
 & $c_{\varphi Q}^{(-)}$& 9.37$\cdot 10^{-3}$& 2.46$\cdot 10^{-2}$& 9.69$\cdot 10^{-3}$& 2.44$\cdot 10^{-2}$ \\ \cline{2-6}
 & $c_{\varphi u}$& 2.99$\cdot 10^{-2}$& 1.54$\cdot 10^{-1}$& 2.99$\cdot 10^{-2}$& 8.18$\cdot 10^{-2}$ \\ \cline{2-6}
 & $c_{\varphi d}$& 4.62$\cdot 10^{-2}$& 3.74$\cdot 10^{-1}$& 4.47$\cdot 10^{-2}$& 1.65$\cdot 10^{-1}$ \\ \cline{2-6}
 & $c_{\varphi t}$& 3.72$\cdot 10^{-1}$& 6.65$\cdot 10^{-1}$& 3.76$\cdot 10^{-1}$& 6.55$\cdot 10^{-1}$ \\ \cline{2-6}
 & $c_{\varphi l_1}$& 2.38$\cdot 10^{-3}$& 4.59$\cdot 10^{-3}$& 2.39$\cdot 10^{-3}$& 4.38$\cdot 10^{-3}$ \\ \cline{2-6}
 & $c_{\varphi l_2}$& 1.03$\cdot 10^{-2}$& 2.55$\cdot 10^{-2}$& 1.01$\cdot 10^{-2}$& 2.56$\cdot 10^{-2}$ \\ \cline{2-6}
 & $c_{\varphi l_3}$& 1.05$\cdot 10^{-2}$& 2.79$\cdot 10^{-2}$& 1.06$\cdot 10^{-2}$& 2.8$\cdot 10^{-2}$ \\ \cline{2-6}
 & $c_{\varphi l_1}^{(3)}$& 8.97$\cdot 10^{-4}$& 7.45$\cdot 10^{-3}$& 8.93$\cdot 10^{-4}$& 7.06$\cdot 10^{-3}$ \\ \cline{2-6}
 & $c_{\varphi l_2}^{(3)}$& 9.79$\cdot 10^{-4}$& 8.1$\cdot 10^{-3}$& 9.73$\cdot 10^{-4}$& 7.8$\cdot 10^{-3}$ \\ \cline{2-6}
 & $c_{\varphi l_3}^{(3)}$& 8.59$\cdot 10^{-3}$& 1.97$\cdot 10^{-2}$& 8.72$\cdot 10^{-3}$& 1.86$\cdot 10^{-2}$ \\ \cline{2-6}
 & $c_{\varphi e}$& 2.52$\cdot 10^{-3}$& 4.62$\cdot 10^{-3}$& 2.54$\cdot 10^{-3}$& 4.62$\cdot 10^{-3}$ \\ \cline{2-6}
 & $c_{\varphi \mu}$& 1.18$\cdot 10^{-2}$& 3.0$\cdot 10^{-2}$& 1.24$\cdot 10^{-2}$& 3.02$\cdot 10^{-2}$ \\ \cline{2-6}
 & $c_{\varphi \tau}$& 1.28$\cdot 10^{-2}$& 3.01$\cdot 10^{-2}$& 1.29$\cdot 10^{-2}$& 2.95$\cdot 10^{-2}$ \\ \cline{2-6}
\hline
\multirow{14}{*}{2L2H}
 & $c_{Qq}^{1,8}$& 2.59$\cdot 10^{-1}$& 2.34& 3.06$\cdot 10^{-1}$& 4.69$\cdot 10^{-1}$ \\ \cline{2-6}
 & $c_{Qq}^{1,1}$& 5.81$\cdot 10^{-1}$& 1.16$\cdot 10^{1}$& 3.64$\cdot 10^{-1}$& 2.82$\cdot 10^{-1}$ \\ \cline{2-6}
 & $c_{Qq}^{3,8}$& 7.95$\cdot 10^{-1}$& 3.96& 6.56$\cdot 10^{-1}$& 5.34$\cdot 10^{-1}$ \\ \cline{2-6}
 & $c_{Qq}^{3,1}$& 1.27$\cdot 10^{-1}$& 1.27$\cdot 10^{-1}$& 1.3$\cdot 10^{-1}$& 1.14$\cdot 10^{-1}$ \\ \cline{2-6}
 & $c_{tq}^{8}$& 4.12$\cdot 10^{-1}$& 2.65& 4.55$\cdot 10^{-1}$& 5.14$\cdot 10^{-1}$ \\ \cline{2-6}
 & $c_{tq}^{1}$& 4.83$\cdot 10^{-1}$& 6.87& 2.83$\cdot 10^{-1}$& 2.19$\cdot 10^{-1}$ \\ \cline{2-6}
 & $c_{tu}^{8}$& 4.38$\cdot 10^{-1}$& 6.97& 5.83$\cdot 10^{-1}$& 7.19$\cdot 10^{-1}$ \\ \cline{2-6}
 & $c_{tu}^{1}$& 8.81$\cdot 10^{-1}$& 1.89$\cdot 10^{1}$& 4.57$\cdot 10^{-1}$& 3.57$\cdot 10^{-1}$ \\ \cline{2-6}
 & $c_{Qu}^{8}$& 7.51$\cdot 10^{-1}$& 6.81& 7.44$\cdot 10^{-1}$& 8.21$\cdot 10^{-1}$ \\ \cline{2-6}
 & $c_{Qu}^{1}$& 6.8$\cdot 10^{-1}$& 8.65& 3.5$\cdot 10^{-1}$& 2.68$\cdot 10^{-1}$ \\ \cline{2-6}
 & $c_{td}^{8}$& 9.38$\cdot 10^{-1}$& 1.32$\cdot 10^{1}$& 1.35& 1.03 \\ \cline{2-6}
 & $c_{td}^{1}$& 1.79& 2.94$\cdot 10^{1}$& 6.2$\cdot 10^{-1}$& 4.61$\cdot 10^{-1}$ \\ \cline{2-6}
 & $c_{Qd}^{8}$& 1.43& 9.3& 1.32& 1.27 \\ \cline{2-6}
 & $c_{Qd}^{1}$& 1.49& 2.32$\cdot 10^{1}$& 5.0$\cdot 10^{-1}$& 3.96$\cdot 10^{-1}$ \\ \cline{2-6}
\hline
\multirow{5}{*}{4H}
 & $c_{QQ}^{1}$& 7.71 & \textemdash & 1.94& 6.75 \\ \cline{2-6}
 & $c_{QQ}^{8}$& 2.15$\cdot 10^{1}$ & \textemdash & 5.82& 2.02$\cdot 10^{1}$ \\ \cline{2-6}
 & $c_{Qt}^{1}$& 2.95$\cdot 10^{2}$ & \textemdash & 1.65& 1.36 \\ \cline{2-6}
 & $c_{Qt}^{8}$& 6.74 & \textemdash & 3.44& 2.75 \\ \cline{2-6}
 & $c_{tt}^{1}$& 3.77 & \textemdash & 9.62$\cdot 10^{-1}$& 8.01$\cdot 10^{-1}$ \\ \cline{2-6}
\hline
\multirow{1}{*}{4l}
 & $c_{ll}$& 6.98$\cdot 10^{-4}$& 8.05$\cdot 10^{-4}$& 6.99$\cdot 10^{-4}$& 8.06$\cdot 10^{-4}$ \\ \cline{2-6}
\hline
\multirow{7}{*}{B}
 & $c_{\varphi G}$& 9.83$\cdot 10^{-4}$& 6.22$\cdot 10^{-3}$& 9.63$\cdot 10^{-4}$& 6.34$\cdot 10^{-3}$ \\ \cline{2-6}
 & $c_{\varphi B}$& 1.97$\cdot 10^{-3}$& 6.65$\cdot 10^{-3}$& 1.97$\cdot 10^{-3}$& 6.67$\cdot 10^{-3}$ \\ \cline{2-6}
 & $c_{\varphi W}$& 5.13$\cdot 10^{-3}$& 2.27$\cdot 10^{-2}$& 5.1$\cdot 10^{-3}$& 1.98$\cdot 10^{-2}$ \\ \cline{2-6}
 & $c_{\varphi WB}$& 2.77$\cdot 10^{-4}$& 1.33$\cdot 10^{-2}$& 2.73$\cdot 10^{-4}$& 1.32$\cdot 10^{-2}$ \\ \cline{2-6}
 & $c_{WWW}$& 7.17$\cdot 10^{-3}$& 7.83$\cdot 10^{-3}$& 7.03$\cdot 10^{-3}$& 7.88$\cdot 10^{-3}$ \\ \cline{2-6}
 & $c_{\varphi \Box}$& 3.91$\cdot 10^{-2}$& 7.0$\cdot 10^{-2}$& 4.0$\cdot 10^{-2}$& 6.72$\cdot 10^{-2}$ \\ \cline{2-6}
 & $c_{\varphi D}$& 6.1$\cdot 10^{-4}$& 2.45$\cdot 10^{-2}$& 6.05$\cdot 10^{-4}$& 2.46$\cdot 10^{-2}$ \\ \cline{2-6}
\hline
\end{tabular}
   }
\caption{\small The 95\% CI upper bounds on the
     EFT coefficients obtained from
     the fits including both the HL-LHC and the FCC-ee (Level-1) pseudo-data,
     see also Figs.~\ref{fig:spider_fcc_nlo_lin_glob} and~\ref{fig:spider_fcc_nlo_quad_glob} for the corresponding graphical representation (relative to the baseline bounds), for  $\Lambda=1$ TeV.
     We present results both for linear and quadratic EFT fits and at the individual and marginalised level.
     }
\label{tab:fcc_all_bounds}
\end{table}
\end{landscape}

\myparagraph{Fisher information analysis}
In order to better scrutinise the interplay between the constraints provided by the (HL-)LHC measurements, on the one hand, and by the FCC-ee ones, on the other hand, it is illustrative to evaluate the Fisher information matrix~\cite{Ethier:2021bye,Ellis:2020unq}, Eq.~\eqref{eq:fisher_smeft_lin}.
%
%
Fig.~\ref{fig:fisher_fcc_lin} displays the diagonal entries of the linear Fisher
information matrix computed over the  complete dataset considered in this work: LEP EWPOs, LHC$_{\rm Run1}$, LHC$_{\rm Run2}$, HL-LHC, and the FCC-ee.
In this plot each row is normalised to 100. 
Note that the LHC Run II datasets and the HL-LHC projections are displayed separately.
For the FCC-ee projections, we evaluate the Fisher matrix individually for each of the relevant $\sqrt{s}$ values.

From the entries of the Fisher information matrix it can be observed how for all LHC processes being considered, the HL-LHC projections display the largest Fisher information, reflecting the expected reduction of statistical and systematic uncertainties. 
The table also confirms that the LEP EWPOs carry a small amount of information once the FCC-ee projections are included in the fit.
Within the global EFT fit, the FCC-ee observables would provide the dominant constraints for all the two-fermion and purely bosonic operators, except for $c_{tG}$, $c_{tW}$, $c_{tZ}$, $c_{\varphi B}$, and $c_{\varphi t}$, where the constraints from HL-LHC processes are still expected to dominate. 
The two-light-two-heavy operators are entirely dominated by HL-LHC, mostly from inclusive $t\bar{t}$ production and by the charge asymmetry measurements $A_C$.

Concerning the impact of the FCC-ee datasets for different center-of-mass energies, Fig.~\ref{fig:fisher_fcc_lin} reveals that the bulk of the constraints on the SMEFT parameter space should be provided by the runs at $\sqrt{s}=91$ GeV and at $\sqrt{s}=240$ GeV, with relevant information arising also from the  $\sqrt{s}=365$ GeV run.
Fig.~\ref{fig:fisher_fcc_lin} highlights the interplay between runs at different values of $\sqrt{s}$ to constrain new physics through the SMEFT interpretation: when their information is combined, the resulting picture is sharper than that of any individual run. 
On the other hand, only moderate information would be provided by the run at the $WW$ threshold ($\sqrt{s}=161$ GeV), and even for the triple gauge operator coefficient $c_{WWW}$ it would be the $Zh$ run that dominates the sensitivity.
Hence, from the viewpoint of EFT analyses, the $\sqrt{s}=161$ GeV run appears to be the less impactful.

The Fisher information matrix shown in Fig.~\ref{fig:fisher_fcc_lin} quantifies the relative sensitivity of various processes to specific operators based on their linear EFT contributions, and corresponds to what one would expect to find in linear individual fits.
However, it cannot capture the complete picture encompassed by the global fit at the marginalised level, and in particular it does not account for the information on the correlations between operators.
Indeed, in the marginalised fits,  these correlations will often modify the picture as compared with expectations based on the Fisher information matrix. 
We also note that the Fisher information matrix, Eq.~(\ref{eq:fisher_smeft_lin}), receives additional contributions when evaluated for quadratic EFT fits.

\myparagraph{Impact of quadratic EFT corrections}
The analog of Fig.~\ref{fig:spider_fcc_nlo_lin_glob} in the case of EFT fits with quadratic corrections is displayed in Fig.~\ref{fig:spider_fcc_nlo_quad_glob}.
Also here one observes a significant reduction of the bounds on the EFT coefficients upon inclusion of the FCC-ee observables.
The improvements observed for some four-quark operators in the linear fit of Fig.~\ref{fig:spider_fcc_nlo_lin_glob} mostly disappear when including quadratic corrections demonstrating that the improvement of the linear fit constraints arise from indirect  correlations with other degrees of freedom.
Indeed, the quadratic fit result illustrates how the FCC-ee observables do not have any direct sensitivity on the four-quark operators, both for the two-light-two-heavy and for the four-heavy ones.

From Table~\ref{tab:fcc_all_bounds} one notices that the differences between linear and quadratic bounds are in general reduced as compared to the case of the {\sc\small SMEFiT3.0} results collected in Tables~\ref{tab:smefit30_all_bounds_p1} and \ref{tab:smefit30_all_bounds_p2} from Chapter \ref{chap:smefit3}. 
This feature is explained by the improved precision of the FCC-ee measurements: since we assume the SM in the pseudo-data,  the best-fit values of the Wilson coefficients move closer to zero with smaller uncertainties, and hence the quadratic terms become less significant. 
We note however that for a subset of operators, such as the two-light-two-heavy ones, which are not constrained by the FCC-ee measurements the discrepancy between linear and quadratic remains large. 

\myparagraph{Disentangling the impact of datasets with fixed $\sqrt{s}$}
As indicated in Table~\ref{tab:FCCee_runs}, the FCC-ee plans to operate sequentially, collecting data at different center-of-mass energies, $\sqrt{s}$, starting at the $Z$-pole and then increasing the energy up to the $t\bar{t}$ threshold.
Plans to define different running scenarios are also being considered, for example directly starting as a Higgs factory with the $\sqrt{s}=240$ GeV run and only later running at the $Z$-pole energy.
It is therefore relevant to disentangle, at the level of the global SMEFT fit, the separate impact of datasets with a given $\sqrt{s}$ value to evaluate the advantages and disadvantages of the proposed running scenarios, see also the Fisher information matrix in Fig.~\ref{fig:fisher_fcc_lin}.

\begin{figure}[t]
    \centering
    \includegraphics[width=0.99\linewidth]{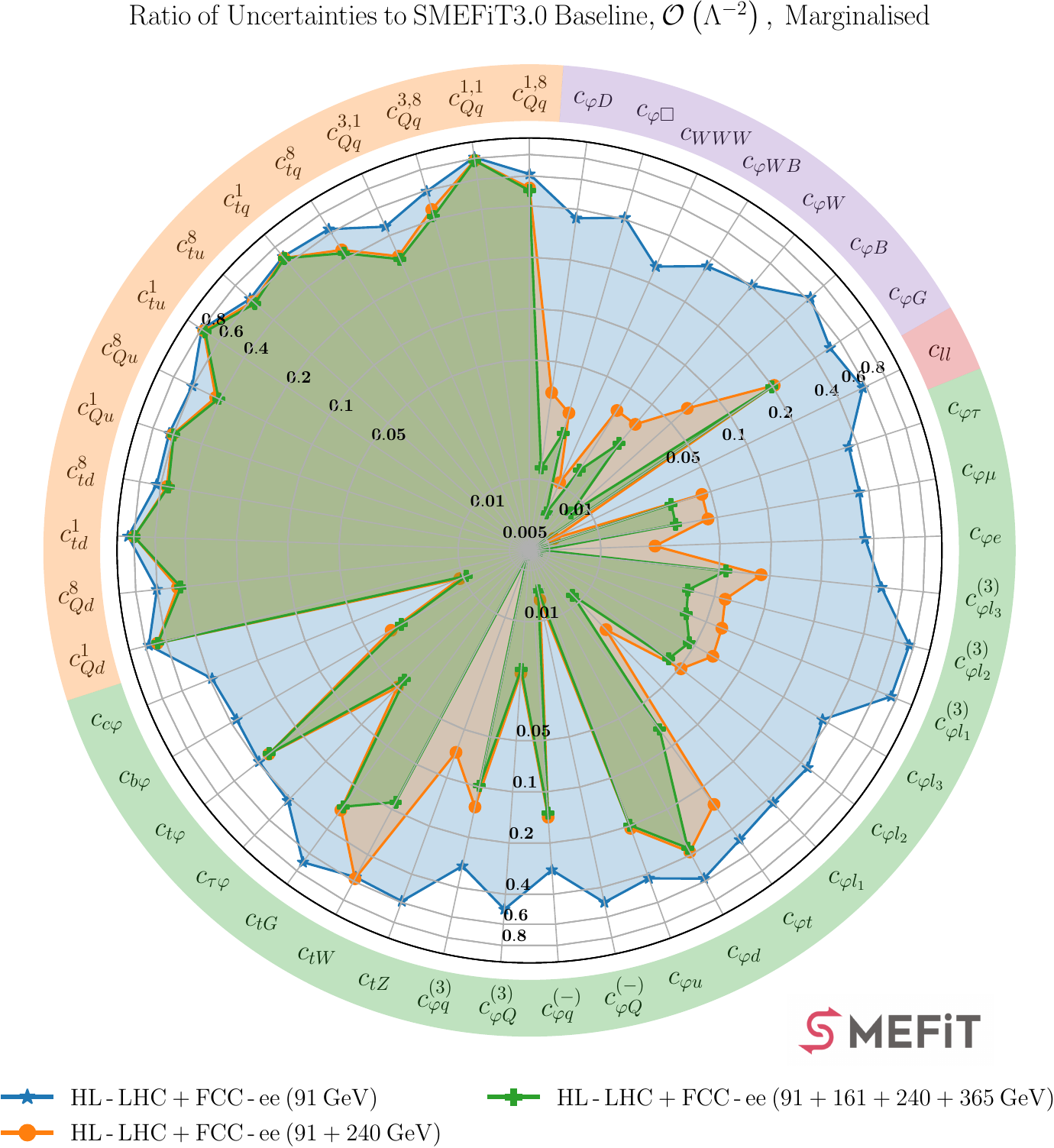}
    \caption{Same as Fig.~\ref{fig:spider_fcc_nlo_lin_glob}, now comparing the sequential impact of the separate $\sqrt{s}$ runs at the FCC-ee with respect to the baseline fit.
    We display the effects of adding the projected FCC-ee dataset at, first, $\sqrt{s}=91$ GeV (blue), followed by adding $\sqrt{s}=240$ GeV (orange) and finally adding both $\sqrt{s}=161$ and 365 GeV (green), which completes the FCC-ee dataset listed in Table~\ref{tab:FCCee_runs}.
    }
\label{fig:spider_fcc_energy_variations}
\end{figure}

Fig.~\ref{fig:spider_fcc_energy_variations} illustrates the sequential impact of the datasets collected at different values of $\sqrt{s}$ at the FCC-ee.
First we show the values of the ratio $R_{\delta c_i}$ when only the $Z$-pole EWPOs at $\sqrt{s}=91$ GeV are included in the fit, then when also the Higgs factory dataset from the $\sqrt{s}=240$ GeV is accounted for, and finally for the full FCC-ee dataset, which includes also the $WW$ run at 161 GeV and the $t\bar{t}$ run at  365 GeV. 
Fig.~\ref{fig:spider_fcc_energy_variations} indicates that the largest impact is obtained when the Higgs, diboson, and fermion-pair production data collected at 240 GeV are included in the fit together with the $Z$-pole run.
We also observe how the measurements from the $\sqrt{s}=161$ GeV and 365 GeV runs are necessary to achieve the ultimate constraining potential of the FCC-ee in the EFT parameter space, with several operators experiencing a marked improvement of the associated bounds. 
This breakdown demonstrates the interplay between the information provided by the FCC-ee runs at the various proposed center-of-mass energies.

\begin{figure}[t]
    \centering
    \includegraphics[width=0.99\linewidth]{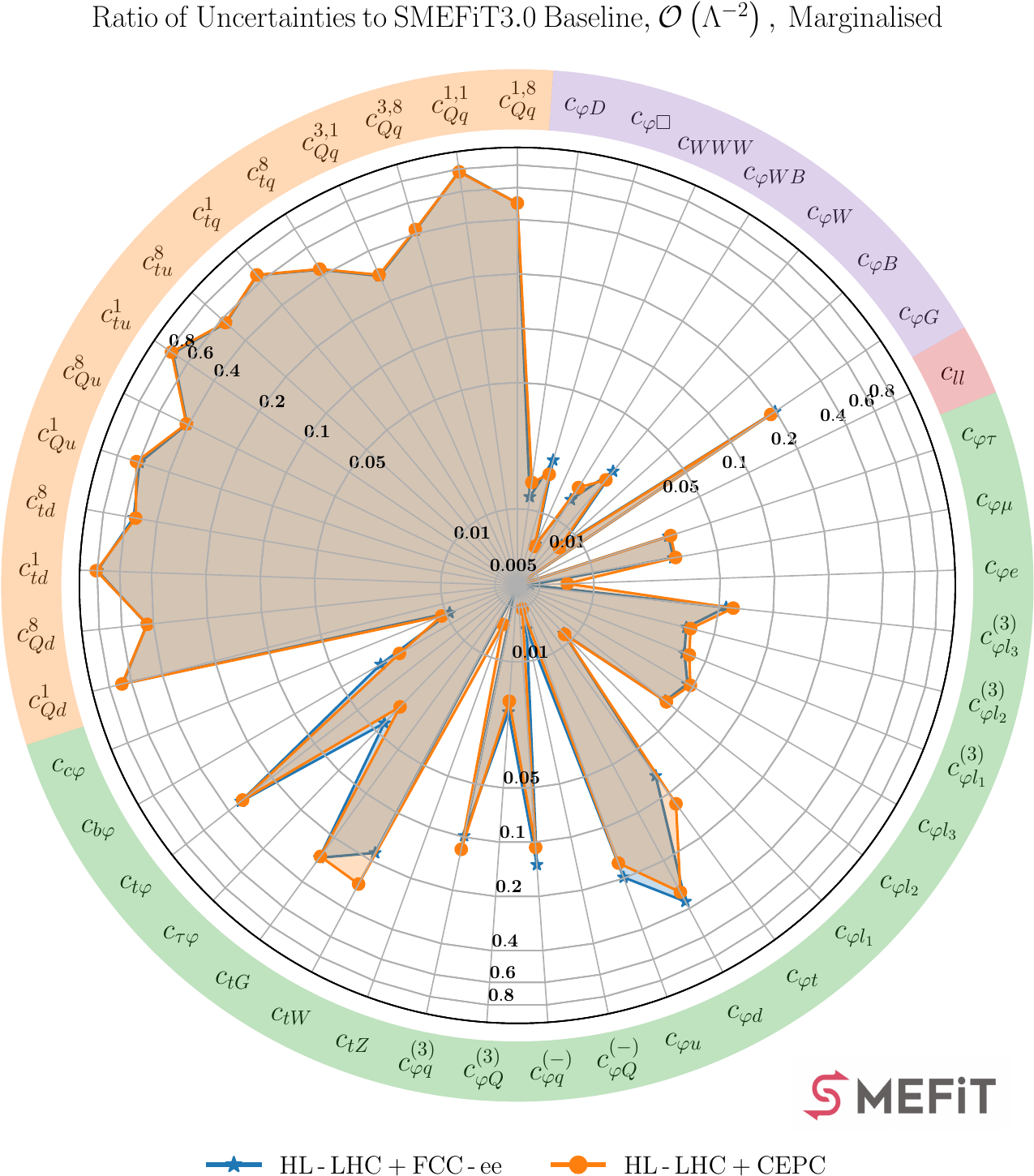}
    \caption{Same as Fig.~\ref{fig:spider_fcc_nlo_lin_glob} now
    comparing the impact of the FCC-ee and CEPC datasets.
    }
    \label{fig:spider_cepc_nlo_lin_glob}
\end{figure}

\myparagraph{FCC-ee impact compared to the CEPC}
We next study the impact of the CEPC measurements on the SMEFT coefficients, relative to that obtained in the FCC-ee  case and shown in Fig.~\ref{fig:spider_fcc_nlo_lin_glob}.
Fig.~\ref{fig:spider_cepc_nlo_lin_glob} displays the constraints provided by both colliders when added on top of the same HL-LHC baseline dataset, always relative to the {\sc\small SMEFiT3.0} baseline fit.
In general, a similar constraining power is obtained, consistent with the lack of major differences between their running plans (see Table~\ref{tab:FCCee_runs}), though the FCC-ee bounds tend to be better than those from CEPC.
The same qualitative conclusions are obtained in the case in which the analysis of Fig.~\ref{fig:spider_cepc_nlo_lin_glob}
is carried with quadratic EFT corrections.

When performing this comparison, we noticed that the total experimental uncertainties provided by the FCC-ee and CEPC collaborations for the Snowmass study, which we adopt in our analysis, differ by more than the scaling of integrated luminosities, as
would be expected in the case of purely statistical uncertainties without further correction factors.
This is illustrated by the Fisher information matrix, defined in Eq.~(\ref{eq:fisher_smeft_lin}) at the linear EFT level, when evaluated in terms of ratios between the two experiments.
If one takes the ratio of the diagonal entries of the  Fisher information matrix between the FCC-ee and the CEPC, given that both colliders share the same theory predictions, one 
obtains
 \begin{equation}
\label{eq:fisherinformation2_ratio_2}
I_{ii}^{\rm (FCC-ee)}/I_{ii}^{\rm (CEPC)} =  \sum_{m=1}^{n_{\rm dat}} \lp \delta_{{\rm exp},m}^2\rp_{\rm CEPC} \Bigg/\sum_{m'=1}^{n_{\rm dat}} \lp \delta_{{\rm exp},m'}^2\rp_{\rm FCC-ee}  \, , \qquad i=1,\ldots,n_{\rm eft},
\end{equation}
namely the ratio of total experimental uncertainties squared.
For observables with only  statistical uncertainties, and assuming that eventual acceptance corrections cancel out, this ratio should reproduce  the corresponding ratios of integrated luminosities reported in Table~\ref{tab:FCCee_runs}, that is,
 \begin{equation}
\label{eq:fisherinformation2_ratio_statistical}
I_{ii}^{\rm (FCC-ee)}/I_{ii}^{\rm (CEPC)}\simeq \lp \mathcal{L}_{\rm FCC-ee}/ \mathcal{L}_{\rm CEPC} \rp \, .
 \end{equation}
 
\begin{figure}[t]
    \centering
    \includegraphics[width=\linewidth]{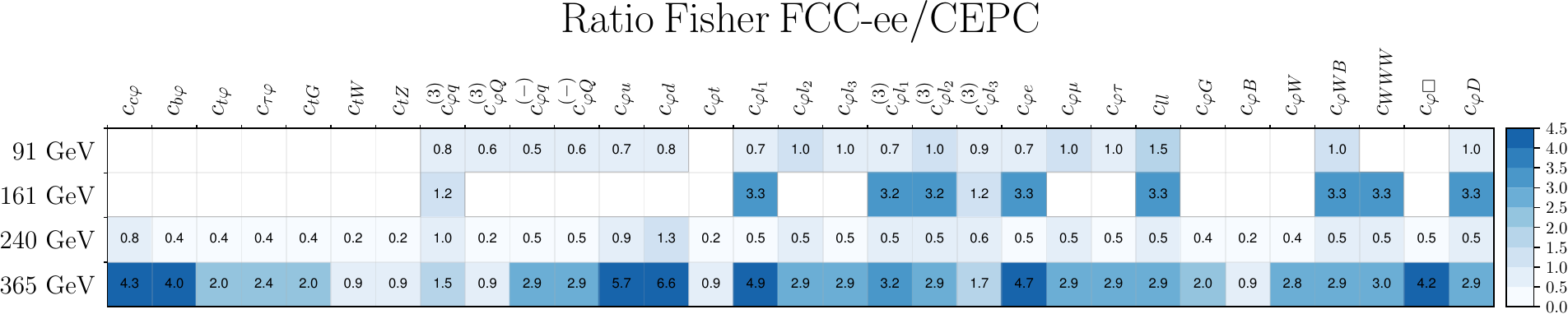}
    \caption{The ratio of the entries of the Fisher information matrix between the FCC-ee and the CEPC, Eq.~(\ref{eq:fisherinformation2_ratio_2}),
    evaluated separately for the observables entering the four centre of mass energies $\sqrt{s}$ considered.
    Since projections for both colliders share theory predictions, this ratio is equivalent to the ratio of total experimental uncertainties squared.
    In turn, if the latter contains only the statistical uncertainties, the entries of the table should match the corresponding ratios of integrated luminosities from Table~\ref{tab:FCCee_runs}.
    }
    \label{fig:fisher_cepc_fccee_energies}
\end{figure}

Fig.~\ref{fig:fisher_cepc_fccee_energies} displays the ratio defined in Eq.~(\ref{eq:fisherinformation2_ratio_2}),  evaluated separately for the observables entering the four data-taking periods considered.
Values of Eq.~(\ref{eq:fisherinformation2_ratio_2}) below unity indicate EFT coefficients for which the CEPC observables should be more constraining than the FCC-ee ones. 
A deviation from the luminosity scaling for the $\sqrt{s}=91$ GeV run is expected, and indeed observed, since each collaboration makes different assumptions for their systematic uncertainties. 
On the other hand, since at $\sqrt{s}= 161, 240$ and 365 GeV the uncertainties considered are purely statistical, for these observables the ratios in Fig.~\ref{fig:fisher_cepc_fccee_energies} are expected to follow Eq.~(\ref{eq:fisherinformation2_ratio_statistical}). 
While in some cases this in indeed true, in particular for the runs at $\sqrt{s}=161$ GeV and $240$ GeV, in other cases there are larger differences. 
Particularly noticeable are those arising in the data-taking period at $\sqrt{s}=365$ GeV. There one expects a ratio of around $3$ purely on the luminosity scaling, but actually one obtains a range of values between 0.9 and 6.6.

It is beyond the scope of this thesis to scrutinise the origin of these differences: they could be explained by different assumptions on the experimental selection procedure and acceptance cuts, for example.
Nevertheless, the analysis of Fig.~\ref{fig:fisher_cepc_fccee_energies} highlights that in general the relative impact in the SMEFT parameter space of the projected FCC-ee and CEPC pseudo-data differs from the expectations based on a pure luminosity scaling.

\section{Constraints on UV-complete models through the SMEFT}
\label{subsec:uvmodels}

We now quantify the constraints that LHC Run II measurements and future collider projections impose on the parameter space of representative UV-complete scenarios.
To this aim, we benefit from the integration of {\sc\small SMEFiT} with {\sc\small matchmakereft} \cite{Carmona:2021xtq} via the {\sc\small Match2Fit} interface presented in Chapter \ref{chap:smefit_uv}.
We consider results for the (indirect) mass reach for new heavy particles at the HL-LHC and FCC-ee obtained from the tree-level matching of a wide range of one-particle extensions of the SM.
We also present results for the reach in the UV couplings for the one-loop matching of a subset of the same one-particle extensions and for the tree-level matching of a multi-particle extension of the SM.
The corresponding results for the CEPC are qualitatively similar to the FCC-ee ones, consistently with Fig.~\ref{fig:spider_cepc_nlo_lin_glob}, and hence are not shown here.

\myparagraph{Tree-level matching of one-particle extensions}
First, we provide results for the UV-complete one-particle models considered in~\mycite{terHoeve:2023pvs}, each of them associated to a different gauge group representation of the new heavy particle, matched at tree-level to the SMEFT.
The one-particle models considered in this work are the same as those considered in Chapter \ref{chap:smefit_uv}, see Table~\ref{tab:UV_particles} for their representations under the SM gauge group.
We restrict the possible UV couplings to ensure consistency with the {\sc\small SMEFiT} flavour assumption after tree-level matching.

\begin{figure}[htbp]
    \centering
    \includegraphics[height=.8\textheight]{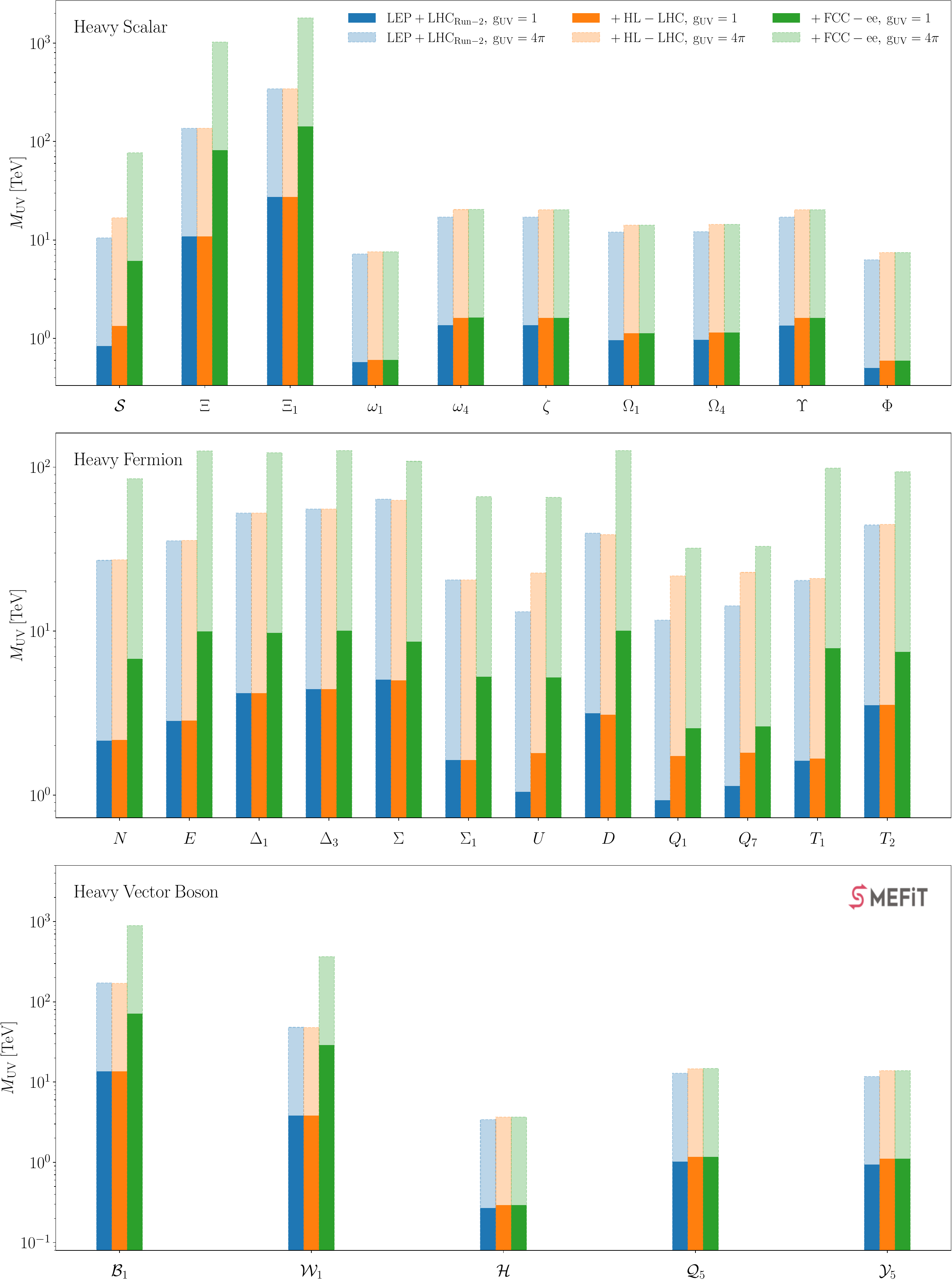}
    \caption{The 95\% CI lower bounds on the heavy particle mass $M_{\rm UV}$ for the one-particle UV-completions of the SM considered in this work, matched to the SMEFT using tree-level relations. 
    In all cases we include corrections up to quadratic order in the EFT expansion.
    From top to bottom, we display results for models with heavy scalars, heavy fermions, and heavy vector bosons, 
    see Table~\ref{tab:UV_particles} for
    the definition of each model.
    We present results for SMEFT analyses based on the current dataset (LEP+LHC$_{\rm RunII}$), then for its extension first with HL-LHC projections, and subsequently with the full set of FCC-ee observables.
    We consider two scenarios for the UV coupling constants, $g_{\rm UV}=1$ (darker) and $g_{\rm UV} = 4\pi$ (lighter). 
    Note the logarithmic scale of the $y$-axis.
    }
    \label{fig:uv-bounds}
\end{figure}

To illustrate the reach of the FCC-ee measurements on the parameter space of these UV models, we derive lower bounds on the mass of the heavy particle $M_{\rm UV}$ for  each of the one-particle extensions listed in Table~\ref{tab:UV_particles} by assuming a given value for the corresponding UV couplings $g_{\rm UV}$.
For simplicity, we consider only models with a single UV coupling.
The projected $95\%$ CI lower bounds on $M_{\rm UV}$ at FCC-ee are shown in Fig~\ref{fig:uv-bounds} for two limiting assumptions on the value of UV couplings: $g_{\rm UV}=1$ and $g_{\rm UV} = 4\pi$.
The chosen values are on the upper edge of what can be considered as weakly and strongly coupled, and in fact, $g_{\rm UV} > 4\pi$ would violate the perturbative limit.
For each model, we present results for a SMEFT analysis based on Level-1 pseudo-data of the current dataset, LEP+LHC$_{\rm Run\,II}$, and then for its extension first with HL-LHC projections, and subsequently with the complete set of FCC-ee observables.

Several observations can be drawn from perusing the results of Fig.~\ref{fig:uv-bounds}.
On the one hand, we find that the HL-LHC projections improve the mass reach for several models, in particular those that include heavy quark partners such as $U$, $Q_1$, and $Q_7$.
On the other hand, the models that are not affected by HL-LHC fall into two distinct categories.
One is composed of models such as $N$, $E$, $\Delta_{1,3}$, $\Sigma$, $\Sigma_1$, and $D$, that generate a subset of operators which display no improved sensitivity in individual fits to HL-LHC pseudo-data, namely purely leptonic and bosonic operators probed by EWPOs as well as $c_{\varphi Q}^{(-)}$ and $c_{\varphi Q}^{(3)}$.
The other class contains the models $\Xi$, $\Xi_1$, $T_2$, $\mathcal{B}_1$, and $\mathcal{W}_1$ that do generate operators that have associated improved bounds at the HL-LHC, but  where the sensitivity to the UV parameters is driven by operators that are instead insensitive to HL-LHC data, such as $c_{\varphi D}$, $c_{\varphi Q}^{(-)}$ or $c_{\varphi Q}^{(3)}$.

The mass reach enabled by FCC-ee measurements increases markedly for several models as compared to the post-HL-LHC limits, in some cases by up to an order of magnitude.
The largest effects are observed for the UV scenarios that modify the interactions of Higgs and electroweak bosons, which are tightly constrained by the FCC-ee data.
These include the $\mathcal{S}$, $\Xi$, and $\Xi_1$ heavy scalar models; the $N$, $E$, $\Delta_1$, $U$, $D$, $T_1$, and $T_2$ vector-like heavy fermion models; and the $\mathcal{B}_1$ and $\mathcal{W}_1$ heavy vector boson models.
For other scenarios, such as those primarily modifying the four-quark interaction vertices, there is only a marginal information gain provided by FCC-ee measurements, consistently with the findings at the Wilson coefficient level in Fig.~\ref{fig:spider_fcc_nlo_lin_glob}.

In terms of the heavy particle mass reach, we observe that FCC-ee measurements will be sensitive to BSM scales of up to around 100 TeV, 10 TeV, and 70 TeV for some of the studied heavy scalar, fermion, and vector boson UV-completions respectively, in the case of $g_{\rm UV}=1$.
This sensitivity increases to around $10^3$ TeV, 200 TeV, and 800 TeV in the case of strongly coupled one-particle extensions of the SM in the upper limit of the perturbative regime, $g_{\rm UV}=4\pi$. 
One also observes how, at least for the one-particle extensions considered here, at the HL-LHC there is not a large difference between the mass reach associated to direct production (with $m_X \sim 7$ TeV at most) and that associated to the indirect bounds obtained in the EFT framework.
On the other hand, at the FCC-ee the production of new heavy (TeV-scale) particles consistent with the LHC exclusion bounds is kinematically impossible due to the limited $\sqrt{s}$ values available, while the EFT bounds instead reach much higher energies, as illustrated by the examples of Fig.~\ref{fig:uv-bounds}.
This result further confirms the powerful sensitivity to heavy new physics enabled by the high-precision electroweak, Higgs, and top quark measurements to be performed at the FCC-ee highlighted by previous studies.

\FloatBarrier
\myparagraph{One-loop matched and multi-particle models}
Following the discussion on single-particle extensions of the SM matched at tree-level, we now evaluate the impact of the FCC-ee data on more general UV completions.
We consider in particular the heavy scalar $\phi$ and the heavy fermion $T_1$ and $T_2$ models, already analysed in Fig.~\ref{fig:uv-bounds}, now matched onto the SMEFT at the one-loop level.
This one-loop matching yields several additional contributions, generally flavour-independent, to bosonic and two-fermion operators as compared to tree-level matching.
One-loop matching contributions can lead to better constraints and, very importantly, allow to constrain otherwise blind directions in the UV parameter space~\mycite{terHoeve:2023pvs}.
In addition, we also provide results for a 3-particle model, matched at tree-level, composed by the heavy vector-like fermions $Q_1$, $Q_7$ and the heavy vector boson $\mathcal{W}$ (see Table~\ref{tab:UV_particles} for their quantum numbers).

\begin{figure}[t]
    \centering
    \includegraphics[width=0.99\linewidth] {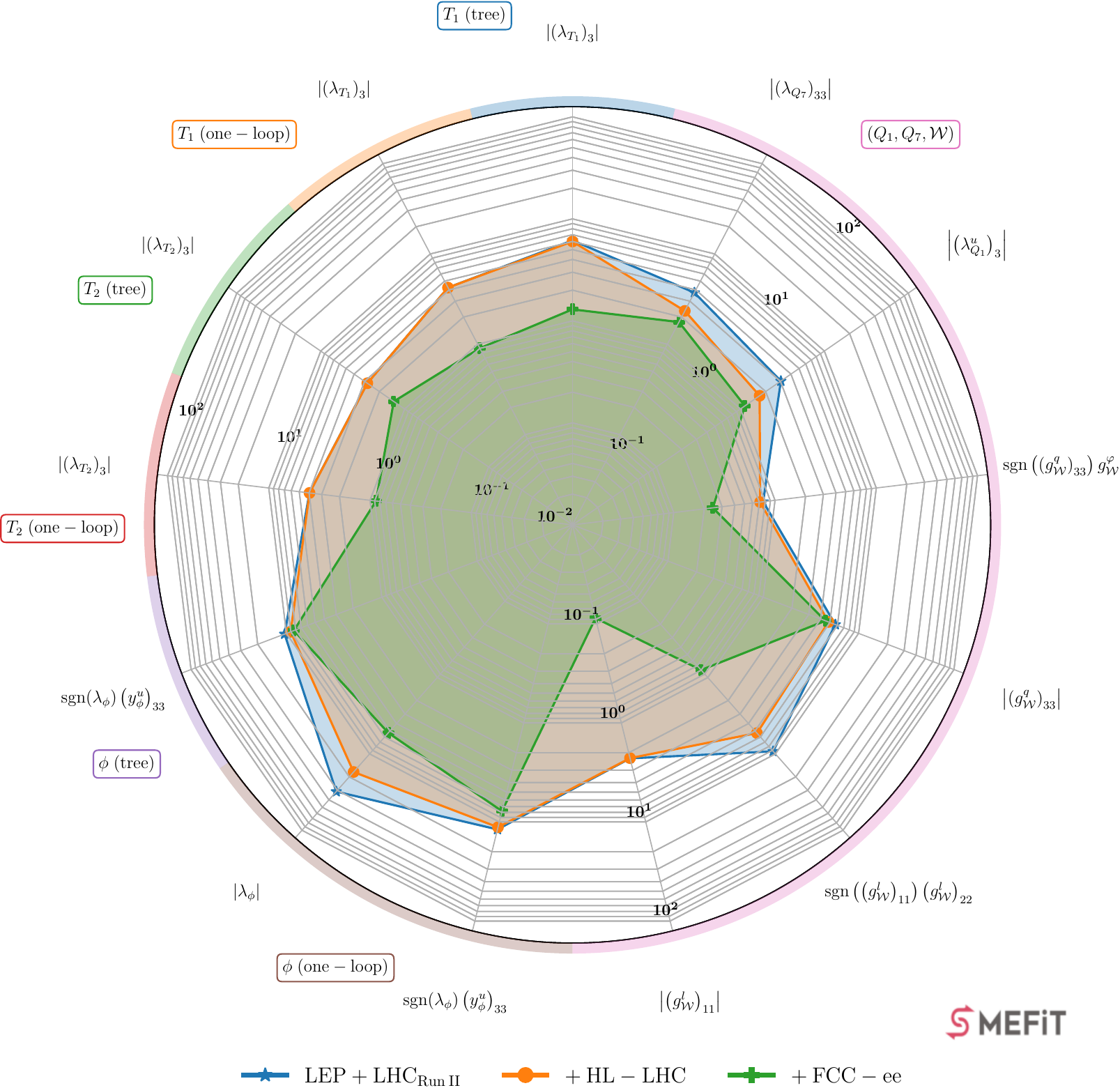}
    \caption{The 95\% CI upper bounds on the UV-invariant couplings of representative models obtained from the {\sc\small SMEFiT3.0} dataset (blue) and from its extension with the HL-LHC (orange) and with both the HL-LHC and FCC-ee (green) projections.
    We consider a 3-particle model, $\lp Q_1,Q_7,\mathcal{W}\rp$, matched at tree-level, and three one-particle models, $T_1$, $T_2$, and $\phi$, matched at one loop. 
    For the multi-particle model, we set the masses $m_{Q_1}= 3$~TeV, $m_{Q_7}= 4.5$~TeV, and $m_{\mathcal{W}}= 2.5$~TeV.
    The one-particle extensions matched at one-loop assume a heavy particle mass of $M_{T_1} = M_{T_2} = 10$~TeV and $M_{\phi} = 5$~TeV, representing the indirect mass reach to be probed at the FCC-ee,
    and for completeness, the associated results from tree-level matching are also displayed.
    }
\label{fig:spider_UVbounds_oneloop_matching}
\end{figure}

Fig.~\ref{fig:spider_UVbounds_oneloop_matching} displays the 95\% CI upper bounds on the UV-invariant combination of couplings~\mycite{terHoeve:2023pvs} of the considered models obtained from the {\sc\small SMEFiT3.0} dataset and from its extension with first the HL-LHC, and then the HL-LHC+FCC-ee projections.
For the multi-particle model matched at tree-level we assume masses of  $m_{Q_1}= 3$~TeV, $m_{Q_7}= 4.5$~TeV, and $m_{\mathcal{W}}= 2.5$~TeV, see also~\mycite{terHoeve:2023pvs}.
For the one-particle extensions matched at one-loop we assume $M_{T_1}= M_{T_2} = 10$~TeV and $M_{\phi} = 5$~TeV, which represent the typical mass reach being probed at the FCC-ee for those kinds of heavy particles, see Fig.~{\ref{fig:uv-bounds}}. 
For reference, the corresponding tree-level results with the same $M_{\rm UV}$ are also provided. 

Consistently with Fig.~{\ref{fig:uv-bounds}}, the sensitivity to the heavy fermions $T_1$ and $T_2$ is not improved at the HL-LHC due to being driven by the constraints from LEP data. Instead, the bounds on these models are significantly tightened after adding the FCC-ee projections. 
The inclusion of one-loop matching results does not alter this picture and has a small and generally positive impact on the bounds.
One-loop matching effects are more marked for the $\phi$ heavy scalar model, where it allows one to constrain the additional UV parameter $|\lambda_\phi|$ in all scenarios and with remarkable improvements both at HL-LHC and FCC-ee.
The bounds on $|\lambda_\phi|$ at LHC Run II and HL-LHC are, however, of limited use since they are beyond the perturbative limit, $|\lambda_\phi|<4 \pi$.
The power of precision measurements at FCC-ee is enough to bring down this bound to a strongly-coupled but perturbative regime, thus providing meaningful information on the UV model.
The bound on sgn$(\lambda_\phi)\left(y_{\phi}^{u}\right)_{33}$ is improved at future colliders for tree-level matching by around $15\%$ at each stage, though the logarithmic scale in the plot hides this improvement. 
The difference diminishes when considering one-loop matching results and in all cases the bound is within the perturbative limit.

Finally, the results of the multiparticle model display several interesting features. 
The UV couplings of the heavy fermions, $|(\lambda_{Q_7})_{33}|$ and $|(\lambda_{Q_1}^{u})_{3}| $, receive improved bounds at both HL-LHC and FCC-ee, as expected since they are related to the coefficients $c_{t\varphi}$ and $c_{\varphi t}$ that show a similar behaviour. 
The UV invariants from the couplings of the heavy spin-1 vector boson with leptons, $|(g_{\mathcal{W}}^{\ell})_{11}|$ and $\text{sgn}((g_{\mathcal{W}}^{\ell})_{11})(g_{\mathcal{W}}^{\ell})_{22}$, see little to no improvement on their bounds from adding HL-LHC projections, but the bounds become at least one order of magnitude tighter once FCC-ee data is considered. 
A similar situation is found for $\text{sgn}((g_{\mathcal{W}}^{q})_{33})g_{\mathcal{W}} ^ {\varphi}$, whose bound is driven mostly by its contribution to $c_{\varphi \square}$ and $c_{\varphi \ell_1}^{(3)}$ but also affected by its appearance in $c_{\varphi Q}^{(-)}$, $c_{\varphi Q}^{(3)}$ and $c_{b \varphi}$. 
The UV coupling $g_{\mathcal{W}}^{\varphi}$ also contributes at tree-level to $c_{t\varphi}$, which connects the fermionic and bosonic sectors in this model.
Such interplay worsens the bound on $\text{sgn} (( g_{\mathcal{W}}^{q} )_{33} )g_{\mathcal{W}} ^ {\varphi}$ by $\sim 30\%$, which highlights the importance of considering models more complex than the one-particle extensions.
The bound on $|(g_{\mathcal{W}}^{q})_{33}|$ is obtained via the Wilson coefficients of four-heavy quark operators such as $c_{QQ}^{1}$ and $c_{QQ}^{8}$. 
Hence, it improves by around $20\%$ when going from LHC to HL-LHC but is not improved further at future lepton colliders.

\section{Summary and outlook}
\label{sec:summary}

In this chapter, we have quantified the impact that measurements at future colliders, first the HL-LHC and then the FCC-ee and the CEPC, would have on the parameter space of the SMEFT in the framework of a global interpretation of particle physics data. These results are enabled by exploiting newly developed functionalities of the {\sc\small SMEFiT} framework, in particular the availability of a projection module which can extrapolate from existing measurements and project them to the settings relevant for other (future) experiments.
This module also makes it possible to carry out global SMEFT analyses based on pseudo-data generated with an arbitrary underlying law.

 We find that the marginalised bounds on the EFT coefficients within the global fit are projected to improve between around 20\% and a factor 3 by the end of the HL-LHC, depending on the operator, with possible further improvements being enabled by optimised analyses not considered here.
 Subsequently, the constraints from the FCC-ee or the CEPC would markedly improve the bounds on most of the purely bosonic and two-fermion operators entering the fit, by up to two orders of magnitude in some cases.
 We have then determined, using the UV matching procedure, how this impact at the EFT coefficient level translates into the parameter space of representative one-particle and multi-particle extensions of the SM matched to the SMEFT.
We find that the stringent constraints on the interactions of the Higgs, $W$ and $Z$ bosons, and top quarks made possible by future $e^+e^-$ circular colliders lead to an indirect mass reach on heavy new particles ranging between a few TeV up to around 100 TeV (for UV couplings $g_{\rm UV}\simeq 1$), depending on the specific UV scenario.
It is hence clear that the precision FCC-ee/CEPC measurements would make possible an extensive indirect exploration of the landscape of UV models containing new heavy particles beyond the direct reach of HL-LHC.

This work could be extended in several directions. 
To begin with, we could consider projections for other proposed future colliders, from the ILC and CLIC to the muon collider at different centre-of-mass energies, to assess what are their strengths and weaknesses as compared to the FCC-ee and the CEPC in the context of a global SMEFT analysis and its matching to UV models.
Second, it would be interesting to further explore whether the EFT impact of Higgs and fermion-pair production at leptonic colliders can be enhanced by including differential information, on the same footing as how it is done for the optimal observables for diboson and top-quark pair production.
Third, one may want to include optimised HL-LHC projections, fully exploiting the increase in statistics in ways which cannot be extrapolated from available Run II measurements, such as extended range in differential distributions or finer binning.
We expect these to improve the constraints set by HL-LHC on both the SMEFT parameter space and that of the UV-complete models. 

Fourth, by extending the analysis of the UV-complete scenarios studied here to other models, both at tree-level and via one-loop matching, and in particular considering more general multi-particle models, one could further scrutinise the indirect constraints that quantum corrections on FCC-ee observables impose on the masses and couplings of new heavy particles beyond the reach of the HL-LHC.
Fifth, albeit on a longer timescale, one may want to generalise the flavour assumptions on the EFT operators entering the global fit baseline and include data from other processes such as Drell-Yan and $B$-meson decays, and subsequently verify the robustness of the obtained projections for future colliders.
In this context, we note that the FCC-ee will also function as a flavour factory, with huge statistics thanks to the large $B$-mesons sample produced in the $Z$-pole run. 

As the global particle physics community moves closer to identifying the next large scientific infrastructure projects that will inform the field for the coming decades, the availability of methodologies quantifying the physics reach of different colliders represents an essential tool to make informed decisions.
The results of this chapter demonstrate that the {\sc\small SMEFiT} open-source framework is suitable to contribute to this endeavour.
Within this framework, progress in the global EFT interpretation of the most updated measurements can be kept synchronised with projections for future colliders, in a way that the baseline always represents the state-of-the-art in terms of experimental constraints. 
For these reasons, we expect that projections based on {\sc\small SMEFiT} will provide a valuable contribution to the upcoming ESPPU and related community studies to take place in the coming few years.

\chapter{Conclusions and future vision}
\label{chap:conclusions}
In this thesis, we performed an extensive quantitative analysis of the theory space beyond the SM in the context of the SMEFT, covering a wide range of particle physics processes both at current and future colliders. We developed an automatised bound setting procedure on explicit SM extensions, and constructed unbinned multivariate observables based on Machine Learning techniques to optimise the sensitivity to EFT effects. These developments resulted in two open-source {\sc\small Python} packages available for the wider particle physics community. We obtained the following deliveries and findings. 

In Chapter \ref{chap:smefit3}, we presented {\sc\small SMEFiT3.0}, a combined SMEFT interpretation up to mass-dimension 6 of the top, Higgs, diboson and electroweak sectors. Our analysis was simultaneously sensitive to  45 (50) Wilson coefficients in the linear (quadratic) fit and included 445 data points, making it the most extensive global SMEFT analyses to date. The LHC cross-section predictions included state-of-the-art theory calculations, both within the SM and the SMEFT where we included corrections up to NLO in the QCD perturbative expansion. Compared to the previous instalment of {\sc\small SMEFiT2.0} \cite{Ethier:2021bye}, we presented two main improvements. First, we removed the approximate implementation of the EWPOs by lifting the hardwired relations among the Wilson coefficients in the EW sector, and replacing it by an exact implementation where all Wilson coefficients were treated dynamically in the fit. This resulted in 14 additional degrees of freedom. We included EWPOs from LEP and SLD, which additionally required recomputing and extending the existing LHC observables from {\sc\small SMEFiT2.0} to properly account for the operators sensitive to the EWPOs. We found that the hardwired implementation gave too stringent constraints as compared to its exact implementation. The second improvement included the addition of LHC datasets partially based on the full Run II luminosity. Here we added the recent STXS1.2 Higgs measurements from ATLAS, diboson production from CMS, as well as a wide range of new top measurements. In total, the number of data points increased by $40\%$ with respect to {\sc\small SMEFiT2.0}, giving a significant tightening of the bounds in the two-light-two-heavy sector as a result. 

We switched viewpoint in Chapter {\ref{chap:smefit_uv}}, where we presented SMEFT-assisted automatised constraints at the level of the masses and couplings of UV-complete models. This was realised by leveraging their matching relations onto the SMEFT that allowed the $\chi^2$ to be reformulated in terms of UV parameters. As a proof of concept, we put constraints on a series of SM extensions with additional heavy scalars, vectors, and heavy bosons. We considered both one-particle, as well as multi-particle extensions and included corrections up to quadratic order in the EFT expansion, while allowing for NLO QCD corrections. Results were shown both at tree-level and one-loop, with the latter bringing in additional sensitivity via loop induced operators. The framework presented in this chapter bridged the gap between the SMEFT and UV landscape by automating, for the first time, the bound-setting procedure on UV-complete models that can be matched onto the SMEFT. We furthermore noted the crucial role played by the EW implementation presented in Chapter \ref{chap:smefit3}.

In Chapter \ref{chap:ml4eft}, we developed and demonstrated the {\sc\small ML4EFT} framework that constructed unbinned multivariate observables based on Machine Learning techniques. Traditional EFT studies rely on SM measurements that are reinterpreted in an EFT context, as opposed to observables that have been explicitly optimised with the EFT in mind. Often, differential cross-sections depend on one or at most two kinematics, thereby losing potentially crucial information, especially in case of complex final states. Moreover, the choice of binning may be non-optimal for EFT purposes. We found that this resulted in suboptimal bounds on the EFT parameters, or equivalently, a reduced reach on new physics scenarios. By contrast, ML optimised observables improved the sensitivity significantly by leveraging the full information encoded in the final state. We also showed that these can be used as benchmark to assess the information loss incurred by adopting a particular set of bins. 

Finally, in Chapter \ref{chap:smefit_fcc} we considered the impact expected from future colliders, analysing first the HL-LHC and then the FCC-ee and CEPC when added on top of the {\sc\small SMEFiT3.0} baseline scenario that we presented in Chapter \ref{chap:smefit3}. We developed a projection module within {\sc\small SMEFiT} to extrapolate existing Run II data to the settings expected at future experiments. We found that the marginalised bounds on the EFT coefficients are expected to improve between 20\% up to a factor of 3 at the end of HL-LHC, depending on the specific operator. Further improvements may be expected by including dedicated bins that target especially the tails of differential distributions as they profit from improved statistics. Adding subsequently projections for future lepton colliders, we found that the marginalised bounds improved by up to two orders of magnitude on nearly all purely bosonic and two-fermion-bosonic EFT coefficients. We then moved to analyse representative one-particle and multi-particle SM extensions using the UV matching procedure developed in Chapter \ref{chap:smefit_uv} and found an indirect mass reach between a few TeV up to around 100 TeV (assuming UV couplings $g_{\rm UV} \simeq 1$). 

All in all, with the global particle physics community moving towards the HL-LHC upgrade and future lepton colliders as proposed in the upcoming ESSPU, the results obtained in this thesis provide a timely and relevant contribution to assess their potential scientific reach.

\subsection*{Future visions}
The work presented and developed in this thesis may be extended along one of the following directions. First, given the upcoming ESPPU that will play a significant role in determining the future of the field, it is especially important to provide a complete overview of the SMEFT sensitivity of the various proposed future colliders such that a well-informed decision may be made. To this end, we are currently working on including additional projections coming from CLIC, ILC and the muon collider (MuCol) into {\sc\small SMEFiT}. Together with our existing projections from HL-LHC, the FCC-ee and the CEPC, we will be in an excellent position to further assess their relative strengths and interplay to contribute to the field's future directions. 

Related to this, we would like to quantify the expected constraints on the Higgs self-coupling coming from proposed future experiments and compare these to current HL-LHC projections. Measuring the Higgs self-coupling is one of the main targets of future colliders as it determines whether the electroweak phase transition that happened in the early universe was first or second order, and it ultimately connects to the matter-antimatter asymmetry puzzle. Here the FCC-ee will be able to provide indirect constraints through higher order corrections to single Higgs production, while the ILC, CLIC or MuCol will also offer a direct probe through Higgs pair production \cite{DiVita:2017vrr}. 

Also within {\sc\small SMEFiT} we aim for several improvements. On the data side, new and more precise measurements may provide a handle on previously ill- or unconstrained directions in the SMEFT parameter space. In particular, high-mass Drell-Yan (DY) tails offer a powerful indirect probe to new physics due to their quadratic growth with energy induced by four-fermion operators. It has been shown in Ref.~\cite{Farina:2016rws} for example that high-mass DY is even competitive with the low energy EWPOs obtained at LEP. Including DY requires extending our fitting basis to include two-lepton-two-quark dim-6 operators. Another class of measurements that should be considered are multijet final states that offer a sensitive probe to anomalous triple gluon interactions \cite{Krauss:2016ely, Hirschi:2018etq, Degrande:2020tno}. Rare SM processes like triboson production also offer a sensitive probe to the triple and quartic gauge couplings and show an interesting complementarity to the Higgs sector \cite{Bellan:2023efn}, and are therefore relevant to add to a global SMEFT fit. 

Furthermore, the connection to low-energy data such as electromagnetic dipole moment (EDM) measurements and flavour observables will provide additional constraints on the SMEFT parameter space, giving rise to richer flavour structures that allow us to further relax our flavour assumptions. This requires taking into account the effect of RG evolution, which can be realised for instance through the public package {\sc\small Wilson} \cite{Aebischer:2018bkb}. Its integration into {\sc\small SMEFiT} is currently work in progress. 

Another natural way of extending our fitting basis is through the addition of dim-8 operators and CP violating ones. Dim-8 effects may become relevant whenever the quadratic corrections coming from dim-6 start to dominate over the linear interference corrections. Given the large number of operators that opens up at dim-8, a case-by-case treatment in which only a subset of operators is considered will be needed. A possibly relevant development in this direction is the {\sc\small geosmeft} framework that associates a geometric interpretation to the SMEFT and enables an efficient scaling of dim-8 effects \cite{Helset:2020yio}. At dim-8, one may also benefit from purely theoretical constraints on the SMEFT via positivity bounds \cite{Adams:2006sv}. Finally, regarding the inclusion of CP violating operators, one may want to include CP sensitive observables such as Higgs production in association with a vector boson, Higgs to four leptons via two $Z$ bosons decaying into two same-flavour and oppositely charged lepton pairs, diboson production and finally, top-pair production \cite{Degrande:2021zpv}. 

Finally, moving on to methodological improvements, an interesting avenue to explore is to gradually start adding full experimental likelihoods to global SMEFT fits as they become publicly available in {\sc\small HistFactory} format \cite{Cranmer:2012sba}. This bypasses the need of the Gaussian approximation and provides a more accurate modelling of systematic uncertainties. More importantly maybe, it enables valuable cross-talk between the theory and experimental communities. In this same direction, cross-correlations between datasets that enters for instance through the PDFs are often neglected, while it would be interesting to study and include this effect. Theory uncertainties on the EFT predictions may also be taken into account \cite{Altmannshofer:2021qrr}, although we expect no significant effects as we already include NLO QCD corrections
to most of our EFT cross-sections. Also, as argued before, simultaneously fitting the PDFs and the EFT parameters along the lines of Refs.~\cite{Costantini:2024xae, Costantini:2024xae, Kassabov:2023hbm} can lead to new insights as to whether PDFs may possibly absorb new physics effects \cite{Hammou:2023heg}, which may be extended by taking into account additionally the developments outlined above.

\end{mainmatter}

\renewcommand{\thechapter}{\Alph{chapter}}
\renewcommand{\thesection}{\thechapter.\arabic{section}}
\renewcommand{\theequation}{\thesection.\arabic{equation}}


\providecommand{\href}[2]{#2}\begingroup\raggedright\endgroup

\begin{backmatter}

\newpage
\markboth{}{} 
\thispagestyle{plain} 

\chapter*{Summary}
\label{chap:summary}
\addcontentsline{toc}{chapter}{Summary}
What are the fundamental building blocks of matter that make up our universe and how do they interact? Despite the great complexity of this question, part of its answer in fact fits on an ordinary coffee mug; the Standard Model (SM) of particle physics, unlike its name might suggest, provides arguably the most successful scientific theory ever constructed by human kind. It has been confirmed by experiment reaching a precision, in some cases, up to the size of a human hair on the earth-to-moon distance. Yet, it leaves many questions unanswered, such as why we observe more matter than antimatter, why neutrinos are massive, how dark matter can be explained and how the SM can be connected to gravity. 

After the discovery of the Higgs boson in 2012 provided the last missing building block of the SM, particle physics has largely entered an era of precision physics. Without any direct signs of physics beyond the SM (BSM) observed so far, new particles, if any, will likely exist at energies beyond our current colliders' energy reach, thus escaping any direct detection. In this scenario, a promising avenue to find new physics nonetheless is provided by the Standard Model Effective Field Theory (SMEFT). The SMEFT formulates indirect effects of BSM physics at energy scales that can be reached in experiments. In this framework, the SM is seen as an approximate theory that is accurate only up to some finite energy, and receives small (unknown) corrections on top of it that capture the effects of new physics. The goal of this thesis is to extract the size of these corrections, called \textit{fingerprints}, from collider data and reinterpret these in terms of new physics models. 

The main advantage of the SMEFT is its ability to analyse many extensions of the SM at once without favouring any of them individually beforehand - it provides a model independent strategy to search for new physics. This makes the SMEFT a highly efficient approach as various SM extensions no longer need to be considered on a case by case basis. Mathematically, the low energy fingerprints are described in terms of Wilson coefficients whose value can be extracted from fits to collider data. At the end of the day, any non-zero value of the Wilson coefficients suggests hints towards BSM physics.

In Chapter \ref{chap:smefit3}, we present {\sc\small SMEFiT3.0}, where we simultaneously analyse a wide range of particle physics processes through a combined treatment of Higgs, top quark, diboson, and electroweak data, adding up to $445$ measurements in total. Due to, first, a more accurate theoretical modelling of the electroweak data, and second, an increase in the number of data points by $40\%$ compared to the predecessor {\sc\small SMEFiT2.0}, we obtain a significant higher precision on the value of the Wilson coefficients. This can be understood as extracting fingerprints at higher resolution, which contributes ultimately to sharpening the landscape of where new physics might hide. 

In Chapter \ref{chap:smefit_uv}, we develop an automatised procedure to analyse a series of specific SM extensions in the language of the SMEFT. Any specific SM extension that introduces new particles beyond our current collider's reach simplifies to the SMEFT at low energies. This can be used to put bounds directly on the masses of the newly hypothesised particles. Here we consider three classes of single particle extensions based on their spin, as well as scenarios where multiple particles are introduced simultaneously. We analyse the effect of the precision of the theoretical modelling on the reach of new physics and argue that the framework presented in this chapter can be used to constrain any user defined SM extension.

So far we have performed SMEFT analyses based on particle physics measurements summarised in terms of either one or at most two variables, such as the momentum of the collision products perpendicular to the incoming protons, or the angle between two outgoing electrons. However, the outcome of a collision (the final state) cannot be fully represented in terms of such few variables. In Chapter \ref{chap:ml4eft} we argue why and go beyond this by adopting machine learning techniques to describe the final state in terms of any number of variables, as well as to obtain a continuous representation of the final state. We show that this leads to an increased sensitivity to the Wilson coefficients as compared to traditional approaches. This chapter comes with a newly developed open source package {\sc\small ML4EFT}. 

Finally, Chapter \ref{chap:smefit_fcc} presents the impact of future colliders on the SMEFT and its reach on new physics. Given that the high luminosity LHC (HL-LHC) scheduled for 2029 will eventually accumulate 25 times as much data as the LHC at the end of 2018, we assess the expected improvement this would give in the context of the SMEFT. In addition, the proposed leptonic future circular collider (FCC-ee) will collide electrons and positrons at an unprecedented rate, giving access to observables with extremely small uncertainties. We quantify the information these expected measurements bring into the parameter space of the SMEFT and advocate that the future of finding new physics looks brighter than ever.

\chapter*{Samenvatting}
\label{chap:samenvatting}
\addcontentsline{toc}{chapter}{Samenvatting}
Wat zijn de fundamentele bouwstenen van de materie waaruit ons universum bestaat en wat is de wisselwerking hiertussen? Ondanks de complexheid van dit vraagstuk, past het antwoord tot op zekere hoogte op een normale koffiemok: het Standaard Model (SM) van de deeltjesfysica. In tegenstelling tot de naam misschien doet vermoeden, is het SM waarschijnlijk de meest succesvolle theorie die de wetenschap tot nu toe gekend heeft. Het is experimenteel bevestigd met een precisie ter grootte van een haar op de afstand van de aarde tot de maan in sommige gevallen, maar toch laat het SM ook veel vragen onbeantwoord. Hoe kan het dat er meer materie is dan antimaterie? Hoe verkrijgen neutrino's massa? Hoe kan donkere materie verklaard worden en hoe kan het SM verenigd worden met de theorie van de zwaartekracht?

Na de ontdekking van het Higgs boson in 2012 is de deeltjesfysica grotendeels in een tijdperk van precisiewaarnemingen terecht gekomen. Bij gebrek aan directe aanwijzingen van nieuwe natuurkunde tot nu toe, is te verwachten dat nieuwe deeltjes zich manifesteren op energie\"en buiten het bereik van onze huidige deeltjesversnellers. Desondanks biedt de zogenaamde ``Standard Model Effective Field Theory'' (SMEFT) een veelbelovend onderzoekspad. SMEFT voorspelt hoe nieuwe natuurkunde zich indirect manifesteert op experimenteel toegankelijke energie schalen. In dit framework wordt het SM gezien als een theorie die alleen accuraat hoeft te zijn tot op een bepaalde energieschaal, terwijl (onbekende) correctietermen worden toegevoegd om het effect van nieuwe natuurkunde beschrijven. Het doel van dit proefschrift is om deze correcties, ook wel vingerafdrukken (\textit{fingerprints}) genoemd, uit deeltjesversneller data te halen en deze vervolgens te interpreteren in termen van modellen die het SM uitbreiden. 

Het voornaamste voordeel van de SMEFT is het vermogen om verschillende uitbreidingen van het SM tegelijkertijd te analyseren zonder er \'e\'en bij voorbaat te prefereren - het biedt een modelonafhankelijke benadering om te zoeken naar nieuwe fysica. Dit maakt SMEFT tot een zeer effici\"ent framework, aangezien modellen niet langer \'e\'en voor \'e\'en geanalyseerd hoeven te worden. Wiskundig gezien worden de \textit{fingerprints} op lagere energie\"en beschreven aan de hand van Wilsonco\"effici\"enten, waarvan de waarde bepaald kan worden aan de hand van fits aan experimentele data. Uiteindelijk suggereert een Wilson co\"effici\"ent ongelijk aan nul waarschijnlijk nieuwe natuurkunde. We hebben het volgende gevonden.

Hoofdstuk \ref{chap:smefit3} behandelde {\sc\small SMEFiT3.0}, een fit aan tegelijkertijd Higgs, top quark, diboson, en elektrozwakke data, wat in totaal $445$ metingen omvat. Dankzij, enerzijds, een nauwkeurigere theoretische beschrijving van de elektrozwakke data, en anderzijds, een toename van $40\%$ in het aantal datapunten ten opzichte van de voorganger {\sc\small SMEFiT2.0}, bepaalden we de Wilson co\"effici\"enten met een aanzienlijk grotere precisie. Dit kan ge\"interpreteerd worden als een scherpere weergave van de vingerafdruk, welke op zijn beurt bijdraagt aan het beter bepalen waar nieuwe natuurkunde zich wel en niet manifesteert.

In hoofdstuk \ref{chap:smefit_uv} ontwikkelden we een geautomatiseerde methode om een reeks specifieke uitbreidingen van het SM te analyseren met behulp van de SMEFT. Vrijwel elke uitbreiding van het SM met nieuwe (zware) deeltjes kan op lagere energie\"en door de SMEFT beschreven worden. Dankzij deze eigenschap kunnen grenzen gesteld worden op de massa’s van nieuwe deeltjes aan de hand van de SMEFT. Hier beschouwden we drie klasse deeltjes op basis van hun spin, evenals scenario’s waarin meerdere deeltjes tegelijkertijd worden toegevoegd. We analyseerden het effect van de precisie van het theoretisch model op de zeggingskracht van nieuwe natuurkunde en lieten zien dat ons framework het automatisch analyseren van vrijwel alle uitbreidingen van het SM toelaat. 

Tot nu toe bevatte onze analyse metingen gebaseerd op ten hoogste twee variabelen, zoals de impuls van de botsingsproducten loodrecht op de inkomende protonen, of de hoek tussen twee uitgaande electronen. De uitkomst van een botsingsreactie kan echter niet in zijn totaliteit worden gevat in zo weinig variabelen. In hoofdstuk \ref{chap:ml4eft} gebruikten we neurale netwerken als hulpmiddel om de botsingsproducten te beschrijven in termen van een willekeurig aantal variabelen, en om hiervan bovendien een continue (in tegenstelling tot een discrete) beschrijving te verkrijgen. We lieten zien dat dit leidde tot een verhoogde gevoeligheid voor de Wilsonco\"effici\"enten ten opzichte van traditionele bepalingen. Dit hoofdstuk kwam met een nieuw open-source pakket {\sc\small ML4EFT}. 

Ten slotte analyseerden we in hoofdstuk \ref{chap:smefit_fcc} de impact van toekomstige deeltjesversnellers op de Wilson co\"effici\"enten. Aangezien de High Luminosity LHC (HL-LHC) in 2030 in gebruik zal worden genomen met een verwachte precisie die tot wel vijf keer zo hoog is als de LHC aan het einde van 2018, analyseerden we de gevolgen hiervan voor de SMEFT. Hiernaast zal, mits goedgekeurd, de Future Circular Collider (FCC-ee) electronen en positronen in ongekende mate met elkaar laten botsen, waardoor metingen verwacht kunnen worden met extreem kleine onzekerheden. We kwantificeerden de informatie die deze verwachte metingen met zich meebrengen vanuit het perspectief van de SMEFT, wat liet zien dat de toekomst van deeltjesfysica er rooskleuriger uitziet dan ooit.

\chapter*{Acknowledgements}
\label{chap:acknowledgements}
\addcontentsline{toc}{chapter}{Acknowledgements}
We have reached the end! It has been an incredible journey and I owe a big thank you to many people that helped me at some point along the way. 

My biggest thank you goes out to my supervisors, Juan and Wouter, for giving me the opportunity to start this PhD and for all the amazing science we did. Juan, thank you for your guidance and trust, and for all the places and conferences I could travel to during my PhD. I learned a lot from you and you helped me grow to become an independent researcher. Wouter, even though we did not interact much, I enjoyed every time our chats, your enthusiasm and support.

Gerhard, Eric, Fabio, Flavia and Lydia, thank you for agreeing to be on my PhD committee and for your valuable time to read and discuss my thesis. 

Next, I would like to deeply thank all current and past members of the Nikhef Theory Group;  Robin, Pieter, Jelle, Max, Giacomo, Coenraad, Heleen, Peter, Avanish, Vaisakh, Anders, Tommaso, Tanishq, Maximilian, Andrea, Sachin, Ankita, Tommaso, Juraj, Johannes, Robert, Luk\'a\v s, Melissa, Marieke, Eric, Jordy and Wouter. Thank you so much for making Nikhef a better place. Thanks also to Andrea and Rahul for exchanging some experimental SMEFT adventures. To the old crew, Rabah, Jake and Emanuele: thank you for guiding me when I started my PhD. A big shout out also to the CT group at Nikhef for their incredible help with stoomboot (love the name!). Of course, to the amazing group in Delft; Abel, Helena, Isabel, Luigi, Sabrya, Sonia, Stijn~-~I wrote my first paper thanks to you. 

I would also like to thank the entire {\sc\small SMEFiT} collaboration: Eleni, Fabio, Juan, Alejo, Eugenia, Giacomo, Marion and Tommaso. I cannot wait to make it to {\sc\small SMEFiT4.0} together. Thank you, Maeve, for all your help and support in the beginning stages of my PhD and showing me around in Cambridge when I visited the group. Manu, thanks for your kindness and I hope at some point you will have a piano as fancy as the one in Benasque. James, I am still impressed by how you managed to cover 80 slides at HEFT in a 12 minute time slot without losing me. Luca, thanks a lot for your patience and sharing your experience in various projects, and most importantly for all the fun we have had as friends at conferences and workshops. I learned a lot from you and I am sure we will continue to collaborate in the future. Hesham, I still crack up when you remind me of the cheese diet in Les Houches. Alejo, Eugenia, I still remember our lovely Turkish commute from the Hay hotel each morning, and Marion, thanks for helping me with our gigantic {\sc\small SMEFiT3.0} madgraph quest. 

Coenraad, it was amazing to have shared our journey together since the Masters, not only as colleagues, but above all else as friends. We even wrote a paper together! Giacomo, I was always impressed by your expertise and I should thank you for being my friend and helping me on so many occasions. Thanks both for being my paranymphs, it means a lot. Maarten, Miguel, I hope we stay TP boys long after physics. Colette, I loved your painting classes and thank you for showing me there is no right or wrong in art. 

Finally, I would like to thank all my family for their continuous love and support. It feels special to have shared Nikhef, CWI and the VU with both my grandparents. 

Jacqueline, Willem, Flip and friends, thank you for sharing so many good moments together and for helping me move to my new place. Maarten, I enjoyed all the good stories you shared and I hope we will spot a few Highland cows soon! Mom, dad, it means a lot to me there will always be a safe place, thank you for that. Thomas, I look forward to having a great time together in sunny SF. Maartje, thank you so much for coming all the way here for my defence, you are right we only need a few words indeed. 

Pauline, thank you for everything, I could never have done this without you.

\begin{flushright}
    Jaco ter Hoeve\\
    Edinburgh, October 2024
\end{flushright}

\end{backmatter}

\end{document}